\documentclass[journal]{IEEEtran}
\usepackage{color}
\usepackage{graphicx} % Required for inserting images
\usepackage{comment}
\usepackage{amsmath}
\usepackage{amssymb}
\usepackage{mathtools}
\usepackage{bm}
\usepackage{url}
\usepackage{cite}  
\usepackage{booktabs}
\usepackage{array}
\usepackage{tabularx}
\usepackage{multirow}
\usepackage{enumitem}
\usepackage[T1]{fontenc}
\usepackage{forest}
\usepackage{xcolor}
\usepackage{makecell}
\usepackage[table]{xcolor}
\usepackage{tabularray}
\usepackage{ragged2e}

\newcolumntype{C}[1]{>{\centering\arraybackslash}m{#1}}
\newcolumntype{L}[1]{>{\raggedright\arraybackslash}m{#1}}

\usepackage{hyperref}
\hypersetup{
    colorlinks=true,
    linkcolor=blue,      % internal links (Fig, Sec, Eq)
    citecolor=blue,       % \cite links
    urlcolor=blue,       % \url{} links
    filecolor=blue,
}

\newcommand{\lj}[1]{\textcolor{red}{\textbf{[#1]}}}
\newcommand{\hz}[1]{\textcolor{red}{\textbf{[#1]}}}

\usepackage[most]{tcolorbox}
\usepackage{xcolor}

\newtcolorbox[auto counter]{definitionbox}[1]{
    breakable,
    colback=gray!10,
    colframe=black,
    boxrule=0.8pt,
    arc=0pt,
    left=3pt,
    right=3pt,
    top=3pt,
    bottom=3pt,
    before={\par\vspace*{8pt}\noindent},
    after skip=8pt,
    before upper={%
        \textbf{Definition \thetcbcounter\ [#1].}\ %
    }
}

\newtcolorbox[auto counter]{theorembox}[1]{
    breakable,
    colback=gray!10,
    colframe=black,
    boxrule=0.8pt,
    arc=0pt,
    width=\linewidth,
    left skip=0pt,
    right skip=0pt,
    left=3pt,
    right=3pt,
    top=3pt,
    bottom=3pt,
    before={\par\vspace*{8pt}\noindent},
    after skip=8pt,
    before upper={%
        \textbf{Theorem \thetcbcounter\ [#1].}\ %
    }
}

\usepackage[table]{xcolor}   % \cellcolor / \rowcolor
\usepackage{graphicx}        % \rotatebox for vertical headers
\usepackage{array}           % robust column types
\usepackage{cite}            % compresses [1],[2],[3] -> [1]-[3]
 
\definecolor{lowcol}{HTML}{C6EFCE}    % Low  -> light green
\definecolor{medcol}{HTML}{BDD7EE}    % Med  -> light blue
\definecolor{highcol}{HTML}{FFE699}   % High -> light amber
\definecolor{hdrcol}{HTML}{E6E6E6}    % header grey
 
\newcommand{\Lc}{\cellcolor{lowcol}L}
\newcommand{\Mc}{\cellcolor{medcol}M}
\newcommand{\Hc}{\cellcolor{highcol}H}

\definecolor{rootc}{HTML}{1F2D3D}
\definecolor{ca}{HTML}{2E5A88}\definecolor{cal}{HTML}{DCE7F2}   % Formulations  (blue)
\definecolor{cb}{HTML}{2E7D4F}\definecolor{cbl}{HTML}{DCEEE3}   % Model-Free    (green)
\definecolor{cc}{HTML}{6C3483}\definecolor{ccl}{HTML}{E9DCF0}   % Model-Based   (purple)
\definecolor{cd}{HTML}{B9770E}\definecolor{cdl}{HTML}{FBEBD2}   % Multi-Agent   (amber)
\definecolor{ce}{HTML}{117A8B}\definecolor{cel}{HTML}{D5EBEF}   % Deployment    (teal)

\title{
Deep Reinforcement Learning for 6G AI-RAN: A Comprehensive Survey
} 
\author{
Jie Lu, 
Peihao Yan,
Qijun Wang,
Ruxin Lin,
and
Huacheng Zeng\\
Department of Computer Science and Engineering, Michigan State University
}

\begin{document}

% \tableofcontents

% \clearpage
\maketitle

 % \hz{Jie: make sure all tables and figures are cited in the text content. Figures and tables are placed close to the spot where they are cited. The format of each reference item is appropriate. Address all latex compile warnings.
 % Make sure all the acronyms are properly defined.}

\begin{abstract}
    The evolution toward sixth-generation (6G) networks is transforming the radio access network (RAN) into a programmable and intelligent control platform that must continuously adapt to heterogeneous services, dynamic environments, and competing performance objectives. Open Radio Access Network (O-RAN) provides the open interfaces, disaggregated architecture, and multi-timescale control loops needed to support this transformation, while deep reinforcement learning (DRL) offers a natural framework for optimizing sequential decisions under uncertainty. However, existing surveys either address artificial intelligence (AI) and machine learning (ML) in O-RAN broadly or focus on isolated DRL use cases, leaving a gap in the systematic connection between DRL methodology, O-RAN architecture, and operational deployment. To the best of our knowledge, this article presents the first dedicated and comprehensive survey of DRL for Open AI-RAN. We review the foundations of model-free, model-based, offline, safe, multi-agent, federated, and transfer learning, and provide an O-RAN-aware framework for formulating RAN control problems through states, observations, actions, rewards, constraints, and temporal structure. We classify DRL applications across radio resource management, mobility management, interference control, traffic steering, energy efficiency, network slicing, integrated sensing and communication, security, and massive MIMO. We further examine multi-agent and federated coordination, foundation models and agentic AI, trustworthy DRL, sim-to-real transfer, continual adaptation, resource-efficient inference, and reinforcement learning operations. Finally, we review experimental platforms, benchmarks, standards, and industry activities, and identify research directions toward sample-efficient, safe, scalable, interoperable, and deployable DRL control for 6G Open AI-RAN.

\end{abstract}

\begin{IEEEkeywords}
    Open Radio Access Network (O-RAN), deep reinforcement learning (DRL), 6G AI-RAN, network intelligence.
\end{IEEEkeywords}

% \clearpage 

\section{Introduction}
\label{1_introduction}

\subsection{Context and Motivation}

% - RAN becomes increasing complicate due to 
%  -- the diversity of demand from diverse UEs, 
%  -- the increasing the data traffic such as VR, 
%  -- reliability requirement of the ...

% - AI has emerged as a fundamental technique for automating the control and optimization of RAN.

The evolution toward sixth-generation (6G) wireless networks is transforming the radio access network (RAN) from a largely static communication infrastructure into a programmable platform that must support diverse services, heterogeneous devices, and rapidly changing operating conditions \cite{wang2023road, jiang2021road}. 
Emerging applications impose different and often competing requirements on throughput, latency, reliability, energy efficiency, sensing capability, and service-level agreement (SLA) satisfaction. At the same time, dense deployments, massive multiple-input multiple-output (MIMO), network slicing, integrated sensing and communication (ISAC), and heterogeneous access technologies substantially increase the dimensionality and coupling of RAN control problems. Conventional optimization, heuristic, and rule-based methods remain effective for well-defined operating conditions, but they can become difficult to design and maintain when network dynamics, objectives, and constraints evolve continuously.

Open Radio Access Network (O-RAN) provides an architectural foundation for addressing this growing complexity through disaggregation, open interfaces, and programmable control \cite{ORAN_OAD_17_2026}.
%\hz{ref, O-RAN spec}
The RAN Intelligent Controller (RIC) introduces multiple control loops spanning different timescales. Non-Real-Time RIC applications, referred to as rApps, support long-term policy optimization and orchestration, whereas Near-Real-Time RIC applications, referred to as xApps, perform sub-second control based on telemetry collected from RAN nodes. Interfaces such as O1, A1, and E2 connect monitoring, policy management, model lifecycle management, and control functions across the Service Management and Orchestration (SMO) framework, the RICs, and the O-RAN Central and Distributed Units. Consequently, O-RAN makes network observations and control actions accessible to independently developed applications, providing a practical substrate for AI-native RAN automation.

Deep reinforcement learning (DRL) is particularly suitable for this environment because many RAN optimization tasks are inherently sequential decision-making problems. Network conditions are stochastic and only partially observable, control actions influence both immediate and future performance, and optimization objectives frequently require balancing several long-term metrics. DRL combines the decision-making framework of reinforcement learning (RL) with the representation capacity of deep neural networks (DNNs), allowing agents to learn control policies from high-dimensional observations without requiring complete analytical models of channel, traffic, mobility, and protocol dynamics. This capability has motivated its application to radio resource management, mobility management, interference mitigation, traffic steering, energy optimization, network slicing, ISAC, security, and massive-MIMO control.

% \hz{ref, seminar/survey papers ones from AI conferences}
Compared with supervised and self-supervised learning \cite{krizhevsky2012imagenet, chen2020simple, grill2020bootstrap}, which primarily learn predictive models or data representations from labeled or automatically constructed training targets, DRL directly learns how control actions affect long-term network performance through interaction with the environment. It does not require explicit labels for the optimal action at every network state or a complete analytical model of the underlying dynamics. Instead, DRL can optimize delayed and cumulative objectives, account for the future consequences of current decisions, and adapt its policy as traffic, mobility, interference, and service demands change. These properties make DRL particularly attractive for closed-loop O-RAN control, where the objective is not merely to estimate the network state, but to continuously select actions that balance multiple operational goals under uncertainty.

Nevertheless, applying DRL to O-RAN involves substantially more than selecting an algorithm (network pipeline) and training it in a simulator. Practical controllers must operate with delayed and aggregated telemetry, changing numbers of users and cells, mixed discrete and continuous actions, and control loops with different temporal resolutions. Multiple independently developed xApps and rApps may interact with the same network resources, creating non-stationarity and potentially conflicting actions. Training data may also be distributed across operators, vendors, and deployment domains, limiting centralized learning. Moreover, production RANs impose strict requirements on latency, safety, robustness, fairness, explainability, and SLA compliance. Policies trained in simulation must therefore address the sim-to-real gap and be supported by lifecycle management, runtime monitoring, rollback mechanisms, reproducible benchmarks, and standards-aligned interfaces.

These characteristics motivate a unified examination of DRL for 6G AI-RAN that connects learning methodology with the O-RAN architecture and its deployment constraints. Such an examination must extend beyond individual algorithms or use cases to address how RAN problems are formulated as Markov decision processes (MDPs), how multiple agents and distributed domains are coordinated, how foundation models and agentic AI can augment sequential control, and how learned policies can be made trustworthy and deployable. This survey accordingly studies DRL for O-RAN across the complete path from foundational principles and problem formulation to applications, training, deployment, benchmarking, and standardization.

% Place this in your preamble (or right before the table) to vertically center the X column
\renewcommand{\tabularxcolumn}[1]{m{#1}} 

\begin{table*}[t]
\centering
\footnotesize 
\renewcommand{\arraystretch}{1.25} % Restored stretch for better vertical padding
\setlength{\tabcolsep}{4pt}        % Slightly adjusted for a cleaner look
\caption{Comparison of this survey paper and existing survey papers. \\ 
(A\&S: Architecture \& Standards; F-Models: Foundational Models; \textbf{L}: Low; \textbf{M}: Medium; \textbf{H}: High.)}
\label{tab:surveys}

% First column changed to m{1.4cm} for vertical center + reduced width
\begin{tabularx}{1\textwidth}{
|>{\raggedright\arraybackslash}m{1.4cm}|*{6}{c|}>{\raggedright\arraybackslash}X|}
\hline
\rowcolor{hdrcol}
\textbf{Work}
 & \makecell{\textbf{RL /} \\ \textbf{DRL}}
 & \makecell{\textbf{O-RAN} \\ \textbf{A\&S}}
 & \makecell{\textbf{O-RAN} \\ \textbf{Use Cases}}
 & \makecell{\textbf{GenAI \&} \\ \textbf{F-Models}}
 & \makecell{\textbf{Trustworthy} \\ \textbf{AI}}
 & \makecell{\textbf{Platforms \&} \\ \textbf{Deployment}}
 & \textbf{Scope} \\
\hline
 
\cite{polese2023understanding, polese2023empowering, agarwal2025open, arnaz2022toward, wani2024open, abdalla2022toward, alam2025comprehensive}
 & \Lc & \Hc & \Mc & \Lc & \Lc & \Mc &
Architecture, open interfaces and general O-RAN overview. \\ \hline
 
\cite{mahmoud2026review , kirana2026ml, brik2022deep}
 & \Mc & \Hc & \Hc & \Lc & \Lc & \Mc &
AI/ML applications, opportunities, and challenges across O-RAN. \\ \hline
 
\cite{oluwaseyi2025llm, cai2025tutorial}
 & \Mc & \Lc & \Lc & \Hc & \Lc & \Lc &
LLM-enhanced RL and LLM-as-operator for wireless and O-RAN control. \\ \hline

\cite{brik2024explainable}
 & \Mc & \Hc & \Hc & \Lc & \Mc & \Mc &
Explainable AI for 6G O-RAN. \\ \hline

\cite{guo2025towards}
 & \Hc & \Lc & \Mc & \Lc & \Hc & \Mc &
Explainable DRL for AI-RAN network slicing. \\ \hline

\cite{liang2026security, mehrban2025integrating, amachaghi2024survey}
 & \Lc & \Mc & \Lc & \Lc & \Mc & \Lc &
Security, privacy, zero-trust, and intrusion detection for O-RAN. \\ \hline
 
\cite{herrera2025tutorial, couto2024survey, masaracchia2023digital}
 & \Lc & \Mc & \Lc & \Lc & \Lc & \Hc &
Deployment, testbeds, datasets, and digital twins for O-RAN. \\ \hline
 
\cite{liang2024energy, abubakar2023energy}
 & \Lc & \Lc & \Lc & \Lc & \Lc & \Mc &
Energy efficiency of (ML-enabled) O-RAN. \\ \hline
 
\cite{deng2026ai}
 & \Lc & \Hc & \Mc & \Lc & \Lc & \Mc &
AI-native O-RAN for non-terrestrial networks. \\ \hline
 
\cite{lee2025survey}
 & \Mc & \Lc & \Mc & \Lc & \Lc & \Lc &
Intelligent traffic steering techniques in O-RAN. \\ \hline
 
\textbf{This paper} & \Hc & \Hc & \Hc & \Hc & \Hc & \Hc &
\textbf{End-to-end DRL for O-RAN: RL fundamentals and MDP modeling,
multi-agent and federated DRL, foundation models, trustworthy DRL, and
the full training-to-deployment lifecycle.} \\ \hline
 
\end{tabularx}
\end{table*}

\subsection{Existing Survey Papers}

% \hz{@Jie: please review the paragraph titles within this section. Especially, how to place Guo \emph{et al.} [14]?}

% \lj{Guo \emph{et al.} is about explainable DRL for network slicing, which is not an LLM paper and it only mentions foundation models as future work. Can we place this paper into Trustworthiness of AI-RAN? }

The convergence of AI and O-RAN has led to a substantial body of survey literature, with individual studies examining this topic from different perspectives. In the following, we first review existing survey papers on AI-RAN and then compare this survey with them in terms of scope and focus.

% \lj{Foundational O-RAN Architecture and System Surveys.}

\textbf{O-RAN Architecture and Systems.}
Polese \emph{et al.} \cite{polese2023understanding} detail O-RAN specifications, open interfaces, and closed-loop control timescales. Their work also surveys data-driven workflows, security models, experimental platforms, and optimization research. A supplementary tutorial~\cite{polese2023empowering} identifies the limitations in current cellular deployments. It then explains how core O-RAN principles integrate into a data-driven, AI-controlled 6G architecture. Subsequent surveys analyze the architecture from specific technical perspectives. For example, Agarwal \emph{et al.} \cite{agarwal2025open} evaluate 6G O-RAN architectures in terms of energy efficiency and latency. They also discuss emerging use cases, AI/ML applications, and digital twin technologies. Arnaz \emph{et al.} \cite{arnaz2022toward} investigate the integration of intelligence and programmability within the RAN. They subsequently review the resulting AI/ML mechanisms. 

Wani \emph{et al.} \cite{wani2024open} outline the O-RAN architecture and its primary components. Abdalla \emph{et al.} \cite{abdalla2022toward} evaluate the practical deployment capabilities and limitations of O-RAN. Alam \emph{et al.} \cite{alam2025comprehensive} present a network slicing tutorial based on O-RAN Alliance and other industry standard specifications. The authors detail open-source projects, experimental platforms, and orchestration options for RAN and transport-network slice subnets. They also analyze the role of AI/ML in slice orchestration, specifically highlighting single- and multi-agent RL. Across these studies, RL is treated generally within the broader AI/ML context. 
Still, existing surveys lack a focused treatment of how RAN control problems can be formulated and solved as sequential decision-making tasks.

% \lj{AI/ML-Enabled O-RAN Intelligence.}

\textbf{AI/ML Applications in O-RAN.}
Some other surveys focus on the intelligence layer. Mahmoud \emph{et al.} \cite{mahmoud2026review } conduct a protocol-driven review of O-RAN intelligence focused on the RIC. Their scope includes the 5G core, the RAN, and the air interface. They also analyze industrial use cases and the broader impacts of RAN intelligence, including model explainability. Kirana \emph{et al.} \cite{kirana2026ml} survey ML integration within the Non-Real-Time and Near-Real-Time RICs. Their work covers resource allocation, mobility management, anomaly detection, traffic optimization, spectrum management, and security. Brik \emph{et al.} \cite{brik2022deep} investigate DL for beyond-5G O-RAN. They present DL case studies deployed on the O-RAN architecture. These studies categorize various AI/ML applications and identify RL as an available method. However, they do not model the control problems as MDPs or propose a dedicated RL methodology.

% \lj{Foundation Models and LLM-Driven O-RAN Control}

\textbf{LLM-Driven O-RAN Control.}
Cai \emph{et al.} \cite{cai2025tutorial} present a tutorial on large language model (LLM)-enhanced RL for wireless networks. They categorize the LLM functions into state perceiver, reward designer, decision-maker, and generator. Each function maps to a specific stage in the RL pipeline. This framework addresses wireless networks generally rather than focusing on the O-RAN architecture. Oluwaseyi \emph{et al.} \cite{oluwaseyi2025llm} model an LLM as a network operator. They propose a formal framework featuring an O-RAN-aligned adapter. In this architecture, the Non-RT RIC generates strategic guidance, while the Near-RT RIC executes reactive control. The authors then analyze the safety and stability of this mechanism. Wei \emph{et al.} \cite{wei2025large} survey the broader application of language models in wireless network management. While these studies demonstrate the applicability of generative methods, they do not integrate foundation models into a comprehensive DRL pipeline for O-RAN.

% \lj{Fine as is; or Explainability and Trustworthiness in AI-RAN.}

\textbf{Explainability and Trustworthiness of AI-RAN.}
Literature on trustworthiness typically addresses two primary domains: explainability and security. Regarding explainability, Brik \emph{et al.} \cite{brik2024explainable} summarize Explainable AI (XAI) methods and metrics for 6G O-RAN. They analyze XAI deployment across O-RAN components and review standardized use cases. Their tutorial also discusses XAI pipeline automation and associated security protocols. Regarding security and privacy, Liang and Zhang~\cite{liang2026security} classify O-RAN vulnerabilities and analyze threat vectors. They review mitigation mechanisms and relevant specifications from standards development organizations. This analysis includes the attack surface introduced by AI/ML components. Mehrban \emph{et al.} \cite{mehrban2025integrating} complement this with an analysis of zero-trust architectures. Amachaghi \emph{et al.} \cite{amachaghi2024survey} review intrusion detection systems for O-RAN. While these studies outline system-level trustworthiness requirements, they do not evaluate the robustness, safety, or privacy of the learning agents.

Guo \emph{et al.} \cite{guo2025towards} survey explainable DRL specifically for intelligent network slicing in AI-RAN. They classify explainability techniques into post-hoc interpretation, surrogate models, and human-in-the-loop steering. The authors also review benchmark environments and O-RAN-compliant testbeds for evaluating these methods. However, the scope of their study is restricted to the network slicing use case and policy explainability. The research utilizes O-RAN primarily as a deployment environment. Additionally, it addresses foundation models, such as LLMs and intent-based networking, strictly as future research directions. Additional research investigates foundation models for the RAN. 

% \lj{Plural forms? Deployment, Testbeds, (Datasets,) and Digital Twins}

\textbf{Deployment, Testbeds, and Digital Twins.}
Several papers have reviewed the practical implementation of O-RAN. Herrera \emph{et al.} \cite{herrera2025tutorial} evaluate deployment solutions using simulation, emulation, and physical testbeds. They detail the procedure for migrating an xApp from simulation to a physical testbed. This process includes deploying the Service Management and Orchestration (SMO) layer and the Non-RT RIC. Couto \emph{et al.} \cite{couto2024survey} catalog public O-RAN datasets and evaluate their applicability for ML model training. Masaracchia \emph{et al.} \cite{masaracchia2023digital} investigate digital twins for configuring an intelligent 6G RAN. These studies concentrate on deployment infrastructure and tooling rather than DRL methodology.

% \lj{Domain-Specific O-RAN Surveys.}

% \paragraph{Energy Efficiency and Others}
\textbf{Domain-Specific O-RAN Surveys.}
Several survey papers have studied O-RAN deployment and operational metrics. Liang \emph{et al.} \cite{liang2024energy} review the energy characteristics of O-RAN. They model the power consumption of the radio, distributed, and centralized units, subsequently surveying applicable ML methods. Abubakar \emph{et al.} \cite{abubakar2023energy} analyze O-RAN energy efficiency from a broader system perspective. Deng \emph{et al.} \cite{deng2026ai} investigate AI-native O-RAN in non-terrestrial networks. They analyze operational challenges and propose an orchestrated framework with defined use cases. Lee \emph{et al.} \cite{lee2025survey} review intelligent traffic steering, noting RL as the primary technique for this application. These studies focus on isolated subsystems and do not address generalized O-RAN control mechanisms.

\textbf{Summary of Existing Surveys.}
The objective of this article is not to exhaustively enumerate all published studies, but rather to position this survey within the most relevant and recent literature at the intersection of learning-based control and RAN optimization. Table~\ref{tab:surveys} summarizes existing survey papers and highlights how this article is positioned in the broader literature. As shown in the table, no existing survey provides a dedicated treatment of DRL methodologies for controlling open AI-RAN systems. This gap motivates the present survey.

% They fail to simultaneously address problem formulation, algorithmic methodology, trustworthiness, and deployment. 

\begin{figure*}
    \centering
    \includegraphics[width=1\linewidth]{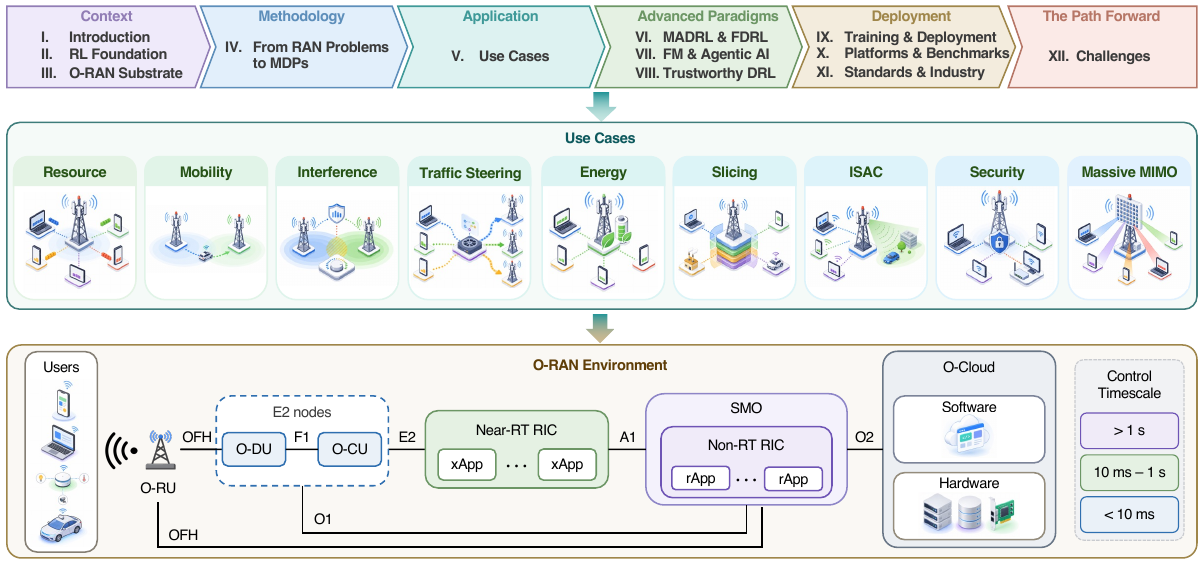}\vspace{-0.1in}
    \caption{Organization of this survey and its relation to the O-RAN control stack. The top row summarizes the paper structure across all the sections, from context and methodology to applications, advanced paradigms, deployment, and open challenges. The middle row highlights representative O-RAN use cases discussed in the survey. The bottom row illustrates the O-RAN environment considered throughout the survey.}\vspace{-0in}
    \label{fig:organization}
\end{figure*}

\subsection{Main Contributions}

To the best of our knowledge, this is the first dedicated and comprehensive survey of DRL for Open AI-RAN. Existing Open AI-RAN surveys (e.g.,  \cite{kirana2026ml, hamdan2023recent, brik2022deep}) review ML techniques broadly but do not provide an in-depth treatment of DRL methodologies. Conversely, existing DRL-oriented surveys are either limited to specific aspects, such as explainable DRL \cite{brik2024explainable}, or centered on deployment artifacts, such as xApps \cite{santos2025managing, hoffmann2023open}. 
The current literature still lacks a unified survey that systematically integrates the following elements for Open AI-RAN: (i) an O-RAN-aware MDP design framework; (ii) single-agent, multi-agent, and federated DRL formulations; (iii) foundation models for DRL-based RAN control; (iv) trustworthy DRL aligned with O-RAN Alliance and the 3rd Generation Partnership Project (3GPP) specifications; and (v) an end-to-end deployment pipeline covering Reinforcement Learning Operations (RLOps), sim-to-real transfer, and benchmark analysis. This survey fills this gap by providing a comprehensive methodological, architectural, and deployment-oriented treatment of DRL for Open AI-RAN. The primary contributions of this survey are summarized below.

% To the best of our knowledge, this is the first dedicated and comprehensive survey of DRL for Open AI-RAN. Prior Open AI-RAN survey papers (e.g., \cite{kirana2026ml, hamdan2023recent, brik2022deep}) review ML broadly without DRL-specific depth, while DRL-focused survey papers are either narrow-scoped (e.g., explainable DRL \cite{brik2024explainable}) or focused on artifacts 
% (e.g., xApps  \cite{santos2025managing, hoffmann2023open}).
% Current literature lacks a comprehensive work that surveys the following elements for AI-RAN: (i) an O-RAN MDP design framework; (ii) single-agent, multi-agent, and federated DRL formulations; (iii) foundation models for DRL; (iv) trustworthy DRL aligned with O-RAN Alliance and 3GPP specifications; and (v) the end-to-end deployment pipeline, encompassing Reinforcement Learning Operations (RLOps), sim-to-real transfer, and benchmark analysis. 
% This survey paper fills this gap.
% The primary contributions of this survey are summarized below.

\begin{itemize}[leftmargin=0.15in]
\item \textit{O-RAN as a designed environment for DRL.}
We map RAN optimization problems to the core elements of an MDP: states, actions, rewards, and constraints. This mapping is contextualized within the O-RAN architecture. The analysis also covers multi-agent and federated formulations specific to O-RAN, including cross-xApp coordination, hierarchical loop control, and multi-vendor federation.

\item \textit{A comprehensive taxonomy of DRL use cases in O-RAN.}
We classify DRL applications systematically by control loop and operational timescale. Analyzed use cases include radio resource management, mobility management, spectrum and interference mitigation, load balancing and traffic steering, energy efficiency, network slicing and QoS enforcement, ISAC, security, and massive MIMO. 

\item \textit{Advanced DRL methods for O-RAN.}
O-RAN environments produce a variable number of users, cells, neighbors, beams, and slices. We organize the encoder design space into concrete architectures, specifically permutation-invariant models, set-based models, and graph neural networks (GNNs). We also analyze foundation models for DRL, an area previously underexplored in O-RAN surveys. We map the foundation-model design space onto O-RAN control loops and interfaces. Investigated architectures include LLMs functioning as policies, reward models, and world models, alongside decision transformers processing O-RAN telemetry. 

\item \textit{Trustworthy and deployable DRL.}
This survey aligns safe DRL, explainability, adversarial robustness, runtime verification, and fairness with O-RAN Alliance AI/ML specifications. It further maps these concepts to 3GPP and European Telecommunications Standards Institute (ETSI) standards, and the Open Testing and Integration Centre (OTIC) certification path. The discussion then details the complete pipeline from training to physical deployment. This encompasses sim-to-real transfer via digital twins, sample efficiency, continual learning, RLOps for O-RAN, and inference within strict latency constraints.

\item \textit{Benchmarks, reproducibility, and open directions.}
We review mainstream testbeds and emulators, including Colosseum~\cite{bonati2021colosseum}, ns-O-RAN~\cite{lacava2023ns}, OpenRAN Gym~\cite{bonati2023openran}, and OAIC~\cite{upadhyaya2022prototyping}, alongside their associated software stacks and datasets. The survey evaluates literature practices regarding the release of code, random seeds, and evaluation protocols to quantify the existing benchmark gap. Based on this, we identify open challenges and future research directions in DRL for Open AI-RAN.

\end{itemize}

\begin{figure*}
    \centering
    \includegraphics[width=1\linewidth]{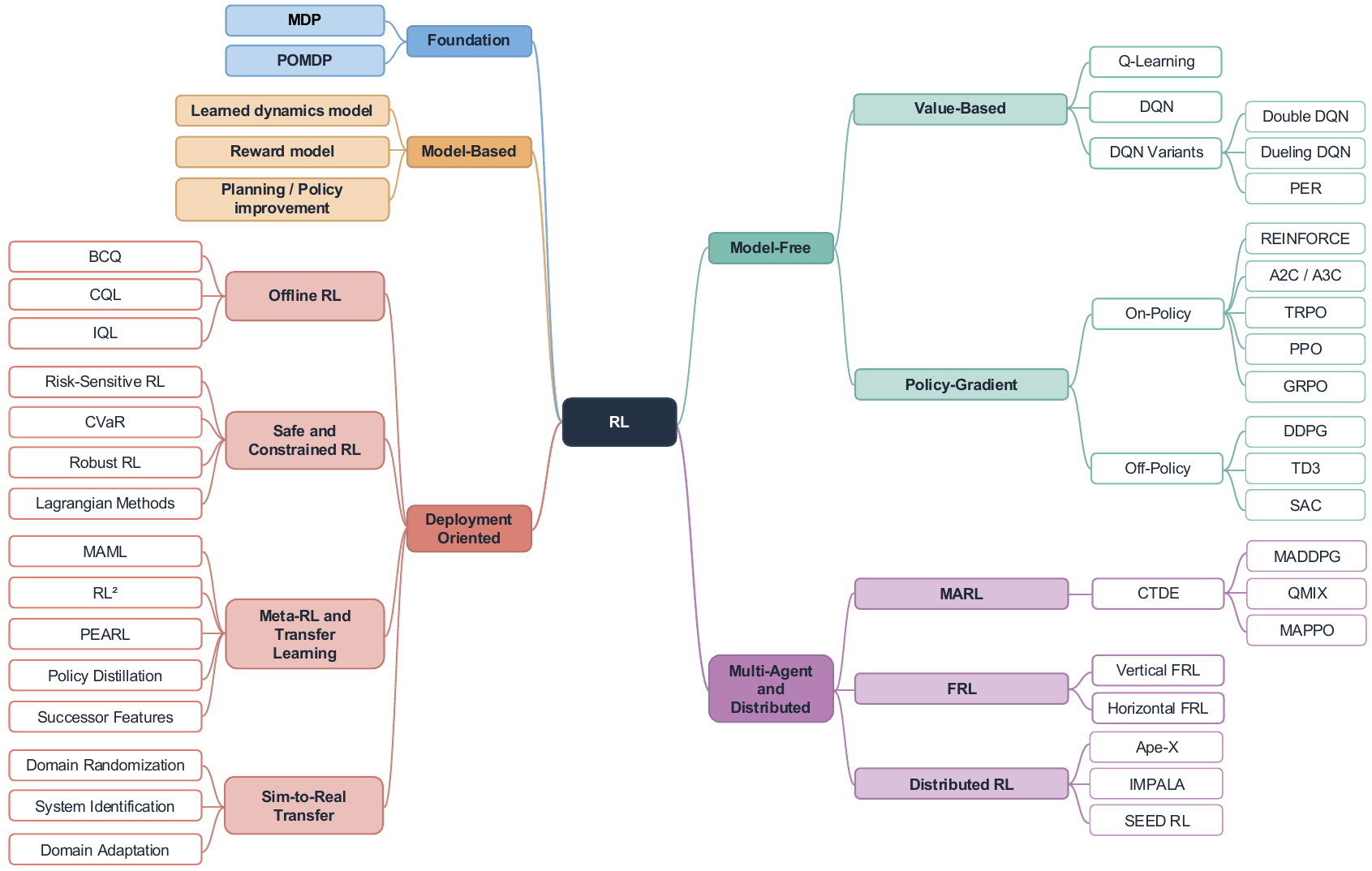}\vspace{-0.1in}
    \caption{Taxonomy of RL methods covered in this survey. RL is organized into five main branches: foundation (MDP and POMDP), model-based RL (learned dynamics, reward modeling, and planning/policy improvement), model-free RL including value-based methods (Q-learning, DQN, and DQN variants) and policy-gradient methods under on-policy and off-policy settings, multi-agent and distributed RL including MARL, federated RL (FRL), and distributed RL, and deployment-oriented RL covering offline RL, safe and constrained RL, meta-RL/transfer learning, and sim-to-real transfer. Representative algorithms listed in each branch illustrate the methodological landscape rather than an exhaustive catalog.}\vspace{-0in}
    \label{fig:rl_taxonomy}
\end{figure*}

\subsection{Paper Organization}

The remainder of this survey is organized as illustrated in Fig.~\ref{fig:organization}. Section~\ref{2_Background} establishes the methodological foundations of DRL, covering value-based and policy-gradient methods, model-based and multi-agent learning, federated and distributed approaches, offline and safe RL, and transfer across simulation and real deployments. Section~\ref{3_ORAN_Architecture_DRL} then introduces the O-RAN environment in which these methods operate, with emphasis on the RIC hierarchy, the SMO and O-Cloud, open interfaces, AI/ML lifecycle management, and control loops at different timescales. Building on these foundations, Section~\ref{4_RAN_to_MDPs} presents a systematic framework for translating RAN control objectives into MDP formulations through the design of observations, actions, rewards, constraints, and temporal structure. Section~\ref{5_Use_Cases} applies this framework to representative use cases, including radio resource and mobility management, interference control, traffic steering, energy efficiency, network slicing, ISAC, security, and massive-MIMO control.

The survey next turns to advanced learning paradigms and the requirements for operational deployment. Section~\ref{6_MADRL_FDRL} examines multi-agent and federated DRL for coordination across applications, control loops, vendors, and network domains. Section~\ref{7_FM_DRL} explores how foundation models, LLMs, trajectory models, and agentic AI can augment O-RAN control, while Section~\ref{8_Trustworthy_DRL} reviews safety, fairness, explainability, robustness, verification, and runtime monitoring. Section~\ref{9_Training_Deployment} discusses training strategies, sim-to-real transfer, continual learning, inference efficiency, and RLOps. Sections~\ref{10_Platforms} and~\ref{11_Standards_Industry} review platforms, benchmarks, reproducibility practices, standards, regulatory considerations, and industry initiatives. Finally, Section~\ref{12_Challenges} identifies open research challenges, and Section~\ref{13_Conclusion} concludes the survey.

% \clearpage 

\section{DRL Foundations}
\label{2_Background}

% \subsection{Overview}
% \label{drl_overview}

DRL combines the sequential decision making framework of RL with the function approximation capabilities of DNNs, enabling autonomous agents to learn control policies from high dimensional observations. 
This section reviews the foundations of DRL for O-RAN control. 
Fig.~\ref{fig:rl_taxonomy} summarizes the taxonomy adopted in this section. It organizes RL into five major categories: foundational decision models, including MDPs and POMDPs; model based RL; model free RL, including value based and policy gradient methods; multi-agent, federated, and distributed RL; and deployment oriented techniques.

%DRL integrates the sequential decision-making framework of reinforcement learning (RL) with the function-approximation capacity of deep neural networks, enabling autonomous agents to learn control policies from high-dimensional observations. This section reviews the principal theoretical and algorithmic building blocks of DRL that underlie the modern learning-based RAN. \lj{Fig.~\ref{fig:rl_taxonomy} summarizes the taxonomy followed in this section, grouping RL methods into model-based RL, model-free value-based and policy-gradient methods, multi-agent/federated/distributed RL, and deployment-oriented techniques.}

\subsection{MDP and POMDP}
\label{MDPs_and_POMDPs}

\begin{definitionbox}{Markov Decision Process (MDP)} 
An MDP is a tuple $\mathcal{M} = (\mathcal{S}, \mathcal{A}, P, r)$, where $\mathcal{S}$ represents the state space, 
$\mathcal{A}$ represents the action space, 
$P: \mathcal{S} \times \mathcal{A} \to \Delta(\mathcal{S})$ is the transition function that maps a state-action pair to a probability distribution over successor states, and $r:\mathcal{S} \times \mathcal{A}  \to  \mathbb{R}$ is the reward function \cite{puterman1990markov}. 
\end{definitionbox}

\textbf{MDP Objective.}
At each discrete time step $t=0,1,2,\ldots$, the agent observes a state $s_t \in \mathcal{S}$ and selects an action $a_t \in \mathcal{A}$. The Markov property assumes that the next state depends only on the current state and action:
$P(s_{t+1} \mid s_t, a_t, s_{t-1}, a_{t-1}, \dots) = P(s_{t+1} \mid s_t, a_t)$.
This assumption makes the MDP analytically tractable. A stationary policy $\pi: \mathcal{S} \to \Delta(\mathcal{A})$ specifies the agent's behavior. The agent aims to find a policy $\pi$ that maximizes the expected discounted return:
\begin{equation}
    J(\pi) = \mathbb{E}_{\tau \sim \pi} \left [ \sum_{t=0}^\infty \gamma^t r(s_t,a_t) \right ] ,
\label{mdp_reward}
\end{equation}
where $\tau=(s_0, a_0, s_1, a_1, \dots)$ denotes a trajectory generated by sampling $s_0 \sim \rho_0$, $a_t \sim \pi(\cdot \mid s_t)$, and $s_{t+1} \sim P(\cdot \mid s_t,a_t)$. The term $\rho_0 \in \Delta(\mathcal{S})$ denotes the initial state distribution, and $\gamma \in [0,1]$ is the discount factor. 

\textbf{Value Functions.}
Two value functions characterize the long-run consequences of a policy:
\emph{state value function} and \emph{state-action value function}.
The state value function, denoted as $V_\pi(s)$, characterizes the expected return starting from $s$ and following $\pi$:
\begin{equation}
    V_\pi(s) = \mathbb{E}_\pi \left[ \sum_{t=0}^\infty \gamma^t r(s_t,a_t) \mid s_0 = s \right].
\label{state_value}
\end{equation}
The state-action value function, denoted as $Q_\pi(s,a)$, characterizes the expected return starting from $s$, taking action $a$, and following $\pi$:
\begin{equation}
    Q_\pi(s,a) = \mathbb{E}_\pi \left[ \sum_{t=0}^\infty \gamma^t r(s_t,a_t) \mid s_0 = s, a_0=a \right].
\end{equation}
Both value functions satisfy the Bellman equations that can recursively decompose the expected long-term return into two parts: the immediate reward received at the current step and the discounted expected value of the subsequent state \cite{wiering2012reinforcement}. 

\textbf{Optimality Equations.}
The objective of MDP is to find an optimal policy $\pi^*$ that receives the highest expected reward. For the optimal state value function $V^* (s)= \max_\pi V_\pi(s)$, the Bellman optimality equation takes the form
\begin{equation}
    V^*(s_t) = \max_{a_t}\left[ r(s_t,a_t) + \gamma \mathbb{E}_{s_{t+1} \sim P(\cdot \mid s_t, a_t) } V^*(s_{t+1}) \right].
\label{bellman_v}
\end{equation}
Similarly, the optimal state-action value function $Q^* (s,a)= \max_\pi Q_\pi(s,a)$ satisfies
\begin{equation}
    \begin{aligned}
        Q^*(s_t,a_t) = & \gamma \mathbb{E}_{s_{t+1} \sim P(\cdot \mid s_t, a_t) } \left[ \max_{a_{t+1}} Q^*(s_{t+1}, a_{t+1}) \right] \\
        & + r(s_t,a_t) .
    \label{bellman_q}
    \end{aligned}
\end{equation}
% \begin{equation}

% Q^(s_t,a_t) = r(s_t,a_t) + \gamma \mathbb{E}{s*{t+1} \sim P(\cdot \mid s_t, a_t) } \left[ \max_{a_{t+1}} Q^*(s_{t+1}, a_{t+1}) \right].

% \label{bellman_q}

% \end{equation}

\begin{definitionbox}{Partially Observable MDP (POMDP)} 
A POMDP can be represented by a tuple $(\mathcal{S}, \mathcal{A}, P, r, {\Omega}, \mathcal{O})$, which extends the MDP with an observation space ${\Omega}$ and an observation distribution $\mathcal{O}: \mathcal{S} \to \Delta({\Omega})$ \cite{kaelbling1998planning}. At each step $t$, the agent receives an observation $o_t \in \Omega$ sampled as $o_t \sim \mathcal{O}(\cdot | s_t)$. 
\end{definitionbox}

\textbf{Partial Observability.}
In a POMDP, the optimal action selection typically conditions on the entire observation-action history $h_t = (o_0, a_0, \dots, o_t)$ or, equivalently, on the belief state $b_t(s) = \Pr (s_t = s \mid h_t)$. Exact solutions are intractable except in the smallest problem instances. 
Therefore, the prevailing approach in DRL is to approximate the belief state via a recurrent neural network (RNN) or a transformer that encodes the observation history \cite{esslinger2022deep}.

\textbf{O-RAN Relevance.}
The distinction between MDP and POMDP  is critical for O-RAN. RICs observe the network through telemetry exposed over the E2 and O1 interfaces, and this telemetry arrives after sampling, aggregation, and transport delay. A DRL agent at the Near-RT RIC therefore faces a POMDP in practice.

\subsection{Value-Based Methods}
\label{Value_Based_Methods}

Value-based methods learn an approximation $Q_\theta$ of the optimal action-value function $Q^*$ and derive the policy by greedy action selection: $\pi^* = \arg\max_a Q_\theta(s, a)$.

\textbf{Tabular Q-Learning.} 
Tabular Q-learning represents the action-value function as a lookup table indexed by state-action pairs \cite{watkins1992q}. Its update rule is
\begin{equation}
\label{q_learning}
\begin{split}
Q(s_t,a_t) \gets{} & Q(s_t,a_t) + \alpha \Big[ r(s_t,a_t) \\
                   & + \gamma\max_{a_{t+1}} Q(s_{t+1},a_{t+1}) - Q(s_t,a_t) \Big],
\end{split}
\end{equation}
where $\alpha \in (0,1]$ is the learning rate. 
In this tabular setting, $Q$ is stored as a lookup table indexed by $(s_t,a_t)$, so each update modifies a single entry without affecting the others. Given sufficient exploration and an appropriate learning rate decay, this update converges to $Q^*$ \cite{watkins1992q}, following the Bellman optimality structure in Eq.~(\ref{bellman_q}). Once $Q \to Q^*$, the optimal policy is given by $\pi^*(s) = \arg\max_a Q^*(s, a)$.

\textbf{Deep Q-Network (DQN).} DQN extends  Q-learning by replacing the tabular action-value function with a parametric function approximator, typically a DNN parameterized by $\theta$. Rather than maintaining an independent value estimate for every state-action pair $(s_t,a_t)$, DQN learns a shared set of parameters $\theta$ that generalizes across the state space. For a discrete action space, the DNN typically takes the state $s_t$ as input and outputs an estimated Q-value $Q_\theta(s_t,a)$ for each possible action $a$.

When nonlinear function approximation is combined with bootstrapping and temporally correlated training samples, Q-learning may exhibit unstable or even divergent training behavior and generally lacks the convergence guarantees available in the tabular setting. To improve training stability, Mnih \emph{et al.}~\cite{mnih2015human} introduced two key mechanisms. First, an experience replay buffer \(\mathcal{D}\) stores previously observed transitions \((s_t,a_t,r_t,s_{t+1})\), and the learner samples mini-batches uniformly from this buffer during training.
Experience replay reduces the temporal correlation among consecutive training samples and improves data efficiency by reusing each transition in multiple updates. Second, a target Q-network with parameters \(\theta^-\) is fixed for multiple training steps and periodically synchronized with the online Q-network. This mechanism reduces the coupling between the bootstrap target and the parameters under update, thereby improving training stability.

With these two mechanisms, DQN commonly minimizes the loss
\begin{equation}
\mathcal{L}_\text{DQN}(\theta) \!=\! \mathbb{E}_{\mathcal{D}} \Big[ \big(  r_t + \gamma \max_{a_{t+1}} Q_{\theta^-}(s_{t+1}, a_{t+1}) 
 - Q_\theta(s_t, a_t) \big)^2 \Big],
\label{dqn_loss}
\end{equation}
where \(\mathcal{D}\) denotes the experience replay buffer, \(\theta\) denotes the parameters of the online Q-network, and \(\theta^-\) denotes the parameters of the target Q-network.

% ====

% DQN has been widely used in DRL applications. It extends classical Q-learning by replacing the tabular Q-function with a parametric function approximator, typically a deep neural network (DNN) parameterized by $\theta$. Instead of maintaining a separate value estimate for each state-action pair $(s_t,a_t)$, DQN learns shared parameters $\theta$ that generalize across the state-action space. 
% To train the DNN within DQN, a typical loss is as follows:
% \begin{equation}
% \mathcal{L}_\text{DQN}(\theta) = \mathbb{E}_{\mathcal{D}} \Big[ \big(  r_t + \gamma \max_{a_{t+1}} Q_{\theta^-}(s_{t+1}, a_{t+1}) 
%  - Q_\theta(s_t, a_t) \big)^2 \Big].
% \label{dqn_loss}
% \end{equation}

% In its original form, DQN often suffers from unstable updates and lacks convergence guarantees.
% To address these issues, Mnih et al. \cite{mnih2015human} introduced two key mechanisms to stabilize deep Q-learning. First, an experience replay buffer $\mathcal{D}$ stores past transitions $(s_t, a_t, r_t, s_{t+1})$, from which mini-batches are sampled uniformly during training. This reduces the temporal correlation among consecutive samples, which would otherwise violate the independent and identically distributed (i.i.d.) assumption commonly used in stochastic gradient descent. Second, a target network with parameters $\theta^-$ is held fixed for a number of training steps and periodically synchronized with the online network. This decouples the bootstrap target from the rapidly changing parameters being optimized, thereby improving training stability.

\textbf{Improvements and Variants.} Several variants address specific weaknesses of DQN. Double DQN \cite{van2016deep} mitigates the systematic overestimation introduced by the $\max$ operator in Eq.~(\ref{dqn_loss}).
It decouples action selection from action evaluation: the next action is selected by the online network, $a^*_{t+1} = \arg \max_{a_{t+1}} Q_\theta(s_{t+1}, a_{t+1}) $, and evaluated by the target network $Q_{\theta^-}(s_{t+1}, a^*_{t+1})$. 
Dueling DQN \cite{wang2016dueling} parameterizes the action-value function by splitting the network after a shared feature extractor into two output heads:
a state-value head $V_\theta(s_t)$ and an advantage head $A_\theta(s_t,a_t)$, 
which are combined as 
\begin{equation}
Q_\theta(s_t, a_t) = V_\theta(s_t) + \Big(A_\theta(s_t, a_t) - \tfrac{1}{|\mathcal{A}|} \sum_{a_{t+1} \in \mathcal{A}} A_\theta(s_t, a_{t+1})\Big),
\end{equation}
where $\mathcal{A}$ denotes the discrete action set and $|\mathcal{A}|$ is its cardinality.
This improves credit assignment when many actions yield similar returns.

Prioritized experience replay (PER) \cite{schaul2015prioritized} samples transitions with probability proportional to their absolute temporal-difference error. This choice focuses updates on transitions where the current estimate is most inaccurate and improves sample efficiency. Distributional RL \cite{bellemare2017distributional} replaces the scalar estimate $Q_\theta(s_t, a_t)$ with a learned distribution over returns. This distribution captures the inherent randomness of returns under the MDP and provides a richer learning signal that often improves empirical performance.

\textbf{Key Takeaways.}
Value-based methods provide a natural fit for discrete and enumerable O-RAN control decisions, such as selecting among scheduling, association, or configuration options. They become impractical when the action space grows continuous or combinatorial, motivating the policy-gradient methods reviewed next.

\subsection{Policy-Gradient Methods}
\label{Policy_Gradient_Methods}

Unlike value-based methods, which first estimate an action-value function and then derive a policy from it, policy-gradient methods directly optimize a parameterized policy \(\pi_\theta(a \mid s)\).
The policy parameters \(\theta\) are adjusted to maximize the expected cumulative return
$J(\pi_\theta)$.

% Different from value-based methods, policy-gradient methods directly optimize a parameterized policy $\pi_\theta(a \mid s)$ by ascending the gradient of the expected return $J(\pi_\theta)$.

\begin{theorembox}{Policy Gradient} 
Sutton \emph{et al.}~\cite{sutton1999policy} established that the gradient of the expected return takes the form
\begin{equation}
    \!\!\!\nabla_\theta J(\pi_\theta) = \mathbb{E}_{s\sim d_{\pi_\theta},\, a \sim \pi_\theta} \left[ \nabla_\theta \log \pi_\theta(a \mid s)\, Q_{\pi_\theta}(s,a) \right],
\label{policy_gradient}
\end{equation}
where $d_{\pi_\theta}(s)=\lim_{t\to \infty}\Pr\{s_t=s \mid s_0, \pi_\theta\}$ is the stationary distribution of states under $\pi_\theta$.
\end{theorembox}

The policy-gradient theorem is foundational because it avoids differentiating the unknown state-transition dynamics. Instead, it expresses the policy gradient as an expectation over states and actions encountered under the current policy, allowing sampled trajectories to estimate the gradient directly. The term
$
    \nabla_\theta \log \pi_\theta(a_t \mid s_t)
$
is commonly referred to as the \emph{score function}. It increases the probability of actions associated with positive returns and decreases the probability of actions associated with negative returns.
In practice, $Q_{\pi_\theta}(s,a)$ in Eq.~(\ref{policy_gradient}) is replaced by an estimator $\hat{A}_t$ of the advantage $A_{\pi_\theta}(s,a) = Q_{\pi_\theta}(s,a) - V_{\pi_\theta}(s)$ to reduce variance without introducing bias.

\textbf{On-Policy Methods.} 
On-policy methods learn from transitions generated by the current policy and update the same policy that collects the data. The REINFORCE algorithm~\cite{williams1992simple} estimates Eq.~(\ref{policy_gradient}) by replacing $Q_{\pi_\theta}(s,a)$ with the empirical Monte Carlo return $G_t = \sum_{k=0}^{\infty} \gamma^k r_{t+k}$, yielding an unbiased but high-variance estimator. 

Actor-critic methods~\cite{konda1999actor} reduce variance by jointly training the \emph{actor} (i.e., the policy $\pi_\theta$) and the \emph{critic} $V_\phi$ (i.e., a learned state-value function) that evaluates the actor's choices. The critic enables a low-variance one-step advantage estimator as follows:
\begin{equation}
    \hat{A}_t = r_t + \gamma V_\phi(s_{t+1}) - V_\phi(s_t).
\label{ac_advantage}
\end{equation}
It serves both as a state-dependent baseline and as a bootstrapped substitute for the Monte Carlo return. 

The asynchronous variant A3C~\cite{mnih2016asynchronous} and its synchronous counterpart A2C~\cite{wu2017baselines} parallelize trajectory collection across environment instances, decorrelating updates without requiring a replay buffer.

Naive policy-gradient updates can produce arbitrarily large policy changes, leading to catastrophic performance collapse. 
Several methods have been proposed to address this issue.

\begin{itemize}[leftmargin=0.15in]
\item 
\emph{Trust Region Policy Optimization (TRPO)~\cite{schulman2015trust}.}
TRPO constrains the Kullback-Leibler (KL) divergence between consecutive policies:
\begin{equation}
    \mathbb{E}_{s \sim d_{\pi_{\theta_\text{old}}}} \left[ D_\text{KL} \left(\pi_{\theta_\text{old}}(\cdot \mid s) \,\|\, \pi_\theta(\cdot \mid s)\right) \right] \leq \delta,
    \label{trpo}
\end{equation} 
where $\theta_\text{old}$ are the policy parameters before the update and $\delta > 0$ is the trust-region radius. This formulation is motivated by a monotonic improvement bound, but enforcing it requires computationally expensive natural-gradient updates via Fisher-vector products.

\item 
\emph{Proximal Policy Optimization (PPO)~\cite{schulman2017proximal}.}
PPO achieves a similar effect with a first-order clipped surrogate objective. 
Its maximization objective function is:
\begin{equation}
\begin{aligned}
\mathcal{L}_{\mathrm{PPO}}(\theta)
&= \mathbb{E}_t \Big[
\min\Big(
\rho_t(\theta)\hat{A}_t, \\
&\qquad
\operatorname{clip}\!\left(
\rho_t(\theta),1-\epsilon,1+\epsilon
\right)\hat{A}_t
\Big)
\Big].
\end{aligned}
\label{loss_ppo}
\end{equation}
% \begin{equation} \mathcal{L}_\text{PPO}(\theta) = \mathbb{E}_t \left[ \min\Big( \rho_t(\theta)\, \hat{A}_t,\; \text{clip}\!\left(\rho_t(\theta), 1-\epsilon, 1+\epsilon\right) \hat{A}_t \Big) \right], \label{loss_ppo} \end{equation}
where $\rho_t(\theta) = \pi_\theta(a_t \mid s_t) / \pi_{\theta_\text{old}}(a_t \mid s_t)$ is the probability ratio, $\epsilon \in (0,1)$ controls the clipping range, and $\hat{A}_t$ is the advantage estimator. PPO typically uses Generalized Advantage Estimation (GAE)~\cite{schulman2017proximal} to compute $\hat{A}_t$ by
\begin{equation}
    \hat{A}_t = \sum_{l=0}^{\infty} (\gamma \lambda)^l\, \delta_{t+l}, \quad \delta_t = r_t + \gamma V_\phi(s_{t+1}) - V_\phi(s_t),
\label{gae}
\end{equation}
where $\lambda \in [0,1]$ trades off bias and variance: $\lambda = 0$ recovers the one-step TD advantage in Eq.~(\ref{ac_advantage}), while $\lambda = 1$ recovers the Monte Carlo advantage. PPO is a widely used policy-gradient algorithm in the DRL-for-O-RAN literature owing to its implementation simplicity and robustness across hyperparameter settings.

\item
\emph{Group Relative Policy Optimization (GRPO)~\cite{shao2024deepseekmath}.}
GRPO was introduced by DeepSeek-AI in DeepSeekMath as a PPO-style 
DRL algorithm for LLM post-training~\cite{shao2024deepseekmath}.
For each prompt \(x\), the behavior policy 
\(\pi_{\theta_{\mathrm{old}}}\) generates a group of \(G\) completions,
\(\mathbf{y}=(y_1,\ldots,y_G)\), where completion \(y_i\) contains
\(T_i\) tokens. Instead of training a separate value network, GRPO
constructs relative advantages by comparing the rewards of completions
generated for the same prompt. Its clipped surrogate objective, which
is maximized with respect to \(\theta\), can be written as
\begin{equation}
\label{grpo_objective}
\begin{aligned}
\mathcal{L}_{\mathrm{GRPO}}(\theta)
= \mathbb{E}_{x,\mathbf{y}}\Bigg[
&\frac{1}{G}\sum_{i=1}^{G}
\min\Big(
\omega_i(\theta)\tilde{A}_i, \\
&\qquad
\operatorname{clip}\!\big(\omega_i(\theta),1-\epsilon,1+\epsilon\big)\tilde{A}_i
\Big) \\
&\qquad
- \beta D_{\mathrm{KL}}\!\big(
\pi_\theta(\cdot \mid x)\,\|\,\pi_{\mathrm{ref}}(\cdot \mid x)
\big)
\Bigg].
\end{aligned}
\end{equation}
Here, $\omega_i(\theta)=\pi_\theta(y_i \mid x)/\pi_{\theta_\text{old}}(y_i \mid x)$ is the importance ratio, $i\in \{1,...,G\}$.
$\pi_\text{ref}$ is a fixed reference policy, and $\beta \ge 0$ controls KL regularization. The normalized relative advantage is formed from the groupwise scalar scores by letting:
$\tilde{A}_i = \frac{u_i-\bar{u}}{\sigma_u + \varepsilon_A}$, 
where
$\bar{u} = \frac{1}{G}\sum_{j=1}^{G} u_j$
and
$\sigma_u = \big(\frac{1}{G}\sum_{j=1}^{G}(u_j-\bar{u})^2\big)^{1/2}$,
and \(\varepsilon_A>0\) is a small constant introduced for numerical
stability. 
Under outcome-level supervision, the same sequence-level
advantage \(\widetilde{A}_i\) is assigned to every token in completion
\(y_i\). 
% More general variants may instead construct token-dependent advantages \(\widetilde{A}_{i,t}\) from process-level rewards.

\ \ 
Because GRPO obtains its baseline through within-group reward normalization rather than through a separately trained critic, it avoids the memory and computational overhead associated with storing and
optimizing a value network. 
This property is particularly attractive
for LLM post-training, where each \(u_i\) may be produced by a reward
model, a verifier, or a rule-based evaluator. 
However, GRPO requires
multiple completions for each prompt, increasing rollout-generation
cost. Its learning signal also depends on meaningful reward variation
within each group; when the completion rewards are identical or nearly
identical, the normalized advantages provide little useful information
for policy improvement.

\end{itemize}

\textbf{Off-Policy Methods.} On-policy methods such as PPO suffer from limited sample efficiency, as each transition is used only once. This limitation matters in real-world RAN deployments, where data collection and exploration are costly. 
In contrast, off-policy methods separate the target policy from the behavior policy, and reuse past transitions for training. 

Deep Deterministic Policy Gradient (DDPG)~\cite{lillicrap2015continuous} is an off-policy actor-critic algorithm.
It extends DQN from discrete to continuous action spaces. The discrete $\max_{a} Q$ in Eq.~(\ref{dqn_loss}) becomes intractable over a continuous $\mathcal{A}$, so DDPG learns a deterministic actor $\mu_\theta: \mathcal{S} \to \mathcal{A}$ that approximates the maximizing action, paired with a critic $Q_\phi(s,a)$. 

DDPG maintains four networks: the online actor $\mu_\theta$ and critic $Q_\phi$ drive learning, while the target actor $\mu_{\theta^-}$ and target critic $Q_{\phi^-}$ provide stable bootstrap targets. DDPG updates the target networks by Polyak averaging $\theta^- \gets \tau\theta + (1-\tau)\theta^-$, $\phi^- \gets \tau\phi + (1-\tau)\phi^-$ with the hyperparameter $\tau \ll 1$. The critic minimizes the mean squared Bellman error:
\begin{equation}
\begin{aligned}
\mathcal{L}_{\mathrm{critic}}(\phi)
&= \mathbb{E}_{\mathcal{D}} \Big[
\big(
r_t + \gamma Q_{\phi^-}\!\left(
s_{t+1},
\mu_{\theta^-}(s_{t+1})
\right)
\\
&\qquad
- Q_\phi(s_t,a_t)
\big)^2
\Big].
\end{aligned}
\label{ddpg_critic}
\end{equation}
 % \begin{equation}
 %    \mathcal{L}_\text{critic}(\phi) = \mathbb{E}_{\mathcal{D}} \left[ \left( r_t + \gamma Q_{\phi^-}\!\left(s_{t+1}, \mu_{\theta^-}(s_{t+1})\right) - Q_\phi(s_t, a_t) \right)^2 \right],
 % \label{ddpg_critic}
 % \end{equation}
where $\mu_{\theta^-}(s_{t+1})$ plays the role of the discrete $\arg\max_{a_{t+1}}$ in Eq.~(\ref{dqn_loss}). The actor is then trained to output this maximizing action by ascending the deterministic policy gradient~\cite{silver2014deterministic},
\begin{equation}
    \nabla_\theta J(\theta) = \mathbb{E}_{s_t \sim \mathcal{D}} \left[ \nabla_\theta \mu_\theta(s_t)\, \nabla_a Q_\phi(s_t, a) \big|_{a=\mu_\theta(s_t)} \right].
\label{ddpg_actor}
\end{equation}
This gradient propagates the critic's gradient through the actor by the chain rule. Exploration is induced by additive noise $\epsilon_t$ on the executed action, $a_t = \mu_\theta(s_t) + \epsilon_t$.

Although DDPG is sample-efficient, it is sensitive to hyperparameters \cite{henderson2018deep} and susceptible to value overestimation \cite{fujimoto2018addressing}.
Twin Delayed DDPG (TD3)~\cite{fujimoto2018addressing} addresses these issues through three modifications. First, \emph{clipped double Q-learning} maintains two critics ($Q_{\phi_1}$ and $Q_{\phi_2}$) and forms the bootstrap target from their minimum, suppressing the positive bias of a single critic queried at the maximizing action:
\begin{equation}
    y_t = r_t + \gamma \min_{i\in\{1,2\}} Q_{\phi^-_i}\!\left(s_{t+1}, \tilde{a}_{t+1}\right).
\label{td3_target}
\end{equation}
Second, \emph{target policy smoothing} perturbs the target action $\tilde{a}_{t+1}$ with clipped Gaussian noise, $\tilde{a}_{t+1} = \mu_{\theta^-}(s_{t+1}) + \mathrm{clip}(\epsilon_t, -c, c)$, $c>0$, preventing the actor from exploiting sharp peaks in $Q_\phi$. Third, \emph{delayed policy updates} refresh the actor and target networks once every $d$ critic steps, allowing the value estimate to stabilize before shaping the policy.

Soft Actor-Critic (SAC)~\cite{haarnoja2018soft} is another off-policy actor–critic algorithm. 
It replaces DDPG's deterministic actor with a stochastic Gaussian policy $\pi_\theta(a \mid s)$ and augments the return with a policy-entropy bonus.
Its return function can be written as:
\begin{equation}
    J(\pi_\theta) = \mathbb{E}_{\tau \sim \pi_\theta} \left[ \sum_{t=0}^\infty \gamma^t \big( r(s_t,a_t) + \beta\, \mathcal{H}(\pi_\theta(\cdot \mid s_t)) \big) \right],
\label{sac_objective}
\end{equation}
where $\mathcal{H}(\pi_\theta(\cdot \mid s_t)) = -\mathbb{E}_{a_t \sim \pi_\theta}[\log \pi_\theta(a_t \mid s_t)]$ is the policy entropy and $\beta > 0$ is a temperature trading off reward and exploration. 

% Like TD3, SAC uses twin critics, but the bootstrap target absorbs the entropy term as $y_t = r_t + \gamma\, \mathbb{E}_{a_{t+1} \sim \pi_\theta}[\min_i Q_{\phi^-_i}(s_{t+1}, a_{t+1}) - \beta \log \pi_\theta(a_{t+1} \mid s_{t+1})]$. The actor is trained via the reparameterization trick $a_t = f_\theta(\epsilon_t; s_t)$, so exploration is built into the policy rather than injected externally as in DDPG and TD3. The temperature $\beta$ is typically tuned automatically against a target entropy~\cite{haarnoja2018soft}. Owing to its sample efficiency, principled exploration, and robustness to hyperparameters, SAC has emerged as a standard continuous-control baseline and is extensively applied for resource allocation in O-RAN.

Like TD3, SAC uses twin critics, but the bootstrap target incorporates
the entropy term:
\begin{equation}
\begin{aligned}
y_t
= r_t + \gamma
\mathbb{E}_{a_{t+1}\sim\pi_\theta}
\Big[
&\min_i Q_{\phi_i^-}(s_{t+1},a_{t+1})
\\
&-\beta\log\pi_\theta(a_{t+1}\mid s_{t+1})
\Big].
\end{aligned}
\label{sac_target}
\end{equation}
The actor is trained via the reparameterization trick
$a_t=f_\theta(\epsilon_t;s_t)$, so exploration is built into the policy
rather than injected externally as in DDPG and TD3. The temperature
$\beta$ is typically tuned automatically against a target
entropy~\cite{haarnoja2018soft}. Owing to its sample efficiency,
principled exploration, and robustness to hyperparameters, SAC has
emerged as a standard continuous-control baseline and is extensively
applied for resource allocation in O-RAN.

\textbf{Key Takeaways.}
Policy-gradient methods support both stochastic discrete control and continuous control. On-policy methods such as PPO emphasize stable updates, while off-policy methods such as DDPG, TD3, and SAC improve sample efficiency, a key requirement when RAN data collection and exploration incur operational cost.

\subsection{Model-Based RL}
\label{Model_Based_Methods}

The methods reviewed in Sections~\ref{Value_Based_Methods} and~\ref{Policy_Gradient_Methods} are \emph{model-free}: they learn value functions or policies directly from interaction data without explicitly modeling the environment transition function \(P\). In contrast, model-based RL learns or assumes an explicit model of the environment and uses it for planning, policy optimization, data generation, or a combination of these purposes~\cite{m2023model}.

\textbf{Learned Model.} A model-based agent typically maintains a parametric dynamics model
\(
\hat{P}_{\psi}(s_{t+1}\mid s_t,a_t)
\),
where \(\psi\) denotes the model parameters. When the reward function is unknown, the agent may additionally learn a parametric reward model
\(
\hat{r}_{\omega}(s_t,a_t)
\),
where \(\omega\) denotes the reward-model parameters. Both models can be trained through supervised learning using observed transitions
\(
(s_t,a_t,r_t,s_{t+1})\in\mathcal{D}
\)
\cite{mnih2015human}.
For example, the dynamics model can be trained to predict \(s_{t+1}\) or its probability distribution conditioned on \((s_t,a_t)\), while the reward model predicts the corresponding immediate reward.

\textbf{Benefits and Limitations.}
Once learned, the environment model can support look-ahead planning, generate synthetic trajectories, or provide imagined rollouts for policy and value-function updates. This ability can substantially improve sample efficiency because the agent can perform multiple simulated updates without repeatedly interacting with the physical environment.

However, model-based RL introduces errors arising from imperfect dynamics and reward models. These errors may accumulate over long imagined rollouts, particularly in high-dimensional, stochastic, or partially observable environments. Consequently, although model-based RL has achieved substantial methodological progress, its application to complex real-world systems remains challenging because of model bias, uncertainty, compounding prediction errors, and the computational cost of planning.

\textbf{Key Takeaways.}
Model-based RL improves sample efficiency by letting agents plan or learn from imagined trajectories. Its relevance to O-RAN depends on model fidelity, because inaccurate channel, mobility, traffic, or protocol dynamics can compound over rollouts and degrade policy learning.

% \subsection{Model-Based RL}
% \label{Model_Based_Methods}

% The methods reviewed in Sections~\ref{Value_Based_Methods} and \ref{Policy_Gradient_Methods} are \emph{model-free}: they estimate value functions or policies directly from interaction data without representing the transition function $P$. 
% In contrast, model-based RL maintains an explicit approximation of the environment dynamics and exploits it for planning, policy improvement, or both \cite{m2023model}. The model-based agent maintains a parametric dynamics model $\hat{P}_\psi(s_{t+1} \mid s_t, a_t)$, parameterized by $\psi$.
% When the reward function is not known a priori, an additional reward model $\hat{r}_\omega(s_t, a_t)$ with parameters $\omega$. Both components can be fit by supervised learning on the transitions stored in $\mathcal{D}$ \cite{mnih2015human}.

% In practice, due to the difficulties in modeling the transition function $P$, the progress of model-based RL methods remains limited in the literature. 

\subsection{Multi-Agent RL}
\label{Multi-Agent_RL}
When multiple decision-makers act concurrently in a shared environment, the single-agent MDP no longer applies. Because state transitions and rewards depend on the joint action of all agents, the environment becomes non-stationary from the perspective of individual agents. Multi-Agent Reinforcement Learning (MARL) addresses this challenge by extending the MDP to $N$ interacting agents. This extension relies on game theory to characterize strategic coupling and leverages partial observability to model bounded informational access.

\begin{definitionbox}{Stochastic Game} 
A stochastic game can be represented by a tuple $(\mathcal{S}, \{ \mathcal{A}_i \}_{i=1}^N, P, \{r\}_{i=1}^N)$, where $N$ is the number of agents, $\mathcal{A}_i$ is agent $i$'s action space, $P: \mathcal{S} \times \mathcal{A}_1 \times \cdots \times \mathcal{A}_N \to \Delta(\mathcal{S})$ is the joint transition function, and $r_i$ is agent $i$'s individual reward function \cite{shapley1953stochastic}. When agents share a common reward, 
%$r_i$ for all $i$, 
the game is fully cooperative.
\end{definitionbox}

\begin{definitionbox}{Decentralized POMDP} 
A decentralized POMDP extends Definition~2 to $N$ cooperating agents, specified by the tuple $\big(\mathcal{S}$, $\{\mathcal{A}_i\}_{i=1}^N$, $P$, $r$, $\{\Omega_i\}_{i=1}^N$, $\{\mathcal{O}_i\}_{i=1}^N\big)$, where $N$ is the number of agents, $\mathcal{A} = \mathcal{A}_1 \times \cdots \times \mathcal{A}_N$ is the joint action space, $P : \mathcal{S} \times \mathcal{A} \to \Delta(\mathcal{S})$ is the joint transition function, $r : \mathcal{S} \times \mathcal{A} \to \mathbb{R}$ is the shared team reward, $\Omega_i$ is the local observation space of agent $i$, $\mathcal{O}_i : \mathcal{S} \to \Delta(\Omega_i)$ is its local observation distribution~\cite{bernstein2002complexity}. 
\end{definitionbox}

Under joint action $\mathbf{a}_t = (a_{1,t}, \dots, a_{N,t})$, the environment transitions to $s_{t+1} \sim P(\cdot \mid s_t, \mathbf{a}_t)$, and each agent $i$ receives a private observation
\begin{equation}
    o_{i,t+1} \sim \mathcal{O}_i(\cdot \mid s_{t+1}).
\end{equation}
Agent $i$ therefore conditions its decentralized policy $\pi_i$ on its local action-observation history $h_{i,t} = (o_{i,0}, a_{i,0}, \dots, o_{i,t})$. The team maximizes the shared discounted return
\begin{equation}
    J(\boldsymbol{\pi}) = \mathbb{E}_{\tau \sim \boldsymbol{\pi}}\!\left[\sum_{t=0}^\infty \gamma^t\, r(s_t, \mathbf{a}_t)\right]
\end{equation}
over the joint policy $\boldsymbol{\pi} = (\pi_1, \dots, \pi_N)$. The decentralized POMDP is the representative formulation for cooperative multi-agent decision-making under partial observability and limited communication~\cite{bernstein2002complexity, oliehoek2016concise}.

\textbf{Centralized Training, Decentralized Execution (CTDE).} 
CTDE is a dominant paradigm for cooperative MARL. During training, a centralized critic conditions on $(s_t, \mathbf{a}_t)$ to mitigate the non-stationarity of concurrently learning agents, while at execution, each agent acts on its local history $h_{i,t}$ alone. MADDPG~\cite{lowe2017multi} applies CTDE to continuous control via deterministic policy gradients, pairing decentralized actors with per-agent centralized critics. QMIX~\cite{rashid2020monotonic} addresses discrete-action cooperation by factorizing the joint action-value function as
% \begin{equation}
%     Q_{\text{tot}}(s_t, \boldsymbol{h}_t, \mathbf{a}_t) = f_\xi\!\left(s_t,\, Q_1(h_{1,t}, a_{1,t}),\, \dots,\, Q_N(h_{N,t}, a_{N,t})\right),
% \end{equation}
\begin{equation}
Q_{\mathrm{tot}}(s_t,\boldsymbol{h}_t,\mathbf{a}_t)
=
f_{\xi}\!\left(
s_t,\!
Q_1(h_{1,t},a_{1,t}),
\dots,\!
Q_N(h_{N,t},a_{N,t})
\right).
\end{equation}
where $\boldsymbol{h}_t = (h_{1,t}, \dots, h_{N,t})$.
QMIX constrains the mixing network $f_\xi$ to be monotonically non-decreasing in each per-agent utility, i.e., $\partial Q_{\text{tot}} / \partial Q_i \geq 0$. This monotonicity enforces the Individual-Global-Max (IGM) property, so decentralized greedy action selection over each $Q_i$ recovers the centralized greedy joint action.
MAPPO~\cite{yu2022surprising} adapts PPO with a centralized value function and matches or exceeds specialized cooperative methods on standard benchmarks, making it a strong default for cooperative MARL.

Competitive and mixed-cooperative settings introduce additional game-theoretic considerations, such as Nash equilibria, regret minimization and opponent modeling, that lie outside the fully cooperative paradigms that are relevant to intra-RAN coordination. We defer discussion of their implications for inter-xApp coordination and multi-vendor deployments to Section~\ref{6_MADRL_FDRL}.

\textbf{Key Takeaways.}
MARL generalizes DRL from one controller to multiple interacting controllers. Stochastic games capture strategic coupling, decentralized POMDPs capture partial local observations, and CTDE provides a practical training pattern for cooperative O-RAN control.

% \subsection{Federated and Distributed RL}
% \label{Federated_RL}
% Federated RL (FRL) combines RL with the federated learning paradigm~\cite{mcmahan2017communication}, in which $K$ participants collaboratively train a shared model under the constraint that raw trajectories never leave the local nodes. Following~\cite{qi2021federated, zhuo2019federated}, FRL approaches are generally divided into two main categories based on how the learning experience is partitioned across participants: In \emph{horizontal} FRL (HFRL), agents share identical state and action spaces but operate within independent and non-overlapping environments; while in \emph{vertical} FRL (VFRL), agents operate within the exact same environment but observe partial views of the state space and may have heterogeneous local action spaces and rewards.

\subsection{Federated and Distributed RL}
\label{Federated_RL}

Federated RL (FRL) integrates RL with the federated learning paradigm~\cite{mcmahan2017communication}. It enables \(K\) participants to collaboratively train a shared model while keeping raw trajectories and local interaction data at their respective nodes. The literature commonly divides FRL methods into two categories according to how participants distribute their learning experience~\cite{qi2021federated,zhuo2019federated}. In \emph{horizontal FRL} (HFRL), participants share the same state and action spaces but interact with independent, non-overlapping environments or environment instances. In \emph{vertical FRL} (VFRL), participants interact with the same underlying environment but observe different subsets or modalities of the global state and may possess heterogeneous local action spaces, reward signals, or control responsibilities.

% \noindent\textbf{Definition 5 (Horizontal FRL).}
\begin{definitionbox}{Horizontal FRL}
Each participant \(k\in\{1,\ldots,K\}\) holds a local MDP
\(
\mathcal{M}_k
=
(\mathcal{S},\mathcal{A},P_k,r_k,\gamma)
\),
where all participants share the state space \(\mathcal{S}\), action space \(\mathcal{A}\), and discount factor \(\gamma\), but may have heterogeneous transition functions \(P_k\) and reward functions \(r_k\). 
\end{definitionbox}

The homogeneous case, in which \(P_k\equiv P\) and \(r_k\equiv r\) for all \(k\), corresponds to parallel experience collection from independent instances of the same MDP. During each communication round, participant \(k\) performs \(E\) local update steps on its parameters \(\eta_k\) using trajectories collected from \(\mathcal{M}_k\). A central coordinator then aggregates the local parameters into a global model according to
\begin{equation}
    \eta_{\mathrm{global}}
    \gets
    \sum_{k=1}^{K}w_k\eta_k,
    \qquad
    w_k
    =
    \frac{n_k}{\sum_{j=1}^{K}n_j},
\label{fed_avg}
\end{equation}
where \(n_k\) denotes the number of local trajectories or transitions contributed by participant \(k\). The resulting global parameters \(\eta_{\mathrm{global}}\) are subsequently broadcast to the participants for the next round of local training.

\textbf{HFRL in O-RAN.} 
HFRL fits cases where geographically or administratively distributed agents perform the same control task in similar environments. For example, a common xApp may be deployed at multiple base stations while being trained from site-specific channel conditions, mobility patterns, and traffic distributions.

% \noindent\textbf{Definition 6 (Vertical FRL).}
\begin{definitionbox}{Vertical FRL}
A VFRL system is defined over a single underlying MDP
\(
\mathcal{M}
=
(\mathcal{S},\mathcal{A},P,r,\gamma)
\),
with the joint action space
\(
\mathcal{A}
=
\prod_{k=1}^{K}\mathcal{A}_k
\).
Each participant \(k\) has a local observation space \(\Omega_k\), an observation kernel
\(
\mathcal{O}_k:
\mathcal{S}
\rightarrow
\Delta(\Omega_k)
\),
and a local action space \(\mathcal{A}_k\). At time step \(t\), participant \(k\) receives a local observation
\(
o_{k,t}
\sim
\mathcal{O}_k(\cdot\mid s_t)
\)
and selects an action
\(
a_{k,t}
\sim
\pi_{\theta_k}(\cdot\mid o_{k,t})
\).
The joint action
\(
\mathbf{a}_t
=
(a_{1,t},\ldots,a_{K,t})
\)
then determines the state transition
\(
s_{t+1}
\sim
P(\cdot\mid s_t,\mathbf{a}_t)
\).
\end{definitionbox}

From the perspective of an individual participant, the decision process is partially observable and therefore corresponds to a POMDP as defined in Definition~2. At the system level, however, the global state evolves according to Markovian dynamics under the joint action \(\mathbf{a}_t\).

Reward access may also be asymmetric. For example, only a subset
\(
\mathcal{K}_r
\subseteq
\{1,\ldots,K\}
\)
may directly observe the global reward \(r(s_t,\mathbf{a}_t)\). Coordination is therefore performed by exchanging model parameters, intermediate representations, embeddings, or gradients of joint value estimates, typically without directly sharing raw local observations or trajectories.

VFRL fits settings where different stakeholders possess complementary information about the same network. For example, VFRL can combine RAN-side telemetry available at the Near-RT RIC with core-network or operations support system (OSS) measurements maintained by a separate administrative domain at the Service Management and Orchestration (SMO) layer~\cite{polese2023understanding}.

\textbf{FRL for O-RAN.}
The motivation for FRL in O-RAN is twofold. First, multi-vendor and multi-operator deployments introduce data-governance, confidentiality, and privacy requirements that may prevent a central learner from collecting raw telemetry and trajectories. Second, continuously transmitting high-volume telemetry from many RAN nodes to a central training facility can impose substantial communication and storage overhead.

FRL also faces several challenges. Statistical heterogeneity across participants, including differences in traffic patterns, channel conditions, hardware capabilities, transition dynamics, and reward distributions, can slow convergence and bias the aggregated model toward participants with larger datasets or dominant operating conditions~\cite{li2020federated}. Moreover, parameter averaging alone does not provide strong privacy guarantees and may still leak information about local training data, motivating its integration with mechanisms such as differential privacy and secure aggregation~\cite{geiping2020inverting}.

FRL inherits an additional difficulty from RL itself. Each participant's policy update may change the distribution of states, actions, and rewards that participants observe in subsequent rounds. The resulting policy-dependent distribution shift creates a non-stationary moving target that classical federated-learning analyses, which typically assume a fixed local data distribution, do not capture~\cite{khodadadian2022federated}.

\textbf{Distributed RL.}
Distributed RL more broadly refers to scaling RL training across multiple computing nodes, independently of privacy or data-locality requirements. Architectures such as Ape-X~\cite{DBLP:journals/corr/abs-1803-00933}, IMPALA~\cite{espeholt2018impala}, and SEED RL~\cite{espeholt2019seed} separate environment actors from one or more centralized learners and exploit parallelism to increase experience-generation throughput and reduce wall-clock training time. Unlike FRL, these architectures generally permit trajectories, gradients, or model parameters to be centrally collected and processed. Their primary objective is computational scalability rather than privacy preservation, data sovereignty, or cross-domain collaboration.

\textbf{Key Takeaways.}
FRL addresses privacy, governance, and data-locality constraints by training across distributed participants without centralizing raw trajectories. Distributed RL addresses a different bottleneck: it accelerates training by parallelizing experience generation and learning when central collection remains acceptable.

\subsection{Offline RL}
\label{Offline_RL}

Offline RL, also known as batch RL, considers the setting in which a policy is trained exclusively from a fixed dataset
\(
\mathcal{D}
=
\{(s_t,a_t,r_t,s_{t+1})\}
\),
collected by one or more behavior policies \(\pi_\beta\), without further interaction with the environment during training~\cite{levine2020offline}. Although conventional off-policy algorithms, such as DDPG, TD3, and SAC, also learn from previously collected transitions, they continually augment their replay buffers with new experience generated by the evolving behavior policy. Offline RL lacks this corrective data-collection loop.

\textbf{Distributional Shift.}
The distributional shift between the learned policy \(\pi_\theta\) and the behavior policy \(\pi_\beta\) creates the central challenge of offline RL. Bellman updates, such as the critic update in Eq.~(\ref{ddpg_critic}), evaluate bootstrap targets using actions that the learned policy selects.
When these actions have little or no representation in \(\mathcal{D}\), the learned critic must extrapolate beyond the support of the behavior-policy distribution. Neural function approximators may assign spuriously high values to such out-of-distribution actions, and repeated bootstrapping can amplify these errors, leading to severe overestimation and policy degradation~\cite{fujimoto2019off}.

\textbf{Representative Methods.}
Representative offline RL methods address this problem. Behavior-Constrained Q-Learning (BCQ) restricts policy improvement to actions that are likely under the behavior policy, thereby limiting extrapolation beyond the dataset support~\cite{fujimoto2019off}. Conservative Q-Learning (CQL) regularizes the critic by lowering the estimated values of actions that are not well supported by the offline dataset~\cite{kumar2020conservative}. Implicit Q-Learning (IQL) avoids explicitly maximizing the value function over potentially out-of-distribution actions and instead performs value learning and policy extraction primarily from actions observed in the dataset~\cite{kostrikov2021offline}.

\textbf{O-RAN Relevance.}
Offline RL is particularly well aligned with the operational constraints of O-RAN. In practice, RIC platforms continuously collect telemetry and control records through interfaces such as E2 and O1, creating large datasets of historical network operation.
In contrast, unconstrained online exploration in a production RAN can cause performance degradation, service disruption, or violations of service-level agreements (SLAs). Consequently, offline pretraining, optionally followed by safe, constrained, or carefully monitored online fine-tuning, provides a practical deployment strategy for DRL-based xApps and rApps.

\textbf{Key Takeaways.}
Offline RL turns historical RAN telemetry and control logs into a training resource while avoiding unsafe online exploration. Its main technical risk comes from distributional shift, so practical offline DRL for O-RAN must constrain or regularize actions that the logged dataset does not support.

\subsection{Safe and Constrained RL}
\label{Safe_Constrained_RL}
The standard MDP objective in Eq.~(\ref{mdp_reward}) maximizes expected return but does not inherently incorporate operational constraints during training and deployment. Such constraints typically include SLA thresholds, hardware safety limits, or fairness criteria. Safe RL optimizes policies subject to such constraints~\cite{garcia2015comprehensive}.
Representative formulations include risk-sensitive objectives that penalize the variance \cite{tamar2015optimizing} or conditional value-at-risk (CVaR) of returns \cite{chow2018risk}, robust optimization frameworks that protect against worst-case environmental dynamics \cite{iyengar2005robust, pinto2017robust}, and constrained formulations that explicitly enforce predefined budget thresholds \cite{altman2021constrained}.
Among these, constrained RL provides a principled theoretical foundation and has emerged as the predominant paradigm.

% \noindent\textbf{Definition 7 (Constrained MDP).} 
\begin{definitionbox}{Constrained MDP}
The standard mathematical framework for Constrained RL is the Constrained MDP (CMDP). A CMDP extends the standard MDP formulation, defined by the tuple $(\mathcal{S}, \mathcal{A}, P, r, \mathcal{C}, \mathbf{d})$. Here, $\mathcal{C} = \{c_1, \dots, c_I\}$ represents a set of cost functions $c_i: \mathcal{S} \times \mathcal{A} \to \mathbb{R}_{\ge 0}$, and $\mathbf{d} = \{d_1, \dots, d_I\}$ denotes the corresponding constraint thresholds \cite{altman2021constrained}. 
The expected discounted cost-return of a policy $\pi$ for the $i$-th constraint is given by
\begin{equation}
    J_{C_i}(\pi) = \mathbb{E}_{\tau \sim \pi} \left[ \sum_{t=0}^\infty \gamma^t c_i(s_t, a_t) \right],
\label{cost_return}
\end{equation}
and the constrained optimization problem is
\begin{equation}
\label{cmdp_objective}
\begin{aligned}
    \max_{\pi}& && J(\pi) \\
    \text{s.t.}& && J_{C_i}(\pi) \le d_i, \quad \forall i \in \{1, \dots, I\}.
\end{aligned}
\end{equation}
\end{definitionbox}
Formulating safety as an explicit constraint in Eq.~(\ref{cmdp_objective}) avoids the sensitivity of reward-shaping approaches, which optimize a scalar surrogate $\tilde{r}(s,a) = r(s,a) - \sum_{i=1}^{I} \nu_i\, c_i(s,a)$ using manually tuned penalty weights $\nu_i \ge 0$.

\textbf{Lagrangian Methods.} 
A prominent class of algorithms reformulates the constrained objective in Eq.~(\ref{cmdp_objective}) into an unconstrained saddle-point problem. The Lagrangian takes the form $\mathcal{L}(\pi_\theta, \boldsymbol{\nu}) = J(\pi_\theta) - \sum_{i=1}^{I} \nu_i \big(J_{C_i}(\pi_\theta) - d_i\big)$, where $\boldsymbol{\nu} \in \mathbb{R}_{\ge 0}^{I}$ denotes the vector of Lagrange multipliers.

Primal-dual optimization methods update these parameters by alternating gradient ascent on the policy parameters $\theta$ and gradient descent on $\boldsymbol{\nu}$:
\begin{equation}
\begin{aligned}
    \theta &\gets \theta + \alpha_\theta \nabla_\theta \mathcal{L}(\pi_\theta, \boldsymbol{\nu}), \\
    \nu_i &\gets \big[ \nu_i + \alpha_\nu (J_{C_i}(\pi_\theta) - d_i) \big]_+, 
\end{aligned}
\label{lagrangian_update}
\end{equation}
where $[\cdot]_+ = \max(\cdot, 0)$. Algorithms such as PPO-Lagrangian and TRPO-Lagrangian~\cite{ray2019benchmarking} integrate this primal-dual framework with the on-policy methods discussed in Section~\ref{Policy_Gradient_Methods} by training a dedicated cost critic $V^{c_i}_{\phi_i}$ concurrently with the standard reward critic $V_\phi$.

Lagrangian methods integrate easily with established policy-gradient architectures, but they guarantee constraint satisfaction, $J_{C_i}(\pi) \le d_i$, only in expectation and strictly upon convergence. Consequently, intermediate policies can violate constraints during training.

\textbf{Key Takeaways.}
Safe and constrained RL extends the MDP objective with operational constraints that matter directly to O-RAN, including SLA, safety, and fairness requirements. CMDPs and Lagrangian methods provide a principled starting point, but deployment still requires care because training-time policies may violate constraints before convergence.

\subsection{Meta-RL and Transfer Learning}
\label{Meta_RL_Transfer}

The methods reviewed so far assume that the task is fixed at training time: the agent learns on the same MDP it will be deployed in. In O-RAN, however, the same xApp or rApp is expected to operate across a class of related but non-identical tasks, such as different cells, slices, and traffic paradigms that share structure but differ in dynamics. Meta-RL and transfer learning address this issue by leveraging shared characteristics across a class of tasks, accelerating adaptation to novel tasks and thereby avoiding the sample complexity of learning from scratch.

% \noindent\textbf{Definition 8 (Meta-RL).}
\begin{definitionbox}{Meta-RL}
Let $\mathcal{P}$ denote a distribution over a class of MDPs that share the same state and action spaces $(\mathcal{S}, \mathcal{A})$, but have distinct transition and reward functions. Meta-RL aims to optimize meta-parameters $\theta$ such that, for any unseen MDP $\mathcal{M}' \sim \mathcal{P}$, a rapid adaptation procedure initialized with $\theta$ yields task-specific parameters $\theta'$. The resulting policy $\pi_{\theta'}$ achieves a high expected return on $\mathcal{M}'$. Formally, this objective is defined as:
\begin{equation}
    \theta^{*} = \arg\max_{\theta}\; \mathbb{E}_{\mathcal{M}' \sim \mathcal{P}}\!\left[\, J_{\mathcal{M}'}\!\left(\pi_{\theta'}\right) \right],
\label{meta_rl_objective}
\end{equation}
where $J_{\mathcal{M}'}$ is the expected discounted return of Eq.~(\ref{mdp_reward}) evaluated on $\mathcal{M}'$.
\end{definitionbox}

Existing approaches mainly differ in their implementation of this adaptation procedure. MAML~\cite{finn2017model} utilizes a few steps of policy-gradient ascent and meta-training $\theta$ to ensure this inner-loop optimization maximizes expected return. RL$^2$~\cite{duan2016rl} encodes the adaptation process within the hidden state of a RNN policy conditioned on the action-observation history, formulating rapid adaptation as in-context learning. PEARL~\cite{rakelly2019efficient} infers a latent task variable $z$ from a small set of trajectories via a learned encoder and conditions the policy as $\pi_\theta(a \mid s, z)$, thereby facilitating sample-efficient, off-policy meta-training.

\textbf{Transfer Learning.}
Transfer learning is a broader paradigm that reuses knowledge acquired from one or more source tasks to improve learning or decision-making on a target task, typically without requiring an explicitly defined distribution over tasks during training~\cite{zhu2023transfer}. Representative approaches include policy distillation, which transfers knowledge from one or more teacher policies to a target policy~\cite{rusu2015policy}, and successor features, which enable the reuse of learned representations and value-function components across tasks with related dynamics or reward structures~\cite{barreto2017successor}.

\textbf{O-RAN Relevance.}
In O-RAN deployments, the primary operational benefit of Meta-RL and transfer learning is their ability to reduce the number of environment interactions required to adapt a policy to a new operating condition. Because the SLA and safety requirements discussed in Section~\ref{Safe_Constrained_RL} tightly constrain online exploration in a production RAN, methods that reduce the data and interaction costs of adapting a deployed policy to a new cell, vendor implementation, network configuration, or traffic pattern are particularly valuable.

\textbf{Key Takeaways.}
Meta-RL and transfer learning address the fact that O-RAN control policies rarely remain confined to one fixed task. They help deployed xApps and rApps adapt across cells, vendors, slices, and traffic regimes with fewer new interactions.

\subsection{Sim-to-Real Gap of DRL}
\label{Sim_to_Real}

Recent studies have demonstrated direct training on small-scale physical testbeds~\cite{yan2025near,lu2026eexapp}, but safety and SLA requirements tightly restrict unconstrained online exploration in a production-grade RAN. Consequently, researchers typically train DRL agents for real-world deployment in simulation, emulation, or digital-twin environments.
However, mismatches in channel dynamics, traffic distributions, user mobility, protocol implementations, hardware characteristics, and communication or computation delays can substantially degrade a policy that learns in a simulated or emulated environment \(\hat{\mathcal{M}}\) and then operates in the corresponding physical environment \(\mathcal{M}\).

% \noindent\textbf{Definition 9 (Sim-to-Real Gap).}

% \begin{definitionbox}{Sim-to-Real Gap}
% Let
% \(
% \pi_{\hat{\mathcal{M}}}^{*}
% =
% \arg\max_{\pi}
% J_{\hat{\mathcal{M}}}(\pi)
% \)
% and
% \(
% \pi_{\mathcal{M}}^{*}
% =
% \arg\max_{\pi}
% J_{\mathcal{M}}(\pi)
% \)
% denote optimal policies for the simulated environment and the physical environment, respectively. The sim-to-real gap is defined as

\begin{definitionbox}{Sim-to-Real Gap}
Let
\[
\begin{aligned}
\pi_{\hat{\mathcal{M}}}^{*}
&= \operatorname*{arg\,max}_{\pi}
J_{\hat{\mathcal{M}}}(\pi), \\
\pi_{\mathcal{M}}^{*}
&= \operatorname*{arg\,max}_{\pi}
J_{\mathcal{M}}(\pi),
\end{aligned}
\]
denote the optimal policies for the simulated and physical
environments, respectively. The sim-to-real gap is defined as
\begin{equation}
    \Delta_{\mathrm{S2R}}
    =
    J_{\mathcal{M}}\!\left(\pi_{\mathcal{M}}^{*}\right)
    -
    J_{\mathcal{M}}\!\left(\pi_{\hat{\mathcal{M}}}^{*}\right),
\label{sim_to_real_gap}
\end{equation}
where \(J_{\mathcal{M}}(\pi)\) is the expected discounted return in Eq.~(\ref{mdp_reward}), evaluated in the physical environment \(\mathcal{M}\). Because \(\pi_{\mathcal{M}}^{*}\) is optimal for \(\mathcal{M}\), \(\Delta_{\mathrm{S2R}}\geq 0\). In practice, \(\pi_{\mathcal{M}}^{*}\) is generally unavailable, and the gap is therefore estimated relative to the strongest available policy or control baseline evaluated on the physical system.
\end{definitionbox}

Three prominent and complementary paradigms have been developed to mitigate the sim-to-real gap.

\begin{itemize}[leftmargin=0.15in]
\item 
\noindent\emph{Domain randomization.}
Domain randomization trains a policy across a distribution of simulated environments whose physical or operational parameters vary systematically~\cite{tobin2017domain,sadeghi2016cad2rl}. Specifically, it replaces a single simulator \(\hat{\mathcal{M}}\) with a parameterized family
\(
\big\{
\hat{\mathcal{M}}_{\xi}
\big\}_{\xi\sim p(\xi)}
\),
where \(\xi\) may represent channel conditions, traffic loads, mobility patterns, processing delays, hardware characteristics, or other uncertain system parameters. The policy is then optimized according to
$
    \max_{\pi}
    \mathbb{E}_{\xi\sim p(\xi)}
    \big[
        J_{\hat{\mathcal{M}}_{\xi}}(\pi)
    \big]
$.
The underlying rationale is that exposure to sufficiently diverse simulated conditions can prevent the policy from overfitting to a single simulator configuration and improve its robustness to previously unseen real-world conditions.

\item 
\noindent\emph{System identification.}
System identification reduces the discrepancy between simulation and reality by calibrating the simulator with measurements from the target physical system~\cite{chebotar2019closing,tan2018sim,zhu2018reinforcement}. Given a parameterized simulator \(\hat{\mathcal{M}}_{\xi}\), the system-identification procedure estimates parameters \(\xi\) that reproduce observed real-world trajectories, transition statistics, or performance metrics as closely as possible.
The calibrated simulator then supports policy retraining or fine-tuning. Because physical environments may evolve over time, system identification may need to run periodically or jointly with policy adaptation.

\item 
\noindent\emph{Domain adaptation.}
Domain adaptation explicitly reduces discrepancies between simulated and real data distributions. A common strategy learns domain-invariant state or observation representations that map simulated and physical measurements to a shared latent space~\cite{ganin2015unsupervised,ganin2016domain,bousmalis2017unsupervised}. Many methods use adversarial training to prevent a domain discriminator from distinguishing features extracted from simulated and real observations.
The resulting representation can allow a policy trained predominantly on simulated data to operate more effectively on physical-system observations. However, aligning observation distributions alone may not eliminate mismatches in transition dynamics or reward functions, so domain adaptation often works best when combined with system identification or limited real-world fine-tuning.
\end{itemize}

\textbf{O-RAN Relevance.}
Sim-to-real transfer is particularly important, yet challenging, for O-RAN. High-fidelity RAN simulators and digital twins are computationally expensive and may not fully capture the stochasticity, partial observability, non-stationarity, implementation-specific behavior, and coupled control loops of a live network.
Effective transfer therefore commonly requires a combination of domain randomization, simulator calibration, representation adaptation, uncertainty estimation, and carefully constrained real-world fine-tuning. Section~\ref{10_Platforms} provides a detailed discussion of available simulation, emulation, and physical evaluation platforms.

\textbf{Key Takeaways.}
The sim-to-real gap measures how much a policy loses when it moves from a training simulator to the physical RAN. Domain randomization, system identification, and domain adaptation reduce this gap from complementary directions, but O-RAN deployments still require cautious validation because live networks combine stochastic dynamics, partial observability, and strict operational constraints.

\section{The O-RAN Environment}

\label{3_ORAN_Architecture_DRL}

\begin{figure}
    \centering
    \includegraphics[width=0.95\linewidth]{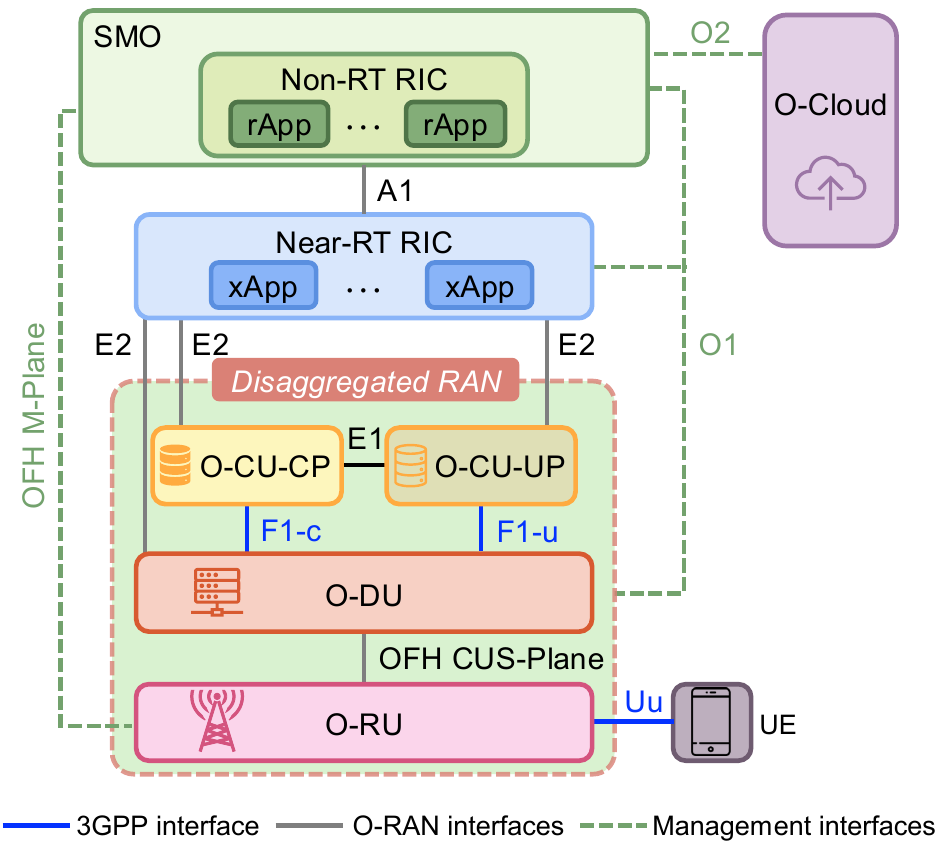}
    \caption{O-RAN architecture for DRL-oriented control. The disaggregated RAN contains the O-CU-CP, O-CU-UP, O-DU, and O-RU; the Non-RT RIC in the SMO hosts rApps, while the Near-RT RIC on the O-Cloud hosts xApps. Interfaces fall into 3GPP-inherited, O-RAN functional, and O-RAN management groups.}
    \label{fig:oran_arc}
\end{figure}

% \subsection{Overview}
% \label{oran_overview}

% A DRL agent does not operate in isolation. Its observation space, action space, decision timing, and compute budget depend on its network location and accessible interfaces. In this sense, the O-RAN architecture is more than just a reference design for a multi-vendor RAN. It also acts as the foundation that shapes the design of every DRL-based controller discussed in this survey. This section reviews the components of this foundation and then maps DRL agents onto it. We begin with the disaggregated data plane (Section~\ref{subsec:disaggregation}) and the RAN Intelligent Controller (RIC) hierarchy (Section~\ref{subsec:ric}). We then illustrate the open interfaces that connect them (Section~\ref{subsec:interfaces}), along with the Service Management and Orchestration (SMO), and the O-Cloud management plane (Section~\ref{subsec:smo-ocloud}). Finally, we map DRL agents onto this foundation (Section~\ref{subsec:drl-mapping}) and give a brief outlook on how 6G will reshape it (Section~\ref{subsec:6g-substrate}).

% % ---------------------------------------------------------------------
% \subsection{RAN Disaggregation}
% \label{subsec:disaggregation}

\subsection{O-RAN Overview}
O-RAN provides the architectural substrate on which DRL-based RAN control operates. It disaggregates traditionally integrated RAN functions, connects them through standardized open interfaces, and introduces programmable RAN Intelligent Controllers (RICs) that host control applications at defined timescales. Fig.~\ref{fig:oran_arc} summarizes these components and their interface groups.

\textbf{Disaggregated Data Plane.}
O-RAN replaces the monolithic base-station stack with a disaggregated framework organized around two functional splits. The \emph{higher-layer split} adopts the 3GPP Option~2 split specified in 3GPP TR~38.801~\cite{3gpp_38_801}, which separates the protocol stack between the Packet Data Convergence Protocol (PDCP) and Radio Link Control (RLC) layers.
Under the higher-layer split, the O-RAN Central Unit (O-CU) hosts Radio Resource Control (RRC), Service Data Adaptation Protocol (SDAP), and PDCP functions, whereas the O-RAN Distributed Unit (O-DU) hosts RLC, Medium Access Control (MAC), and High-PHY functions. The O-CU and O-DU communicate through the F1 interface, which includes F1-C for control-plane signaling and F1-U for user-plane traffic.
The O-CU can be further separated into an O-CU Control Plane (O-CU-CP) and an O-CU User Plane (O-CU-UP). The E1 interface standardized in 3GPP TS~38.463 connects these two entities and supports independent scaling, deployment, and lifecycle management of signaling and user-plane functions~\cite{3gpp_38_463}.

\textbf{Open Fronthaul Split.}
The \emph{lower-layer split} follows the 7.2x functional split defined by O-RAN Alliance Work Group~4~\cite{oran_wg4_cus}. It separates High-PHY functions at the O-DU from Low-PHY and radio-frequency (RF) functions at the O-RAN Radio Unit (O-RU). The O-DU and O-RU exchange control, user-plane, synchronization, and management information through the Open Fronthaul (OFH) interface.

\textbf{DRL Relevance.}
RAN disaggregation makes agent placement part of the DRL problem formulation. An agent associated with the O-CU observes higher-layer signaling and user-plane context; an agent associated with the O-DU observes MAC and High-PHY dynamics; and an in-node or radio-side controller can access lower-latency measurements closer to the scheduler or RF chain. These locations shape the observation space, action space, and inference budget available to a DRL controller.

% \textbf{Key Takeaways.}
O-RAN decomposes the RAN into functional units whose interfaces expose different control and observation opportunities. This decomposition gives DRL studies a concrete architectural basis for defining where an agent runs, what it observes, and which actions it can execute.

% ---------------------------------------------------------------------
\subsection{The RIC Hierarchy}
\label{subsec:ric}

The RIC hierarchy organizes O-RAN programmability by control timescale. O-RAN introduces two RICs~\cite{polese2023understanding}: the Non-Real-Time RIC (Non-RT RIC) for slower policy and model-management loops, and the Near-Real-Time RIC (Near-RT RIC) for edge control loops.

\begin{itemize}[leftmargin=0.15in]
\item 
\emph{Non-RT RIC.}
The Non-RT RIC runs within the Service Management and Orchestration (SMO) framework and supports control loops longer than one second. It hosts \emph{rApps}, which process aggregated telemetry, generate policy decisions, provide enrichment information, and manage AI and ML model lifecycles~\cite{oran_wg2_nonrtric_arch}. 

\item 
\emph{Near-RT RIC.}
The Near-RT RIC runs at the network edge or regional cloud and supports control loops from 10\,ms to 1\,s. It hosts \emph{xApps}, which subscribe to telemetry from E2 nodes, such as O-CU and O-DU, and send control actions through the E2 interface~\cite{oran_wg3_ricarch}. Both rApps and xApps follow plug-in software models with standardized deployment methods, versioned container images, and managed lifecycles.
\end{itemize}

\textbf{In-node Control.}
Existing dApp work addresses control loops below the Near-RT RIC timescale. The community has proposed \emph{dApps} as lightweight in-node control applications co-located with the O-CU or O-DU~\cite{d2022dapps}. These applications can act on real-time Key Performance Measurements (KPMs)~\cite{oran_wg3_e2sm_kpm,3gpp_28_552}, such as physical resource block (PRB) usage counters, transmitted data volumes, or Channel Quality Indicator (CQI) samples, as well as proprietary user-plane data whose latency or volume does not fit the E2 interface budget.

Current O-RAN specifications do not yet treat dApps as a normative control tier, so the literature generally presents them as a research-stage extension. Nevertheless, 6G-oriented O-RAN studies discuss dApps as a possible third tier for link adaptation, MAC scheduling, beam management, and other fast radio-control functions~\cite{d2022dapps,oran_ngrg}.

\textbf{DRL Relevance.}
The RIC host defines the boundary of a DRL agent. An rApp agent typically works with aggregated, slower-timescale telemetry; an xApp agent works with E2-based subscriptions and near-real-time control actions; and a dApp agent can target sub-10\,ms decisions close to the scheduler. This host choice determines the telemetry streams that form the observation space, the interfaces that execute actions, and the compute budget that constrains inference time.

\textbf{Key Takeaways.}
The RIC hierarchy maps O-RAN control applications to distinct timescales. This hierarchy offers a structured way to place DRL within rApp, xApp, and emerging dApp agents according to latency, observability, and control authority.

% ---------------------------------------------------------------------
\subsection{SMO, O-Cloud, and the AI/ML Lifecycle}
\label{subsec:smo-ocloud}

\textbf{SMO Services.}
The SMO provides the operator-centric management and orchestration plane for O-RAN. As shown in Fig.~\ref{fig:oran_arc}, it hosts the Non-RT RIC, Data Management and Exposure services, and AI/ML workflow functions~\cite{polese2023understanding}. These services connect management-plane telemetry, analytics, policy generation, and model lifecycle operations.

The Data Management and Exposure services follow a producer-consumer model. Producers within the SMO and Non-RT RIC publish performance reports, Key Performance Indicator (KPI) predictions, and analytical outputs, while consumers, mainly rApps, discover and subscribe to the required data types. A KPI~\cite{3gpp_28_554} is a higher-level metric derived from one or more KPMs and summarizes service quality or operator objectives, such as handover success rate, per-slice throughput percentile, or latency SLA margin.

\textbf{AI/ML Lifecycle.}
The AI/ML workflow provides the management structure for model training and deployment. It covers data collection, training, validation, deployment, execution, and lifecycle management. It also gives operators a standardized mechanism for integrating custom training pipelines, including offline pre-training pipelines, into the SMO.

\textbf{O-Cloud Execution Substrate.}
O-Cloud provides the cloud-native execution platform for virtualized RAN functions and the Near-RT RIC. The SMO manages the logically separated O-Cloud through the O2 interface~\cite{oran_wg6_o2}. Through O2, O-Cloud allocates heterogeneous hardware resources, including CPUs, GPUs, FPGAs, and accelerator Network Interface Cards (NICs), to support containerized workloads.

O-Cloud placement affects the latency and compute budget of intelligent control applications. Central sites provide higher computational capacity and therefore fit offline training or complex rApp policies. Regional and edge tiers place the Near-RT RIC and xApps closer to O-CU and O-DU nodes, reducing E2 latency but imposing tighter inference and resource constraints.

\textbf{DRL Relevance.}
SMO and O-Cloud services connect DRL training, model versioning, and policy execution. The SMO can aggregate O1 telemetry and relayed E2 KPM reports for offline DRL pre-training, while O-Cloud hardware heterogeneity influences whether a learned policy runs as a central rApp, an edge xApp, or a lightweight dApp. As a result, deployment studies often combine model placement with compression, distillation, or staged policy updates.

\textbf{Key Takeaways.}
The SMO supplies management-plane data, AI/ML lifecycle functions, and Non-RT RIC services, while the O-Cloud supplies the execution substrate. Together they determine how DRL models move from historical data and training pipelines to runtime control applications.

% \begin{table*}[t]
% \centering
% \scriptsize
% \caption{Mapping of DRL agents onto the O-RAN substrate.}
% \label{tab:drl-mapping}
% \renewcommand{\arraystretch}{1.0}
% \setlength{\tabcolsep}{5pt}
% \begin{tabularx}{\textwidth}{@{}
%     >{\raggedright\arraybackslash}p{1.5cm}
%     >{\raggedright\arraybackslash}p{1.5cm}
%     >{\raggedright\arraybackslash}p{1.6cm}
%     >{\raggedright\arraybackslash}X
%     >{\raggedright\arraybackslash}X
%     >{\raggedright\arraybackslash}X
%     @{}}
% \toprule
% \textbf{Agent host} & \textbf{Controller} & \textbf{Timescale} &
% \textbf{Observation path} & \textbf{Control path} &
% \textbf{Representative DRL Problems} \\
% \midrule
% rApp & Non-RT RIC & $>1$\,s &
% R1 (SMO data services), aggregated O1 telemetry &
% A1-P policies, A1-EI, A1-ML model updates, O1 configurations &
% Long-term resource planning, slice SLA management, policy guidance \\
% \addlinespace
% xApp & Near-RT RIC & $10$\,ms--$1$\,s &
% E2 (KPM service model), A1-EI &
% E2 (RC service model) &
% Mobility management, resource management, traffic steering \\
% \addlinespace
% dApp & O-CU/O-DU & $<10$\,ms &
% Local node state (MAC/PHY counters, HARQ buffers) &
% Local control functions (scheduler, link adaptation) &
% Scheduling, beam management, link adaptation \\
% \bottomrule
% \end{tabularx}
% \end{table*}

\begin{table*}[!t]
\centering
\caption{Mapping of DRL agents onto the O-RAN substrate.}
\label{tab:drl-mapping}
\scriptsize
\setlength{\tabcolsep}{1.2pt}
\renewcommand{\arraystretch}{1.18}
\begin{tabular}{|>{\centering\arraybackslash}m{0.08\textwidth}|
                  >{\centering\arraybackslash}m{0.10\textwidth}|
                  >{\centering\arraybackslash}m{0.09\textwidth}|
                  >{\raggedright\arraybackslash}m{0.21\textwidth}|
                  >{\raggedright\arraybackslash}m{0.22\textwidth}|
                  >{\raggedright\arraybackslash}m{0.25\textwidth}|}
\hline
\rowcolor[gray]{0.9}
\multicolumn{1}{|c|}{\parbox[c][6ex][c]{0.08\textwidth}{\centering\textbf{Agent Host}}} &
\multicolumn{1}{c|}{\parbox[c][6ex][c]{0.10\textwidth}{\centering\textbf{Controller}}} &
\multicolumn{1}{c|}{\parbox[c][6ex][c]{0.09\textwidth}{\centering\textbf{Timescale}}} &
\multicolumn{1}{c|}{\parbox[c][6ex][c]{0.21\textwidth}{\centering\textbf{Observation Path}}} &
\multicolumn{1}{c|}{\parbox[c][6ex][c]{0.22\textwidth}{\centering\textbf{Control Path}}} &
\multicolumn{1}{c|}{\parbox[c][6ex][c]{0.25\textwidth}{\centering\textbf{Representative DRL Problems}}} \\
\hline

\textbf{rApp}
& Non-RT RIC
& $>1$\,s
& R1 (SMO data services), aggregated O1 telemetry.
& A1-P policies, A1-EI, A1-ML model updates, and O1 configurations.
& Long-term resource planning, slice SLA management, and policy guidance. \\
\hline

\textbf{xApp}
& Near-RT RIC
& $10$\,ms--$1$\,s
& E2 (KPM service model) and A1-EI.
& E2 (RC service model).
& Mobility management, resource management, and traffic steering. \\
\hline

\textbf{dApp}
& O-CU/O-DU
& $<10$\,ms
& Local node state, including MAC/PHY counters and HARQ buffers.
& Local control functions, including scheduling and link adaptation.
& Scheduling, beam management, and link adaptation. \\
\hline

\end{tabular}
\end{table*}

% ---------------------------------------------------------------------
\subsection{Open Interfaces: Observation and Control Paths}
\label{subsec:interfaces}

Open interfaces define the observation and control paths available to O-RAN agents. Fig.~\ref{fig:oran_arc} groups these interfaces into three categories: 3GPP-inherited interfaces, O-RAN functional interfaces, and O-RAN management interfaces. For DRL, these interfaces specify which telemetry can become state information and which protocol mechanisms can carry actions.

\textbf{3GPP-Inherited Interfaces.}
3GPP-inherited interfaces carry user- and control-plane traffic along the RAN data path. The Uu interface connects the User Equipment (UE) to the Next Generation NodeB (gNB)~\cite{3gpp_38_300,3gpp_38_331}. In O-RAN, the physical layer terminates at the O-RU, while higher-layer signaling extends through the O-DU and O-CU. The F1 interface connects the O-DU and O-CU, with F1-C carrying control signaling and F1-U carrying user-plane traffic~\cite{3gpp_38_473}. The E1 interface separates the O-CU-CP and O-CU-UP, enabling independent control- and user-plane functions~\cite{3gpp_38_463}.

\textbf{O-RAN Functional Interfaces.}
O-RAN functional interfaces provide runtime programmability for disaggregated RAN components. The OFH interface connects the O-DU and O-RU through Control (C), User (U), Synchronization (S), and Management (M) planes. The C, U, and S planes, collectively called the CUS-plane, carry real-time scheduling commands, IQ samples, and timing information~\cite{oran_wg4_cus}. The M-plane carries non-real-time O-RU management traffic and therefore aligns with the management-interface group below.

The E2 interface creates the main Near-RT RIC observation and control loop. It connects the Near-RT RIC to E2 nodes, including O-CU-CP, O-CU-UP, and O-DU, and uses service models such as KPM and RAN Control (RC) to support subscription-based telemetry and control~\cite{oran_wg3_ricarch}. In many DRL formulations, KPM service models define observable performance features, while RC service models define feasible control operations.

The A1 and R1 interfaces connect slower-timescale intelligence with near-real-time control. A1 connects the Non-RT RIC inside the SMO to the Near-RT RIC and carries policy guidance (A1-P), enrichment information (A1-EI), and machine learning models (A1-ML)~\cite{oran_wg2_a1}. R1 gives rApps access to SMO and Non-RT RIC framework services, including data access, ML services, configuration, and A1 policy creation~\cite{oran_wg2_r1}. The Y1 interface exposes RAN analytics from the Near-RT RIC to authorized internal and external users~\cite{oran_wg3_y1}.

\textbf{O-RAN Management Interfaces.}
O-RAN management interfaces handle lifecycle and infrastructure orchestration at the SMO. The O1 interface provides Fault, Configuration, Accounting, Performance, and Security (FCAPS) management and carries high-volume telemetry from the O-CU, O-DU, and Near-RT RIC to the SMO~\cite{oran_wg1_o1}. This telemetry often supplies aggregated metrics for offline DRL training.
The OFH M-plane complements O1 by managing the O-RU, while the O2 interface connects the O-Cloud to the SMO for infrastructure and workload management~\cite{oran_wg6_o2}. These interfaces support model deployment and resource allocation decisions that affect where DRL training and inference workloads can run.

\textbf{Key Takeaways.}
Open interfaces translate the O-RAN architecture into concrete \emph{observation} and \emph{control} paths, making O-RAN well suited for DRL-based operation. At the same time, E2 service models, A1 policy mechanisms, O1 telemetry, and O2 workload management impose practical timing and operational constraints on DRL, as they determine the state information, action space, training data, and deployment options available to DRL agents.

% ---------------------------------------------------------------------
\subsection{DRL in O-RAN: Agents, Loops, and Timescales}
\label{subsec:drl-mapping}

Existing DRL formulations for O-RAN usually map agents to control loops by host, interface, and timescale. The disaggregated data plane, RIC hierarchy, SMO/O-Cloud management plane, and open interfaces together separate radio control across loops ranging from scheduler-level decisions to multi-second policy updates. Table~\ref{tab:drl-mapping} summarizes how common DRL agent hosts align with this substrate.

\begin{itemize}[leftmargin=0.15in]
\item 
\emph{Deployment locations structure the learning problem.}
An O-RAN DRL agent typically runs as an rApp at the Non-RT RIC, an xApp at the Near-RT RIC, or a dApp inside an E2 node. Each host brings a different control period, observation path, action path, and compute budget.

\item 
\emph{Interfaces define observations and actions.}
O-RAN exposes state and control through standardized service models rather than through vendor-specific internal variables. The KPM service model defines observable performance metrics, the RC service model defines supported control operations, and the A1 interface defines how higher-level agents shape lower-level behavior through policy. This interface discipline gives DRL studies a shared vocabulary for state and action design.

\item
\emph{The hierarchy creates multi-agent and federated settings.}
The same E2 node may receive near-real-time xApp control and slower rApp policy guidance. Multiple xApps may also share an E2 node and interact through coupled radio resources. In geographically or administratively separated deployments, RICs can train local agents while exchanging model updates rather than raw telemetry, matching the federated RL paradigm~\cite{mcmahan2017communication,qi2021federated}.
\end{itemize}

\textbf{Key Takeaways.}
DRL studies can map O-RAN agents by answering three architectural questions: where the agent runs, which interface supplies its observations, and which interface carries its actions. This mapping helps keep later MDP, multi-agent, and deployment formulations tied to concrete O-RAN control loops.

\begin{table*}[!t]
\centering
\caption{Taxonomy of representative DRL-for-O-RAN studies by MDP formulation dimensions. Lightly shaded cells mark dimensions emphasized by the corresponding research problem.}
\label{tab:oran_drl_formulation_taxonomy}
\scriptsize
\setlength{\tabcolsep}{1.2pt}
\renewcommand{\arraystretch}{1.18}
\begin{tabular}{|>{\centering\arraybackslash}m{0.12\textwidth}|>{\raggedright\arraybackslash}m{0.15\textwidth}|>{\raggedright\arraybackslash}m{0.16\textwidth}|>{\raggedright\arraybackslash}m{0.13\textwidth}|>{\raggedright\arraybackslash}m{0.16\textwidth}|>{\raggedright\arraybackslash}m{0.15\textwidth}|>{\centering\arraybackslash}m{0.07\textwidth}|}
\hline
\rowcolor[gray]{0.9}
\multicolumn{1}{|c|}{\parbox[c][6ex][c]{0.12\textwidth}{\centering\textbf{Research Problem}}} &
\multicolumn{1}{c|}{\parbox[c][6ex][c]{0.13\textwidth}{\centering\textbf{Decision Model Selection}}} &
\multicolumn{1}{c|}{\parbox[c][6ex][c]{0.16\textwidth}{\centering\textbf{Observation/State Design}}} &
\multicolumn{1}{c|}{\parbox[c][6ex][c]{0.13\textwidth}{\centering\textbf{Action-Space Design}}} &
\multicolumn{1}{c|}{\parbox[c][6ex][c]{0.16\textwidth}{\centering\textbf{Reward Design}}} &
\multicolumn{1}{c|}{\parbox[c][6ex][c]{0.15\textwidth}{\centering\textbf{Temporal Structure \& Evaluation Scope}}} &
\multicolumn{1}{c|}{\parbox[c][6ex][c]{0.07\textwidth}{\centering\textbf{Refs.}}} \\
\hline

\textbf{Radio resource management}
& Single-agent MDP or POMDP; some studies use multi-timescale MDPs for slice-aware resource control.
& \cellcolor{highcol!35}CQI, SINR, queue length, traffic demand, slice KPI, and QoS/SLA indicators.
& \cellcolor{highcol!35}Scheduler choice, resource share, rate target, or radio-resource quota.
& Weighted network utility over throughput, latency, fairness, energy, and QoS satisfaction.
& Near-RT control for scheduling and allocation; non-RT policy guidance for coarse slice-resource management.
& \cite{mhatre2024aiaas1,tan2025deep,villegas2025drl2,sohaib2024drl} \\
\hline

\textbf{Power, spectrum, and interference control}
& Continuous-control MDP / POMDP; multi-cell studies introduce coupled or coordinated decision processes.
& Channel quality, interference measurements, spectrum occupancy, traffic load, and multi-cell context.
& \cellcolor{highcol!35}Transmit-power level, spectrum-sharing decision, coordination parameter, or interference-mitigation action.
& \cellcolor{highcol!35}Spectral efficiency, coexistence, energy saving, interference reduction, and robustness terms.
& Mostly near-RT evaluation; multi-cell studies evaluate generalization under dynamic interference and traffic.
& \cite{eskandari2025network,gopal2025adapshare,xdiff,reinders2026aiim,ergu2025radar} \\
\hline

\textbf{Mobility and traffic steering}
& POMDP or hierarchical MDP; multi-RAT steering studies decompose long-term access selection and short-term association.
& \cellcolor{highcol!35}RSRP/RSRQ, serving and neighboring-cell load, UE distribution, mobility state, service type, and handover history.
& Target cell, target RAT, handover trigger, steering bias, or routing/offloading decision.
& Load balance, URLLC/QoE satisfaction, handover success, ping-pong reduction, and latency.
& \cellcolor{highcol!35}Near-RT steering over seconds to minutes; hierarchical studies separate slow RAT selection from fast association.
& \cite{tamim2023intelligent,habib2023hierarchical,lacava2023programmable,kavehmadavani2024empowering,dai2024intelligent} \\
\hline

\textbf{Network slicing and SLA control}
& MDP, CMDP, constrained MARL, or federated DRL depending on whether studies model SLA limits, agents, and data locality.
& Slice requests, utilization, per-slice KPIs, SLA margin, queue state, and violation history.
& Admission, slice quota, scaling, reservation, priority, or A1-style policy guidance.
& \cellcolor{highcol!35}SLA compliance, revenue, isolation, utilization, latency cost, and explicit constraint violations.
& \cellcolor{highcol!35}Multi-timescale control across rApps and xApps; studies evaluate time-varying demand and SLA violations.
& \cite{abedin2022elastic,ghafouri2024multi,zangooei2023flexible,nagib2025safeslice,hazarika2024enhancing} \\
\hline

\textbf{Multi-agent O-RAN control}
& \cellcolor{highcol!35}Stochastic game or Dec-POMDP; many works adopt centralized training with decentralized execution.
& Local cell, slice, UAV, or access-point observations; some studies add neighbor information or exchanged messages.
& Agent-specific resource allocation, task scheduling, precoding, xApp coordination, or conflict-mitigation action.
& Shared network utility, local reward with cooperation terms, constrained team reward, or fairness-aware objective.
& Distributed near-RT control; studies evaluate non-stationarity, communication overhead, and xApp/resource coupling.
& \cite{iturria2022multi,rezazadeh2023multi,he2025heterogeneous,shokouhi2025distributed,seid2025multiagent} \\
\hline

\textbf{Federated and distributed DRL}
& \cellcolor{highcol!35}Horizontal or vertical FRL over local MDP/POMDP instances; some works combine federation with hierarchy or MARL.
& Local RAN traces, slice telemetry, edge-compute load, jamming indicators, and site-specific traffic distributions.
& Same control variables as the local task, such as slice resources, VNF split, task offloading, or mitigation action.
& Local reward plus global model aggregation objective; privacy and communication costs shape the training design.
& \cellcolor{highcol!35}Cross-site training over communication rounds; studies evaluate heterogeneous sites without centralizing raw data.
& \cite{abouaomar2022federated,amiri2023edge,ndikumana2023federated,hazarika2024enhancing,kouchaki2025federated} \\
\hline

\textbf{Trustworthy and evaluation-aware formulations}
& CMDP, risk-aware MDP, robust/adversarial MDP, or explainability-augmented DRL formulation.
& Standard telemetry plus safety margins, adversarial perturbations, uncertainty estimates, or explanation features.
& Projected safe action, defense action, resource-control action, or action replacement after explanation.
& \cellcolor{highcol!35}Constraint cost, risk-sensitive reward, robustness objective, or explanation-guided performance criterion.
& \cellcolor{highcol!35}Offline, simulated, or testbed evaluation; studies stress latency budgets, adversarial shifts, and reproducibility.
& \cite{tsampazi2024pandora,fiandrino2023explora,nagib2023safe,nagib2025safeslice,hassan2025advo} \\
\hline
\end{tabular}
\end{table*}

% ---------------------------------------------------------------------
\subsection{Toward 6G: AI-Native O-RAN}
\label{subsec:6g-substrate}

Research work on 6G intends to extend O-RAN from AI-enabled control toward more AI-native networking. The 5G O-RAN architecture introduced AI/ML mainly through controller applications and lifecycle services, while 6G visions place learning, inference, and adaptation closer to the core network design~\cite{hexax_ai_native,ngmn_6g_position}. Several directions in the literature affect how future DRL agents may observe, decide, and coordinate.

\textbf{AI-Native Design.}
AI-native 6G designs treat AI as a network function rather than only an add-on application. The O-RAN Alliance next Generation Research Group (nGRG) and 3GPP Release~20 study items explore native AI/ML support in the RAN, including standardized model exchange, distributed training procedures, and in-node intelligent functions~\cite{oran_ngrg}. For DRL, these developments can tighten the link between inference and the data path, expose richer telemetry, and support online adaptation under managed lifecycle procedures.

\textbf{Integrated Sensing and Communication (ISAC).}
ISAC expands the radio interface from communication-only operation to joint communication and environmental sensing~\cite{liu2022integrated}. ISAC can enlarge the observation space of DRL agents with radar-like measurements, including range, Doppler, and angle information. It can also enlarge the action space in beamforming and waveform design, where DRL formulations often balance sensing utility with throughput, latency, and interference objectives.

\textbf{Non-Terrestrial Networks.} Non-Terrestrial Networks (NTNs) bring satellites, high-altitude platforms, and drones into the RAN \cite{kodheli2020satellite}. NTN-aware O-RAN studies examine xApps and rApps that account for orbital dynamics, long propagation delays, and intermittent connectivity. They also investigate corresponding adaptations to E2
and A1 service models \cite{oran_ntn_2025,baena2025space}. These settings challenge DRL formulations that assume stationary dynamics, short control horizons, or dense feedback.

\textbf{Edge and Federated Intelligence.}
Edge and federated intelligence push learning and inference closer to radio nodes. The distributed O-Cloud can place DRL inference at regional or edge tiers, while federated learning can train policies across sites without centralizing raw telemetry~\cite{mcmahan2017communication,qi2021federated}. This direction aligns with O-RAN's multi-site and multi-vendor setting, where local adaptation and data governance both affect deployable DRL designs.

\textbf{Key Takeaways.}
AI-native 6G O-RAN expands the design space for DRL through native model lifecycle support, sensing-rich observations, NTN-aware control, and edge or federated learning. These trends do not change the architectural need for clear agent placement; instead, they make the mapping among host, interface, timescale, and data source more important for future DRL studies.

\section{DRL in O-RAN: MDP Formulation}
\label{4_RAN_to_MDPs}
This section reviews how existing DRL-for-O-RAN studies formulate RAN control problems as sequential decision-making problems. We compare the literature along five recurring formulation dimensions: (i) decision model, (ii) observation and state design, (iii) action-space design, (iv) reward and constraint design, and (v) temporal structure and evaluation scope. Table~\ref{tab:oran_drl_formulation_taxonomy} summarizes representative studies along these dimensions, with shaded cells highlighting the formulation aspects emphasized by each research problem.

The appropriate formulation depends strongly on the control objective and the corresponding O-RAN control loop. Resource-management studies primarily characterize radio conditions and resource-allocation decisions; mobility and traffic-steering studies emphasize user mobility, cell load, and handover outcomes; slicing studies place greater emphasis on SLA constraints and multi-timescale control; and multi-agent and federated studies additionally account for coupling among controllers, cells, or deployment sites. The following subsections examine these formulation choices systematically.

\subsection{Decision Model Selection}
\label{subsec:formulation_class}

\textbf{Single-Controller Formulations.}
When a single xApp or rApp is responsible for a control task, existing studies typically formulate the problem as an MDP or POMDP. The specific choice depends primarily on whether the information available to the controller sufficiently represents the underlying network state. Such formulations have been used for resource allocation, scheduler reconfiguration, power control, and traffic steering \cite{tan2025deep,villegas2025drl2,tamim2023intelligent}. For example, Tan \emph{et al.} \cite{tan2025deep} formulate downlink resource allocation as a DRL problem in which the controller adapts PRB allocation to jointly optimize throughput and fairness. Villegas \emph{et al.} \cite{villegas2025drl2} use a DRL-based xApp to dynamically configure parameters of a QoS-aware MAC scheduler according to changing network conditions. Tamim \emph{et al.} \cite{tamim2023intelligent} formulate traffic steering as a sequential control problem in which congestion information guides proactive traffic redirection for URLLC services. These studies illustrate how a conventional single-controller MDP can capture a broad range of O-RAN control tasks when one logical controller owns the corresponding action space.

\textbf{Constraint-Aware and Coupled Formulations.}
More complex decision models are required when operational constraints must be enforced explicitly or when multiple controllers share network resources. Zangooei \emph{et al.} \cite{zangooei2023flexible} formulate flexible RAN slicing as constrained multi-agent RL, while Nagib \emph{et al.} \cite{nagib2025safeslice} study SLA-compliant slicing through safe DRL. Iturria-Rivera \emph{et al.} \cite{iturria2022multi}, Rezazadeh \emph{et al.} \cite{rezazadeh2023multi}, and He \emph{et al.} \cite{he2025heterogeneous} model interacting O-RAN controllers as multiple learning agents to capture coupling among cells, slices, or xApps. Collectively, these studies extend the single-controller formulation by explicitly representing either operational constraints or interactions among multiple decision makers.

\textbf{Federated Formulations.}
Federated formulations address a different aspect of the problem: how policies can be learned across multiple sites without centralizing raw network trajectories. 
For example, Abouaomar \emph{et al.} \cite{abouaomar2022federated} study federated DRL for O-RAN slicing, Amiri \emph{et al.} \cite{amiri2023edge} apply federated DRL to dynamic virtual network functions (VNF) splitting, and Ndikumana \emph{et al.} \cite{ndikumana2023federated} combine federated learning with deep Q-learning for offloading and routing. Unlike multi-agent formulations, which primarily change how the control problem is decomposed among interacting agents, federation primarily changes how policy parameters are trained and shared while keeping raw observations and trajectories local.

\subsection{Observation and State Design}
\label{subsec:state_space}

The observation or state representation determines what information the DRL agent uses to make control decisions. Existing studies generally select these variables according to the physical resources, service objectives, and network entities affected by the control task.

\textbf{Radio-Resource and Interference States.}
Existing RRM and scheduling studies commonly use CQI, SINR, buffer or queue length, traffic demand, PRB utilization, and QoS indicators to characterize current radio-resource conditions \cite{tan2025deep,villegas2025drl2,sohaib2024drl}. Power, spectrum, and interference-control studies further incorporate channel gains, interference measurements, neighboring-cell load, and spectrum occupancy to capture interactions between the controlled cell and its radio environment \cite{eskandari2025network,gopal2025adapshare,xdiff,reinders2026aiim}.

\textbf{Mobility, Service, and Slicing States.}
Mobility and traffic-steering studies require observations that describe both radio quality and service continuity. Tamim \emph{et al.} \cite{tamim2023intelligent}, Habib \emph{et al.} \cite{habib2023hierarchical}, and Lacava \emph{et al.} \cite{lacava2023programmable}, for example, use combinations of serving- and neighboring-cell measurements, cell load, UE distribution, mobility context, service type, and handover history. Network-slicing studies instead emphasize slice demand, per-slice utilization, latency and throughput KPIs, SLA margins, and violation history to represent resource isolation and service satisfaction \cite{abedin2022elastic,ghafouri2024multi,zangooei2023flexible,nagib2025safeslice}.

\textbf{Representation Granularity.}
State representations also differ in their granularity. Single-agent RRM studies commonly construct per-UE, per-cell, or per-scheduler state vectors, whereas slicing studies operate on per-slice or per-service summaries \cite{mhatre2024aiaas1,tan2025deep,filali2024open,lotfi2023attention}. Multi-agent studies expose local observations to individual cells, slices, UAVs, access points, or edge nodes, and may augment them with neighboring states or exchanged messages to capture inter-agent coupling \cite{iturria2022multi,rezazadeh2023multi,shokouhi2025distributed,seid2025multiagent}.

\textbf{Representation Mechanisms.}
Although many studies concatenate KPIs into flat state vectors, more structured representations have also emerged. Orhan \emph{et al.} \cite{orhan2021connection} combine graph neural networks with RL for an O-RAN connection-management xApp, allowing network topology to be incorporated directly into the state representation. Lotfi \emph{et al.} \cite{lotfi2023attention} employ attention-based processing for O-RAN slice management, while Fiandrino \emph{et al.} \cite{fiandrino2023explora} use feature-attribution analysis to relate DRL resource-allocation decisions to individual state features. Overall, state design follows the structure of the underlying control problem: RRM emphasizes radio and queue conditions, mobility emphasizes radio quality and movement context, slicing emphasizes SLA and utilization information, and multi-agent control emphasizes local observations together with inter-agent relationships.

\subsection{Action-Space Design}
\label{subsec:action_space}

% \hz{what about dividing work to: continuous actions, discrete actions, and hybrid actions?}

% \lj{Revised as follows:}

Existing O-RAN studies characterize action spaces according to both the controlled network decision and its mathematical representation. The decisions reviewed in this subsection include virtual network function (VNF) redirection, flow admission to a radio access technology (RAT), UE-to-serving-cell selection, transmit-power control, PRB partitioning, RU sleep scheduling, and scheduling-policy configuration. According to their mathematical representation, these action spaces can be organized into discrete, continuous, and hybrid formulations.

\textbf{Discrete Action Spaces.}
Discrete action spaces enumerate a finite set of categorical control decisions. Tamim \emph{et al.} \cite{tamim2023intelligent} define a traffic-steering action space that contains a no-steering action and all source--destination VNF redirection pairs. Their formulation produces $1+V^2$ actions, where $V$ denotes the number of VNFs. Habib \emph{et al.} \cite{habib2023hierarchical} adopt a hierarchical discrete formulation in which the meta-controller selects a queue-load threshold from a finite set and the lower-level controller admits each traffic flow to either Long-Term Evolution (LTE) or 5G New Radio (NR).

UE-to-serving-cell selection provides another discrete formulation. In this context, user association refers to selecting the serving cell or RAN node for a UE. Truong \emph{et al.} \cite{truong2025reinforcement} train a traffic-steering actor offline in the non-real-time RIC and deploy the trained actor as an xApp in the near-real-time RIC to infer user-association decisions. These studies use discrete actions to represent VNF redirection, flow admission, and UE-to-cell selection, while the formulation of Tamim \emph{et al.} \cite{tamim2023intelligent} also illustrates how the number of enumerated actions can increase with network size.

\textbf{Continuous Action Spaces.}
Continuous action spaces represent numerical control variables over bounded domains. Eskandari \emph{et al.} \cite{eskandari2025network} employ SAC to generate an $M$-dimensional transmit-power vector, where each component specifies the transmit power of one access point within the range of $10$ to $40$~dBm. Gopal \emph{et al.} \cite{gopal2025adapshare} formulate AdapShare as a contextual-bandit problem in which a continuous action partitions a shared PRB pool between LTE and NR under an available-resource constraint. These studies use continuous actions to represent direct adjustments of radio parameters and resource shares.

Continuous actions can also encode scheduling policies that are executed by lower-layer controllers. Yan \emph{et al.} \cite{xdiff} design xDiff to generate a preference value in $[-1,1]$ for each DU, UE, and PRB combination. The xApp sends these preference values to the corresponding DUs through the E2 interface. At each DU, a modified proportional fair (PF) scheduler first incorporates the preference information into its UE-priority metric and then assigns RBs with higher preference values to the prioritized UEs. Therefore, the xDiff action represents a continuous scheduling policy that is incorporated into the DU scheduler rather than a fixed RB-assignment vector generated by the near-real-time RIC.

\textbf{Hybrid Action Spaces.}
Hybrid action spaces jointly contain discrete- and continuous-valued components in the action executed at one decision step. Either component may contain multiple variables, and the continuous component does not need to be conditioned on a single categorical choice. Lu \emph{et al.} \cite{lu2026eexapp} design EExAPP to combine a discrete RU sleep-scheduling vector with continuous per-slice PRB-allocation ratios in a joint action generated by coordinated PPO actor--critic pairs. Ammar \emph{et al.} \cite{ammar2026towards} develop ASH-MARL, in which a D3QN agent selects discrete VNF-scaling actions and a TD3 agent selects continuous transmit-power and CPU-capacity allocations before the two outputs are executed as one action.

\textbf{Transformations of Hybrid Actions.}
Some O-RAN studies transform mixed actions into a homogeneous representation to use DRL algorithms with a single action type or to reduce combinatorial action-space growth. Ammar \emph{et al.} \cite{ammar2025maritime} start from discrete VNF scaling and migration decisions and continuous CPU-capacity, transmit-power, and UAV-trajectory controls, but map UAV trajectories to five grid movements and quantize resource allocations into two levels to obtain a fully discrete action space for A2C and PPO. Qazzaz \emph{et al.} \cite{qazzaz2026oreo} take the opposite approach in OREO: PPO produces continuous RU-activation scores and continuous weights for the user-association scoring rule. The RU-activation scores are then thresholded to determine the binary RU states, while the association weights are normalized via softmax to ensure feasibility.

\textbf{Action Granularity and Feasibility.}
Action granularity describes the network entity and control scope associated with each decision. Mhatre \emph{et al.} \cite{mhatre2024aiaas1}, Tan \emph{et al.} \cite{tan2025deep}, and Filali \emph{et al.} \cite{filali2024open} formulate resource-management actions at the scheduler, resource-allocation, or slice level, respectively. In multi-agent settings, Iturria-Rivera \emph{et al.} \cite{iturria2022multi}, He \emph{et al.} \cite{he2025heterogeneous}, and Shokouhi \emph{et al.} \cite{shokouhi2025distributed} distribute the joint action across cells, xApps, or access points, respectively. These formulations distinguish action granularity from the discrete, continuous, or hybrid domain of the action itself.

Existing formulations also encode feasibility through bounds and coupled constraints. The reviewed power-control \cite{eskandari2025network}, spectrum-sharing \cite{gopal2025adapshare}, interference-management \cite{xdiff}, and hybrid-control \cite{lu2026eexapp} studies impose transmit-power bounds, total-resource constraints, bounded preference values, or coupled slot and slice-allocation constraints, respectively. Tsampazi \emph{et al.} \cite{tsampazi2024pandora} further evaluate how action-space design interacts with reward definition and decision period in PandORA, showing that these formulation choices can affect comparisons among DRL agents.

\subsection{Reward Design}
\label{subsec:reward}

The reward function translates operator objectives into learning feedback. Because RAN optimization generally involves multiple competing objectives, reward design is often one of the most application-specific aspects of the DRL formulation.

\textbf{KPI-based Rewards.}
A common approach is to construct a scalar reward as a weighted combination of multiple network KPIs. RRM studies combine throughput, latency, fairness, QoS, and energy terms; mobility and traffic-steering studies combine load balancing, service continuity, latency, and handover penalties; and slicing studies combine utilization, revenue, isolation, and SLA satisfaction \cite{mhatre2024aiaas1,tan2025deep,tamim2023intelligent,abedin2022elastic}. In this way, higher-level operator objectives and standardized KPIs can be translated into scalar feedback for policy optimization \cite{3gpp_28_554}.

\textbf{Constraint-Aware Reward.}
Reward shaping alone, however, does not guarantee that operational requirements are satisfied. Constraint-aware studies therefore separate the optimization objective from explicit limits on latency, SLA violations, safety, or resource feasibility. 
For example, Zangooei \emph{et al.} \cite{zangooei2023flexible} formulate slice limits through constrained multi-agent RL, while Nagib \emph{et al.} \cite{nagib2025safeslice} design SafeSlice to reduce SLA violations during O-RAN slicing. Such formulations treat critical requirements as costs or constraints rather than merely assigning them additional penalty weights in the reward.

\textbf{Reward Sensitivity.}
The choice and weighting of reward terms can substantially affect the learned behavior. 
To investigate the reward sensitivity, Tsampazi \emph{et al.} \cite{tsampazi2024pandora} evaluate O-RAN agents under alternative rewards and action spaces, illustrating how formulation choices affect comparative performance. Fiandrino \emph{et al.} \cite{fiandrino2023explora} further use explanation tools to connect learned resource-allocation actions to input features. 
To address the reward sensitivity, robust and adversarial studies introduce attack-aware or risk-aware objectives when policies must operate under perturbed telemetry or malicious behavior \cite{ergu2025radar,hassan2025advo}. These results highlight that reward and constraint design determines not only the optimization target but also important properties such as fairness, robustness, and SLA compliance.

\begin{figure*}[!t]
\centering
\includegraphics[width=1\linewidth,trim=0 0 0 0, clip]{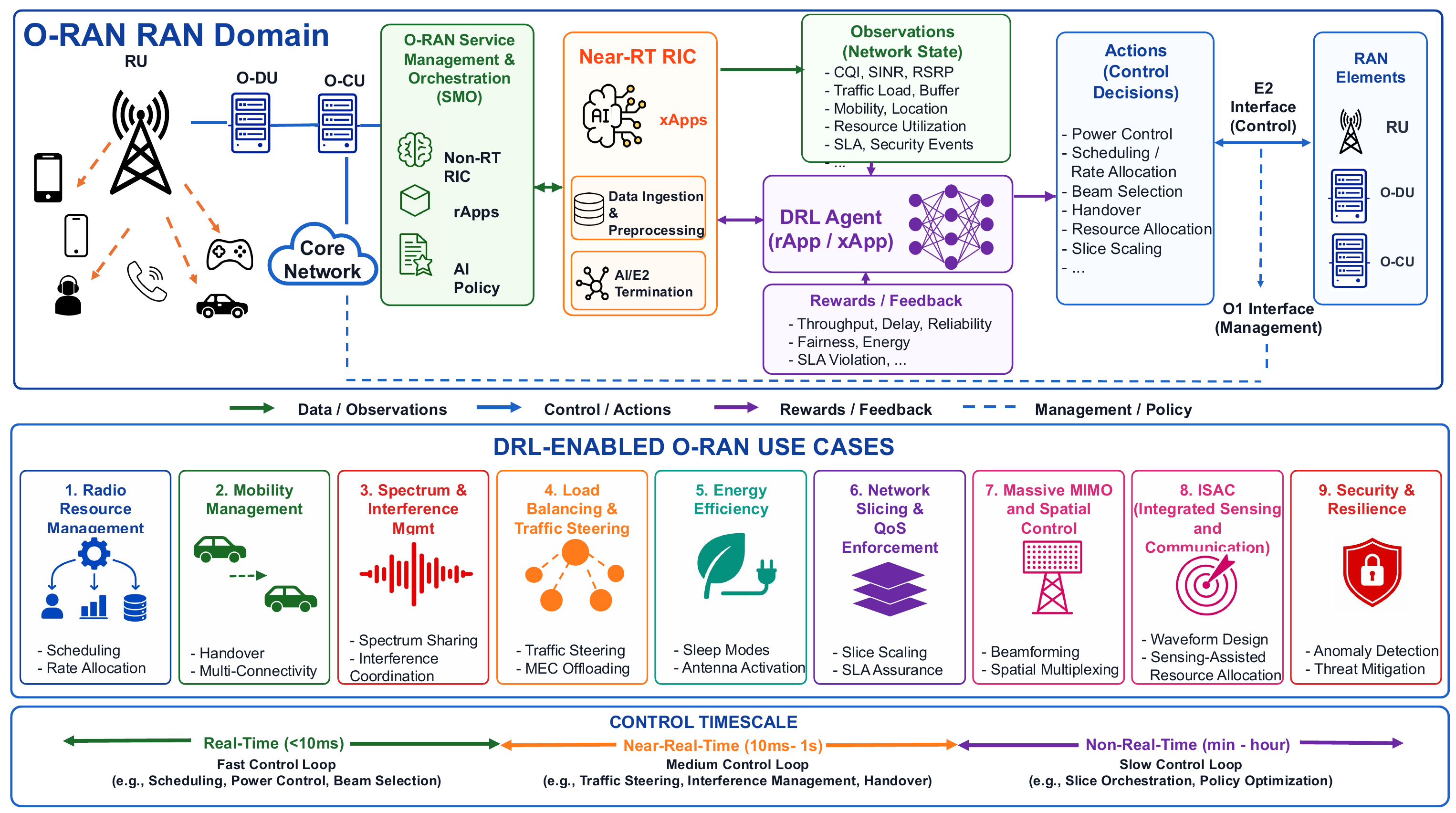}
\caption{An overview of DRL use cases in O-RAN. The top subfigure shows the DRL components within an O-RAN system, and the bottom subfigure illustrates some use cases with different optimization objectives.}
\label{fig:DRL_UseCases}
\end{figure*}

\subsection{Temporal Structure and Evaluation Scope}
\label{subsec:dynamics_timescales}

The final formulation dimension concerns when decisions are made and over what temporal horizon their effects are evaluated. This aspect is particularly important in O-RAN because different controllers operate at different timescales and network conditions evolve continuously.

\textbf{Control-loop Timescale.}
The decision interval is typically aligned with the O-RAN control loop hosting the DRL agent. Scheduler reconfiguration and radio-resource control are commonly evaluated at near-real-time timescales, whereas slice orchestration and policy guidance operate over slower non-real-time loops \cite{villegas2025drl2,mhatre2024aiaas1,hazarika2024enhancing}. Hierarchical traffic-steering approaches further decompose a single optimization problem across multiple timescales. Habib \emph{et al.} \cite{habib2023hierarchical}, for example, separate slower access-network selection from faster user association.

\textbf{Delay and Non-stationarity.}
The temporal formulation must also account for changing traffic, mobility, channel quality, and interference. Accordingly, existing studies evaluate policies under time-varying load, UE mobility, adversarial perturbations, or heterogeneous deployment conditions \cite{xdiff,kavehmadavani2024empowering,ergu2025radar,kouchaki2025federated}. 
In particular, federated and meta-learning approaches address related forms of non-stationarity by adapting policies across sites, local data distributions, or tasks instead of assuming a single stationary training environment \cite{hazarika2024enhancing,kouchaki2025federated}.

\textbf{Evaluation Horizon.}
Simulation and testbed studies also reveal the importance of specifying the temporal scope of evaluation. 
For instance, PandORA \cite{tsampazi2024pandora} jointly evaluates the effects of reward design, action space, decision interval, and policy architecture, while ns-O-RAN \cite{lacava2023ns} exposes xApps through an O-RAN-compliant E2 interface for simulation-based evaluation. Such studies typically specify the observation and action spaces, reward definition, control interval, episode or horizon length, and evaluation scenario. Explicit reporting of these parameters is essential because two studies using the same DRL algorithm may effectively solve different control problems when their decision periods or evaluation horizons differ.

% ---------------------------------------------------------------------
\subsection{Summary}
\label{subsec:formulation_summary}

The reviewed literature does not converge on a single generic MDP formulation for DRL-based O-RAN control. Instead, the formulation is determined by five closely related choices: who makes the decision, what network information is observable, which control variables are exposed, how operator objectives and constraints are represented, and when decisions are executed and evaluated. These choices are in turn shaped by the target RAN function and the O-RAN control loop in which the agent operates.

\section{DRL in O-RAN: Use Cases}
\label{5_Use_Cases}
% \textbf{The following structures are not fixed. Please feel free to adjust based on your understanding and current works.}

% \lj{A figure in each section to summarize all references.}

% \subsection{Overview}

% \textbf{Comparative table needed: problem, timescale, algorithm class, state, action, reward, scale tested, representative refs}

% \textbf{Organize problems by objective and timescale; for each problem, present the typical state, action, reward, constraints.}

\subsection{Overview}

\noindent\textbf{Why DRL Fits O-RAN.}
DRL is well suited to the control and optimization of O-RAN for three main reasons.

\begin{itemize}[leftmargin=0.15in]
\item 
First, O-RAN provides a programmable and data-driven control architecture. The Non-RT RIC and Near-RT RIC host rApps and xApps, respectively, enabling learning-based policies to observe network conditions and optimize multiple RAN functions through standardized interfaces. This architecture provides the telemetry, control hooks, and hierarchical control loops required to deploy and manage DRL agents.

\item
Second, many RAN control problems are naturally sequential decision-making tasks. Decisions such as resource allocation, scheduling, handover, traffic steering, and energy management influence not only the immediate network performance but also future traffic loads, interference conditions, queue states, and user associations. These problems are further characterized by stochastic channel and traffic dynamics, partial observability, coupled control variables, and long-term performance objectives, making them well suited to MDP- or POMDP-based formulations.

\item 
Third, DRL can optimize a control policy directly from environmental feedback without requiring labeled examples of optimal actions. This property is valuable in RAN environments, where obtaining optimal labels through exhaustive optimization or expert intervention is often prohibitively expensive. Moreover, DRL can continually adapt its policy as network conditions evolve. Nevertheless, DRL does not eliminate the sim-to-real gap: policies trained in simulators or digital twins may still experience performance degradation when deployed in previously unseen physical networks. Addressing this gap requires techniques such as domain randomization, transfer learning, offline-to-online adaptation, safe exploration, and continual learning, as discussed later in this survey.
\end{itemize}

\begin{table*}[!t]
\centering
\caption{Representative DRL formulations for O-RAN use cases.}
\label{tab:drl-usecase-formulations}
\scriptsize
\setlength{\tabcolsep}{1.2pt}
\renewcommand{\arraystretch}{1.18}

\begin{tabular}{|
    >{\centering\arraybackslash}m{0.14\textwidth}|
    >{\centering\arraybackslash}m{0.09\textwidth}|
    >{\raggedright\arraybackslash}m{0.20\textwidth}|
    >{\raggedright\arraybackslash}m{0.16\textwidth}|
    >{\raggedright\arraybackslash}m{0.22\textwidth}|
    >{\centering\arraybackslash}m{0.12\textwidth}|}
\hline

\rowcolor[gray]{0.9}
\multicolumn{1}{|c|}{
    \parbox[c][6ex][c]{0.14\textwidth}{
        \centering\textbf{Use Case}}} &
\multicolumn{1}{c|}{
    \parbox[c][6ex][c]{0.09\textwidth}{
        \centering\textbf{Timescale}}} &
\multicolumn{1}{c|}{
    \parbox[c][6ex][c]{0.20\textwidth}{
        \centering\textbf{State}}} &
\multicolumn{1}{c|}{
    \parbox[c][6ex][c]{0.16\textwidth}{
        \centering\textbf{Action}}} &
\multicolumn{1}{c|}{
    \parbox[c][6ex][c]{0.22\textwidth}{
        \centering\textbf{Reward}}} &
\multicolumn{1}{c|}{
    \parbox[c][6ex][c]{0.12\textwidth}{
        \centering\textbf{DRL}}} \\
\hline

\textbf{Power Control}
& ms--sec
& SINR, CQI, and interference.
& TX power level.
& Throughput and energy efficiency.
& DQN, PPO, SAC \\
\hline

\textbf{Scheduling}
& ms
& Buffer status, CQI, and queue length.
& RB assignment.
& Throughput and latency.
& DQN, PPO \\
\hline

\textbf{Resource Allocation}
& sec--min
& Traffic demand and load.
& Resource share.
& QoS and resource utilization.
& PPO, SAC \\
\hline

\textbf{Slice Orchestration}
& min--hour
& Slice KPIs and SLA status.
& Slice scaling.
& SLA satisfaction.
& Hierarchical RL \\
\hline

\textbf{Handover}
& ms--sec
& RSRP and mobility state.
& HO trigger.
& Handover success rate.
& DQN \\
\hline

\textbf{Traffic Steering}
& sec--min
& Cell load and UE distribution.
& Target cell or RAT.
& Load balancing.
& PPO \\
\hline

\textbf{MEC Offloading}
& ms--sec
& CPU load and latency.
& Offloading decision.
& Delay reduction.
& Actor--Critic \\
\hline

\textbf{Energy Saving}
& min--hour
& Traffic load.
& Sleep-mode activation.
& Energy saving.
& DQN, PPO \\
\hline

\textbf{Security Mitigation}
& sec--min
& Attack indicators.
& Defense policy.
& Attack resilience.
& MARL \\
\hline

\textbf{ISAC Resource Control}
& ms--sec
& Sensing and communication demands.
& Resource partition.
& Joint sensing--communication utility.
& PPO, SAC \\
\hline

\end{tabular}
\end{table*}

These architectural and methodological properties have made DRL one of the most widely investigated AI paradigms for O-RAN control and optimization.

\noindent\textbf{A Glance at DRL in O-RAN.}
Although the literature contains a large and diverse body of DRL studies, their optimization objectives generally fall within a relatively small set of network-wide goals. These include radio resource utilization (e.g., \cite{mhatre2024aiaas1}), mobility robustness (e.g., \cite{yan2026tarmm,lacava2023programmable}), spectrum and interference mitigation (e.g., \cite{ergu2025radar,xdiff}), load balancing and traffic steering (e.g., \cite{tamim2023intelligent,habib2023hierarchical,lacava2023programmable}), energy efficiency (e.g., \cite{abedin2022elastic}), network slicing and service-level agreement satisfaction (e.g., \cite{yan2025near,nagib2025safeslice}), sensing-communication integration (e.g., \cite{nikbakht2024memory,villegas2026isac}), massive-MIMO and spatial control (e.g., \cite{an2024dragon4,eskandari2025network,shokouhi2025distributed}), and network security and resilience (e.g., \cite{ergu2025radar,abou2024federated,motalleb2023moving}). Organizing the literature according to these objectives reveals common decision structures and recurring challenges across seemingly different RAN functions.

Fig.~\ref{fig:DRL_UseCases} presents an overview of DRL use cases in the O-RAN architecture. 
Table~\ref{tab:drl-usecase-formulations} summarizes representative DRL applications in O-RAN, highlighting their optimization objectives, control timescales, state representations, action spaces, reward formulations, evaluation scales, and commonly adopted DRL algorithms. 
In the remainder of this section, we review these studies according to the major operational domains.

\subsection{Radio Resource Management (RRM)}
\label{sec:usecase_rrm}

RRM is one of the most extensively investigated applications of DRL in O-RAN, as conventional optimization, heuristic, and rule-based methods often struggle to adapt to rapidly varying channel conditions, traffic demands, interference, and service requirements. By exploiting the programmable control loops of O-RAN, DRL agents can observe CQI/SINR measurements, traffic loads, queue states, resource utilization, and QoS or SLA indicators and then adapt transmit powers, scheduling policies, radio-resource allocations, and slice configurations to balance throughput, latency, fairness, energy efficiency, and service satisfaction. Existing studies cover four main directions. For power control, Eskandari \emph{et al.}~\cite{eskandari2025network} employ SAC to generate continuous per-AP power allocations for network slicing in O-RAN-enabled cell-free massive MIMO, while Ghafouri \emph{et al.}~\cite{ghafouri2024multi} and Shokouhi and Wong~\cite{shokouhi2025distributed} extend cell-free resource control through multi-level learning and distributed multi-agent precoding, respectively. For scheduling, Villegas \emph{et al.}~\cite{villegas2025drl2} dynamically reconfigure MAC schedulers according to network conditions, whereas Sohaib \emph{et al.}~\cite{sohaib2024drl} jointly schedule eMBB and URLLC traffic to balance throughput, latency, and reliability. Related interference-aware approaches generate scheduling guidance rather than direct per-slot assignments: xDiff produces DU-, UE-, and RB-specific preference values for proportional-fair schedulers~\cite{xdiff}, while AIIM coordinates overlapping PRB allocations across heterogeneous gNBs~\cite{reinders2026aiim}. For rate and radio-resource allocation, Hammami and Nguyen~\cite{hammami2022policy18} compare on-policy and off-policy DRL, Tan \emph{et al.}~\cite{tan2025deep} optimize the throughput--fairness tradeoff, and Rezazadeh \emph{et al.}~\cite{rezazadeh2023multi} and Seid \emph{et al.}~\cite{seid2025multiagent} distribute allocation decisions among multiple agents. This direction has also been extended through risk-aware and meta-reinforcement learning for adaptation under uncertainty and previously unseen conditions~\cite{kasi2025risk,lotfi2025meta30}, as well as robust learning against manipulated observations~\cite{ergu2025radar,ergu2024efficient14}. Finally, slice-resource orchestration coordinates admission, association, resource quotas, and SLA enforcement over longer or multiple timescales. 
% Representative studies investigate elastic industrial slicing~\cite{abedin2022elastic}, multi-timescale and QoS-aware resource management~\cite{mhatre2024aiaas1,mhatre2024intelligent22}, intelligible and near-real-time allocation~\cite{rezazadeh2024intelligible10,yan2025near}, federated and hierarchical orchestration~\cite{abouaomar2022federated,rezazadeh2022specialization16,hazarika2024enhancing,qiao2025resource}, and constrained, latency-aware, and safe slicing under dynamic SLAs~\cite{zangooei2023flexible,raftopoulos2024drl,nagib2025safeslice}. Collectively, these studies show that DRL-based RRM in O-RAN has evolved from isolated scheduling and allocation decisions toward distributed, multi-timescale, adaptive, and trustworthy resource control.

\textbf{Power Control.}
Power control adapts the transmit powers of cells or distributed access points (APs) to improve spectral and energy efficiency while mitigating interference and satisfying heterogeneous QoS requirements. DRL formulations generally use channel quality, user rates, interference, and traffic demand as states; transmit-power levels or power profiles as actions; and spectral efficiency, energy consumption, and QoS satisfaction as rewards.

Eskandari \emph{et al.}~\cite{eskandari2025network} employ SAC for power control in O-RAN-enabled cell-free massive MIMO, generating continuous per-AP power allocations to balance eMBB rates and URLLC requirements. Ghafouri \emph{et al.}~\cite{ghafouri2024multi} propose a multi-level DRL framework that coordinates slice admission and radio-resource allocation in cell-free O-RAN networks.
Shokouhi and Wong~\cite{shokouhi2025distributed} use multi-agent reinforcement learning to coordinate distributed precoding for eMBB and URLLC traffic. Oh \emph{et al.}~\cite{oh2023decentralized} study decentralized pilot assignment to reduce pilot contamination, complementing power and spatial-resource control in cell-free O-RAN.
Related studies address inter-cell interference through resource coordination. Yan \emph{et al.}~\cite{xdiff} propose xDiff, which generates DU-, UE-, and RB-specific preference values to guide proportional-fair scheduling. Reinders \emph{et al.}~\cite{reinders2026aiim} develop AIIM, a PPO-based xApp that coordinates cross-cell PRB allocation to reduce interference and QoS violations.
Robustness has also received attention. Ergu and Nguyen~\cite{ergu2025radar} propose RADAR to protect DRL-based resource allocation against manipulated observations. Ergu \emph{et al.}~\cite{ergu2024efficient14} show that falsified signal-power observations can mislead DRL agents and degrade V2X resource allocation. Overall, direct DRL-based power control is primarily demonstrated in cell-free O-RAN~\cite{eskandari2025network}, while related studies address slicing, precoding, pilot assignment, interference coordination, and robustness.

%%%%%%%%%%%%%%%%%%%%%%%%%%%%%%%%%%%%%%%%%%%
% \hz{multi-agent should be next section.}
% Multi-agent DRL (MADRL) has further been explored to coordinate power control decisions across neighboring cells while maintaining scalability in large-scale deployments. Because learned RRM policies directly affect radio resources, recent work also studies adversarially robust and attack-aware resource allocation, including RADAR and efficient adversarial attacks against DRL-based O-RAN resource allocation \cite{ergu2025radar,ergu2024efficient14}.

\textbf{Scheduling.}
Scheduling determines how radio resources and transmission opportunities are assigned among users and services under varying channel, traffic, and QoS conditions. DRL formulations commonly observe CQI reports, buffer status, packet-delay budgets, historical throughput, and service requirements. Their actions may select scheduling policies, configure scheduler parameters, generate user or RB preferences, or determine resource assignments, while rewards balance throughput, latency, fairness, reliability, and resource utilization.

Villegas \emph{et al.}~\cite{villegas2025drl2} deploy a DRL-based xApp that dynamically configures a Lyapunov-based MAC scheduler to satisfy QoS requirements while limiting resource consumption. Their subsequent work incorporates sensing information into scheduler reconfiguration, enabling context-aware scheduling decisions~\cite{villegas2026isac}. Sohaib \emph{et al.}~\cite{sohaib2024drl} propose a distributed Thompson-sampling-based DRL framework for jointly scheduling eMBB and URLLC users while balancing their distinct throughput, latency, and reliability requirements. Wu \emph{et al.}~\cite{wu2025mogul} propose MOGUL, which uses model-based optimization to reduce the scheduling space before applying online deep multi-agent reinforcement learning to select scheduling decisions.

Other studies retain the real-time O-DU scheduler and provide DRL-generated guidance. Yan \emph{et al.}~\cite{xdiff} generate DU-, UE-, and RB-specific preference values that guide proportional-fair scheduling at each O-DU. Reinders \emph{et al.}~\cite{reinders2026aiim} develop a PPO-based xApp that coordinates PRB partitions across neighboring gNBs to reduce interference and QoS violations. Overall, DRL-enabled O-RAN scheduling ranges from scheduler reconfiguration and reduced-space learning to policy-guided O-DU execution and cross-cell coordination.

\textbf{Rate Allocation.}
Rate allocation determines how transmission opportunities are distributed among users, cells, or services while satisfying QoS requirements. Unlike scheduling, which focuses on resource assignment decisions, rate allocation addresses the optimization of achievable data rates under limited radio resources.
A typical formulation seeks to maximize a network utility function:
$\max \sum_{u=1}^{U} U(R_u)$,
where $U(\cdot)$ is a utility function representing throughput, fairness, or QoS satisfaction. A commonly adopted utility function is
$
U(R_u)=\log(R_u)
$,
which naturally promotes proportional fairness among users.

In DRL-based rate allocation, the state space usually contains user QoS requirements, queue lengths, channel conditions, and historical allocation decisions. The action space specifies resource fractions or target transmission rates assigned to users or services. Recent studies have modeled rate allocation as a sequential decision-making problem and applied PPO, DDPG, and SAC algorithms to learn adaptive allocation policies. Tan \emph{et al.}~\cite{tan2025deep} present DRL-based resource allocation in O-RAN, and Seid \emph{et al.}~\cite{seid2025multiagent} extend this direction to multi-agent dynamic resource allocation for O-RAN-enabled TN-NTN metaverse services. 
% To improve scalability in large O-RAN deployments, multi-agent DRL approaches distribute decision-making across multiple cells or RIC instances \hz{ref}. 
Furthermore, risk-aware DRL and meta-reinforcement learning techniques \cite{kasi2025risk} have been proposed to enhance robustness under traffic bursts and non-stationary environments, enabling rapid adaptation to changing network conditions. Related RRM work \cite{abdelmoaty2025enabling} also connects resource adaptation with mobility robustness in scalable cell-free massive MIMO handover under O-RAN.

\textbf{Resource Slicing and Orchestration.}
Network slicing introduces an additional dimension to RRM by requiring resource isolation and service-level guarantees across multiple logical networks. In O-RAN, slice management can be implemented through coordinated optimization between the non-RT RIC and near-RT RIC.
Assume that radio resources are divided among $N$ slices. The allocation ratio for slice $i$ is denoted by $\rho_i$, satisfying
$
\sum_{i=1}^{N}\rho_i = 1
$.
The objective is to maximize aggregate slice utility:
$
\max \sum_{i=1}^{N} U_i(\rho_i)
$, 
subject to SLA and resource constraints.
DRL-based slice orchestration frameworks typically use slice-level traffic demands, QoS violations, and resource utilization metrics as state inputs. Actions correspond to adjusting slice resource quotas, admission thresholds, or scheduling priorities. Rewards are designed to reflect SLA compliance, resource utilization efficiency, and operator revenue.

Several studies have demonstrated that DRL can improve slice admission and resource-allocation efficiency compared with static partitioning. Rezazadeh \emph{et al.}~\cite{rezazadeh2024intelligible10} investigate intelligible protocol learning for O-RAN slicing, while Mhatre \emph{et al.} study multi-timescale and QoS-aware slice-resource management~\cite{mhatre2024aiaas1,mhatre2024intelligent22}. Sohaib \emph{et al.}~\cite{sohaib2025optimizing} further examine the latency, reliability, and throughput tradeoffs involved in URLLC resource optimization. Hierarchical DRL is particularly suitable for O-RAN because it decomposes slice orchestration across different decision levels and timescales. Hazarika \emph{et al.}~\cite{hazarika2024enhancing} combine hierarchical O-RAN slicing with federated DRL for vehicular networks, whereas Qiao \emph{et al.}~\cite{qiao2025resource} decompose slice-resource allocation into multiple learning levels to improve scalability. Together with the multi-timescale framework in~\cite{mhatre2024aiaas1}, these studies show how slower slice-level policies can coordinate faster resource adaptation, making hierarchical DRL a promising approach to scalable slice-aware RRM.

\subsection{Mobility Management}
\label{sec:mobility}
Mobility management aims to maintain service continuity and QoS as users move across the network. In O-RAN, programmable control loops allow DRL agents to make adaptive mobility decisions based on radio measurements, traffic conditions, and user mobility patterns. Existing studies mainly focus on handover optimization, mobility-aware traffic steering, and intelligent resource management.

Traditional handover mechanisms rely on fixed thresholds and often struggle with dynamic mobility, traffic, and QoS requirements. By observing signal quality, cell load, user trajectories, and service demands, DRL-based xApps can learn adaptive policies that improve service continuity and resource utilization. Dai \emph{et al.}~\cite{dai2024intelligent} present an O-RAN-compliant handover xApp that combines PPO with reservoir computing to capture sequential user- and cell-level states while reducing training overhead. Their ns-3 evaluation demonstrates improved cell performance and user experience under dynamic conditions. Li \emph{et al.}~\cite{li2025toward} propose a MEC--O-RAN orchestration framework for deploying DRL agents across multiple control timescales. They further identify asynchronous traffic, topology generalization, and costly online exploration as key challenges to practical DRL operation at O-RAN edges.

Recent research extends mobility management beyond conventional handover optimization. Kalntis \emph{et al.}~\cite{kalntis2026meta} employ meta-learning to enable rapid adaptation to changing environments and mobility patterns, making their approach suitable for large-scale NextG O-RAN deployments. Wadud \emph{et al.}~\cite{wadud2026ai} propose a multimodal AI framework that integrates heterogeneous network information to optimize handover decisions. Yan \emph{et al.}~\cite{yan2026tarmm} incorporate user mobility prediction into traffic steering and delay-critical edge-AI offloading. Qazzaz \emph{et al.}~\cite{qazzaz2026xapp} develop O-RAN xApps for mobility and resource management in non-terrestrial networks, highlighting the convergence of handover control, traffic steering, and intelligent network orchestration.

\subsection{Spectrum and Interference Management}
\label{sec:spectrum_split}
Spectrum and interference management are essential for maintaining network performance in dense O-RAN deployments. Leveraging the programmability of the RIC, DRL-enabled xApps can dynamically adapt spectrum allocation, interference coordination, and resource-sharing policies according to traffic demands and radio conditions.

Several representative works illustrate this trend. Gopal \emph{et al.} \cite{gopal2025adapshare} present AdapShare, an RL-based dynamic spectrum sharing framework that enables O-RAN controllers to adapt spectrum access policies in shared-spectrum environments. Yan \emph{et al.} \cite{xdiff} proposes xDiff, which combines online learning, DRL, and diffusion models to perform collaborative inter-cell interference management, achieving substantial performance gains in dynamic multi-cell O-RAN deployments. Similarly, Reinders \emph{et al.} \cite{reinders2026aiim} developed AIIM, an adaptive interference mitigation framework for heterogeneous multi-vendor O-RAN networks, highlighting the feasibility of AI-driven interference coordination in practical deployments.

Beyond spectrum sharing and interference mitigation, DRL has also been integrated with network slicing and QoS management. Existing studies jointly optimize spectrum utilization, slice isolation, and service reliability under diverse traffic demands, including industrial IoT \cite{abedin2022elastic}, cell-free networks \cite{ghafouri2024multi}, and URLLC services \cite{sohaib2025optimizing}. 
Practical deployment issues such as online adaptation and real-world operation of DRL agents in O-RAN environments have also received increasing attention \cite{li2025toward}.

\subsection{Load Balancing and Traffic Steering}
\label{sec:traffic_qoe}
Load balancing and traffic steering aim to improve network-wide resource utilization by dynamically distributing users and traffic flows across available radio access resources. In O-RAN, these functions can be implemented through xApps that continuously monitor cell load, user distribution, and QoS indicators and make adaptive steering decisions.

\textbf{Inter-Cell Offloading.}
Inter-cell offloading alleviates congestion by redirecting users or traffic flows to neighboring cells or access networks with available capacity. DRL-based traffic steering jointly considers cell load, channel quality, traffic demand, and QoS requirements. Kavehmadavani \emph{et al.}~\cite{kavehmadavani2024empowering} combine LSTM-based traffic prediction with multi-agent DRL for slice-aware traffic steering and resource allocation across multiple timescales. Tamim \emph{et al.}~\cite{tamim2023intelligent} integrate congestion prediction with deep Q-learning to proactively redirect URLLC traffic and reduce queuing delay. Habib \emph{et al.}~\cite{habib2023hierarchical} propose hierarchical RL for multi-RAT traffic steering, where a meta-controller selects load thresholds and a lower-level controller assigns traffic to LTE or NR. These studies demonstrate the benefits of predictive, multi-agent, and hierarchical learning for adaptive inter-cell offloading.

\textbf{Inter-RAT Steering.}
Inter-RAT steering is becoming increasingly important in O-RAN as operators integrate heterogeneous access technologies, including sub-6 GHz, mmWave, Wi-Fi, satellite, and future 6G systems. Leveraging the programmable RIC architecture, xApps and rApps can dynamically steer users across RATs to balance load, improve QoS, and optimize network utilization \cite{lacava2023programmable,nguyen2023network}.

Recent studies have extensively explored AI-driven traffic steering. 
Results in \cite{tamim2023intelligent,kavehmadavani2024empowering} show that DRL-based approaches outperform conventional policies in terms of throughput, latency, and load balancing. 
Per \cite{habib2024machine}, hierarchical learning frameworks improve scalability in multi-RAT environments.
At the same time, \cite{erdol2022federated} shows that federated and meta-learning techniques enable distributed adaptation across multiple domains with limited information exchange. 
In \cite{sharma2025adaptive,truong2025reinforcement}, hybrid policy-RL solutions have also been proposed to enhance reliability and deployment feasibility in operational O-RAN systems.

Beyond terrestrial networks, traffic steering is being extended to emerging scenarios such as non-terrestrial networks, vehicular communications, and mobility-aware edge computing \cite{qazzaz2026xapp,sroka2024policy,yan2026tarmm}. Recent works further incorporate mobility prediction, handover optimization, and traffic forecasting to support proactive steering decisions \cite{kalntis2026meta,kefalas2025traffic}. Experimental O-RAN testbeds have also demonstrated the practicality of DRL-enabled steering xApps for real-time closed-loop optimization \cite{barker2025real}. These advances indicate a clear trend toward intelligent and autonomous multi-RAT traffic management in future O-RAN deployments.

\textbf{Multi-access Edge Computing (MEC)  Offload Decision.}
MEC offloading determines whether computation tasks are processed locally or transferred to edge or cloud servers under latency, communication, and computing constraints. Ndikumana \emph{et al.}~\cite{ndikumana2023federated} combine federated learning with deep Q-learning to jointly optimize task offloading and fronthaul routing among distributed O-RAN edge clouds, minimizing communication and computation delays. Filali \emph{et al.}~\cite{filali2023communication7} propose a two-level DRL framework that jointly allocates radio blocks and MEC computing resources to satisfy URLLC requirements. Mart\'inez-Morfa \emph{et al.}~\cite{martinez2024drl} develop DRL-based xApps for dynamic RAN--MEC resource allocation and slicing, coordinating radio and computing resources under changing demands. Yan \emph{et al.}~\cite{yan2026tarmm} further integrate mobility prediction, proactive resource preparation, and multi-agent RL for delay-critical edge-AI offloading. These studies show that O-RAN-oriented DRL is extending traffic steering from radio access selection toward joint communication, computing, routing, and mobility optimization.

\subsection{Energy Efficiency}
\label{sec:energy}

Energy efficiency is becoming increasingly important as O-RAN deployments scale in size and complexity. Existing surveys identify the programmability and disaggregation of O-RAN as key enablers for adapting radio, computing, and infrastructure resources to traffic demand \cite{abubakar2023energy,liang2024energy}. Current DRL research mainly considers RU sleep control, spatial resource optimization, and functional placement or splitting.

\textbf{Base Station Sleep Modes.}
Lu \emph{et al.}~\cite{lu2026eexapp} present EExAPP, a DRL-based xApp that jointly optimizes RU sleep scheduling and DU resource slicing. Its dual-actor--dual-critic PPO architecture balances energy savings against QoS, and is evaluated on a real O-RAN testbed with commercial RUs and smartphones. Qazzaz \emph{et al.}~\cite{qazzaz2026oreo} propose OREO, which employs PPO within a hierarchical rApp--xApp architecture to jointly determine RU activation and user association in terrestrial and non-terrestrial O-RAN deployments. Together, these studies demonstrate traffic-aware RU sleep control at different O-RAN control layers.

\textbf{Dynamic Antenna Activation.}
Direct DRL-based antenna activation remains largely unexplored in O-RAN. Related work by Shokouhi and Wong~\cite{shokouhi2025distributedkasi2025risk} applies multi-agent DRL to distributed precoding in cell-free massive-MIMO O-RAN, demonstrating adaptive spatial-domain control across distributed radio nodes. However, its actions concern precoding rather than antenna on/off states; thus, energy-aware antenna selection remains an open research direction rather than an established O-RAN use case.

\textbf{Functional Split Adaptation.}
Amiri \emph{et al.}~\cite{amiri2023edge} propose a federated DRL framework for dynamically splitting VNFs across O-RAN edge sites, jointly improving resource utilization and limiting reconfiguration overhead. Joda \emph{et al.}~\cite{joda2022deep} develop a DQN-based method for joint user association and CU--DU placement, reducing end-to-end delay and deployment cost. Although the latter does not explicitly optimize energy consumption, it provides a decision framework that could incorporate processing and transport energy into future split- and placement-adaptation policies.

\subsection{Network Slicing and QoS Enforcement}
\label{sec:slicing}

Network slicing has become one of the most active application domains of DRL in O-RAN, enabling multiple logical networks with heterogeneous QoS and SLA requirements to coexist on shared physical infrastructure. Compared with traditional optimization methods, DRL naturally captures the sequential and coupled nature of slice orchestration, allowing resource allocation policies to continuously adapt to traffic dynamics, service priorities, and network conditions. Recent studies investigate a broad spectrum of slicing problems, including admission control \cite{rezazadeh2024sliceops}, resource allocation \cite{filali2023communication7,mhatre2024intelligent22,qiao2025resource}, slice scaling \cite{rezazadeh2024sliceops,lotfi2023attention}, SLA-aware orchestration \cite{dai2024ran6,raftopoulos2024drl}, federated learning \cite{rezazadeh2022specialization16,abouaomar2022federated,zhang2022federated}, and LLM-assisted management \cite{lotfi2025llm,lotfi2025prompt,lotfi2025oran,lotfi2026scalable}.

% Network slicing\cite{ghafouri2024multi, abedin2022elastic, rezazadeh2022specialization16, rezazadeh2024sliceops} is one of the most extensively studied DRL applications in O-RAN. By enabling the creation of isolated virtual networks with distinct service requirements, slicing introduces a complex resource management problem that naturally aligns with reinforcement learning.

\noindent\textbf{Slice Admission Control.}
Slice admission control determines whether newly arriving slice requests should be accepted according to available radio and computing resources while minimizing SLA violations. DRL enables adaptive admission policies that jointly optimize long-term revenue, resource utilization, and QoS satisfaction under highly dynamic traffic conditions. Filali \emph{et al.}~\cite{filali2023communication7} formulate admission control as a sequential decision-making problem and demonstrate that DRL significantly outperforms heuristic policies under varying traffic loads. Related O-RAN slicing studies (e.g., \cite{filali2024open,filali2026drl}) further consider slice provisioning for MVNOs and service-specific URLLC requirements, showing that admission decisions must be coordinated with computation, communication, and isolation constraints.

\noindent\textbf{Slice Resource Allocation.}
Resource allocation is the core control problem in O-RAN slicing. Instead of allocating resources only at the user level, slice-aware DRL agents dynamically distribute radio, computing, and transport resources among multiple slices according to their service priorities and QoS requirements. 
Ghafouri \emph{et al.}~\cite{ghafouri2024multi,ghafouri2024multi} propose a multi-level DRL framework for network slicing and resource management in O-RAN-based 6G cell-free networks, demonstrating the benefit of hierarchical control for coordinating slice-level and radio-level decisions. 
Mhatre \emph{et al.}~\cite{mhatre2024intelligent22} introduce a QoS-aware slice resource allocation framework with user-association parameterization, where DRL is used to jointly optimize slice resources and user association for beyond-5G O-RAN architectures. 
Other works \cite{abedin2022elastic,dai2024ran6,yan2025near,martinez2024drl} extend this resource-allocation view to industrial monitoring, near-real-time QoS optimization, SLA-driven intelligent slicing, and joint RAN--MEC resource allocation. 
Qiao \emph{et al.}~\cite{qiao2025resource} further develop a hierarchical learning approach for network slicing in O-RAN, highlighting the importance of decomposing slice orchestration into multiple learning levels to improve scalability.

% Slice admission control determines whether incoming slice requests should be accepted or rejected based on available resources and expected QoS impact. DRL-based approaches can learn admission policies that balance revenue generation, resource utilization, and SLA compliance under dynamic demand conditions \cite{filali2023communication7}.

\noindent\textbf{Dynamic Slice Scaling.}
Dynamic slice scaling adjusts slice resources according to time-varying traffic demand and service requirements. Recent studies have investigated hierarchical DRL, federated DRL, and attention-based slice management to improve scalability and adaptability in large-scale O-RAN deployments. 
Hazarika \emph{et al.}~\cite{hazarika2024enhancing} propose a hierarchical O-RAN slicing framework with federated DRL for vehicular networks, enabling distributed slice optimization while preserving local data privacy. 
Rezazadeh \emph{et al.}~\cite{rezazadeh2022specialization16} investigate the specialization of federated DRL agents for scalable and distributed 6G RAN slicing orchestration, showing that specialized agents can improve distributed slice management. 
In \cite{abouaomar2022federated} and \cite{amiri2023edge}, Federated and edge-AI-enabled slicing has also been used for O-RAN slicing and dynamic VNF splitting, further illustrating the role of distributed learning in elastic slice scaling. 
Lotfi \emph{et al.}~\cite{lotfi2023attention} introduce an attention-based DRL framework for O-RAN slice management, using attention mechanisms to capture the relative importance of slice states and improve resource adaptation under dynamic conditions.

% Slice scaling dynamically adjusts resource allocations according to traffic fluctuations and service requirements. This has become one of the most active research directions in O-RAN slicing. Representative studies include hierarchical DRL, federated DRL, attention-based management, and LLM-assisted orchestration frameworks. These solutions demonstrate significant improvements in adaptability and resource efficiency compared with static provisioning approaches.

\noindent\textbf{SLA-aware QoS Enforcement.}
Maintaining SLA guarantees is a central objective of network slicing. In DRL-based slicing, SLA requirements are often incorporated into reward functions, constraints, or penalty terms so that the agent can balance resource efficiency against latency, reliability, isolation, and service-priority requirements. Dai \emph{et al.}~\cite{dai2024ran6} explicitly study O-RAN-enabled intelligent network slicing to meet SLAs, while Yan \emph{et al.}~\cite{yan2025near} focus on near-real-time resource slicing for QoS optimization. Raftopoulos \emph{et al.}~\cite{raftopoulos2024drl} further examine latency-aware slicing under time-varying SLAs, and safe-DRL formulations such as SafeSlice incorporate safety constraints to reduce SLA violations during learning and operation \cite{nagib2023safe,nagib2025safeslice}. Rezazadeh \emph{et al.}~\cite{rezazadeh2023multi} study multi-agent DRL for RAN resource allocation in O-RAN, where multiple agents coordinate resource decisions under shared network constraints. These multi-agent formulations are especially relevant for SLA-aware slicing because slice isolation and QoS enforcement require coordinated decisions across radio resources, users, and services.

\noindent\textbf{Intelligent Slice Orchestration.}
Recent research further extends DRL-based slicing toward automation-native and AI-native orchestration. 
Rezazadeh \emph{et al.}~\cite{rezazadeh2024sliceops} propose SliceOps, an explainable MLOps framework for automation-native 6G networks, emphasizing the need for lifecycle management, explainability, and operational automation in slice control. 
In \cite{lotfi2025llm,lotfi2025prompt,lotfi2026scalable}, Lotfi \emph{et al.} introduce LLM-augmented DRL for dynamic O-RAN network slicing, where language-model guidance and context-aware prompt learning assist DRL-based slice orchestration. Security-aware slicing has also been explored. 
Motalleb \emph{et al.}~\cite{motalleb2023moving} propose a moving-target-defense-based secured slicing system in O-RAN, showing that slice orchestration must consider not only QoS and resource efficiency but also resilience against adversarial attacks. 
Overall, existing studies indicate that O-RAN slicing is evolving from single-objective resource allocation toward hierarchical, federated, explainable, safe, and secure closed-loop orchestration.

% \noindent\textbf{SLA-Aware Resource Reservation}
% Maintaining SLA guarantees is a central objective of most slicing frameworks. Recent studies increasingly incorporate SLA-awareness directly into reward design and resource orchestration decisions. Such approaches seek to balance resource efficiency against the risk of QoS degradation, particularly under highly dynamic traffic conditions.

\subsection{Integrated Sensing and Communication (ISAC)}
\label{sec:isac}
ISAC is widely regarded as a key enabling technology for future 6G networks. By jointly utilizing radio resources for communication and environmental sensing, ISAC introduces new optimization challenges involving sensing accuracy, communication performance, and resource efficiency. These highly dynamic and multi-objective decision problems naturally align with reinforcement learning, making DRL a promising approach for real-time ISAC control.

Recent studies have begun exploring the convergence of ISAC, AI, and O-RAN architectures. 
In \cite{nikbakht2024memory}, Nikbakht \emph{et al.} demonstrated that reinforcement learning can effectively optimize the communication-sensing tradeoff through adaptive waveform selection and memory-based policy learning. From an architectural perspective, Hamidi-Sepehr \emph{et al.} discussed the RAN requirements needed to support native ISAC functionalities, highlighting the importance of programmable control and AI-driven resource management \cite{hamidi2025ran}. Building upon the O-RAN framework, Baena \emph{et al.} proposed native ISAC support through sensing-aware service models and new control interfaces, enabling the deployment of sensing-oriented xApps and closed-loop optimization mechanisms \cite{baena2026toward}. Similarly, Polese \emph{et al.} introduced programmable inference and sensing applications based on O-RAN dApps, extending the AI-native RAN vision toward distributed sensing and inference services at the network edge \cite{polese2026enabling}.

While the literature has a considerable volume of work on the convergence of AI and ISAC, the research on DRL-enabled ISAC in O-RAN remains at an early stage.
% Recent results are promising. 
In \cite{villegas2026isac}, Villegas \emph{et al.} developed an ISAC-assisted DRL framework for dynamic MAC scheduler reconfiguration, where sensing information is incorporated into the decision-making process of O-RAN controllers to improve scheduling performance under dynamic network conditions. 
These initial efforts suggest that future O-RAN deployments may employ sensing-aware xApps and multi-agent DRL frameworks to jointly optimize communication, sensing, and computing resources in 6G networks.
% \subsubsection{Joint Waveform Design}
% Joint waveform design aims to optimize transmission parameters that simultaneously support communication and sensing objectives. Although DRL-based waveform optimization has attracted growing interest in the broader ISAC literature, explicit O-RAN implementations remain limited.
% \subsubsection{Beam Allocation for Sensing and Communication}
% Beam allocation is particularly important in highly directional ISAC systems. Learning-based approaches can dynamically allocate spatial resources between sensing and communication tasks according to changing operational requirements. However, existing O-RAN research has only begun to explore such capabilities.
% \subsubsection{Resource Trade-offs}
% ISAC fundamentally involves balancing competing sensing and communication objectives. Future DRL-enabled O-RAN controllers may play an important role in dynamically managing these trade-offs across multiple timescales. At present, however, ISAC remains one of the least explored use cases in the O-RAN DRL literature and represents a significant research opportunity.

\subsection{Security and Resilience}
\label{sec:security}

The openness, disaggregation, and programmability of O-RAN significantly expand the attack surface of radio access networks. Compared with conventional RAN architectures, O-RAN introduces new vulnerabilities through open interfaces, virtualized network functions, distributed intelligence, and AI-driven control loops. Consequently, security and resilience have become essential requirements for DRL-enabled O-RAN control, where learning agents must not only optimize network performance but also maintain reliable operation under faults, attacks, and highly dynamic environments \cite{ergu2025radar,ranslicejamming2026,ergu2024unmasking,chiejina2024systemlevel,blackboxoran2025,fggm2025}.

From a DRL perspective, security mechanisms can be viewed as sequential decision-making problems in which the agent continuously observes network states, identifies potential threats, and selects mitigation actions that balance security, service continuity, and resource utilization. 
Existing studies mainly investigate four complementary directions. 
\begin{itemize}[leftmargin=0.15in]
\item 
First, DRL enables adaptive anomaly detection by learning normal operational patterns from O-RAN telemetry, including traffic statistics, radio KPIs, and service performance indicators, allowing abnormal behaviors to be detected without relying on static thresholds \cite{ranslicejamming2026,ergu2024unmasking}.

\item 
Second, DRL has been applied to automated mitigation and anti-jamming, where agents dynamically adjust spectrum access, transmission parameters, or resource allocation policies to maintain communication reliability under adversarial interference \cite{chiejina2024systemlevel}. 
\item 
Third, moving-target defense and adaptive network reconfiguration leverage reinforcement learning to continuously modify network configurations, making attacks more difficult while minimizing the impact on legitimate users \cite{blackboxoran2025}. 
\item 
Finally, recent work has begun to investigate adversarially robust DRL, incorporating adversarial training and robust policy optimization to improve the resilience of AI-driven xApps against manipulated observations, poisoned training data, and malicious control actions \cite{ergu2025radar,oranevasion2024,maloran2025,kouchaki2025federated}.
\end{itemize}

Existing research \cite{habler2025adversarial,chiejina2024systemlevel,ergu2025radar,hassan2025advo} indicates that security should be treated as an integral component of the DRL control loop rather than as an independent protection mechanism. Instead of performing security analysis only after attacks are detected, future O-RAN systems are expected to integrate anomaly perception, adaptive mitigation, robust policy learning, and continuous risk assessment into a unified closed-loop control framework. Such architectures will become increasingly important as AI-native O-RAN evolves toward large-scale multi-vendor deployments involving distributed xApps, federated learning, and autonomous network management.

\subsection{Massive MIMO and Spatial Control}

Massive MIMO and spatial control are becoming increasingly important in O-RAN as future networks rely on large antenna arrays, cell-free deployments, and fine-grained PHY/MAC adaptation. Compared with conventional radio resource management, spatial control introduces additional decisions over MIMO layers, beams, antenna activation, pilot assignment, precoding, and distributed access-point coordination. These decisions are highly coupled with channel quality, interference, traffic load, and service requirements, making them natural candidates for DRL-enabled xApps and emerging low-latency control applications.

\begin{table*}[!t]
\centering
\caption{Taxonomy of DRL use cases in O-RAN.}
\label{tab:use_cases}
\scriptsize
\setlength{\tabcolsep}{1.5pt}
\renewcommand{\arraystretch}{1.06}

\begin{tabular}{|
>{\centering\arraybackslash}m{0.13\textwidth}|
>{\raggedright\arraybackslash}m{0.155\textwidth}|
>{\centering\arraybackslash}m{0.145\textwidth}|
>{\raggedright\arraybackslash}m{0.34\textwidth}|
>{\centering\arraybackslash}m{0.17\textwidth}|}
\hline

\rowcolor[gray]{0.9}
\multicolumn{1}{|c|}{
    \parbox[c][5ex][c]{0.13\textwidth}{
        \centering\textbf{Use-Case Domain}}} &
\multicolumn{1}{c|}{
    \parbox[c][5ex][c]{0.155\textwidth}{
        \centering\textbf{Main Control Target}}} &
\multicolumn{1}{c|}{
    \parbox[c][5ex][c]{0.145\textwidth}{
        \centering\textbf{O-RAN Scope}}} &
\multicolumn{1}{c|}{
    \parbox[c][5ex][c]{0.34\textwidth}{
        \centering\textbf{Compact DRL Formulation}}} &
\multicolumn{1}{c|}{
    \parbox[c][5ex][c]{0.17\textwidth}{
        \centering\textbf{Representative Refs.}}} \\
\hline

\textbf{Radio resource management} (\S\ref{sec:usecase_rrm})
&
Power control, scheduling, rate allocation, and slice-level
radio-resource orchestration
&
Mainly near-RT xApps; non-RT policies for coarse orchestration
&
State: CQI/SINR, load, interference, queue, and QoS/SLA indicators.
Action: power, RB/scheduler, rate, or quota.
Reward: throughput--latency--fairness--energy tradeoff.
&
\cite{mhatre2024aiaas1,ergu2025radar,seid2025multiagent,
rezazadeh2024intelligible10,ergu2024efficient14,tan2025deep,
villegas2025drl2,sohaib2024drl,abdelmoaty2025enabling,
mhatre2024intelligent22,kasi2025risk,sohaib2025optimizing}
\\
\hline

\textbf{Massive MIMO and spatial control}
&
MIMO-layer adaptation, antenna/resource activation,
cell-free coordination, and spatial interference control
&
Near-RT xApps for adaptive spatial control; emerging dApps
for fast PHY/MAC decisions
&
State: channel/rank quality, beam or layer status, antenna usage,
traffic load, and interference.
Action: active antennas, MIMO layers, beam/spatial resource,
or power profile.
Reward: spectral efficiency, energy saving, and interference reduction.
&
\cite{an2024dragon4,oh2023decentralized,eskandari2025network,
shokouhi2025distributed,hazarika2024enhancing,villegas2025drl2}
\\
\hline

\textbf{Mobility management} (\S\ref{sec:mobility})
&
Handover, user association, mobility-aware steering,
and mobility prediction
&
Near-RT xApps via E2; non-RT mobility policy
and enrichment information
&
State: serving/neighbor signal quality, cell load,
trajectory/speed, QoS, and handover history.
Action: target cell, offset, time-to-trigger, or steering decision.
Reward: service continuity with fewer failures and ping-pongs.
&
\cite{yan2026tarmm,lacava2023programmable,dai2024intelligent,
li2025toward,kalntis2026meta,wadud2026ai,qazzaz2026xapp,
abedin2022elastic}
\\
\hline

\textbf{Spectrum and interference management}
(\S\ref{sec:spectrum_split})
&
Dynamic spectrum sharing, coexistence,
and inter-cell interference mitigation
&
Near-RT xApps for fast adaptation; non-RT policies
for sharing rules
&
State: spectrum occupancy, interference, traffic/slice demand,
and multi-cell context.
Action: access policy, sharing decision, or coordination parameter.
Reward: spectrum efficiency, coexistence, reliability, and QoS.
&
\cite{ergu2025radar,xdiff,gopal2025adapshare,reinders2026aiim,
abedin2022elastic,ghafouri2024multi,sohaib2025optimizing,
li2025toward}
\\
\hline

\textbf{Load balancing and traffic steering}
(\S\ref{sec:traffic_qoe})
&
Inter-cell offloading, inter-RAT steering,
traffic forecasting, and MEC offloading
&
Near-RT xApps for cell/RAT/flow steering;
rApps for long-term policies
&
State: cell load, channel quality, UE distribution,
service type, mobility, and edge/cloud load.
Action: target cell/RAT, route, bias, or offloading placement.
Reward: balanced load, latency, QoE, and steering stability.
&
\cite{tamim2023intelligent,habib2023hierarchical,
lacava2023programmable,kavehmadavani2024empowering,
nguyen2023network,habib2024machine,erdol2022federated,
sharma2025adaptive,truong2025reinforcement,
sroka2024policy,kefalas2025traffic}
\\
\hline

\textbf{Energy efficiency} (\S\ref{sec:energy})
&
Base-station sleep, antenna activation, transmit-power saving,
and split/placement adaptation
&
Non-RT rApps and SMO/O-Cloud for long-term energy policy;
xApps for short-term load response
&
State: traffic forecast, load, user distribution,
radio/compute usage, and fronthaul status.
Action: sleep/wake, antenna level, carrier/power profile,
or split/placement choice.
Reward: energy saving under coverage and SLA constraints.
&
\cite{abedin2022elastic,abou2024federated,lotfi2025meta30}
\\
\hline

\textbf{Network slicing and QoS enforcement}
(\S\ref{sec:slicing})
&
Admission control, scaling, SLA-aware reservation,
isolation, and QoS-aware resource management
&
Non-RT rApps for slice policy; near-RT xApps
for short-term slice adaptation
&
State: slice requests, KPIs/KPMs, utilization, queues,
SLA margin, and violation history.
Action: admit/reject, quota, reservation, priority,
or A1 guidance.
Reward: SLA compliance, revenue, isolation, and utilization.
&
\cite{ghafouri2024multi,abedin2022elastic,
rezazadeh2022specialization16,rezazadeh2024sliceops,
filali2023communication7,hazarika2024enhancing,
amiri2023edge,raftopoulos2024drl,nagib2023safe,
lotfi2025prompt,mhatre2024intelligent22,martinez2024drl,
filali2026drl,nagib2025safeslice,lotfi2026scalable,
dai2024ran6,yan2025near,abouaomar2022federated,
filali2024open}
\\
\hline

\textbf{Integrated sensing and communication}
(\S\ref{sec:isac})
&
Communication--sensing tradeoff, waveform/resource adaptation,
and sensing-aware scheduling
&
Near-RT xApps and emerging dApps;
sensing-aware service models/interfaces
&
State: communication demand, sensing target,
radio/sensing measurements, beam/waveform status,
and environment context.
Action: waveform, beam, sensing/communication split,
or scheduler policy.
Reward: joint communication utility, sensing accuracy,
latency, and efficiency.
&
\cite{nikbakht2024memory,hamidi2025ran,baena2026toward,
polese2026enabling,villegas2026isac}
\\
\hline

\textbf{Security and resilience} (\S\ref{sec:security})
&
Anomaly detection, automated mitigation, anti-jamming,
moving-target defense, and adversarially robust DRL
&
Near-RT xApps for rapid mitigation; non-RT/SMO
for analytics, governance, and policy enforcement
&
State: abnormal KPM/traffic patterns, attack indicators,
action history, uncertainty, and logs.
Action: isolate, reconfigure, block, update model,
or escalate alarm.
Reward: attack suppression with low false alarms,
overhead, and SLA degradation.
&
\cite{ergu2025radar,ranslicejamming2026,
ergu2024unmasking,chiejina2024systemlevel,
blackboxoran2025,fggm2025}
\\
\hline

\end{tabular}
\end{table*}

Existing studies mainly investigate three complementary directions. 

\begin{itemize}[leftmargin=0.15in]
\item 
First, DRL has been applied to link-level and MAC-layer spatial adaptation. 
In  \cite{an2024dragon4}, An \emph{et al.} proposed DRAGON, a DRL-based MIMO layer and MCS adapter for O-RAN 5G networks, showing that learning agents can adjust transmission configuration according to changing radio conditions. 
In  \cite{villegas2025drl2}, Villegas \emph{et al.} further studied DRL-based dynamic MAC scheduler reconfiguration in O-RAN, which is closely related to spatial control because scheduler decisions interact with rank, channel quality, and resource-block allocation. These works indicate that near-RT control can move beyond cell-level resource allocation toward more adaptive PHY/MAC configuration.

\item 
Second, cell-free massive MIMO creates new coordination problems for O-RAN because distributed radio units must jointly manage pilots, power, and spatial resources. 
In \cite{oh2023decentralized}, Oh \emph{et al.} investigated decentralized pilot assignment for scalable O-RAN cell-free massive MIMO, highlighting the need for distributed coordination under limited signaling overhead. 
In \cite{eskandari2025network}, Eskandari \emph{et al.} studied DRL-based power control for network slicing in O-RAN-enabled cell-free massive MIMO, demonstrating how slice requirements, cell-free coordination, and spatial resource allocation can be jointly optimized. 
Recent multi-agent approaches (e.g., \cite{shokouhi2025distributed}) further extend this direction to distributed precoding and cell-free O-RAN control, where neighboring radio units or agents cooperate to improve spectral efficiency and support heterogeneous eMBB and URLLC traffic.

\item 
Third, massive MIMO spatial control increasingly overlaps with slice-aware and mobility-aware orchestration. Hierarchical O-RAN slicing with federated DRL can exploit spatial resources and distributed observations to support vehicular networks under dynamic traffic and mobility conditions \cite{hazarika2024enhancing}. Overall, the current literature suggests that massive MIMO control in O-RAN is shifting from isolated beam or power decisions toward joint spatial, slicing, scheduling, and distributed coordination problems. However, explicit DRL-based beam management, antenna activation, and real-time spatial interference coordination remain less mature than conventional RRM and slicing, leaving substantial room for future work on scalable multi-agent learning, fast inference, and O-RAN-compliant PHY/MAC control loops.
\end{itemize}

\subsection{Summary}
This section reviewed the major use cases of DRL in O-RAN, including radio resource management, mobility management, interference control, traffic steering, energy efficiency, network slicing, ISAC, security, and massive MIMO. Across these domains, DRL enables adaptive control by mapping dynamic network observations to actions that optimize long-term performance under varying traffic, channel, mobility, and service conditions. Table~\ref{tab:use_cases} summarizes the representative DRL use cases in O-RAN.

% \subsubsection{Power Control}

% \subsubsection{Scheduling}

% \subsubsection{Rate Allocation}

% \subsubsection{Slice Resource Orchestration}

% \subsection{Mobility Management}

% \subsubsection{Handover}

% \subsubsection{Multi-Connectivity}

% \subsubsection{Beam Switching}

% \subsection{Spectrum and Interference Management}

% \subsubsection{Dynamic Spectrum Sharing}

% \subsubsection{Incumbent Coexistence}

% \subsubsection{Beam Coordination}

% \subsection{Load Balancing and Traffic Steering}

% \subsubsection{Inter-Cell Offloading}

% \subsubsection{Inter-RAT Steering}

% \subsubsection{MEC Offload Decisions}

% \subsection{Energy Efficiency}
% \subsubsection{Base Station Sleep Modes}

% \subsubsection{Dynamic Antenna Activation}

% \subsubsection{Functional Split Adaptation}

% \subsection{Network Slicing and QoS Enforcement}
% \subsubsection{Slice Admission Control}

% \subsubsection{Slice Scaling}

% \subsubsection{SLA-Aware Resource Reservation}

% \subsection{ISAC}
% \subsubsection{Joint Waveform Design}

% \subsubsection{Beam Allocation for Sensing and Communication}

% \subsubsection{Resource Trade-offs}

% \subsection{Security and Resilience}
% \subsubsection{Anomaly Detection}

% \subsubsection{Automated Mitigation}

% \subsubsection{Adversarial Defense}

% \subsection{Lessons Learned}
% \textbf{Cross-cutting observations: which formulations recur, where reward shaping consistently fails, which use cases are over-studied vs. neglected, what reproducibility looks like in this literature.}

% \clearpage 
\section{Multi-Agent and Federated DRL for O-RAN}
\label{6_MADRL_FDRL}

% \lj{A figure in each section to summarize all references.}

\begin{figure*}[!t]
    \centering
\includegraphics[width=1\linewidth,trim=0 0 0 0, clip]{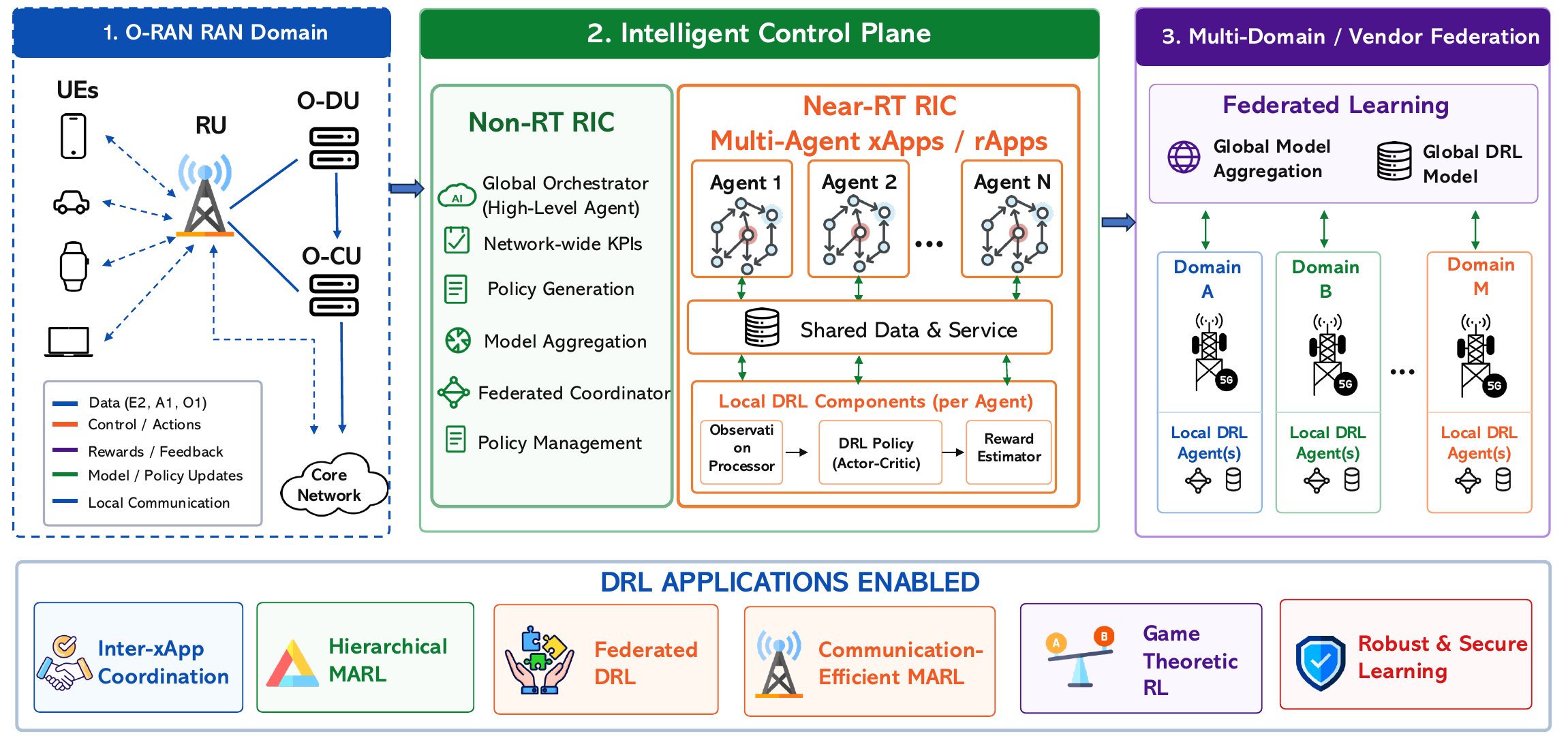}
    \caption{Multi-agent and federated DRL architecture for O-RAN, illustrating hierarchical RIC coordination and distributed model aggregation across multiple network domains and vendors.}
    \label{fig:multi}
\end{figure*}

Future O-RAN deployments are expected to involve a large number of distributed control entities, including xApps, rApps, base stations, slices, edge clouds, and user devices, all interacting across different timescales and administrative domains.
Such large-scale environments introduce several challenges for single-agent DRL approaches in the following aspects. 
First, a single agent often cannot observe the complete network state due to the inherently distributed architecture of O-RAN. Second, centralized training becomes increasingly difficult as network size grows, resulting in excessive computational complexity and signaling overhead. Third, independent DRL agents may generate conflicting actions when multiple optimization objectives coexist.

To address these challenges, recent research has increasingly adopted multi-agent reinforcement learning (MARL), hierarchical reinforcement learning (HRL), and federated reinforcement learning (FRL). These paradigms aim to distribute intelligence across multiple control entities while preserving scalability, coordination efficiency, and privacy. 
As illustrated in Fig.~\ref{fig:multi}, current research on DRL-enabled O-RAN applications can be broadly categorized into six directions: inter-xApp coordination, hierarchical MARL, federated DRL, communication-efficient MARL, game-theoretic RL, and robust \& secure learning.
% \hz{the description of Fig.~\ref{fig:multi} here does not agree with the figure itself.}

\subsection{Inter-xApp Coordination/Orchestration}
\label{sec:inter_xapp_coordination}

A key characteristic of O-RAN is the coexistence of multiple intelligent xApps within the Near-RT RIC. These xApps are often designed independently to optimize specific functions such as mobility management, traffic steering, spectrum allocation, scheduling, and energy efficiency. Since they operate on shared network resources, their actions may interact and even conflict, creating what is commonly known as the \emph{xApp conflict problem}. 
For example, traffic-steering and energy-saving xApps may issue contradictory decisions competing for the same radio resources in O-RAN systems \cite{zafar2024ric}.

To address this challenge, researchers increasingly model inter-xApp coordination as a multi-agent learning problem. Instead of independently optimizing local objectives, multiple intelligent agents cooperate to maximize network-wide utility while accounting for mutual interactions. 
Early work by Orhan \emph{et al.} \cite{orhan2021connection} demonstrated the integration of graph neural networks and reinforcement learning into an O-RAN connection-management xApp, showing the benefits of exploiting network topology information for coordinated decision making. 
More broadly, machine learning has been recognized as a key enabler of intelligent O-RAN operation in Beyond-5G systems \cite{erdol2025machine}. 

Recent studies have explored more sophisticated coordination frameworks. 
For example, He \emph{et al.} \cite{he2025heterogeneous} proposed a heterogeneous-agent PPO architecture in which xApps with different objectives and action spaces jointly learn coordinated policies. Their results indicate that cooperative learning significantly outperforms independently trained agents in dynamic O-RAN environments. Digital twins further enhance this capability by providing a safe virtual environment for policy training and evaluation. In particular, the digital-twin-assisted framework in \cite{he2025digital} enables xApps to test coordination strategies before deployment, improving learning efficiency and reducing operational risks.

Beyond coordination among Near-RT RIC applications, recent work has also investigated interoperability across O-RAN control layers. 
Giannopoulos \emph{et al.} \cite{giannopoulos2026interoperable} demonstrated coordinated rApp/xApp control for mobility-aware spectrum allocation.
% , while \cite{traka2025enabling} \lj{empty journal in this citation} extended O-RAN closed-loop control to heterogeneous access networks. 
Another promising direction is conflict mitigation through AI-based orchestration. The xApp Distillation framework proposed in \cite{erdol2024distillation} identifies and resolves conflicting actions among independently developed applications without requiring a fully unified learning architecture.

Practical deployment aspects are also receiving increasing attention. Platforms such as MANATEE \cite{montebugnoli2026manatee} provide testing and lifecycle-management capabilities for large-scale xApp ecosystems, facilitating the validation of coordinated control strategies. Overall, existing studies demonstrate that effective inter-xApp coordination can substantially improve network-wide performance and robustness. However, challenges including non-stationarity, communication overhead, and multi-vendor interoperability remain open research issues, motivating further investigation into MARL, digital twins, and AI-native orchestration frameworks for future O-RAN systems.
% One of the defining features of O-RAN is the deployment of multiple xApps on a shared Near-RT RIC platform. Each xApp is typically designed to optimize a specific network function, such as traffic steering, load balancing, mobility management, interference mitigation, or energy optimization. Although these applications are developed independently, they often operate on shared network resources and therefore influence one another’s performance.

\subsection{Hierarchical MARL Across Loops}
\label{sec:hierarchical_marl}

% The hierarchical control architecture of O-RAN naturally aligns with hierarchical reinforcement learning principles. O-RAN defines multiple control loops operating at different timescales, ranging from sub-second radio optimization in the Near-RT RIC to long-term policy generation and orchestration in the Non-RT RIC.

In O-RAN, many network optimization problems involve both strategic and operational decisions. For example, network slicing requires long-term admission and resource provisioning decisions as well as short-term scheduling and traffic management actions. Attempting to learn these decisions using a single DRL agent often results in prohibitively large state and action spaces.
Hierarchical reinforcement learning addresses this issue by decomposing complex decision-making processes into multiple levels. High-level agents typically generate long-term objectives, policies, or resource budgets, while lower-level agents execute fine-grained operational actions.

Habib \emph{et al.} \cite{habib2023hierarchical} propose a hierarchical DRL framework for traffic steering in multi-RAT O-RAN environments. Their architecture separates strategic access-network selection from short-term user-association decisions, significantly reducing learning complexity. Similar hierarchical approaches have been investigated in  \cite{mhatre2024intelligent22} for resource orchestration in massive MIMO systems and dynamic spectrum management scenarios.

Network slicing has emerged as one of the most prominent application domains for hierarchical DRL. The SliceOps framework proposed in \cite{hazarika2024enhancing} demonstrates how multi-level learning agents can jointly perform slice admission control, resource scaling, and SLA enforcement. By separating strategic orchestration decisions from operational resource allocation, the framework improves both scalability and service reliability.

More recently, researchers have begun combining HRL with MARL to improve the scalability of O-RAN control. Lotfi \emph{et al.} \cite{lotfi2024open} present a distributed hierarchical DRL framework in which multiple agents cooperate to perform scalable radio resource allocation across O-RAN domains. Lotfi \emph{et al.} \cite{metahrl2025} further introduce a meta-hierarchical RL framework that incorporates meta-learning to enable faster adaptation to dynamic traffic patterns and previously unseen network environments. Bao \emph{et al.} \cite{bao2026llm} propose LLM-hRIC, where an LLM-assisted hierarchical controller coordinates long-term decision making in the Non-RT RIC while the Near-RT RIC performs fine-grained real-time optimization. Riggio \cite{riggio2026deployable} presents a deployable hierarchical ML framework for traffic steering that maps high-level policies to real-time xApp execution in practical O-RAN deployments. 
These hierarchical architectures improve both temporal and spatial scalability while naturally aligning with the layered control structure of the O-RAN architecture.

\subsection{Federated DRL Across Vendors}
\label{sec:federated_drl}

One of the defining characteristics of O-RAN is its multi-vendor and geographically distributed architecture. While this openness promotes innovation and interoperability, it also introduces significant challenges for AI-native network optimization. Operational data are often distributed across operators, vendors, edge domains, and infrastructure providers, making centralized DRL training difficult due to privacy requirements, data ownership constraints, communication overhead, and security concerns.

To address these issues, Federated DRL has emerged as a promising paradigm for distributed intelligence in O-RAN. Instead of collecting raw data at a central controller, individual network entities train local DRL agents using their own observations and periodically exchange model parameters through a federation server or aggregation mechanism. This enables collaborative policy learning while preserving data locality and reducing information exposure.

% \hz{add pargraph titles to better organize the survey}

\textbf{Federated DRL for Network Slicing.}
Network slicing has become one of the most actively studied applications of federated DRL in O-RAN. 
Early work by Abouaomar \emph{et al.} \cite{abouaomar2022federated} introduced a federated DRL framework for O-RAN slicing in 6G networks, demonstrating how distributed RAN domains can jointly learn slice-management policies without sharing proprietary operational data. 
Similar ideas were explored by Zhang \emph{et al.} \cite{zhang2022federated}, who applied federated DRL to resource allocation across multiple network slices, showing improved scalability and privacy preservation compared with centralized learning. More recently, Hazarika \emph{et al.} \cite{hazarika2024enhancing} combined hierarchical slicing architectures with federated DRL to support vehicular O-RAN environments characterized by highly dynamic traffic conditions and heterogeneous service requirements.

\textbf{Federated DRL for Service Orchestration.}
Federated DRL has also been adopted for edge computing and service orchestration. Ndikumana \emph{et al.} \cite{ndikumana2023federated} proposed a federated DQN framework for joint task offloading and fronthaul routing optimization in O-RAN, where distributed edge nodes collaboratively improve decision policies without exchanging local traffic information. Similarly, Amiri \emph{et al.} \cite{amiri2023edge} developed a federated DRL solution for dynamic VNF splitting in O-RAN slicing, enabling edge-cloud coordination while reducing data transfer overhead. These studies demonstrate the suitability of federated learning for MEC-enabled O-RAN deployments, where computation and decision making are naturally distributed across multiple edge domains.

\textbf{Federated DRL for Emerging Use Cases.}
Beyond traditional RAN optimization, federated DRL has recently expanded into several emerging O-RAN scenarios. Alsenwi \emph{et al.} \cite{alsenwi2025ran} proposed a distributed learning framework for multi-RIS-assisted vehicular networks based on O-RAN principles, while Ahmed \emph{et al.} \cite{ahmed2025federated} leveraged federated DRL to support autonomous multirobot reconfiguration through O-RAN-controlled wireless infrastructures. These studies illustrate the growing role of federated intelligence in integrating O-RAN with cyber-physical systems, edge robotics, and future 6G applications.

\textbf{Federated DRL for Network Security.}
The open and heterogeneous nature of O-RAN further motivates federated DRL approaches for security and robustness enhancement. Abou El Houda \emph{et al.} \cite{abou2024federated} employed federated DRL to mitigate jamming attacks without requiring centralized collection of sensitive radio measurements. Similarly, El-Hajj \cite{el2025secure} proposed a zero-trust O-RAN optimization framework combining federated learning and security-aware orchestration to enhance trustworthiness in multi-vendor deployments. In addition, Kouchaki \emph{et al.} introduced OpenAI-DApp \cite{kouchaki2023openai}, an open platform for distributed federated reinforcement learning applications in O-RAN, and later proposed Federated Neuroevolution O-RAN \cite{kouchaki2025federated}, which improves the robustness and adaptability of DRL-based xApps through federated evolutionary optimization.

% Compared with centralized DRL, federated DRL offers several advantages for O-RAN. It preserves data privacy, reduces raw-data transfer, improves scalability across geographically distributed deployments, and naturally aligns with O-RAN’s disaggregated architecture. Moreover, federated training enables local agents to adapt to site-specific traffic characteristics while still benefiting from knowledge acquired across the federation.

\textbf{Outlook on Federated DRL.}
Despite the progress in the past years, several challenges remain unresolved. First, O-RAN environments often exhibit highly non-IID data distributions, leading to slower convergence and degraded policy quality. Second, periodic parameter synchronization can introduce substantial communication overhead, particularly for large DNN models and Near-RT control applications. Third, open multi-vendor ecosystems remain vulnerable to model-poisoning, backdoor, and Byzantine attacks, making secure aggregation and trust management critical research issues. 
Consequently, research is increasingly exploring personalized federated DRL, hierarchical federation architectures, asynchronous model aggregation, federated neuroevolution, and secure zero-trust learning frameworks. 
These directions are expected to play a central role in enabling scalable and trustworthy AI-native O-RAN deployments in future 6G networks.

\subsection{Communication-Efficient MARL}
\label{sec:communication_efficient_marl}

Although MARL enhances the scalability of DRL, efficient coordination among distributed agents remains a major challenge in O-RAN. In practical deployments, AI agents operating at different cells (e.g., DUs, CUs, edge clouds, and xApps) must exchange observations, actions, policies, or model updates to achieve coordinated behavior. As the network scale grows, communication overhead can become a critical bottleneck, especially for Near-RT RIC applications where decision latency is tightly constrained.

\textbf{Inter-agent Communication Efficiency.}
Several recent studies have therefore focused on reducing inter-agent communication while preserving coordination performance.
One approach is to exploit localized interactions and distributed decision making. 
Early work on team learning in virtualized O-RAN by Iturria-Rivera \emph{et al.} \cite{iturria2022multi} demonstrated that distributed agents can learn cooperative resource-management policies using only partial local observations, avoiding the need for global state dissemination. 
Similarly, the MARL-based resource allocation framework proposed by Rezazadeh \emph{et al.} \cite{rezazadeh2023multi} distributes decision making among multiple RAN entities, thereby reducing centralized signaling requirements within the O-RAN control architecture.

Another direction is the use of hierarchical coordination structures, where communication occurs only between adjacent layers rather than among all agents. 
In \cite{ghafouri2024multi}, multi-level DRL architectures for network slicing and resource management decompose decisions across multiple control layers, enabling local optimization while exchanging only aggregated information with upper-level controllers. Likewise, a hierarchical MARL framework for edge-cloud mobility management in \cite{giarre2025hierarchical} organize agents into multiple decision levels, significantly reducing coordination overhead compared with fully connected MARL schemes. Such designs align naturally with O-RAN’s hierarchical control loops spanning Non-RT RIC, Near-RT RIC, and distributed network elements.

\textbf{Federated Learning for Communication Efficiency.}
Communication efficiency is also increasingly addressed through federated and distributed learning paradigms. Rather than sharing raw network measurements or user data, federated DRL approaches exchange only model parameters or gradients, thereby reducing the communication overhead. 
For example, Hazarika \emph{et al.} \cite{hazarika2024enhancing} combine hierarchical network slicing with federated DRL for vehicular O-RAN environments, reducing inter-node signaling while preserving privacy. Similar ideas have emerged in distributed O-RAN optimization scenarios where local agents periodically synchronize models instead of continuously exchanging observations.

\textbf{Task Decomposition for Communication Efficiency.}
Several studies further reduce coordination overhead by exploiting task decomposition and agent specialization. In constrained MARL-based RAN slicing \cite{zangooei2023flexible}, agents optimize slice-level decisions using local resource information while coordination is enforced through constrained optimization objectives rather than extensive message exchange. Heterogeneous-agent designs for xApp coordination \cite{he2025heterogeneous} similarly assign different responsibilities to specialized agents, allowing only selected information to be shared among xApps. Such architectures become increasingly important as the number of AI-native xApps deployed in Near-RT RIC continues to grow.

\textbf{Communication Efficiency in Distributed Systems.}
Recent work has also explored communication-efficient coordination in highly distributed scenarios such as cell-free massive MIMO and edge-cloud O-RAN systems. 
In \cite{shokouhi2025distributed}, the authors proposed a distributed precoding framework based on MARL, enabling access points to learn coordinated transmission policies using primarily local channel information.
In \cite{zhang2026optimized}, the authors proposed edge-cloud traffic scheduling solutions to employ distributed agents that cooperate through limited state exchange to balance latency and resource utilization across heterogeneous computing tiers.

\textbf{Key Takeaways.}
Overall, existing research suggests that communication-efficient MARL in O-RAN is evolving along three complementary directions: (i) hierarchical coordination to limit information exchange across control layers, (ii) federated synchronization to replace raw-data sharing with model sharing, and (iii) localized cooperation mechanisms that rely on partial observations and sparse inter-agent communication. These approaches are expected to become increasingly important as future O-RAN deployments scale toward dense multi-cell, multi-vendor, and AI-native network environments.

% Available width for the contents of the six columns
\newlength{\oranTableWidth}

% ============================================================
% Table
% ============================================================

\begin{table*}[!t]
\centering
\caption{Taxonomy of multi-agent and federated DRL studies for O-RAN.}
\label{tab:marl_fdrl_oran_taxonomy}

\scriptsize

% Horizontal padding inside cells
\setlength{\tabcolsep}{2pt}

% Vertical spacing
\renewcommand{\arraystretch}{1.12}

% ============================================================
% Calculate the actual available width
%
% 6 columns -> 12\tabcolsep
% 7 vertical rules -> 7\arrayrulewidth
% 1pt additional safety margin
% ============================================================

\setlength{\oranTableWidth}{%
    \dimexpr
    \linewidth
    -12\tabcolsep
    -7\arrayrulewidth
    -1pt
    \relax
}

% Original column-width ratio:
% 0.14 : 0.14 : 0.165 : 0.21 : 0.205 : 0.09
%
% Normalized:
% 0.147368 : 0.147368 : 0.173684 :
% 0.221053 : 0.215789 : 0.094738

\begin{tabular}{
    |C{0.147368\oranTableWidth}
    |L{0.147368\oranTableWidth}
    |L{0.173684\oranTableWidth}
    |L{0.221053\oranTableWidth}
    |L{0.215789\oranTableWidth}
    |C{0.094738\oranTableWidth}|
}

% ============================================================
% Header
% ============================================================

\hline
\rowcolor{gray!20}

\textbf{Research Direction}
&
\textbf{Research Focus}
&
\makecell[c]{
    \textbf{Learning/Coordination}\\
    \textbf{Technique}
}
&
\makecell[c]{
    \textbf{O-RAN Scope}
}
&
\makecell[c]{
    \textbf{Role in Multi-Agent/Federated}\\
    \textbf{DRL}
}
&
\textbf{Refs.}
\\
\hline

% ============================================================
% I. Inter-xApp coordination
% 3 rows
% ============================================================

\multirow[c]{3}{=}[-1.5\baselineskip]{
    \centering
    \bfseries
    \makecell[c]{
        Inter-xApp\\
        coordination\\
        (\S\ref{sec:inter_xapp_coordination})
    }
}
&
xApp conflict mitigation
&
Conflict-aware coordination and orchestration
&
Multiple xApps sharing Near-RT RIC resources
&
Resolves inconsistent actions among independently developed xApps.
&
\cite{
    zafar2024ric,
    erdol2024distillation,
    montebugnoli2026manatee
}
\\

\cline{2-6}

&
Graph/RL xApp control
&
Graph neural networks with reinforcement learning
&
Connection management and topology-aware xApp decisions
&
Uses graph structure and RL for coordinated network control.
&
\cite{
    orhan2021connection
}
\\

\cline{2-6}

&
Cross-layer learning
&
Heterogeneous-agent PPO, digital twins, and rApp/xApp coordination
&
Multi-objective xApps and non-RT/near-RT closed loops
&
Supports coordinated learning, validation, and cross-loop interoperability.
&
\cite{
    he2025heterogeneous,
    he2025digital,
    giannopoulos2026interoperable
}
\\

\hline

% ============================================================
% II. Hierarchical MARL across loops
% 3 rows
% ============================================================

\multirow[c]{3}{=}[-1.5\baselineskip]{
    \centering
    \bfseries
    \makecell[c]{
        Hierarchical MARL\\
        across loops\\
        (\S\ref{sec:hierarchical_marl})
    }
}
&
Multi-RAT traffic steering
&
Hierarchical DRL
&
Strategic access-network selection and short-term user association
&
Decomposes steering into high-level and low-level decisions.
&
\cite{habib2023hierarchical}
\\

\cline{2-6}

&
Slice/resource hierarchy
&
Multi-timescale DRL and hierarchical resource management
&
Non-RT RIC policies and near-RT RIC slice/resource adaptation
&
Separates long-term orchestration from short-term radio control.
&
\cite{mhatre2024intelligent22}
\\

\cline{2-6}

&
Hierarchical federated slicing
&
Hierarchical slicing with federated DRL
&
Vehicular O-RAN slicing under dynamic traffic and mobility
&
Coordinates admission, scaling, and SLA enforcement.
&
\cite{hazarika2024enhancing}
\\

\hline

% ============================================================
% III. Federated DRL across vendors
% 4 rows
% ============================================================

\multirow[c]{4}{=}[-2.5\baselineskip]{
    \centering
    \bfseries
    \makecell[c]{
        Federated DRL across\\
        vendors\\
        (\S\ref{sec:federated_drl})
    }
}
&
Federated network slicing
&
Federated DRL and distributed slice learning
&
Multi-domain O-RAN slicing and slice resource allocation
&
Learns slice policies without exposing local data.
&
\cite{
    abouaomar2022federated,
    zhang2022federated,
    hazarika2024enhancing
}
\\

\cline{2-6}

&
Edge/MEC federated control
&
Federated DQN and federated DRL
&
Task offloading, fronthaul routing, and dynamic VNF splitting
&
Coordinates edge-cloud decisions with limited data transfer.
&
\cite{
    ndikumana2023federated,
    amiri2023edge
}
\\

\cline{2-6}

&
Vehicular and robotic systems
&
Distributed and federated learning
&
Multi-RIS vehicular networks and O-RAN-controlled multirobot systems
&
Extends federated learning to mobile cyber-physical scenarios.
&
\cite{
    alsenwi2025ran,
    ahmed2025federated
}
\\

\cline{2-6}

&
Secure federated platforms
&
Zero-trust FL, federated RL apps, and neuroevolution
&
Jamming mitigation, O-RAN dApps, xApps, and distributed learning infrastructure
&
Improves robustness and provides practical federation mechanisms.
&
\cite{
    abou2024federated,
    el2025secure,
    kouchaki2023openai,
    kouchaki2025federated
}
\\

\hline

% ============================================================
% IV. Communication-efficient MARL
% 4 rows
% ============================================================

\multirow[c]{4}{=}[-2.5\baselineskip]{
    \centering
    \bfseries
    \makecell[c]{
        Communication-\\
        efficient MARL\\
        (\S\ref{sec:communication_efficient_marl})
    }
}
&
Local team learning
&
Multi-agent team learning with partial observations
&
Virtualized and disaggregated O-RAN controllers
&
Reduces reliance on global state exchange.
&
\cite{iturria2022multi}
\\

\cline{2-6}

&
Distributed RAN allocation
&
MARL with decentralized execution
&
Multi-agent RAN resource allocation in O-RAN
&
Distributes allocation decisions across RAN entities.
&
\cite{rezazadeh2023multi}
\\

\cline{2-6}

&
Hierarchical communication
&
Multi-level DRL and hierarchical MARL
&
Network slicing, resource management, and edge-cloud mobility management
&
Limits exchange to adjacent layers or aggregate policies.
&
\cite{
    ghafouri2024multi,
    giarre2025hierarchical,
    hazarika2024enhancing
}
\\

\cline{2-6}

&
Sparse radio/edge control
&
Constrained MARL and local-information cooperation
&
Slice control, cell-free massive MIMO, distributed precoding,
and edge scheduling
&
Enforces coordination through constraints or selective information sharing.
&
\cite{
    zangooei2023flexible,
    he2025heterogeneous,
    shokouhi2025distributed,
    zhang2026optimized
}
\\

\hline

% ============================================================
% V. Game-theoretic perspectives
% 4 rows
% ============================================================

\multirow[c]{4}{=}[-2.5\baselineskip]{
    \centering
    \bfseries
    \makecell[c]{
        Game-theoretic\\
        perspectives\\
        (\S\ref{sec:game_theoretic})
    }
}
&
Matching-game slicing
&
Distributed matching game with DRL
&
Elastic O-RAN slicing for industrial monitoring and control
&
Matches service demands with radio-computing resources.
&
\cite{abedin2022elastic}
\\

\cline{2-6}

&
Auction/resource pricing
&
Auction and cost-sharing mechanisms
&
Multi-tenant x-haul and cloud resource allocation
&
Supports fair and truthful multi-tenant allocation.
&
\cite{mondal2023fairauction}
\\

\cline{2-6}

&
Stackelberg spectrum trading
&
Stackelberg game, blockchain trading, and MADDPG
&
Spectrum trading and network slicing
&
Captures leader--follower spectrum trading.
&
\cite{boateng2022consortium33}
\\

\cline{2-6}

&
Cooperative and constrained MARL
&
Team games, constrained MARL, and equilibrium-aware evaluation
&
Distributed RAN allocation, TN--NTN services,
distributed precoding, slicing, and xApp evaluation
&
Connects O-RAN MARL with stability, fairness, and incentives.
&
\cite{
    iturria2022multi,
    rezazadeh2023multi,
    seid2025multiagent,
    shokouhi2025distributed,
    lotfi2025task,
    zangooei2023flexible,
    tsampazi2024pandora
}
\\

\hline

\end{tabular}
\vspace{-2mm}

\end{table*}

\subsection{Game-Theoretic Perspectives}
\label{sec:game_theoretic}

Game-theoretic perspectives offer a useful lens for analyzing multi-agent and federated DRL in O-RAN because many control problems involve multiple decision makers with partially aligned or conflicting objectives. xApps, rApps, slices, tenants, radio units, edge-cloud nodes, and service classes may simultaneously compete for spectrum, computing, fronthaul capacity, scheduling priority, and slice quotas. In such settings, a single-agent MDP formulation can optimize one controller's policy, but it often hides strategic interactions among distributed entities. Game-theoretic models complement MARL by making these interactions explicit through cooperative games, non-cooperative games, matching games, auctions, Stackelberg games, and constrained stochastic games.

Existing O-RAN studies already reflect this connection in several concrete forms. 
In \cite{abedin2022elastic}, Abedin \emph{et al.} combine a distributed matching game with DRL for elastic O-RAN slicing in industrial monitoring and control, where the matching model captures the association between service demands and available radio-computing resources while DRL adapts slicing decisions under dynamic traffic conditions.
In \cite{mondal2023fairauction}, Mondal and Ruffini study fairness-guaranteed and auction-based x-haul and cloud resource allocation in multi-tenant O-RANs, using auction and cost-sharing mechanisms to support truthful resource allocation among tenants with different demands. 
In \cite{boateng2022consortium33}, Boateng \emph{et al.} further develop a consortium-blockchain-based spectrum trading framework for network slicing, where Stackelberg-game-based resource trading is combined with multi-agent DRL to coordinate strategic spectrum allocation. These works show that game-theoretic structure is particularly valuable when O-RAN control involves tenant fairness, pricing, resource trading, or incentive compatibility.

A second line of work connects game-theoretic reasoning with cooperative and constrained MARL. Team-learning formulations in virtualized O-RAN treat distributed controllers as agents that must coordinate under partial observations and shared network objectives \cite{iturria2022multi}. MARL-based RAN resource allocation, TN-NTN service management, distributed precoding, and task-specific O-RAN resource management further illustrate how local agents can learn policies whose collective behavior determines system-level efficiency and fairness \cite{rezazadeh2023multi,seid2025multiagent,shokouhi2025distributed,lotfi2025task}. Constrained MARL-based slicing also introduces game-like coupling among slices, because each slice-level decision affects the feasible resource region and SLA satisfaction of the others \cite{zangooei2023flexible}.

Overall, game-theoretic analysis exposes an important limitation of reward-driven DRL: \emph{locally high rewards do not necessarily imply stable, fair, or incentive-compatible multi-agent behavior.} 
Comparative O-RAN xApp studies show that different reward definitions and action spaces can produce conflicting or unfair resource-allocation outcomes even when individual agents appear effective in isolation \cite{tsampazi2024pandora}. Future work should therefore integrate DRL with equilibrium analysis, mechanism design, auction models, contract theory, mean-field approximations, and incentive-aware reward shaping. Such hybrid approaches are especially important for multi-vendor and multi-tenant O-RAN deployments, where autonomous agents must not only improve performance but also remain stable, interpretable, and fair under strategic interaction.

\subsection{Summary}

% This section reviewed multi-agent and federated DRL for distributed O-RAN control, including inter-xApp coordination, hierarchical MARL across control loops, federated learning across vendors and domains, communication-efficient coordination, and game-theoretic approaches. These paradigms distribute learning and decision making across multiple network entities, improving scalability, coordination, and data privacy while addressing the limitations of centralized DRL in large-scale O-RAN deployments. Table~\ref{tab:marl_fdrl_oran_taxonomy} summarizes representative multi-agent and federated DRL studies in the literature.

Table~\ref{tab:marl_fdrl_oran_taxonomy} summarizes representative multi-agent and federated DRL studies for distributed O-RAN control. This section reviewed inter-xApp coordination, hierarchical MARL, federated learning across vendors and domains, communication-efficient coordination, and game-theoretic approaches. Collectively, these paradigms distribute learning and decision making across network entities, improving scalability, coordination, and data privacy in large-scale O-RAN deployments.

\begin{figure*}[!t]
    \centering
    \includegraphics[width=\textwidth]{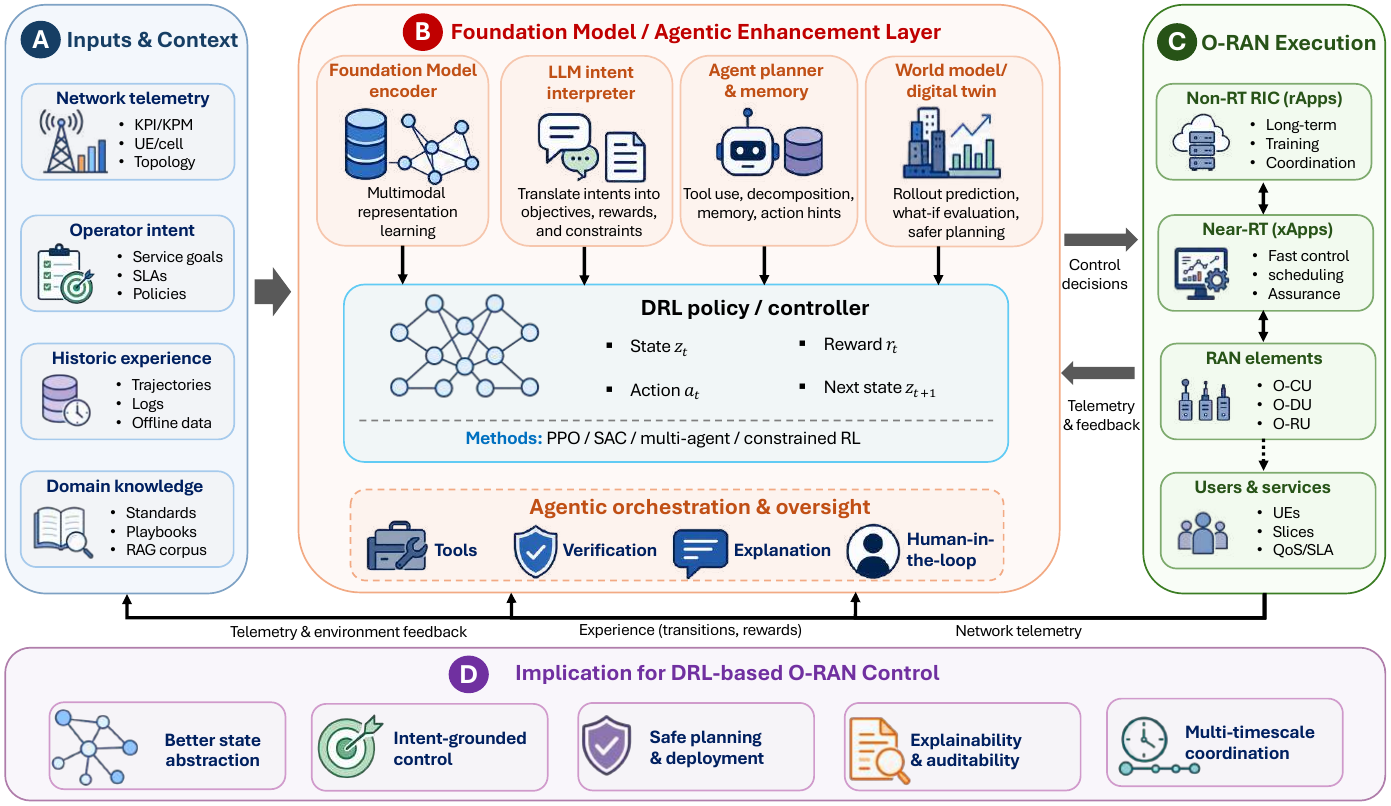}
    \caption{Foundation-model- and agentic-AI-assisted DRL pipeline for O-RAN control.}
    \label{fig:fm_agentic_drl_pipeline}
\end{figure*}

% \lj{A figure in each section to summarize all references.}

% \subsection{Inter-xApp Coordination}

% \subsection{Hierarchical MARL Across Loops}

% \subsection{Federated DRL Across Vendors}

% \subsection{Communication-Efficient MARL}

% \subsection{Game-Theoretic Perspectives}

% \clearpage 

\section{Foundation Models and Agentic AI for DRL-Based O-RAN Control}
\label{7_FM_DRL}

\newlength{\fmTableWidth}

% ============================================================
% Table VII
% ============================================================

\begin{table*}[!t]
\centering
\caption{Taxonomy of studies on foundation-model- and agentic-AI-assisted DRL for O-RAN control.}
\label{tab:fm_agentic_drl_taxonomy}

\scriptsize

% Horizontal padding inside cells
\setlength{\tabcolsep}{2pt}

% Vertical spacing
\renewcommand{\arraystretch}{1.12}

% ============================================================
% Calculate actual available width
%
% 6 columns -> 12\tabcolsep
% 7 vertical rules -> 7\arrayrulewidth
% 1pt safety margin
% ============================================================

\setlength{\fmTableWidth}{%
    \dimexpr
    \linewidth
    -12\tabcolsep
    -7\arrayrulewidth
    -1pt
    \relax
}

% ============================================================
% Normalized column-width ratios
%
% Original:
% 0.112 : 0.129 : 0.148 : 0.207 : 0.235 : 0.090
%
% Normalized:
% 0.121607 : 0.140065 : 0.160695 :
% 0.224756 : 0.255157 : 0.097720
% ============================================================

\begin{tabular}{
    |C{0.121607\fmTableWidth}
    |L{0.140065\fmTableWidth}
    |L{0.160695\fmTableWidth}
    |L{0.224756\fmTableWidth}
    |L{0.255157\fmTableWidth}
    |C{0.097720\fmTableWidth}|
}

% ============================================================
% Header
% ============================================================

\hline
\rowcolor{gray!20}

\textbf{Control Dimension}
&
\textbf{Research Focus}
&
\makecell[c]{
    \textbf{Enabling}\\
    \textbf{Model/Technique}
}
&
\textbf{O-RAN Use Case}
&
\textbf{DRL Integration}
&
\textbf{Refs.}
\\
\hline

% ============================================================
% I. Representation and grounding
% 4 rows
%
% Important:
% calibrated vertical shift = -4.5\baselineskip
% ============================================================

\multirow[c]{4}{=}[-2\baselineskip]{%
    \centering
    \bfseries
    \makecell[c]{%
        Representation\\
        and grounding\\
        (\S\ref{sec:fm_state_representation})
    }
}
&
Wireless/RAN encoders
&
Signal-level foundation model
&
PHY-aware RAN intelligence, beam/angle prediction, RF fingerprinting
&
Builds reusable state encoders for PHY-aware DRL and improves transfer across radio tasks.
&
\cite{
    mashaal2026iqfm,
    liurobust
}
\\

\cline{2-6}

&
Latent slicing representation
&
Generative and contrastive learning
&
O-RAN slicing with limited labeled data
&
Improves latent state abstraction and data efficiency for O-RAN slicing control.
&
\cite{
    nouri2024generative
}
\\

\cline{2-6}

&
Domain-specific LLM/FM
&
O-RAN knowledge foundation
&
O-RAN knowledge grounding and control support
&
Supplies domain knowledge, explanations, and reusable priors for DRL agents.
&
\cite{
    gajjar2025oransight,
    soliman2025foundation
}
\\

\cline{2-6}

&
LLM-assisted operation
&
Network operator intelligence
&
6G RAN operation and wireless network management
&
Supports intent interpretation, diagnosis, monitoring, and high-level control guidance.
&
\cite{
    oluwaseyi2025llm,
    navidan2025closed,
    alkuwaiti6577052distributed
}
\\

\hline

% ============================================================
% II. Intent-driven DRL and resource control
% 4 rows
%
% Important:
% calibrated vertical shift = -4.8\baselineskip
% ============================================================

\multirow[c]{4}{=}[-3\baselineskip]{%
    \centering
    \bfseries
    \makecell[c]{%
        Intent-driven DRL\\
        and resource\\
        control\\
        (\S\ref{sec:fm_intent_drl_control})
    }
}
&
LLM-enabled RL
&
Reward and policy guidance
&
Wireless resource allocation and network optimization
&
Assists reward design, action recommendation, and sequential resource optimization.
&
\cite{
    zheng2026large,
    sun2026large
}
\\

\cline{2-6}

&
Intent processing
&
Semantic task specification
&
Intent-driven network optimization and 6G control
&
Maps high-level intents into states, rewards, constraints, and control workflows.
&
\cite{
    habib2025llm,
    liu2025lameta,
    zaidi2026reasoning
}
\\

\cline{2-6}

&
O-RAN slicing
&
Prompt- and RAG-assisted DRL
&
Dynamic O-RAN network slicing
&
Structures context, prompts, reward hints, and learning signals for DRL-based slicing.
&
\cite{
    lotfi2025llm,
    lotfi2025prompt,
    lotfi2025oran,
    lotfi2026scalable
}
\\

\cline{2-6}

&
Service-aware control
&
Intent-driven scheduling and prioritization
&
Inter-slice prioritization, edge scheduling, and cell-free O-RAN optimization
&
Supports inter-slice prioritization, edge scheduling, and cell-free O-RAN optimization.
&
\cite{
    chiaranillm,
    sun2026igaa,
    shokouhi2026agentic
}
\\

\hline

% ============================================================
% III. Agentic and hierarchical control
% 6 rows
%
% calibrated vertical shift = -6.1\baselineskip
% ============================================================

\multirow[c]{6}{=}[-5\baselineskip]{%
    \centering
    \bfseries
    \makecell[c]{%
        Agentic and\\
        hierarchical\\
        control\\
        (\S\ref{sec:fm_agentic_hierarchical_control})
    }
}
&
RIC-layer architecture
&
Hierarchical agentic O-RAN
&
Autonomous O-RAN control across non-RT and near-RT RIC layers
&
Embeds DRL policies within hierarchical, intent-driven, and auditable O-RAN control loops.
&
\cite{
    elkael2026agentran,
    bao2026llm,
    li2026agentic
}
\\

\cline{2-6}

&
Multi-scale autonomy
&
Self-evolving AI-RAN agents
&
Real-time O-RAN control and AI service provisioning
&
Coordinates perception, reasoning, adaptation, and service provisioning across RAN loops.
&
\cite{
    navidan2026toward,
    he2026agentic,
    natanzi2026advanced
}
\\

\cline{2-6}

&
Agent state and memory
&
Representation-, intention-, and memory-driven agents
&
AI-RAN state understanding, intent grounding, and long-term memory
&
Links telemetry, historical context, and high-level reasoning for DRL-based control.
&
\cite{
    ding2025ridas,
    wang2026bridging
}
\\

\cline{2-6}

&
AI-native 6G agents
&
Agent-based network autonomy
&
6G AI-RAN orchestration and autonomous network operation
&
Defines system-level requirements for agentic coordination around learning-based policies.
&
\cite{
    ferrag20266g,
    gajjar2026agents,
    altintas2026beyond
}
\\

\cline{2-6}

&
AI-native O-RAN operation
&
Real-time adaptation and conflict resolution
&
6G O-RAN resource management and conflict resolution
&
Connects low-latency adaptation, conflict handling, and resource control with learning-based policies.
&
\cite{
    salmi2026ai
}
\\

\cline{2-6}

&
Multi-agent orchestration
&
Negotiation and safe automation
&
Network function orchestration, rApp coordination, and non-RT RIC automation
&
Adds coordination, explainability, safety checks, and conflict handling around xApp/rApp and DRL decisions.
&
\cite{
    gemayel2025network,
    oranconflictllm2026,
    baena2026knows,
    ngo2026llm
}
\\

\hline

% ============================================================
% IV. Trajectory and world modeling
% 4 rows
%
% calibrated vertical shift = -4.0\baselineskip
% ============================================================

\multirow[c]{4}{=}[-2.5\baselineskip]{%
    \centering
    \bfseries
    \makecell[c]{%
        Trajectory and\\
        world modeling\\
        (\S\ref{sec:fm_trajectory_world_models})
    }
}
&
Decision Transformer
&
Return-conditioned sequence modeling
&
Intent-driven RAN management and RAN automation
&
Enables offline trajectory modeling, long-horizon policy learning, and intent-conditioned action generation.
&
\cite{
    habib2025harnessing,
    habib2026reimagining
}
\\

\cline{2-6}

&
Mamba-based decision model
&
Efficient long-context temporal modeling
&
Intent-driven 6G RAN management and RAN slicing
&
Processes long telemetry histories for efficient sequential decision making.
&
\cite{
    habib2026generative,
    habib2026hierarchical
}
\\

\cline{2-6}

&
World model and digital twin
&
Predictive planning and simulation-in-the-loop control
&
Predictive planning, digital-twin validation, RAN synthesis, and network testing
&
Supports model-based evaluation, synthetic rollouts, and safer validation before live deployment.
&
\cite{
    rezazadeh2025agentic,
    zou2026telecom,
    hu2025reflection,
    aghayev2026genesis
}
\\

\cline{2-6}

&
Agentic DRL foundation
&
RL for agent behavior and communication
&
Wireless agent control and future industrial AI-agent communication
&
Clarifies how DRL supports autonomous agent behavior, reliable communication, and coordination.
&
\cite{
    zheng2026advanced,
    zhou2026convergence
}
\\

\hline

\end{tabular}

\end{table*}

\subsection{Overview}
\label{sec:fm_agentic_drl_overview}
The openness, programmability, and disaggregation of O-RAN create a demanding decision-making environment for DRL agents. A practical O-RAN controller must reason over heterogeneous radio measurements, traffic dynamics, slice-level KPIs, service intents, topology changes, and control-loop feedback while adapting to changing numbers of UEs, cells, slices, links, xApps, and rApps.

Most conventional DRL methods target specific tasks and deployment settings, which limits their transferability and increases their dependence on costly online interaction. Foundation models and agentic AI complement these methods by introducing large-scale pretraining, reusable representations, contextual reasoning, tool use, memory, and sequence-level decision modeling into the DRL pipeline \cite{zhang2025multi,soliman2025foundation,wei2025large,cai2025tutorial}. Recent studies on LLM-enabled network operation, LLMs for 6G RAN, decision-making LLMs, and agentic AI-native networks reinforce this direction \cite{oluwaseyi2025llm,ccimen2025overview,yang2025decision,altintas2026beyond,ferrag20266g,gajjar2026agents}.

Foundation models and agentic AI augment rather than replace DRL in O-RAN control. DRL remains suitable for sequential optimization, while foundation models provide representation learning, semantic grounding, prior knowledge, trajectory-level reasoning, and human-interpretable feedback. Agentic AI further supports task decomposition, tool use, memory, and multi-step orchestration \cite{alkuwaiti6577052distributed,zheng2026large,zhou2024large}. The following generic pipeline summarizes these complementary roles:
\begin{equation}
    z_t = f_{\phi}(s_t,I_t,h_t,\mathcal{K}), \quad
    a_t \sim \pi_{\theta}(a_t \mid z_t,I_t),
    \label{eq:fm_agentic_control_pipeline}
\end{equation}
where $s_t$ is the O-RAN state/observation, $I_t$ is the intent or task context, $h_t$ is historical context, $\mathcal{K}$ is domain knowledge, $z_t$ is the learned representation, and $\pi_{\theta}$ is the DRL policy. 

This abstraction shows how foundation models can enter before policy learning as encoders, during learning as reward or policy guides, and after action execution as explanation, verification, or planning modules. Agentic components can orchestrate these modules and coordinate their interactions with xApps, rApps, and RIC control loops.
Fig.~\ref{fig:fm_agentic_drl_pipeline} expands this abstraction into an end-to-end O-RAN control pipeline, linking heterogeneous inputs, foundation-model and agentic enhancement modules, the DRL controller, and RIC-layer execution.

% \begin{figure*}[!t]
%     \centering
%     \includegraphics[width=\textwidth]{figures/7_telemetry1.pdf}
%     \caption{Foundation-model- and agentic-AI-assisted DRL pipeline for O-RAN control.}
%     \label{fig:fm_agentic_drl_pipeline}
% \end{figure*}

This section reviews foundation models and agentic AI for DRL-based O-RAN control from four technical perspectives. First, foundation models can learn reusable state representations from multimodal wireless observations, I/Q streams, topology information, and variable-size network entities. Second, LLMs can augment DRL by translating intents into objectives, constraints, prompts, rewards, or action guidance. Third, agentic AI architectures can organize LLMs, tools, xApps, rApps, and hierarchical RIC control loops for autonomous O-RAN operation. Fourth, trajectory and world models can exploit historical control sequences, digital twins, and learned dynamics for offline learning and predictive planning. The section concludes by synthesizing the implications of these directions for DRL-based O-RAN control.

Table~\ref{tab:fm_agentic_drl_taxonomy} summarizes representative studies along the same four directions and highlights their control dimensions, enabling techniques, O-RAN use cases, and DRL integration roles.

\subsection{Foundation Models for Network State Representation}
\label{sec:fm_state_representation}

Network state representation is a critical interface between O-RAN environments and DRL agents. Classical DRL often assumes a fixed-dimensional state vector, but O-RAN observations are heterogeneous and deployment-dependent. They include physical-layer signals, traffic demand, UE mobility, slice KPIs, topology information, xApp/rApp feedback, and service intents. Moreover, the number of UEs, cells, slices, links, and control applications can change over time. Foundation models address this mismatch by learning reusable representations from large-scale wireless data and multimodal telemetry \cite{zhang2025multi,mashaal2026iqfm,zou2026telecom}.

\textbf{Wireless and Multimodal Foundation Models.}
Several studies examine foundation models as reusable wireless encoders. Zhang \emph{et al.} \cite{zhang2025multi} survey multimodal data-enhanced foundation models for wireless prediction and control and explain how these models can jointly encode radio signals, traffic traces, topology information, and service-level context. Mashaal and Abou-Zeid \cite{mashaal2026iqfm} introduce IQFM, a wireless foundation model that operates directly on I/Q streams. Their results show how pretraining supports modulation classification, beam prediction, angle-of-arrival estimation, and RF fingerprinting without relying heavily on handcrafted features.

Nouri \emph{et al.} \cite{nouri2024generative} propose a generative semi-supervised VAE-contrastive learning framework for O-RAN slicing and show that generative and contrastive objectives improve representation robustness when labels remain limited. 
Liu \emph{et al.} \cite{liurobust} explore robust foundation-model-empowered RAN intelligence in embodied robot scenarios. Although their setting extends beyond O-RAN control, it illustrates how pretrained radio representations may transfer across RAN tasks.

\textbf{Latent and Task-conditioned State Abstraction.}
For DRL, the central question is how these pretrained representations are integrated into the policy. 
In the generic control pipeline of Eq.~\eqref{eq:fm_agentic_control_pipeline}, a pretrained or fine-tuned encoder $f_{\phi}$ maps the raw state/observation $s_t$, service intent or task context $I_t$, temporal context $h_t$, and domain knowledge $\mathcal{K}$ into the compact latent state $z_t$. 
The DRL policy and value function can then operate on $z_t$ instead of raw observations. This abstraction is consistent with LLM-enabled RL studies that view foundation models as modules for reducing state complexity and improving transfer across wireless optimization tasks \cite{zheng2026large,cai2025tutorial}. It is also aligned with telecom world models, which treat latent states as the basis for prediction and planning \cite{zou2026telecom}.

\textbf{Variable-size, Set-based, and Topology-aware Representations.}
O-RAN states/observations often consist of sets or graphs rather than fixed vectors. For example, a state may include a variable number of UEs, slices, cells, or xApps. Set encoders, attention pooling, and permutation-invariant aggregation can make the representation independent of entity ordering and input size. In parallel, graph representations can capture interference relations, UE-cell associations, fronthaul dependencies, and interactions among control applications. Surveys on multimodal foundation models and decision-making LLMs identify GNNs and topology-aware learning as important tools for wireless prediction and control \cite{zhang2025multi,yang2025decision}. These mechanisms are especially important for O-RAN because a policy trained in one cell cluster or slicing configuration should not depend on a fixed input dimension.

\textbf{Temporal Representation.}
O-RAN states/observations are also temporally dependent because traffic bursts, mobility, congestion, and control-loop effects evolve over time. Temporal encoders can summarize recent observation-action histories, while world models and trajectory-based models learn latent dynamics for prediction and planning.
In this research line, Rezazadeh \emph{et al.} \cite{rezazadeh2025agentic} develop a causal O-RAN world model that compresses KPI and physical resource block histories into a stochastic latent state and produces forecasts conditioned on candidate actions for counterfactual planning. Zou \emph{et al.} \cite{zou2026telecom} formulate the Telecom World Model as a factored POMDP that separates controllable variables from exogenous processes and combines field prediction, network dynamics, and a telecom foundation model for KPI trajectory prediction under uncertainty. Habib \emph{et al.} \cite{habib2026reimagining} propose a hierarchical online decision transformer that uses intents and trajectory context and combines offline pretraining with online fine-tuning.

These studies show how historical context can strengthen the latent state passed to DRL. Section~\ref{sec:fm_trajectory_world_models} will examine how trajectory and world models use this context for action selection and planning.

\subsection{LLM-Augmented DRL for Intent-Driven O-RAN Control}
\label{sec:fm_intent_drl_control}

Intent-driven O-RAN control allows operators or vertical applications to specify high-level objectives, such as maximizing slice throughput, reducing URLLC latency, improving energy efficiency, or maintaining SLA compliance, without manually configuring low-level radio parameters. Operators often express these intents in natural language or service-level terms, whereas DRL agents require states, actions, rewards, and constraints. LLM-augmented DRL bridges this semantic gap by interpreting intents, retrieving domain knowledge, generating task specifications, and guiding DRL agents.

\textbf{LLM Roles and Representative Frameworks.}
Cai \emph{et al.} \cite{cai2025tutorial} identify four roles for LLMs in DRL: state perception, reward design, decision making, and generation. 
Zheng \emph{et al.} \cite{zheng2026large} demonstrate LLM-based state representation and semantic extraction in multi-agent RL for service migration and request routing across UAV and satellite networks. 
Yang \emph{et al.} \cite{yang2025decision} organize wireless decision making around prompt learning, chain-of-thought reasoning, inference, decision frameworks, and multi-agent coordination.

Tageldien \emph{et al.} \cite{tageldien2025large} map LLMs to the five stages of the intent lifecycle. 
Sidhu and Sharma \cite{sidhu2025artificial} review AI support for intent interpretation, predictive assurance, anomaly detection, and resource optimization. 
Hong \emph{et al.} \cite{hong2025comprehensive} examine LLM applications in network design, configuration, fault management, security, and orchestration, together with hallucination and domain adaptation risks. 
For O-RAN control, Oluwaseyi \emph{et al.} \cite{oluwaseyi2025llm} formalize an LLM-RAN operator that maps intents and network states to validated commands, with strategic guidance in the non-RT RIC and reactive execution in the near-RT RIC.

\textbf{Intent-to-Control Translation.}
An LLM can translate a high-level intent $I$ and O-RAN knowledge $\mathcal{K}$ to a task specification, i.e.,
\begin{equation}
    \tau_I = g_{\psi}(I,\mathcal{K})
    =(\mathcal{S}_I,\mathcal{A}_I,r_I,\mathcal{C}_I,\Omega_I),
\end{equation}
where $\mathcal{S}_I$, $\mathcal{A}_I$, $r_I$, $\mathcal{C}_I$, and $\Omega_I$ represent intent-conditioned state variables, actions, reward components, constraints, and auxiliary control knowledge. 
Based on this formulation, Habib \emph{et al.} \cite{habib2025llm} study LLM-based intent processing with attention-based hierarchical reinforcement learning and show how the resulting workflow maps intents into network optimization tasks. 
Liu \emph{et al.} \cite{liu2025lameta} propose LAMeTA, in which a large AI model supports intent-aware network optimization through a two-stage design. 
Sun \emph{et al.} \cite{sun2026igaa} study intent-driven agentic scheduling for edge services, while Shokouhi and Wong \cite{shokouhi2026agentic} investigate intent-driven optimization in cell-free O-RAN. Together, these studies position LLMs and large AI models as semantic interfaces between service objectives and control policies.

\textbf{LLM-Augmented DRL for O-RAN Slicing.}
Network slicing is one of the most direct use cases for LLM-augmented DRL. 
For this use case, Lotfi \emph{et al.} \cite{lotfi2025llm} propose LLM-augmented DRL for dynamic O-RAN slicing, where language models structure control information for DRL-based resource allocation. 
Follow-up studies extend this design through prompt-tuned LLM-augmented DRL \cite{lotfi2025prompt}, ORAN-GUIDE with RAG-driven prompt learning \cite{lotfi2025oran}, and scalable LLM-augmented DRL with context-aware prompt learning \cite{lotfi2026scalable}. 
In this family of methods, the LLM does not act as the low-level controller; instead, it interprets intents, summarizes context, produces reward hints, or provides prompt-conditioned guidance to the DRL agent. 
Chiarani \emph{et al.} \cite{chiaranillm} study LLM-guided reinforcement learning for adaptive inter-slice resource prioritization and further show how language guidance supports slicing decisions under service differentiation.

\textbf{Reward, Constraint, and Policy Guidance.}
Reward design is a major bottleneck in DRL-based O-RAN because a controller must balance throughput, latency, reliability, fairness, energy efficiency, and SLA satisfaction. LLMs can help decompose intents into measurable reward terms and constraints, reducing manual trial-and-error. 
Large-language-model-enabled reinforcement learning for wireless optimization \cite{zheng2026large} and the tutorial on LLM-enhanced RL \cite{cai2025tutorial} both highlight this role. LLMs can also provide policy priors, action preferences, or explanations, while DRL retains responsibility for sequential optimization. Sun \emph{et al.} \cite{sun2026large} study LLM-empowered resource allocation in intent-driven wireless networks and provide a non-DRL reference point in which LLMs act as direct resource-allocation assistants.

\textbf{Retrieval and Grounding.}
Prompting alone is fragile when the LLM lacks domain grounding. Retrieval-augmented generation can condition LLM outputs on O-RAN documentation, historical cases, and policy databases, thereby reducing hallucination and improving consistency. 
ORAN-GUIDE \cite{lotfi2025oran} provides a representative example of RAG-driven prompt learning for LLM-augmented RL in O-RAN slicing. 
Related studies on ReAct-style multimodal LLMs \cite{zaidi2026reasoning}, closed-loop LLM intelligence \cite{navidan2025closed}, and O-RAN-specific LLMs such as ORANSight-2.0 \cite{gajjar2025oransight} further establish grounding, reasoning, and explanation as central requirements for LLM-assisted O-RAN control. Together, these techniques ground the semantic guidance that enters the DRL pipeline and reduce the risk that unverified LLM outputs influence O-RAN actions.

\subsection{Agentic and Hierarchical AI for O-RAN Control}
\label{sec:fm_agentic_hierarchical_control}

Agentic AI extends foundation models from passive encoders or advisors to active components in multi-step control workflows. Agentic systems integrate LLMs, tools, memory, planning mechanisms, and specialized agents to decompose objectives, coordinate functions, and execute control procedures. This paradigm is relevant to O-RAN, where autonomous control is distributed across the Non-RT RIC, Near-RT RIC, xApps, rApps, and domain-specific control loops. Recent surveys and position papers on agentic AI-native networks argue that future 6G RAN control will require intelligent systems capable of perceiving, reasoning, coordinating, and adapting across these layers
\cite{altintas2026beyond,ferrag20266g,gajjar2026agents,he2026agentic}.

\textbf{Agentic O-RAN Architectures.}
Some research efforts have been invested in the study of agentic O-RAN architecture.
For example, Elkael \emph{et al.} \cite{elkael2026agentran} propose AgentRAN, an agentic AI architecture for autonomous control of open 6G networks. AgentRAN uses natural-language intents to generate and orchestrate distributed AI agents that interact with O-RAN control functions. Bao \emph{et al.} \cite{bao2026llm} propose LLM-hRIC, an LLM-empowered hierarchical RAN intelligent control framework that emphasizes cooperation between RIC layers and domain-specific fine-tuning. 
Li \emph{et al.} \cite{li2026agentic} propose an agentic O-RAN framework that separates LLM reasoning from deterministic and auditable radio control. 
Together, these studies establish a key design principle: LLMs and agents should support reasoning, orchestration, and policy generation, while latency-critical radio actions should remain constrained, auditable, and verifiable.

\textbf{Representation-Driven and Intention-Driven Agents.}
Several agentic frameworks explicitly connect representation, intent, and control. RIDAS \cite{ding2025ridas} introduces representation-driven and intention-driven agents for AI-RAN and bridges state representation with intent grounding in a unified agent design. Natanzi \emph{et al.} \cite{natanzi2026advanced} study advanced AI service provisioning in O-RAN through LLM engine integration, while Navidan \emph{et al.} \cite{navidan2026toward} propose a multi-scale agentic AI framework for real-time network control and management. Wang \emph{et al.} \cite{wang2026bridging} introduce a memory-centric paradigm for 6G agentic AI-RAN and argue that memory bridges low-level telemetry with high-level reasoning. These studies shift the discussion from single-agent DRL toward layered and memory-enabled autonomous control.

\textbf{Multi-agent Orchestration and Conflict Handling.}
O-RAN naturally involves multiple control applications. As xApps and rApps become agentic, conflicts may arise from inconsistent objectives, overlapping control scopes, or incompatible policies. 
Li \emph{et al.} \cite{oranconflictllm2026} study conflict-aware rApp policy orchestration in O-RAN, while Baena \emph{et al.} \cite{baena2026knows} investigate semantic negotiation for human-supervised RAN agentic coordination. Gemayel and Mokh \cite{gemayel2025network} explore LLM-based multi-agent systems for network function orchestration. Ngo \emph{et al.} \cite{ngo2026llm} propose an LLM-based Net Analyzer rApp for explainable and safe automation in the non-RT RIC. These studies complement DRL by addressing coordination, diagnosis, explanation, and safety functions that are difficult to capture with a single reward-maximizing policy.

\textbf{Relation to DRL.}
From a DRL perspective, these agentic functions surround rather than replace the policy learner: they translate intents, maintain memory, coordinate xApps and rApps, resolve conflicts, and verify candidate actions. The DRL controller remains responsible for sequential optimization within the resulting objectives and constraints.

\subsection{Trajectory and World Models for Sequential Decision Making}
\label{sec:fm_trajectory_world_models}

O-RAN control is inherently sequential. Resource allocation, slicing, mobility management, interference coordination, and xApp/rApp policy adaptation all exhibit delayed effects. An action taken at one time may influence future congestion, SLA satisfaction, handover behavior, and resource availability. This motivates trajectory models and world models, which learn from sequences of observations, actions, rewards, and outcomes. In foundation-model- and agentic-AI-assisted DRL, these models enable offline learning from logs, long-horizon dependency modeling, and predictive planning before the live RAN executes candidate actions \cite{habib2025harnessing,zou2026telecom,rezazadeh2025agentic}.

\textbf{Trajectory Modeling and Decision Transformers.}
Decision Transformers reformulate RL as conditional sequence modeling and predict actions from desired return-to-go and historical trajectories. Habib \emph{et al.} \cite{habib2025harnessing} combine LLMs, Informers, and Decision Transformers for intent-driven RAN management and show how their sequence-modeling framework connects service intents with historical trajectories. Habib \emph{et al.} \cite{habib2026reimagining} further propose hierarchical online Decision Transformer ideas for RAN automation and emphasize agentic decision making over temporal network contexts. These studies show how offline logs can initialize trajectory-conditioned policies, while online variants can adapt through stochastic exploration and continual trajectory-level updates.

\textbf{Mamba and Efficient Long-context Decision Models.}
Transformer-based trajectory models can be expensive for long telemetry sequences. Mamba and state-space sequence models provide a more efficient alternative for capturing long-range temporal dependencies. Habib \emph{et al.} \cite{habib2026generative} study generative AI for intent-driven 6G RAN management using the Mamba model, and Habib \emph{et al.} \cite{habib2026hierarchical} propose Hierarchical Decision Mamba for RAN slicing. These studies represent a broader trend toward sequence architectures that process long network histories more efficiently.

\textbf{World Models, Digital Twins, and Simulation-in-the-Loop Planning.}
World models aim to learn network dynamics rather than only imitate past actions. A generic world model takes the form
\begin{equation}
    \hat{s}_{t+1},\hat{r}_t = W_{\omega}(s_t,a_t,h_t),
\end{equation}
where $W_{\omega}$ predicts the future state and reward under a candidate action.

Based on such a world model, Zou \emph{et al.} \cite{zou2026telecom} introduce telecom world models that unify digital twins, foundation models, and predictive planning. Rezazadeh \emph{et al.} \cite{rezazadeh2025agentic} propose agentic world modeling for near-real-time generative state-space reasoning in 6G. Hu \emph{et al.} \cite{hu2025reflection} study reflection-driven self-optimization with simulation-in-the-loop workflows. Together, these studies show how predictive models, uncertainty estimates, and digital twins can evaluate candidate actions before the live RAN executes them. GENESIS \cite{aghayev2026genesis} further highlights how AI agents can support autonomous 6G RAN synthesis, research, and testing, thereby providing environments for validating learned control strategies.

\textbf{Relation to DRL.}
Trajectory and world models are not substitutes for DRL. Instead, they provide pretrained sequence representations, offline policy initialization, synthetic rollouts, or planning modules that can improve sample efficiency and safety. This view also connects with broader studies on advanced DRL for agentic AI \cite{zheng2026advanced} and reinforcement learning for future industrial AI-agent communication \cite{zhou2026convergence}, where sequential decision making and reliable agent communication are treated as coupled problems.

\subsection{Key Takeaways}
\label{sec:fm_agentic_drl_takeaways}

Existing studies position foundation models and agentic AI as complementary layers around DRL rather than direct replacements for DRL-based optimization. Foundation models primarily improve the information interface of DRL by learning reusable representations, grounding intents, extracting temporal context, and supporting predictive reasoning. Agentic AI primarily improves the control workflow by decomposing objectives, invoking tools, coordinating xApps and rApps, and linking high-level reasoning with auditable radio-control procedures. This division is consistent with recent studies on wireless foundation models, LLM-augmented reinforcement learning, O-RAN-specific LLMs, hierarchical RIC control, and agentic AI-native RAN architectures \cite{zhang2025multi,mashaal2026iqfm,zheng2026large,cai2025tutorial,gajjar2025oransight,bao2026llm,elkael2026agentran,salmi2026ai}.

From a DRL perspective, the most immediate benefits are improved state abstraction, intent-conditioned reward and constraint design, offline policy learning, and safer policy evaluation through trajectory and world models. These capabilities are especially relevant to O-RAN because the environment is heterogeneous, partially observable, topology-dependent, and governed by multiple control loops. At the same time, existing studies show that O-RAN designers must integrate explainability, grounding, conflict handling, and verifiability with learning performance rather than add these deployment-oriented properties after policy training \cite{lotfi2025oran,habib2025harnessing,zou2026telecom,rezazadeh2025agentic,guo2025towards,oranconflictllm2026,baena2026knows,ngo2026llm}.

Overall, foundation-model- and agentic-AI-assisted DRL provides an emerging bridge between data-driven O-RAN optimization and autonomous network operation. 
%The paper-level challenges and future research section examines the broader, cross-cutting issues of data availability, real-time inference, safety assurance, cross-deployment generalization, and benchmarking.

% \clearpage 

\section{Trustworthy DRL for O-RAN}
\label{8_Trustworthy_DRL}
\begin{figure*}
    \centering
    \includegraphics[width=1\linewidth]{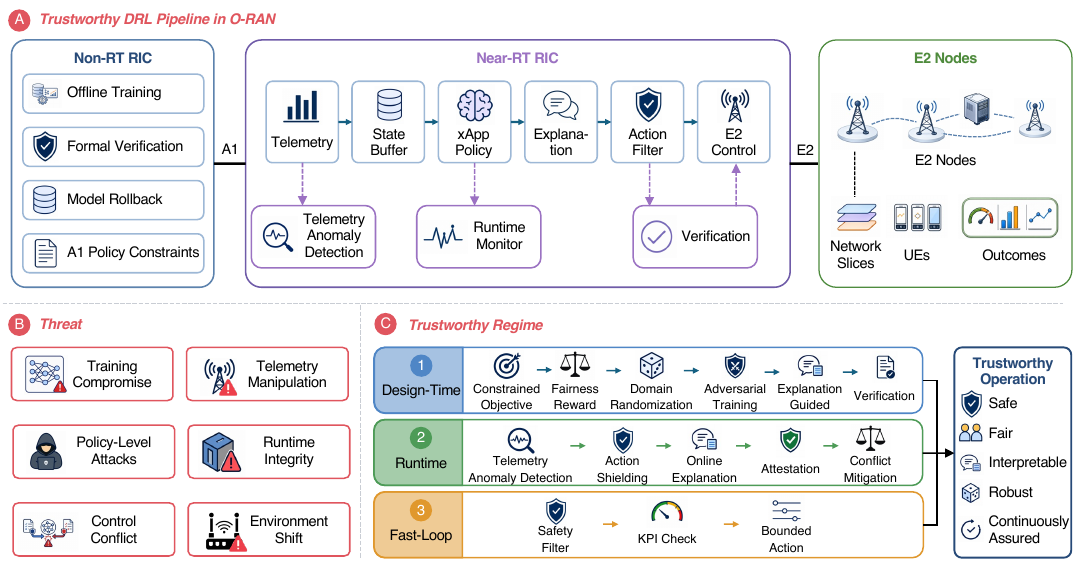}
    \caption{Trustworthy DRL pipeline in O-RAN. \textit{Part A} maps trustworthy DRL onto the O-RAN environment, \textit{Part B} summarizes the main threats, and \textit{Part C} groups the main trustworthy regimes by control loop.}
    \label{fig:trust}
\end{figure*}

 \newlength{\trustTableWidth}

% ============================================================
% Trustworthy DRL Mapping Table
% ============================================================

\begin{table*}[!t]
\centering
\caption{Trustworthy DRL dimensions mapped onto the O-RAN control loops.}
\label{tab:trust_mapping}

\scriptsize

% Horizontal padding inside cells
\setlength{\tabcolsep}{2pt}

% Vertical spacing
\renewcommand{\arraystretch}{1.12}

% ============================================================
% Calculate actual available width
%
% 4 columns -> 8\tabcolsep
% 5 vertical rules -> 5\arrayrulewidth
% 1pt safety margin
% ============================================================

\setlength{\trustTableWidth}{%
    \dimexpr
    \linewidth
    -8\tabcolsep
    -5\arrayrulewidth
    -1pt
    \relax
}

% ============================================================
% Normalized column-width ratios
%
% Original:
% 0.09 : 0.29 : 0.32 : 0.19
%
% Normalized:
% 0.101124 : 0.325843 : 0.359551 : 0.213483
% ============================================================

\begin{tabular}{
    |C{0.101124\trustTableWidth}
    |L{0.325843\trustTableWidth}
    |L{0.359551\trustTableWidth}
    |L{0.213483\trustTableWidth}|
}

% ============================================================
% Header
% ============================================================

\hline
\rowcolor{gray!20}

\textbf{Dimension}
&
\textbf{Failure Mode Addressed}
&
\textbf{Representative Mechanisms}
&
\textbf{Primary O-RAN Host}
\\
\hline

% ============================================================
% Safety
% ============================================================

\textbf{Safety}
&
Constraint violations during training and deployment (e.g., SLA, power, and interference limits).
&
Constrained MDP, Lagrangian methods, risk-sensitive objectives, barrier functions, and action projection.
&
Non-RT RIC (design), Near-RT RIC (enforcement).
\\
\hline

% ============================================================
% Fairness
% ============================================================

\textbf{Fairness}
&
Systematic resource starvation for UEs, slices, or cells.
&
Welfare objectives, fairness constraints, and fairness-aware rewards.
&
Near-RT RIC, Non-RT RIC.
\\
\hline

% ============================================================
% Explainability
% ============================================================

\textbf{Explainability}
&
Lack of transparency hinders accountability and network troubleshooting.
&
Feature attribution, reward decomposition, and interpretable surrogate policies.
&
Near-RT RIC (online), Non-RT RIC (offline).
\\
\hline

% ============================================================
% Robustness
% ============================================================

\textbf{Robustness}
&
Distribution shifts, non-stationarity, and adversarial telemetry manipulation.
&
Domain randomization, robust optimization, adversarial training, policy smoothness, and anomaly detection.
&
Non-RT RIC (training), Near-RT RIC (monitoring).
\\
\hline

% ============================================================
% Verification and monitoring
% ============================================================

\textbf{Verification and monitoring}
&
Undetected runtime faults, lack of behavioral guarantees, and conflicting agents.
&
Formal policy verification, runtime shielding, fallback controllers, and conflict mitigation.
&
Non-RT RIC (verification), Near-RT RIC (runtime).
\\
\hline

\end{tabular}

\end{table*}

Trustworthy DRL extends performance-oriented RAN control by requiring a learned policy to be not only effective, but also dependable under operational constraints. Conventional DRL evaluations often emphasize cumulative reward, convergence, and sample efficiency. Production O-RAN deployments, however, must additionally consider SLA violations, unfair resource allocation, behavior under unseen conditions, manipulated telemetry, and whether operators can inspect and trust automated decisions. These requirements are particularly important in O-RAN because learned policies may directly influence radio resources, mobility, slicing, and other network functions through closed-loop control.

We therefore view trustworthiness as a lifecycle property that spans policy design, validation, deployment, and runtime operation. Fig.~\ref{fig:trust} illustrates this perspective. 
% Part A places assurance mechanisms such as policy constraints, formal verification, action filtering, anomaly detection, runtime monitoring, and rollback within the O-RAN control architecture. Part B summarizes representative threats, including training compromise, telemetry manipulation, policy-level attacks, control conflicts, and environmental shifts. Part C organizes the corresponding assurance mechanisms across design-time, runtime, and fast-loop operation. 
This section reviews trustworthy DRL through five complementary dimensions: \emph{safety, fairness, explainability, robustness, and verification with runtime monitoring}. Table~\ref{tab:trust_mapping} maps these dimensions to their failure modes, representative mechanisms, and primary O-RAN hosts.

% ---------------------------------------------------------------------
\subsection{Safe DRL}
\label{Safe_DRL_ORAN}

Safe DRL augments reward maximization with mechanisms that limit harmful behavior during training and deployment. In O-RAN, safety requires a learned controller to respect service, physical, and operational limits, including SLA requirements, transmit-power bounds, interference constraints, and restrictions on actions that could destabilize a live service.

\textbf{Design-Time Safety Mechanisms.}
Design-time safety specifies the constraints and risk requirements that a policy should satisfy before it is admitted to the operational control loop. The standard objective in Eq.~(\ref{mdp_reward}) maximizes expected return but does not encode such limits by itself. Constrained MDPs, risk-sensitive objectives, and certificate-based methods provide complementary mechanisms for incorporating operational requirements into policy learning or action selection. Three representative approaches are as follows.

\begin{itemize}[leftmargin=0.15in]
\item
\emph{Constrained policy optimization.}
These methods maximize return subject to explicit cost or safety constraints, thereby separating performance objectives from operational limits. Constrained policy optimization solves the problem in Eq.~(\ref{cmdp_objective}) directly in the policy space. 
Following this approach, Achiam \emph{et al.} \cite{achiam2017constrained} restrict policy updates to a trust region so that expected costs remain below a predefined budget. Ray \emph{et al.} \cite{ray2019benchmarking} report that such methods typically satisfy constraints in expectation and most reliably near convergence, which motivates additional runtime safeguards during training and deployment.

\item
\emph{Risk-sensitive optimization.}
Risk-sensitive methods account for the severity or probability of unfavorable outcomes rather than optimizing only average return. This is useful when rare SLA or latency violations are operationally significant even if mean performance is high. 
Following this approach, Tamar \emph{et al.} \cite{tamar2015optimizing} optimize variance-aware objectives, while Chow \emph{et al.} \cite{chow2018risk} use conditional value-at-risk to reduce rare but severe failures.

\item
\emph{Certificate-based safety.}
Certificate-based methods associate the learned policy with a mathematical condition that characterizes a safe operating region. Examples include Lyapunov functions \cite{berkenkamp2017safe} and control barrier functions \cite{cheng2019end}. The controller can then reject or modify actions that violate the certificate. Related safety-layer methods project a proposed action onto a nearby feasible action that satisfies instantaneous constraints.
\end{itemize}

\textbf{O-RAN Deployment Considerations.}
Safe DRL is especially important in RAN environments because exploratory actions can immediately affect users. A common deployment pattern therefore trains a policy offline or in simulation and permits only constrained, monitored adaptation in the live network. This separation also clarifies different types of constraints: trajectory-level constraints can capture quantities such as accumulated latency or SLA violations over a time window, whereas instantaneous constraints can enforce per-step limits on transmit power, resource allocation, or interference.

\textbf{Examples in O-RAN.}
% \hz{why is it called ``evidence''?}
Recent O-RAN studies apply these safety mechanisms to slicing, handover, and coverage control. 
For examaple, Zangooei \emph{et al.} \cite{zangooei2023flexible} encode SLA limits as explicit constraints in cooperative multi-agent O-RAN slicing and report fewer SLA violations during training under dynamic slice counts. Yungaicela-Naula \emph{et al.} \cite{yungaicela2026rslaq} integrate SLA targets into a QoS xApp objective and translate operator policies into resource-allocation decisions.

Risk-sensitive O-RAN studies focus on rare but costly SLA failures. Nagib \emph{et al.} \cite{nagib2025safeslice} propose SafeSlice for inter-slice resource allocation, using a sigmoid-based risk-sensitive reward for cumulative latency and a learned cost model to project actions onto a safe set. Tuerxun and Nakao \cite{tuerxun2026safe} combine model-based RL with online KPI quantile estimation and uncertainty-aware safety margins. Kasi \emph{et al.} \cite{kasi2025risk} apply a risk-aware soft actor-critic rApp to coverage optimization and allow operators to tune the risk parameter according to confidence in the offline model.

Offline entry strategies provide another way to reduce unsafe live exploration. Navarro \emph{et al.} \cite{navarro20252offran} propose 2OffRAN, which trains a handover agent from logged data and evaluates the policy with off-policy methods before deployment. Collectively, these studies illustrate an important design principle for O-RAN: operational limits should remain explicit constraints or safety mechanisms, while the reward primarily captures the performance objective to be optimized.

\textbf{Key Takeaways.}
Safe DRL for O-RAN makes SLA, power, interference, and exploration limits explicit in the learning and control process. Existing work mainly combines constrained optimization, risk-sensitive learning, certificate-inspired action filtering, and offline entry strategies to reduce unsafe behavior before and during deployment.

% ---------------------------------------------------------------------
\subsection{DRL for Network Fairness}
\label{Fairness_DRL}

Safety alone does not guarantee that network benefits are distributed equitably. A throughput-oriented policy can favor UEs and cells with favorable channel conditions because these allocations yield high immediate reward, while cell-edge UEs, congested nodes, or low-priority slices receive systematically weaker service. Fairness therefore concerns how performance and resources are distributed across users, slices, cells, tenants, and operators.

\textbf{Metrics and DRL Formulations.}
The networking literature provides several fairness criteria that can be incorporated into DRL. The $\alpha$-fairness family provides a tunable trade-off between efficiency and equality, includes proportional fairness as a special case, and approaches max-min fairness as the fairness parameter increases \cite{mo2002fair}. Jain's index provides a scalar measure of allocation uniformity \cite{jain1984quantitative}. In DRL, these criteria are commonly introduced through reward shaping, explicit constraints, or welfare objectives. Reward shaping adds a fairness term to the scalar reward but can be sensitive to weight selection. Constraint-based formulations enforce minimum fairness levels using constrained-RL machinery, whereas welfare-based formulations optimize a concave function of per-entity returns and connect fairness with multi-objective RL \cite{siddique2020learning}. A recent fairness survey further notes that per-step fairness can differ from fairness accumulated over an entire trajectory \cite{fairrlsurvey2024}.

\textbf{Examples in O-RAN.}
% \hz{why is it `evidence'? ``Examples in O-RAN''?}
Several O-RAN studies have used reward design to expose the throughput-fairness trade-off. 
For example, Tan \emph{et al.} \cite{tan2025deep} train PPO and TD3 downlink resource-allocation xApps with a reward that combines total throughput and Jain's index. Their results show that TD3 distributes physical resources more evenly, whereas PPO achieves higher aggregate throughput by favoring heavily loaded UEs. Giwa \emph{et al.} \cite{giwa2025hetnet} jointly consider power, bandwidth slicing, and scheduling in heterogeneous O-RAN and report fairness gains over a throughput-greedy baseline under Near-RT RIC inference constraints. Gopal \emph{et al.} \cite{gopal2025adapshare} propose AdapShare, in which DDPG and TD3 agents partition resource blocks between LTE and NR networks using Jain's index and an intent weight. Tsampazi \emph{et al.} \cite{tsampazi2024pandora} evaluate competing xApps and show that reward weights become a central mechanism for balancing slices with similar objectives.

Other work couples network fairness with QoS targets or distributes fairness decisions across multiple controllers. 
For example, Comsa \emph{et al.} \cite{comsa2025fairq} propose FAIR-Q, which uses dual RL controllers to select scheduling rules and tune a generalized proportional-fair algorithm. The framework treats fairness as maintaining delay distributions within an acceptable operating region, consistent with latency-fairness models in wireless scheduling \cite{lopezsanchez2022latency}. Hashemi Nezhad \emph{et al.} \cite{hasheminezhad2025derric} propose DERRIC-SO, where self-organizing PPO agents place RICs using a reward that combines user throughput and local Jain fairness. 
Classical optimization also provides useful baselines and complementary mechanisms.
For instance, Mondal \emph{et al.} \cite{mondal2023fairauction} use Vickrey-Clarke-Groves auctions for multi-tenant x-haul and cloud-resource leasing, while Aslan \emph{et al.} \cite{aslan2024fairvran} use online convex optimization to enforce two-sided $\alpha$-fairness in virtualized RAN platforms.

\textbf{Temporal and Multi-level Fairness.}
Fairness in O-RAN depends on both the entity being protected and the time horizon over which fairness is measured. A scheduler may make an unequal allocation in one slot while remaining fair over a longer observation window. O-RAN also introduces several nested fairness levels, including UEs within a cell, slices sharing a cell, cells sharing spectrum, and tenants sharing infrastructure. When multiple xApps or rApps optimize overlapping resources, their fairness objectives may conflict, linking fairness design to the runtime conflict-mitigation mechanisms discussed later in this section.

\textbf{Key Takeaways.}
Fairness in O-RAN DRL concerns how performance gains and resources are distributed, not only whether operational constraints are satisfied. Existing studies mainly encode fairness through reward terms, constraints, welfare functions, QoS-aware objectives, and distributed coordination across multiple agents.

% ---------------------------------------------------------------------
\subsection{Explainable DRL}
\label{Explainable_DRL}

Explainable DRL addresses the opacity of neural policies in automated RAN control. A deep policy maps telemetry observations to actions through a neural network, but operators may not be able to determine why a particular action was selected. In O-RAN, this opacity complicates fault attribution across xApps, troubleshooting of automated control loops, and post-deployment accountability.

\textbf{Explanation Mechanisms and O-RAN Placement.}
Existing XRL surveys organize explanations into feature-level, learning-process, and policy-level approaches \cite{milani2024explainable}. Feature-level methods identify which observations most strongly influence an action, using tools such as LIME \cite{ribeiro2016should} or Shapley-value attribution \cite{lundberg2017unified}. Learning-process explanations interpret behavior through rewards, value functions, or training dynamics, while policy-level explanations summarize a learned policy using an interpretable surrogate such as a decision tree or rule set. Heuillet \emph{et al.} \cite{heuillet2021explainability} similarly observe that many deep-RL explanations analyze a trained policy after learning rather than making the policy interpretable by design.

O-RAN adds a system-level requirement: the explanation mechanism must match the control timescale. Computationally expensive explanations can run offline at the Non-RT RIC to support auditing, debugging, and model improvement. Near-RT RIC explanations, in contrast, may need to accompany control actions within a 10~ms to 1~s budget. This distinction motivates three practical categories in the O-RAN literature: post-hoc action explanation, explanation-guided training, and interpretable policy representation.

\begin{itemize}[leftmargin=0.15in]
\item 
\textit{Post-hoc action explanation.}
Post-hoc methods explain decisions produced by an already trained agent. Fiandrino \emph{et al.} \cite{fiandrino2023explora} propose EXPLORA, which constructs an attributed graph connecting a DRL resource-allocation xApp's actions to input-state features and then uses the explanations to detect and replace performance-degrading actions. Fatehi \emph{et al.} \cite{fatehi2026interpretable} propose AE-MAPPO, which embeds attention modules into a multi-agent PPO controller for RAN slicing and uses attention weights as low-cost action-level explanations. Sun \emph{et al.} \cite{sun2025explainable} train a vehicular-slicing DRL agent whose attention layer receives supervision from offline Shapley values, combining fast attention-based inference with attribution guidance.

\item 
\textit{Explanation-guided training.}
Explanation-guided methods incorporate interpretability signals directly into policy learning. Rezazadeh \emph{et al.} \cite{rezazadeh2023explanation} use XAI attributions to guide a DRL agent for 6G RAN slicing toward decisions with higher confidence and interpretability. Rezazadeh \emph{et al.} \cite{rezazadeh2024sliceops} extend this idea in SliceOps, an explainable MLOps loop that monitors confidence during training and tracks policy transparency as the model converges. Rezazadeh \emph{et al.} \cite{rezazadeh2024intelligible10} further propose STEP, which combines multi-agent DRL with an information-bottleneck mechanism to learn compact and interpretable inter-slice communication.

\item 
\textit{Interpretable policy representation.}
A third direction makes the policy itself more transparent. Lu \emph{et al.} \cite{lu2026demystifying} propose DeRAN, which maps high-dimensional telemetry into meaningful features and synthesizes symbolic rules for continuous and discrete control. Duttagupta \emph{et al.} \cite{duttagupta2025symbxrl} propose SymbXRL, which uses first-order logic to describe concepts and relations behind actions and supports intent-based steering. Jabbari \emph{et al.} \cite{jabbari2026sia} extend symbolic interpretability to anticipatory control across RAN slicing and massive MIMO scheduling tasks. Brik \emph{et al.} \cite{brik2024explainable} provide a broader O-RAN XAI tutorial that maps reactive and proactive explanation mechanisms onto learning-based network control.
\end{itemize}

\textbf{Key Takeaways.}
Explainable DRL in O-RAN supports transparency, troubleshooting, and operator accountability. Existing work mainly follows three paths: explaining trained actions, incorporating explanations into policy learning, and representing policies through interpretable symbolic or surrogate structures.

% ---------------------------------------------------------------------
\subsection{Robustness of DRL Policy}
\label{Robustness_DRL}

Robustness concerns whether a learned policy remains reliable when deployment conditions differ from those encountered during training. In O-RAN, two sources of mismatch are particularly important. \emph{Non-adversarial shifts} arise from normal changes in traffic, channel quality, mobility, interference, or deployment conditions, whereas \emph{adversarial shifts} arise when an attacker deliberately manipulates telemetry, training data, models, or the operating environment.

\textbf{Robustness to Non-adversarial Shift.}
Policies trained in simulators, digital twins, or logged datasets can encounter live states that are underrepresented or absent during training. This includes sim-to-real mismatch, offline-RL support mismatch, and non-stationary traffic or channel dynamics. Generic robust-RL methods address these effects by exposing policies to broader training distributions or by explicitly optimizing against uncertainty. Domain randomization trains across a distribution of simulated environments \cite{tobin2017domain}, while robust MDP formulations optimize against worst-case dynamics within an uncertainty set \cite{iyengar2005robust}. Pinto \emph{et al.} \cite{pinto2017robust} train policies against learned disturbances, and Vinitsky \emph{et al.} \cite{vinitsky2020robust} extend adversarial training with population-based adversaries.

O-RAN studies evaluate robustness under sim-to-real mismatch, non-stationarity, transfer, and generalization. Li \emph{et al.} \cite{sim2real_oran} study RL-driven next-generation networks with digital twins and report that shifts in UE distributions degrade simulator-trained policies. Alves Esteves \emph{et al.} \cite{robustsliceplacement} combine an actor-critic agent with a graph convolutional network for slice placement and evaluate robustness under stair-stepped load changes. Several slicing studies further target adaptation to unseen network conditions. Nagib \emph{et al.} \cite{nagib2023safe} combine policy reuse and distillation to reduce reward variance and accelerate convergence. Zeng and Niu \cite{meta_oran_slicing} propose a multi-task meta-initialization scheme for rapid adaptation to unseen slicing tasks, while Lotfi \emph{et al.} \cite{lotfi2023attention} design an attention-based distributed slicing agent and evaluate it across diverse traffic conditions.

\textbf{Adversarial Robustness.}
Adversarial robustness addresses deliberate manipulation of the information or components used by the policy. O-RAN broadens this attack surface because telemetry and control information can pass through E2 nodes, shared RIC data layers, xApps, rApps, and management services before influencing a decision. State-adversarial MDPs model one form of this threat by allowing bounded perturbations to observations \cite{zhang2020robust}. Smoothness regularization and adversarial losses can then train policies whose actions vary less under small observation perturbations \cite{oikarinen2021robust}.

Existing O-RAN studies demonstrate attacks against both inference components and DRL controllers. Balakrishnan \emph{et al.} \cite{oranevasion2024} study evasion attacks against a graph-based connection-management xApp. Sapavath \emph{et al.} \cite{sapavath2023experimental} manipulate a shared Near-RT RIC database and reduce the accuracy of an interference-classifier xApp. Chiejina \emph{et al.} \cite{chiejina2024systemlevel} deploy a malicious xApp that poisons spectrograms and key performance metrics in the RIC database and compare defensive distillation with adversarial training under Near-RT RIC latency constraints.

For DRL controllers, Ergu \emph{et al.} \cite{ergu2024unmasking} study compromised users or jammers that falsify signal-power reports and corrupt the DRL state. Tashman and Cherkaoui \cite{ranslicejamming2026} study a budget-constrained jamming adversary against DRL-based slicing and report slice-dependent SLA violations. Aizikovich \emph{et al.} \cite{aizikovich2025rogue} examine false telemetry from a malicious cell in traffic steering and propose an LSTM-based telemetry monitor. Lacava \emph{et al.} \cite{maloran2025} report backdoor and poisoning attacks on a live O-RAN deployment, while Gajjar \emph{et al.} \cite{blackboxoran2025} study black-box attacks that clone the target model and apply universal perturbations against xApps and rApps.

\textbf{Defense Mechanisms.}
O-RAN defenses must address both training-time compromise and runtime manipulation. Groen \emph{et al.} \cite{groen2024securing} discuss adversarial training, policy-smoothness regularization, telemetry anomaly detection, and moving-target defense at the Near-RT RIC. Population-based learning, including neuroevolution, can diversify policies and reduce dependence on a single brittle model \cite{kouchaki2025federated}. Hoang \emph{et al.} \cite{hoang2024security} survey attacks and defenses across O-RAN and 6G systems. Certified robustness provides a complementary direction by bounding worst-case behavior over a perturbation set; recent grey-box attacks on DRL MU-MIMO scheduling illustrate the type of worst-case behavior such certificates target \cite{fggm2025}.

\textbf{Key Takeaways.}
Robustness in O-RAN DRL covers both natural distribution shift and adversarial manipulation. Existing work studies sim-to-real mismatch, non-stationarity, transfer, telemetry attacks, poisoning, black-box perturbations, and defenses based on robust training, monitoring, diversification, and certification.

% ---------------------------------------------------------------------
\subsection{Verification and Runtime Monitoring}
\label{Verification_Monitoring}

Verification and runtime monitoring provide the assurance layer around a deployed controller. Safety, fairness, explainability, and robustness influence how a policy is designed and trained, but deployment still requires evidence that the policy behaves acceptably, mechanisms that detect failures after admission, and safeguards that prevent harmful or conflicting actions. In the O-RAN hierarchy, formal verification is naturally associated with offline and Non-RT RIC workflows, whereas monitoring, enforcement, and conflict mitigation are more closely coupled to Near-RT RIC operation.

\textbf{Formal Verification.}
Formal verification asks whether a trained policy satisfies a stated property over a specified input or state region rather than only on sampled test cases. 
Katz \emph{et al.} \cite{katz2017towards} introduce Reluplex for neural networks with rectified-linear units, and Katz \emph{et al.} \cite{katz2019marabou} later extend this line through Marabou. A DRL verification survey reviews how such tools can certify properties such as bounded action variation under bounded input perturbations \cite{drlverification2023}. Related work verifies probabilistic policies \cite{bacci2022verified} and examines verification of learning-based systems in networking contexts \cite{dethise2021analyzing}.

Direct verification of neural xApp policies remains less developed than verification of simplified behavioral models. Metere \emph{et al.} \cite{metere2025formal} encode a cell-switching xApp and its environment in PRISM and verify quantitative thresholds that characterize the trade-off between energy saving and service availability. Their work positions formal analysis as a design and admission-control mechanism for xApps. A recurring limitation is scalability because exact verification becomes expensive for high-dimensional telemetry, continuous actions, and large neural policies.

% ======================================================================
%  TABLE: Taxonomy of trustworthy DRL for O-RAN (O-RAN-specific studies)
%  Fits a two-column IEEE (IEEEtran) layout as a full-width table*.
%  Vertical centering: every column is m{} (middle-aligned); each trust
%  dimension label is placed on the MIDDLE row of its block, so it sits
%  centered in the merged grid by geometry (no \multirow height guess).
%  Requires: \usepackage{array}, \usepackage[table]{xcolor}  % \rowcolor
%  (\multirow is no longer used; harmless if your preamble still loads it)
% ======================================================================
\begin{table*}[!t]
\centering
\caption{Taxonomy of O-RAN-specific studies on trustworthy DRL for RAN control.}
\label{tab:trust_drl_taxonomy}
\scriptsize
\setlength{\tabcolsep}{2.2pt}
\renewcommand{\arraystretch}{1.18}
% \begin{tabular}{|>{\centering\arraybackslash}m{0.115\textwidth}|m{0.165\textwidth}|m{0.225\textwidth}|m{0.33\textwidth}|m{0.095\textwidth}|}
\begin{tabular}{|>{\centering\arraybackslash}m{0.115\textwidth}|m{0.165\textwidth}|m{0.225\textwidth}|m{0.33\textwidth}|>{\centering\arraybackslash}m{0.095\textwidth}|}
\hline
\rowcolor[gray]{0.9}
\multicolumn{1}{|c|}{\parbox[c][6ex][c]{0.115\textwidth}{\centering\textbf{Trust Dimension}}} &
\multicolumn{1}{c|}{\parbox[c][6ex][c]{0.165\textwidth}{\centering\textbf{Research Focus}}} &
\multicolumn{1}{c|}{\parbox[c][6ex][c]{0.225\textwidth}{\centering\textbf{Technique}}} &
\multicolumn{1}{c|}{\parbox[c][6ex][c]{0.33\textwidth}{\centering\textbf{Key Idea}}} &
\multicolumn{1}{c|}{\parbox[c][6ex][c]{0.095\textwidth}{\centering\textbf{Refs.}}} \\
\hline

% ===================== SAFE DRL (3) — label on row 2 =====================
& Constrained SLA control
& Constrained MARL; QoS-driven policy
& Encodes SLA targets as explicit constraints/objectives, cutting violations while scaling with slice count.
& \cite{zangooei2023flexible,yungaicela2026rslaq} \\
\cline{2-5}
\textbf{Safety} (\S\ref{Safe_DRL_ORAN})
& Risk-sensitive objectives
& Risk-aware and model-based safe RL
& Bounds tail risk via learned cost/quantile models; a tunable risk knob keeps SLA violations low.
& \cite{nagib2025safeslice,tuerxun2026safe,kasi2025risk} \\
\cline{2-5}
& Offline / safe exploration
& Offline RL with off-policy evaluation
& Trains fully offline and validates via OPE, avoiding unsafe live exploration.
& \cite{navarro20252offran} \\
\hline

% ===================== FAIRNESS (3) — label on row 2 =====================
& Fairness reward shaping
& Jain-index / utility-augmented reward (PPO, TD3, DDPG)
& Adds a fairness term to the reward; reward weights become the main equity-vs-throughput lever.
& \cite{tan2025deep,giwa2025hetnet,gopal2025adapshare,tsampazi2024pandora} \\
\cline{2-5}
\textbf{Fairness} (\S\ref{Fairness_DRL})
& Constraint-based fair RL
& Dual RL; online convex $\alpha$-fairness
& Keeps the delay distribution in a fair region with horizon-level regret guarantees.
& \cite{comsa2025fairq,lopezsanchez2022latency,aslan2024fairvran} \\
\cline{2-5}
& Welfare and auction mechanisms
& Self-organizing agents; VCG fair auction
& Allocates via fair auction / SO agents and bounds the worst-off tenant's cost.
& \cite{hasheminezhad2025derric,mondal2023fairauction} \\
\hline

% ================== EXPLAINABLE DRL (3) — label on row 2 ==================
& Post-hoc action explanation
& Attribution graphs; attention; Shapley supervision
& Explains and repairs degrading actions at low inference cost while preserving fidelity.
& \cite{fiandrino2023explora,fatehi2026interpretable,sun2025explainable} \\
\cline{2-5}
\textbf{Explainability} (\S\ref{Explainable_DRL})
& Explanation-guided training
& Attribution-steered learning; explainable MLOps; information bottleneck
& Uses explanations during training to favor confident, transparent, lower-conflict policies.
& \cite{rezazadeh2023explanation,rezazadeh2024sliceops,rezazadeh2024intelligible10} \\
\cline{2-5}
& Interpretable policy representation
& Symbolic distillation; first-order-logic concepts
& Distills policies into symbolic/logical rules for auditability with little reward loss.
& \cite{lu2026demystifying,duttagupta2025symbxrl,jabbari2026sia,brik2024explainable} \\
\hline

% ===================== ROBUSTNESS (6) — label on row 3 =====================
& Sim-to-real and robust architectures
& Digital twin; actor-critic + GCN; attention agent
& Sustains performance under distribution shift via twins and robust network architectures.
& \cite{sim2real_oran,robustsliceplacement,lotfi2023attention} \\
\cline{2-5}
& Transfer and meta adaptation
& Policy transfer/distillation; meta-initialization
& Reuses and adapts policies for safer, faster, lower-variance convergence.
& \cite{nagib2023safe,meta_oran_slicing} \\
\cline{2-5}
\textbf{Robustness} (\S\ref{Robustness_DRL})
& Evasion and poisoning attacks
& Perturbed-input evasion; RIC-DB poisoning
& Show that perturbed inputs and poisoned shared state break DRL xApps.
& \cite{oranevasion2024,sapavath2023experimental,chiejina2024systemlevel} \\
\cline{2-5}
& Telemetry and PHY attacks
& Signal spoofing; jamming; rogue base station
& Falsified telemetry corrupts the agent state, causing SLA loss and recovery lag.
& \cite{ergu2024unmasking,ranslicejamming2026,aizikovich2025rogue} \\
\cline{2-5}
& Backdoor and black/grey-box attacks
& Training-time triggers; limited-access gradient attacks
& Plant backdoors or attack with little model knowledge on a live RAN.
& \cite{maloran2025,blackboxoran2025,fggm2025} \\
\cline{2-5}
& Adversarial defenses and hardening
& Adversarial training; smoothness; anomaly detection; population diversity
& Combine adversarial training, moving-target defense, and diverse populations to cut brittleness.
& \cite{groen2024securing,kouchaki2025federated,hoang2024security} \\
\hline

% ============== VERIFICATION \& MONITORING (5) — label on row 3 ==============
& Formal verification
& Probabilistic model checking (PRISM)
& Model-checks safety thresholds, but verifies a model rather than the trained weights.
& \cite{metere2025formal} \\
\cline{2-5}
& Runtime monitoring and attestation
& Unsupervised KPI/security detectors; hash-based attestation
& Detect KPI anomalies and verify xApp integrity before actions take effect.
& \cite{argos2025,xsec2024,orandefence2025} \\
\cline{2-5}
\textbf{Verification and Monitoring} (\S\ref{Verification_Monitoring})
& Conflict detection
& Rule-based framework; GCN graph; causal/Shapley attribution
& Identify direct/indirect/implicit conflicts and attribute KPI degradation.
& \cite{adamczyk2023conflict,oranconflictgcn2025,conflictcausal2025} \\
\cline{2-5}
& Conflict resolution (cooperative / game-theoretic)
& Team learning/distillation; Nash/E-G arbitration; digital-twin check
& Merge or arbitrate xApp policies, testing actions on a twin before applying.
& \cite{zhang2022team,erdol2024distillation,oranxapps2025,comix2025} \\
\cline{2-5}
& Conflict resolution (orchestration)
& LLM-driven rApp pipelines
& Composes conflict-free xApp sets at the orchestration layer; least standardized.
& \cite{oranconflictllm2026} \\
\hline

\end{tabular}
\end{table*}

\textbf{Runtime Monitoring and Enforcement.}
Runtime assurance addresses failures that emerge after deployment, including telemetry drift, model degradation, unexpected interactions, and adversarial manipulation. Shielding wraps a learned policy with a filter that blocks or replaces actions violating a temporal-logic property \cite{alshiekh2018safe}. Related runtime-assurance schemes monitor whether an agent remains inside a certified safe region and switch to a verified baseline controller when the policy leaves that region.

Telemetry-based monitoring complements action enforcement. Anomaly detectors over E2 and O1 telemetry can flag operational faults and telemetry manipulation. 
Argos scores UE-level key performance indicators to identify rogue or downgrade-inducing cells \cite{argos2025}. XSec reads security telemetry from the shared data layer and provides explanation-equipped detection within a Near-RT RIC budget \cite{xsec2024}. O-RAN Defense checks xApp integrity through execution-time hash challenges and detects tampered or substituted xApps before their actions affect the RAN \cite{orandefence2025}.

\textbf{Conflict Detection and Mitigation.}
Runtime assurance must also address interactions among independently developed xApps and rApp policies. Multiple xApps can issue actions to the same E2 node or manipulate coupled radio resources, creating direct, indirect, or implicit conflicts. Broader taxonomies further distinguish horizontal conflicts among peer xApps from vertical conflicts between xApps and rApp-level policies \cite{oranxapps2025}.

Conflict detection combines rule-based and data-driven approaches. Adamczyk and Kliks \cite{adamczyk2023conflict} instantiate a Near-RT RIC Conflict Mitigation Framework with per-type detection logic and message flows among RIC components. Al Shami \emph{et al.} \cite{oranconflictgcn2025} reconstruct conflict graphs among xApps from observed behavior using a graph neural network. Sharma \emph{et al.} \cite{conflictcausal2025} combine Shapley attribution with causal treatment-effect estimation to assign responsibility for KPI degradation.

Conflict resolution can occur through cooperative learning, runtime arbitration, or orchestration-level composition. 
For example, Zhang \emph{et al.} \cite{zhang2022team} allow power-control and resource-allocation xApps to share intended actions during team learning so that each DQN can condition on the others' choices. Erdol \emph{et al.} \cite{erdol2024distillation} merge several trained xApps into a distilled agent that selects among their policies. Elyasi \emph{et al.} \cite{oranxapps2025} study QoS-aware runtime mitigation using game-theoretic controllers based on Nash social welfare and Eisenberg-Gale solutions. Giannopoulos \emph{et al.} \cite{comix2025} use a network digital twin to evaluate conflicting power-control actions before applying them to the live RAN. Li \emph{et al.} \cite{oranconflictllm2026} move arbitration to the orchestration layer by synthesizing conflict-aware rApp pipelines with a language-model-driven multi-agent framework.

\textbf{Key Takeaways.}
Verification and runtime monitoring complement training-time trustworthiness with admission checks, continuous observation, action enforcement, fallback control, and conflict mitigation. Together, these mechanisms provide assurance when a deployed policy encounters failures or interactions that were not fully captured during training.

\subsection{Summary}

Trustworthy DRL for O-RAN requires assurance throughout the complete control lifecycle. Safety constrains harmful actions, fairness governs how benefits and resources are distributed, explainability makes learned decisions inspectable, robustness addresses distribution shifts and attacks, and verification with runtime monitoring provides admission and operational assurance. These dimensions are complementary rather than independent, and practical O-RAN controllers must combine design-time mechanisms with runtime monitoring, enforcement, and fallback strategies. Table~\ref{tab:trust_drl_taxonomy} summarizes representative trustworthy DRL studies in the literature, including their trustworthiness objectives, mechanisms, O-RAN scope, and evaluation focus.

% \clearpage 

\section{Training and Deployment}
\label{9_Training_Deployment}
Training and deployment are tightly coupled in DRL-based O-RAN control. A policy that performs well during training may still fail after deployment if the training data, simulator, control interface, inference platform, or update process does not match the conditions of the operational network. Therefore, deploying DRL in O-RAN requires more than selecting and training an algorithm. It requires an end-to-end lifecycle that connects data collection and training with validation, deployment, monitoring, and adaptation.

% \textbf{Lifecycle.} 
A typical lifecycle of an AI agent begins with data collection and policy training, followed by validation, model optimization, deployment, runtime monitoring, rollback, and retraining. These stages are interdependent. The training process determines the conditions that the policy has experienced, whereas the deployment environment determines what the policy can observe, how quickly it must act, and which control variables it can modify. For example, an xApp operating in a near-real-time control loop must be validated using realistic E2 measurements, and its deployed policy must satisfy the corresponding inference-latency budget.

% \textbf{Section roadmap.} 
This section follows the policy lifecycle from training to operation. We first compare offline, online, and hybrid training strategies. We then discuss sim-to-real transfer and digital twins, sample-efficient learning, continual adaptation under changing network conditions, efficient inference at the network edge, and reinforcement learning operations (RLOps) for managing deployed policies.

% =====================================================================
\subsection{Training Strategies: Offline, Online, and Hybrid}
\label{subsec:where_train}

Existing literature organizes DRL training for O-RAN into three main strategies: offline training from previously collected data, online training through interaction with a simulator, emulator, or controlled testbed, and hybrid training that combines these sources. These strategies differ in their exploration opportunities, sample-collection costs, safety risks, and ability to represent the deployment environment~\cite{navarro20252offran,bonati2023openran,nagib2023safe}.

The intended deployment host is a cross-cutting consideration for all three training strategies. rApps can use SMO-level data and comparatively long control intervals, whereas xApps rely on Near-RT RIC behavior and E2 telemetry, and dApps operate closer to MAC- and PHY-level functions under tighter timing constraints~\cite{polese2023understanding,d2022dapps}. Li \emph{et al.}~\cite{li2025toward} similarly connect the DRL lifecycle to the placement and timing constraints of MEC and O-RAN components. Training and validation environments therefore commonly reproduce the observations, actions, delays, and compute limits of the intended host.
% \hz{you mentioned three classes. The paragraph titles in the following should only have these three.}
% \textbf{\textcolor{blue}{[This section has been revised. ]}}

\textbf{Offline Training.} 
Offline training learns a policy from a fixed dataset collected before deployment and avoids exploratory control of the operational network \cite{fujimoto2019off}. Historical KPI, KPM, configuration, and control records can therefore support policy learning without exposing users to an immature policy. Algorithms such as BCQ~\cite{fujimoto2019off}, CQL~\cite{kumar2020conservative}, and IQL~\cite{kostrikov2021offline} address this setting by constraining policy improvement or value estimation when candidate actions have weak support in the logged data.

Operational telemetry often requires trajectory reconstruction before it can be used for offline RL. The offline-RL literature represents training data as aligned state, action, reward, and next-state transitions~\cite{levine2020offline}, whereas operational logs may record telemetry, control commands, and outcomes through separate services and at different times. RLOps studies therefore identify behavior-policy provenance, delay alignment, reward reconstruction, and data-quality filtering as parts of the data-management pipeline~\cite{li2022rlops}. In O-RAN, 2OffRAN reconstructs logged handover trajectories and combines them with off-policy evaluation before deployment~\cite{navarro20252offran}.

Deployment-oriented O-RAN studies use offline-first training when exploratory actions could disrupt service. Navarro \emph{et al.}~\cite{navarro20252offran} introduce 2OffRAN for offline handover optimization, while SafeSlice~\cite{nagib2025safeslice} and risk-aware coverage optimization~\cite{kasi2025risk} limit the exposure of operational networks to poorly evaluated actions. These studies indicate that logged-data coverage and offline evaluation are central considerations when offline policies are prepared for deployment.
% \hz{add refs if possible.}

\textbf{Online Training.} Online training allows the agent to collect new trajectories while learning and therefore supports broader exploration of the state and action spaces. In O-RAN research, this interaction is usually performed in simulators, emulators, or controlled testbeds rather than in production networks, where an immature policy could reduce throughput, violate slice SLAs, or cause unstable handovers. Representative environments include ns-O-RAN for ns-3-based O-RAN simulation~\cite{lacava2023ns}, ColO-RAN for programmable xApp experimentation~\cite{polese2022colo}, and OpenRAN Gym for AI/ML development, data collection, and testing~\cite{bonati2023openran}.

\textbf{Hybrid Training.}
Hybrid training combines logged experience with controlled online interaction to address limitations of either source alone. Offline-only policies can encounter distribution shift when they select actions that are uncommon in the dataset \cite{fujimoto2019off, kumar2020conservative}, while simulator-only policies can depend on simplified traffic, channel, mobility, or control-delay models \cite{yan2020learning, bonati2021colosseum}. Hybrid pipelines therefore use logged data or simulation for pretraining, followed by emulator or testbed validation and limited policy adaptation under controlled conditions \cite{zheng2022online, wagenmaker2023leveraging }.

Existing safe-RL studies provide mechanisms for constraining the interactive stages of hybrid training. Nagib \emph{et al.}~\cite{nagib2023safe,nagib2025safeslice} combine learning and safety mechanisms for RAN slicing, while shielding and action-projection methods restrict actions that violate predefined constraints~\cite{alshiekh2018safe,cheng2019end}. These mechanisms support staged exploration, although their effectiveness depends on the accuracy and coverage of the specified safety constraints.

\textbf{Key Takeaways.} 
The three strategies provide complementary design options rather than a universal training prescription. Offline learning reduces live exploration, controlled online learning expands the available experience, and hybrid training combines offline data reuse with subsequent online adaptation. In O-RAN, the appropriate choice is shaped by the availability \cite{polese2022colo} and representativeness of training data \cite{polese2024colosseum}, the fidelity of simulation or emulation \cite{polese2024colosseum, alomar2023causalsim}, the control timescale of the target function \cite{tsampazi2024pandora}, and the cost or operational risk associated with live adaptation \cite{polese2022colo, polese2024colosseum}.

%\hz{is this part for Hybrid training? If yes, remove the title of this paragraph. If not, it is not clear to me why we need to discuss this here.}

% =====================================================================
\subsection{Sim-to-Real Transfer}
\label{subsec:sim2real}

The sim-to-real problem arises when a policy trained in an approximate environment that encounters different observations, dynamics, or control interfaces after deployment in the physical RAN. O-RAN simulators may simplify channel dynamics, traffic and mobility, E2/control-loop timing, measurement noise, partial observability, hardware behavior, or interactions among xApps \cite{polese2024colosseum, bonati2023openran}. 

The literature addresses this gap through four complementary mechanisms: domain randomization broadens the training distribution~\cite{tobin2017domain, demirel2026generalization, mehta2020active,lien2023revisiting}, system identification calibrates the training environment using target-system measurements~\cite{tan2018sim, chebotar2019closing, ruah2024calibrating}, digital twins maintain an updated virtual representation of the deployed network~\cite{masaracchia2023digital, he2025digital, hossen2026opentwin}, and shadow deployment evaluates a candidate policy using live observations before broader activation~\cite{gassert2024stepping}. The first two mechanisms modify the training environment, whereas the latter two provide progressively more deployment-specific validation.

\textbf{Domain Randomization.}
Domain randomization improves policy robustness by varying the simulated conditions encountered during training. Tobin \emph{et al.}~\cite{tobin2017domain} introduce this approach by sampling environment parameters from predefined distributions rather than training in one fixed environment. For RAN control, the randomized parameters can include traffic load, packet arrivals, user locations, mobility, channel statistics, interference, network topology, UE characteristics, slice composition, and control timing~\cite{demirel2026generalization}. Exposure to these variations reduces dependence on a single simulated configuration.

The effectiveness of domain randomization depends on how the parameter distributions are selected. Demirel \emph{et al.}~\cite{demirel2026generalization} examine generalization across changing RAN conditions and show that the selection of training scenarios affects performance under unseen conditions. Broader sim-to-real studies find that uniform or excessively adverse randomization can enlarge the training space, increase policy variance, and reduce training efficiency~\cite{mehta2020active,lien2023revisiting}. Domain randomization therefore improves coverage when its ranges represent plausible deployment variation, but it does not directly calibrate the simulator to a particular target network.

\textbf{System Identification.}
System identification reduces the deployment gap by estimating simulator parameters from measurements of the target system. Tan \emph{et al.}~\cite{tan2018sim} identify actuator behavior from physical measurements and incorporate the resulting model into simulation before transferring the learned controller. Chebotar \emph{et al.}~\cite{chebotar2019closing} further adapt simulation parameters using real-world experience, thereby iteratively aligning the training environment with observed physical behavior. These studies establish measurement-driven calibration as an alternative to selecting a broad parameter distribution in advance.

RAN-oriented calibration applies the same principle to radio and network models. Ruah \emph{et al.}~\cite{ruah2024calibrating} calibrate an RF-emulation environment using measurements that characterize the target wireless conditions. Such calibration can cover channel behavior, traffic and mobility models, processing delays, or control-interface timing. The resulting environment is more representative of the measured deployment, although its validity can decrease when operating conditions move beyond the calibration data.
%\hz{this is a survey paper, aiming to convey results in the literature. So referneces are needed.}

\textbf{Digital Twins.}
Digital twins extend one-time system identification by maintaining a virtual representation that is updated as the deployed network changes. Masaracchia \emph{et al.}~\cite{masaracchia2023digital} describe an O-RAN digital-twin architecture in which network observations support modeling, optimization, and validation. He \emph{et al.}~\cite{he2025digital} integrate a digital twin with reinforcement learning for intelligent xApp management. These architectures position the twin alongside SMO, O-Cloud, and RIC functions so that candidate policies or configurations can be evaluated before operational activation.

O-RAN prototypes provide initial evidence for continuously synchronized twins. Hossen \emph{et al.}~\cite{hossen2026opentwin} present OpenTwin, which infers configuration-dependent behavior and updates its KPM estimates from live measurements. Their experiments report up to 96\% KPM-mirroring accuracy and use the twin to evaluate an energy-saving xApp without disrupting live operation. This result demonstrates twin-assisted validation for a particular xApp and experimental platform, while evidence across multiple sites, vendors, and closed-loop DRL tasks remains limited.

\textbf{Shadow Deployment.}
Shadow deployment evaluates policy integration under live observations without immediately replacing the incumbent controller. In passive shadow operation, the candidate and incumbent policies process the same observations, but only the incumbent action controls the network \cite{li2022rlops}. The candidate outputs can then be examined for timing failures, unsupported actions, constraint violations, and disagreement with the deployed controller. This stage evaluates the observation and software paths more realistically than offline or digital-twin evaluation.

Existing shadow-mode RL work also studies restricted forms of policy activation. Gassert and Althoff~\cite{gassert2024stepping} retain an incumbent policy and progressively allow the learned policy to control selected states. Their approach differs from purely passive shadow execution because some learned actions affect the environment, but it shows how incumbent control can support a gradual transition toward policy activation. Passive shadow evaluation remains unable to observe the counterfactual outcomes of unapplied actions and therefore does not fully establish post-deployment policy performance.
%\hz{the position of this paragraph is not clear to me. Also, references are needed.}

\textbf{Key Takeaways.} 
The four techniques address different parts of the sim-to-real tranfer gap. Domain randomization broadens the simulated training distribution, system identification aligns simulation with measurements from a target network, digital twins maintain an updated validation environment for repeated validation, and shadow deployment evaluates policy integration using live observations. The literature consequently presents these mechanisms as complementary stages rather than equivalent solutions.

% =====================================================================
\subsection{Sample-Efficient Training}
\label{subsec:sample_efficiency}

Sample efficiency is especially important in cellular networks because realistic interaction data can be expensive to obtain \cite{polese2022colo, qin2022cooperative}. Simulation can generate large datasets but may lack realism, emulation improves fidelity but limits scale and parallelism, and physical-network exploration can affect service quality \cite{alomar2023causalsim, villa2024colosseum}. As a result, O-RAN deployment favors methods that extract more value from existing data and reduce the number of new interactions required for training or adaptation \cite{li2022rlops, bonati2023openran, polese2022colo }.

\textbf{Model-Based and Model-Guided Learning.}
Model-based RL improves data reuse by learning network dynamics and using the learned model for planning or synthetic rollouts. Dreamer, for example, learns a world model and optimizes behavior through imagined trajectories~\cite{hafner2023mastering}. Applied to O-RAN, related models can represent slicing, mobility, scheduling, or coverage dynamics, although prediction errors can accumulate over long rollouts.

Model-guided approaches can also reduce the learning problem without constructing a complete world model. Wu \emph{et al.}~\cite{wu2025mogul} present MOGUL, which first uses model-based optimization to restrict the 5G O-RAN scheduling action space and then applies online multi-agent DRL; the reported experiments outperform the compared model-based and model-free baselines. Tuerxun and Nakao~\cite{tuerxun2026safe} combine model-based optimization with KPI-quantile estimates for safe slicing. These studies show how domain models guide exploration, but their reported benefits remain specific to the evaluated scheduling and slicing environments.

\textbf{Reuse of Logged Data.}
Logged trajectories improve sample efficiency by replacing part of the new interaction process with historical experience. BCQ constrains learned actions toward those represented in the behavior data~\cite{fujimoto2019off}, CQL estimates conservative values for distribution-shifted actions~\cite{kumar2020conservative}, and IQL avoids explicitly evaluating unseen actions during policy extraction~\cite{kostrikov2021offline}. In O-RAN, 2OffRAN applies logged-data learning and off-policy evaluation to handover control~\cite{navarro20252offran}. These methods reduce new data collection but remain sensitive to trajectory coverage, reward reconstruction, and behavior-policy provenance.

\textbf{Federated RL.}
Federated RL increases the diversity of reusable experience by coordinating learning across distributed RAN sites without centralizing their raw trajectories. Abouaomar \emph{et al.}~\cite{abouaomar2022federated} study federated DRL for O-RAN slicing, while Zhang \emph{et al.}~\cite{zhang2022federated} coordinate distributed resource management through federated learning. Ndikumana \emph{et al.}~\cite{ndikumana2023federated} and Hazarika \emph{et al.}~\cite{hazarika2024enhancing} further examine federated DRL in disaggregated and vehicular RAN settings. These studies report benefits from distributed experience, while also identifying communication cost, heterogeneous data distributions, and local-policy divergence as practical constraints.
%\hz{there are many references for Federated RL in O-RAN.}

\textbf{Meta-RL and Transfer Learning.}
Meta-learning and transfer learning reuse prior policies or shared structure when adapting to a new cell, slice, traffic pattern, or network configuration. Erdol \emph{et al.}~\cite{erdol2022federated} combine federated and meta-learning concepts for O-RAN traffic steering, while Nagib \emph{et al.}~\cite{nagib2023safe} transfer learned behavior through a staged safe-slicing pipeline. Lotfi \emph{et al.}~\cite{lotfi2025meta30} and Kalntis \emph{et al.}~\cite{kalntis2026meta} respectively study meta-learning for resource allocation and handover adaptation. The evaluations support adaptation across their tested conditions, although broader transfer across independently operated sites and heterogeneous vendor stacks is not yet established.
%\hz{refs are needed. This is a survey paper.}

\textbf{Imitation and Warm Starting.}
Demonstrations and teacher policies provide an initial behavior before extensive RL interaction. Hester \emph{et al.}~\cite{hester2018deep} introduce Deep Q-learning from Demonstrations and report better early learning performance than the compared demonstration-free DQN on 41 of 42 evaluated Atari tasks. Policy distillation similarly transfers action or value information from trained teacher policies into a student policy~\cite{rusu2015policy}. These broader DRL results support initialization from rule-based or optimization-based RAN controllers, but O-RAN-specific comparisons among imitation, random initialization, and offline pretraining remain sparse.
%\hz{refs needed.}

\textbf{Key Takeaways.} 
The literature improves sample efficiency through both data reuse and knowledge reuse. Model-based methods reuse learned dynamics, offline RL reuses historical trajectories, federated learning combines distributed experience, transfer methods reuse policies across related tasks, and demonstrations supply an informed initialization. Each approach reduces some new interaction while introducing assumptions about model fidelity, dataset coverage, site similarity, or teacher quality.

%=====================================================================
\subsection{Continual Learning and Non-Stationarity}
\label{subsec:continual}

Deployed RAN policies encounter non-stationarity when changes in traffic, mobility, interference, or service requirements shift the observation process, network dynamics, or control objective beyond the training setting~\cite{padakandla2021survey,khetarpal2022towards}. Existing work addresses these changes through three complementary approaches: transfer and meta-learning adapt pretrained policies to identifiable new tasks \cite{nagib2023safe, lotfi2025meta30}, scheduled or drift-triggered retraining produces new policy versions \cite{gama2014survey}, and continual learning updates models incrementally as new data arrive \cite{gain2024open, benzaid2024federated, niknami2026ran}. These approaches rely on lifecycle mechanisms to validate and deploy their updates \cite{oran-aiml-workflow, rezazadeh2024sliceops}.

\textbf{Transfer- and Meta-Learning-Based Adaptation.}
Representative O-RAN DRL studies formulate adaptation across explicitly defined source and target tasks. Nagib \emph{et al.}~\cite{nagib2023safe} combine policy reuse and policy distillation to initialize an O-RAN slicing agent under previously unseen network conditions. Compared with the evaluated baselines, their method improves initial performance and convergence while reducing reward variance by 64.6\%. Lotfi and Afghah~\cite{lotfi2025meta30} use model-agnostic meta-learning to adapt resource-block and downlink-power allocation to new operating conditions. These studies demonstrate efficient policy adaptation when task changes are identifiable, but they do not address the detection of unannounced drift during operation.

\textbf{Scheduled and Drift-Triggered Retraining.}
Scheduled and drift-triggered retraining determine when a new policy version is produced. The data-stream literature distinguishes updates performed without an explicit drift alarm from informed adaptation initiated after detecting a change~\cite{gama2014survey}. Scheduled approaches retrain periodically using recent data, whereas triggered approaches initiate retraining after detecting changes in the input distribution or monitored performance \cite{lu2018learning, gama2014survey}. The current O-RAN DRL studies do not compare these alternatives or demonstrate a complete pipeline from drift detection to policy validation, leaving the selection of update cadence and trigger criteria insufficiently evaluated. 
%\hz{need references.} 

\textbf{Continual Learning.}
Existing O-RAN continual-learning studies primarily address traffic analysis and intrusion detection rather than DRL control. Gain \emph{et al.}~\cite{gain2024open} introduce Incremental O-RAN Traffic Categorization to learn new encrypted-traffic classes while retaining previously learned classes, reporting approximately 98\% average accuracy. Benza\"id \emph{et al.}~\cite{benzaid2024federated} combine federated learning with replay-based continual learning for distributed anomaly detection and report retention of more than 98.8\% of previously learned knowledge. Niknami and Wu~\cite{niknami2026ran} further integrate novelty detection, replay-driven updates, teacher--student distillation, and federated continual learning for O-RAN intrusion detection. These studies establish continual-learning evidence for predictive models, while evidence for continually updated O-RAN control policies remains limited.
%\hz{needs references...}

\textbf{Lifecycle Management.}
Lifecycle management provides the operational layer that connects the preceding adaptation approaches to deployment. The O-RAN AI/ML workflow covers model training, validation, publication, deployment, inference, and continuous operation~\cite{oran-aiml-workflow}. SliceOps similarly connects runtime monitoring with model retraining and redeployment~\cite{rezazadeh2024sliceops}. Within this lifecycle, transfer or continual-learning methods determine how a model is updated, scheduled or drift-triggered rules determine when the update begins, and lifecycle procedures determine how the resulting candidate is validated, released, and monitored.
%\hz{what is the position of this paragraph? what is its relationship with previous paragraphs?}

\textbf{Key Takeaways.}
Existing O-RAN DRL studies provide the examples for transfer- and meta-learning-based adaptation to identifiable task changes, whereas continual learning has mainly been demonstrated for traffic classification and intrusion detection. Scheduled or drift-triggered retraining and lifecycle management can initiate, validate, and deploy model updates, but end-to-end evidence for continually adapting deployed O-RAN DRL controllers remains limited.

% =====================================================================
\subsection{Efficient Inference and Edge Deployment}
\label{subsec:inference_cost}

O-RAN control timescales determine the latency budget and feasible execution location of a deployed DRL policy. Polese \emph{et al.}\ describe non-real-time control loops operating above 1\,s and near-real-time xApp control loops operating between 10\,ms and 1\,s~\cite{polese2023understanding}. D'Oro \emph{et al.}\ introduce dApps at the CU or DU to support inference and control below 10\,ms~\cite{d2022dapps}. Efficient deployment therefore involves two related questions: how to reduce policy inference cost and where to execute the resulting policy.

%\hz{if you cannot find sufficient references for DRL in O-RAN, you can use the reference to DRL itself.}

\textbf{Quantization.} Quantization reduces the numerical precision of policy weights and activations, although the resulting speedup depends on support for low-precision operations on the target platform. Krishnan \emph{et al.} \cite{krishnan2019quarl}  evaluate quantized DRL policies in QuaRL and report that aggressively quantized policies can accelerate inference on quantization-compatible, resource-constrained edge devices without reducing performance in their evaluated tasks. These results support quantization as a candidate for resource-constrained DRL deployment, but they do not establish latency or control performance for an O-RAN xApp or dApp.
%\hz{need references}

\textbf{Pruning and Structured Sparsity.} Pruning reduces policy-network redundancy, whereas practical acceleration depends on whether the runtime and hardware can exploit the resulting sparsity. Livne and Cohen \cite{livne2020pops} introduce Policy Pruning and Shrinking (PoPS), which combines iterative pruning, policy shrinking, and transfer learning to obtain compact DRL policies while retaining performance on CartPole, Lunar Lander, Pong, and Pacman. Hoefler \emph{et al.} \cite{hoefler2021sparsity} survey sparse neural-network execution and show that parameter reduction alone does not guarantee hardware acceleration because the achievable gains also depend on the sparsity structure, computational kernels, and target platform. An O-RAN evaluation therefore needs to distinguish model-size reduction from measured inference-latency reduction.
%\hz{need references}

\textbf{Hardware- and Platform-Aware Design.} Hardware-aware design incorporates target-platform costs into architecture or precision selection before deployment. Wang \emph{et al.} \cite{wang2019haq} introduce HAQ, which uses accelerator-derived latency and energy feedback, rather than model size or floating-point operations (FLOPs) alone, to select layer-wise numerical precision for neural networks. For DRL policies, Kosta \emph{et al.} \cite{kosta2022rapid} introduce RAPID-RL, an early-exit DQN architecture that adjusts computation according to input difficulty and evaluates the resulting performance-operation trade-off on Atari and drone-navigation tasks. These studies provide transferable methods for edge policy design, although neither evaluates an O-RAN controller.
%\hz{need references}

\textbf{O-Cloud and RAN Placement.} Execution placement affects inference efficiency because communication, virtualization, and model-execution delays share the same end-to-end control-loop budget. Polese \emph{et al.} \cite{polese2023understanding} characterize the O-Cloud as pooled computing and virtualization infrastructure managed through the O2 interface. D'Oro \emph{et al.} \cite{d2022orchestran} introduce OrchestRAN to select data-driven algorithms and their execution locations according to service objectives and timing requirements. For tighter control loops, their dApp architecture instead places inference and control closer to the CU or DU~\cite{d2022dapps}. These studies organize placement according to control timescale and available resources rather than identifying a single execution platform as preferable for all controllers.
%\hz{need references}

\textbf{Deployment Validation.} Deployment-oriented validation evaluates control utility and implementation efficiency after compression. QuaRL \cite{krishnan2019quarl} compares policy performance with edge-device inference efficiency, PoPS \cite{livne2020pops} evaluates policy performance together with model compactness, and RAPID-RL \cite{kosta2022rapid} reports both task performance and computational operations, respectively. For an O-RAN controller, a corresponding evaluation reports the original and compressed model specifications, RAN control KPIs, memory and compute requirements, model-inference latency, and end-to-end control-loop latency on the intended execution platform~\cite{d2022orchestran,d2022dapps}. Such target-platform measurements avoid treating parameter or FLOP reduction as sufficient evidence of deployment-time acceleration~\cite{hoefler2021sparsity,wang2019haq}.
%\hz{need references}

\textbf{Key Takeaways.} Published DRL studies show that quantization, pruning, and adaptive computation can reduce policy cost while retaining benchmark performance, although the realized gains depend on the target hardware and runtime support. O-RAN studies further show that feasibility depends on execution placement and end-to-end control-loop latency. The reviewed O-RAN examples primarily addresses timing and placement, whereas direct evaluations of compressed DRL policies within xApps or dApps remain limited.

% =====================================================================
\subsection{RLOps: The DRL Lifecycle in O-RAN}
\label{subsec:rlops}

%\hz{if you cannot find sufficient references for DRL in O-RAN, you can use the reference to DRL not specific to O-RAN.}

RLOps connects policy development and network operation through a repeated lifecycle of training, validation, deployment, monitoring, and update. Li \emph{et al.} \cite{li2022rlops} formulate RLOps for O-RAN around model specification, development, production serving, operational monitoring, versioning, and safety and security. This formulation complements the O-RAN AI/ML workflow, which organizes data collection and processing, training, validation and publication, deployment, inference, and continuous operation~\cite{oran-aiml-workflow,polese2023understanding}.

\textbf{Data and Model Versioning.} Reproducible DRL operation depends on linking interaction data to the policy and environment that generated it. Li \emph{et al.} \cite{li2022rlops} distinguish RL data collected from a live network from data generated by a digital twin and propose joint versioning of datasets, environment configurations, models, hyperparameters, and code. This linkage is particularly relevant to real-world RL, where non-stationarity, delayed effects, partial observability, and safety constraints complicate policy evaluation~\cite{dulac2021challenges}. In O-RAN, aligning observations, actions, rewards or network outcomes, timing information, and policy identifiers as versioned interaction records supports both data reconstruction and attribution of network KPIs to a controller version.
%\hz{reference needed.}

\textbf{Testing and Validation.} Existing work supports evaluating candidate policies across progressively more realistic environments before wider deployment. Li \emph{et al.}~\cite{li2022rlops} place testing and validation before production serving and discuss limited alpha--beta deployment as a means of measuring application performance while restricting its initial impact. Bonati \emph{et al.}~\cite{bonati2023openran} develop OpenRAN Gym to collect data and design, train, and test xApps on the Colosseum emulator before transferring them to experimental platforms including Arena, POWDER, and COSMOS. 

Polese \emph{et al.}~\cite{polese2022colo} further demonstrate DRL-based xApps for RAN slicing, scheduling, and online model training within the ColO-RAN experimental framework. These studies motivate a progression from offline or digital-twin evaluation to O-RAN-compliant testbed experiments and limited live deployment, although they do not define a common policy-promotion protocol.
%\hz{reference needed.}

\textbf{Deployment and Monitoring.} Monitoring, action-level safeguards, and rollback address different post-deployment failure modes. Li \emph{et al.}~\cite{li2022rlops} monitor accumulated reward, application-specific network KPIs, inference time, throughput, and memory use, and associate sustained underperformance with policy retraining or replacement. Polese \emph{et al.}~\cite{polese2023understanding} similarly describe continuous O-RAN operations in which inference outputs and network performance are monitored to identify models that require refinement or retraining. 

\textbf{Runtime Assurance and Rollback.}
At the action level, Alshiekh \emph{et al.}~\cite{alshiekh2018safe} introduce shielding to monitor policy actions and correct those that violate a temporal-logic safety specification. Versioning then provides a separate recovery mechanism through which an underperforming policy can be replaced by a previously validated version.
%\hz{reference needed.}

\textbf{Current O-RAN Coverage and Gap.} The principal gap lies in RL-specific operational conventions rather than the absence of a generic O-RAN lifecycle. The O-RAN AI/ML workflow defines the principal lifecycle stages and an AI/ML model catalog, although several model-management procedures remain under development~\cite{oran-aiml-workflow,polese2023understanding}. Li \emph{et al.}\ extend this lifecycle with RL-specific concerns, including agent--environment data dependencies, synchronized artifact versioning, digital-twin development, limited rollout, model decay, and safety monitoring~\cite{li2022rlops}. 

Experimental platforms demonstrate data collection, policy training, and closed-loop testing~\cite{bonati2023openran,polese2022colo}, but the reviewed literature does not yet establish common O-RAN conventions for interaction-record schemas, policy-promotion gates, or rollback interfaces.
%\hz{reference needed.}

\textbf{Key Takeaways.} The literature positions RLOps as the operational link between DRL development and continuous O-RAN control. The O-RAN workflow provides generic model-lifecycle functions, while DRL operation adds policy-dependent data provenance, staged validation, runtime assurance, and versioned policy replacement; common O-RAN conventions for these RL-specific elements have not yet been established.

\section{Platforms and Benchmarks}
\label{10_Platforms}
%\lj{A figure in each section to summarize all references.}

% \subsection{Testbeds and Emulators}

% \subsection{Software Stacks}

% \subsection{The Benchmark Gap}

% \subsection{Datasets}

% \subsection{Reproducibility}

\begin{figure*}
\centering
\includegraphics[width=0.98\linewidth]{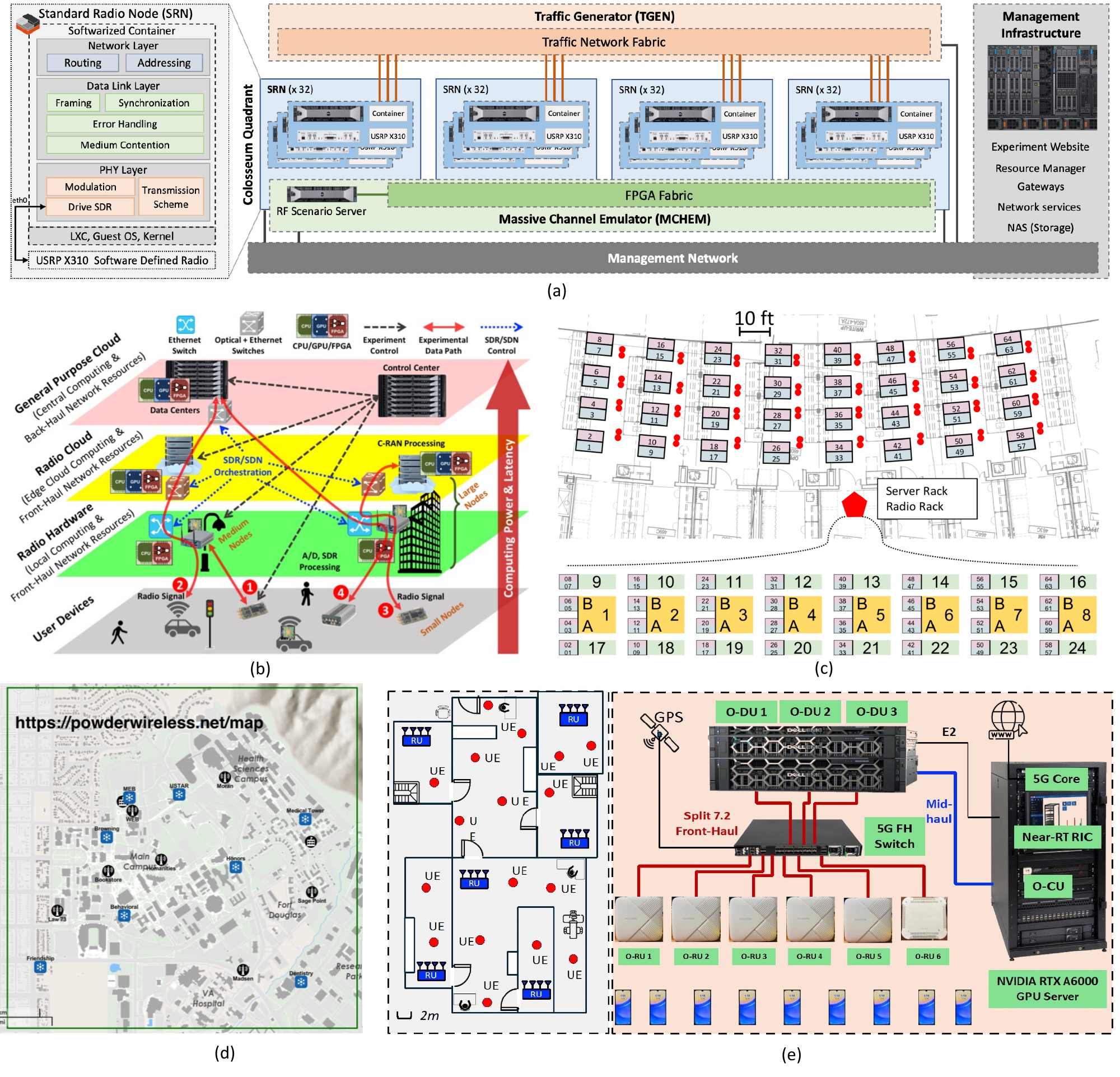}\vspace{-0.2in}
\caption{Some representative experimental platforms and testbeds for O-RAN and DRL-enabled RAN research.
\textbf{(a) Colosseum}, a large-scale wireless network emulator supporting repeatable RF experiments \cite{bonati2021colosseum};
\textbf{(b) COSMOS}, an urban-scale advanced wireless testbed supporting cloud, edge, and programmable RAN research \cite{raychaudhuri2020challenge};
\textbf{(c) Arena}, an indoor programmable wireless testbed for reproducible radio and networking experiments \cite{bertizzolo2020arena};
\textbf{(d) POWDER}, a city-scale programmable wireless testbed for advanced cellular and O-RAN experimentation \cite{breen2020powder};
and \textbf{(e) the MSU INSS Lab Private 5G Network}, a campus-deployed private 5G testbed with commercial O-RUs, smartphones, OpenAirInterface/srsRAN support, and GPU-enabled Near-RT RIC infrastructure \cite{msu_private_5g_platform}.}
\label{fig:testbeds}
\end{figure*}

Experimental platforms and benchmarks are the bridge between algorithmic claims and deployable intelligent control in O-RAN. A DRL agent is not a static predictor: it learns a policy through repeated interaction, and the policy later changes the behavior of the wireless system it controls. This creates a sharper evaluation problem than in conventional supervised learning or offline radio optimization. A policy that appears effective in a simplified simulator may become unstable when it receives delayed measurements, when users move according to unmodeled mobility patterns, when the channel statistics shift, or when the action has to be transported through an actual RIC-to-RAN control loop. Unconstrained exploration in a production RAN can create service and SLA risks; consequently, pre-deployment training and controlled simulation/emulation/testbed evaluation are important for learning-based controllers\cite{herrera2025tutorial}.

% The O-RAN architecture makes this evaluation problem more structured, but also more demanding. A DRL controller is commonly implemented as an xApp in the near-RT RIC, as an rApp in the non-RT RIC, or as a lower-level application close to the RAN node. Each placement defines an observation path, a control path, and a timing budget. The near-RT RIC, for example, operates at the timescale from 10 ms to 1 s and uses the E2 interface to collect KPMs and send control actions. The non-RT RIC operates above 1 s and guides the near-RT RIC through policies, enrichment information, and model-management functions. As a result, evaluating only the final network reward is insufficient. A complete evaluation must also check whether the measurements are actually exposed by the intended service model, whether the action can be expressed through the intended control interface, whether inference fits the loop deadline, whether the xApp or rApp can be onboarded and monitored, and whether the RIC--RAN interaction remains stable under realistic traffic and channel conditions.

This section surveys the experimental platforms for DRL in O-RAN from five perspectives. 
First, we review hardware-in-the-loop and wireless-emulation platforms. Second, we survey software stacks that enable xApp and rApp development, closed-loop control, and trajectory logging. Third, we discuss datasets and trace-generation pipelines, emphasizing that DRL requires decision trajectories. Fourth, we identify the benchmark gap that prevents fair comparison across studies. Finally, we propose reporting guidelines that make DRL-based O-RAN experiments reproducible, robust, and useful for future standardization and certification discussions.

\begin{table*}[!t]
\centering
\caption{Representative platforms for the evaluation of DRL-based control and optimization in O-RAN.}
\label{tab:platforms}
\scriptsize
\setlength{\tabcolsep}{2.2pt}
\renewcommand{\arraystretch}{1.18}

\begin{tabular}{
|>{\centering\arraybackslash}m{0.11\textwidth}
|>{\RaggedRight\arraybackslash}m{0.145\textwidth}
|>{\RaggedRight\arraybackslash}m{0.235\textwidth}
|>{\RaggedRight\arraybackslash}m{0.205\textwidth}
|>{\RaggedRight\arraybackslash}m{0.255\textwidth}|}
\hline

\rowcolor[gray]{0.9}
\multicolumn{1}{|c|}{
    \parbox[c][6ex][c]{0.11\textwidth}{
        \centering\textbf{Platform Class}
    }
}
&
\multicolumn{1}{c|}{
    \parbox[c][6ex][c]{0.145\textwidth}{
        \centering\textbf{Examples}
    }
}
&
\multicolumn{1}{c|}{
    \parbox[c][6ex][c]{0.235\textwidth}{
        \centering\textbf{Main Strength}
    }
}
&
\multicolumn{1}{c|}{
    \parbox[c][6ex][c]{0.205\textwidth}{
        \centering\textbf{Typical DRL Tasks}
    }
}
&
\multicolumn{1}{c|}{
    \parbox[c][6ex][c]{0.255\textwidth}{
        \centering\textbf{Main Limitation}
    }
}
\\
\hline

% ===================== LARGE-SCALE WIRELESS EMULATION =====================

\textbf{Large-scale wireless emulation}
&
Colosseum~\cite{polese2024colosseum},\newline
ColO-RAN~\cite{polese2022colo},\newline
OpenRAN Gym~\cite{bonati2023openran}
&
Repeatable wireless channels with softwarized protocol stacks and RIC components.
&
Slicing, scheduling, resource allocation, traffic steering, and robustness testing.
&
Access may be limited; experiments still depend on platform-specific configuration and container images.
\\
\hline

% ===================== INDOOR / LOCAL OTA TESTBEDS =====================

\textbf{Indoor or local over-the-air testbeds}
&
Arena~\cite{bertizzolo2020arena}
&
Real indoor propagation with synchronized programmable SDRs and repeatable over-the-air experimentation.
&
Indoor cellular control, multi-cell coordination, spectrum sharing, and robustness evaluation.
&
Smaller scale than city-scale platforms and harder to reproduce than controlled RF emulation.
\\
\hline

% ===================== CITY-SCALE PUBLIC PLATFORMS =====================

\textbf{City-scale public platforms}
&
POWDER~\cite{breen2020powder},\newline
COSMOS~\cite{kohli2020open,raychaudhuri2020challenge},\newline
AERPAW~\cite{panicker2021aerpaw}
&
Outdoor propagation, mobility, mmWave, aerial, and urban scenarios.
&
Traffic steering, handover, beam control, and UAV/aerial RAN control.
&
More expensive experiment scheduling and less freedom for unsafe exploration.
\\
\hline

% ===================== FEDERATED / MULTI-SITE INFRASTRUCTURES =====================

\textbf{Federated and multi-site research infrastructures}
&
SLICES~\cite{fdida2022slices},\newline
PAWR platforms~\cite{bonati2023openran}
&
Access to heterogeneous shared research infrastructures across multiple sites and deployment environments.
&
Cross-site generalization, multi-site training and evaluation, federated DRL, and cross-domain orchestration.
&
Common benchmark definitions and uniform telemetry across heterogeneous platforms remain limited.
\\
\hline

% ===================== OPEN AI CELLULAR FRAMEWORKS =====================

\textbf{Open AI cellular frameworks}
&
OAIC~\cite{upadhyaya2022prototyping},\newline
OpenRAN Gym extensions~\cite{bonati2023openran}
&
AI-enabled control workflows, xApp prototyping, and security and explainability hooks.
&
Explainable control, adversarial evaluation, and rapid xApp prototyping.
&
Often requires integration with a separate simulator, emulator, or SDR stack for radio realism.
\\
\hline

% ===================== PRIVATE 5G CAMPUS TESTBED =====================

\textbf{Private 5G campus testbed}
&
MSU INSS Lab Private 5G Network~\cite{msu_private_5g_platform}
&
Commercial indoor/outdoor O-RUs, smartphones,
OpenAir\allowbreak Interface and srsRAN support,
and a GPU-enabled Near-RT RIC.
&
Real-network xApp and DRL evaluation, private 5G control, and site-aware experimentation.
&
Single-campus deployment with more limited scale than large shared testbeds.
\\
\hline

\end{tabular}
\end{table*}

\subsection{Hardware-in-the-Loop and Wireless Emulation Platforms}
\label{subsec:hil-platforms}

\textbf{Bridging Simulation and Real-world Deployment.}
Hardware-in-the-loop (HIL) and wireless-emulation platforms provide an intermediate stage between mathematical simulation and commercial-network deployment. They expose DRL agents to realistic radio, traffic, protocol-stack, and timing behavior while retaining greater repeatability and experimental control than production networks. This capability is particularly important in O-RAN, where a learning agent must interact with a disaggregated RAN, telemetry pipelines, RIC components, and practical control interfaces rather than an idealized resource-allocation model. Fig.~\ref{fig:testbeds} shows some representative platforms in the research community.

\textbf{Large-scale Emulation and Closed-loop xApp Experimentation.}
Colosseum \cite{bonati2021colosseum} is one of the most widely used environments for this purpose, providing a large-scale wireless network emulator in which software-defined radio nodes operate through a configurable channel emulator. ColO-RAN and OpenRAN Gym extend this infrastructure toward O-RAN experimentation by integrating RAN nodes, RIC components, traffic generation, data collection, and xApps for closed-loop control \cite{bonati2023openran,bonati2022intelligent,polese2022colo}. 
These platforms are particularly suitable for DRL because policies can be trained and evaluated under repeatable channel and traffic conditions while interacting with softwarized protocol stacks and realistic control loops. OpenRAN Gym\cite{bonati2023openran} further structures the AI/ML workflow from RAN data collection and model development to xApp deployment and closed-loop evaluation, enabling reproducible comparison of learning-based control policies.

\textbf{Wireless Testbeds for Complementary Deployment Scenarios.}
Public experimental infrastructures provide different forms of wireless and system realism. 
For example, Arena \cite{bertizzolo2020arena} offers an indoor SDR-based environment for reproducible cellular and networking experiments, while POWDER \cite{breen2020powder} provides city-scale programmable sub-6~GHz infrastructure with support for O-RAN experimentation. 
COSMOS \cite{raychaudhuri2020challenge,kohli2020open} targets dense urban, mmWave, cloud/edge, and xHaul experimentation, whereas AERPAW \cite{panicker2021aerpaw} extends programmable wireless experimentation to UAVs and aerial mobility scenarios. SLICES \cite{fdida2022slices} aims to federate European networking, cloud, orchestration, and research infrastructures, and OAIC integrates AI-enabled cellular control with O-RAN-oriented experimentation and security research \cite{upadhyaya2022prototyping}. Table~\ref{tab:platforms} summarizes the roles of these platforms in DRL-for-O-RAN evaluation.

\textbf{Exposing Deployment Constraints Beyond Simulation.}
HIL and over-the-air platforms expose practical deployment constraints that are often abstracted away in simplified simulators. When moving from simulation to emulated or real testbeds, inference and communication delays can affect whether a policy satisfies its control-loop budget, while KPM reporting periodicity and control-path latency may differ from the idealized timing assumed by the MDP. In addition, practical deployments may encounter compatibility, data-formatting, and orchestration issues across RICs, RAN components, and software stacks~\cite{herrera2025tutorial,lacava2023ns}. Multiple xApps operating on the same RAN functions may also introduce direct or indirect conflicts when their control objectives or actions interact~\cite{polese2023understanding}. Real radio environments can further expose operating conditions and data distributions that were not represented during training. At the same time, platform configurations, software and container versions, traffic profiles, channel scenarios, random seeds, and other experimental parameters should therefore be documented and, when possible, released together with the implementation and evaluation artifacts~\cite{henderson2018deep,agarwal2021deep}. We therefore recommend using HIL evaluation together with open software stacks, public configurations, trace-driven experiments, and well-defined benchmark tasks to improve the realism, transparency, and comparability of DRL evaluations for O-RAN.
% \hz{all these statements need to be backed by references. if you cannot find references for this, remove it.}

\textbf{Campus-scale Private 5G Experimentation.}
The MSU Private 5G Network provides a complementary deployment environment based on commercial infrastructure \cite{msu_private_5g_platform}. Deployed in the MSU Engineering Building, it includes six indoor O-RUs, one outdoor O-RU, and more than 20 commercial smartphones, providing realistic indoor-and-outdoor conditions for O-RAN control loops, xApp development, and DRL-based RAN optimization. The platform supports both OpenAirInterface and srsRAN, enabling studies of policy portability across different RAN implementations, while four NVIDIA RTX PRO 6000 Blackwell Max-Q GPUs support Near-RT RIC inference, monitoring, and learning workloads. Its deployment visualization in Fig.~\ref{fig:testbeds}(e) can further support mobility, topology-aware optimization, radio-map-assisted control, and digital-twin-style evaluation. The platform therefore complements large-scale emulation environments by enabling DRL-for-O-RAN methods to be studied under practical campus private-5G conditions.

\subsection{Software Stacks for xApp/rApp Development and Closed-Loop Control}
\label{subsec:software-stacks}

% \textbf{From DRL algorithms to deployable O-RAN applications.}
Software stacks determine whether a DRL policy remains an abstract algorithm or can be integrated into an operational xApp or rApp. A useful stack should provide a RAN environment, a RIC or RIC-like controller, telemetry collection, action delivery, and experiment logging. For DRL, logging is particularly important because observations, actions, rewards, next observations, timestamps, KPMs, KPIs, and control-loop states are required to reproduce training and evaluation. The software stack therefore acts as the interface between the learning algorithm and the practical O-RAN control loop.

\textbf{Softwarized Implementations of RAN and RIC.}
Open-source RAN stacks such as srsRAN/OCUDU \cite{srsran} and OpenAirInterface \cite{kaltenberger2020openairinterface} enable 3GPP-compliant network functions to run on general-purpose hardware, often together with SDR front ends. SCOPE \cite{bonati2021scope} provides another softwarized prototyping environment and has been used extensively in the Colosseum ecosystem. 
At the control layer, platforms such as the O-RAN Software Community (OSC) Near-RT RIC, ONF SD-RAN, FlexRIC, and 5G-EmPOWER provide different tradeoffs among standards alignment, modularity, deployment complexity, and service-model support \cite{polese2023understanding,foukas2016flexran,onos-sdran}. 
In particular, FlexRIC \cite{schmidt2021flexric} is convenient for rapid prototyping because of its lightweight RIC and agent architecture, whereas OSC\cite{oransc2026} and SD-RAN\cite{onf_sdran_opensource} provide a broader integration with O-RAN-oriented software ecosystems.

\textbf{Interface-aware Simulation and Experimentation Scalability.}
Simulation-oriented software complements physical RAN implementations by enabling large numbers of controlled experiments. 
For example, ns-O-RAN \cite{lacava2023ns} connects an ns-3-based 4G/5G simulated RAN to a real Near-RT RIC through an O-RAN-compliant E2 interface. This allows xApps to be developed and evaluated using realistic control interfaces before deployment on experimental infrastructure, while supporting substantially larger traffic, mobility, and topology sweeps than many physical testbeds. More generally, such platforms illustrate an important principle for DRL evaluation: simulators should expose measurements and execute actions through interfaces resembling those used by actual xApps rather than hiding the entire control loop behind an internally computed reward.

\textbf{DRL Automation, Explainability, and Robustness.}
Recent frameworks extend software support beyond basic xApp execution toward systematic AI/ML experimentation. PandORA \cite{tsampazi2024pandora} automates the design and training of DRL agents, packages them as xApps, and evaluates them on Colosseum, enabling controlled comparisons across architectures, rewards, action spaces, and decision timescales. EXPLORA  \cite{fiandrino2023explora} incorporates explainability into DRL-based O-RAN control and evaluates explanations in an O-RAN-compliant Near-RT RIC environment, while AdvO-RAN \cite{hassan2025advo} studies adversarial DRL, robust training, and evasion attacks against learning-based xApps. Together, these efforts show that O-RAN software stacks increasingly need to support not only policy execution, but also design-space exploration, interpretability, robustness, and safety analysis.

% \textbf{Fragmentation and the need for reproducible software interfaces.}
\textbf{Fragmentation and Reproducibility.}
Current O-RAN experimentation platforms differ substantially in their support for interfaces, service models, RIC integration, and application lifecycle management. For example, existing platforms may implement only a subset of the E2 functionality or support different combinations of standardized and custom E2 service models, while the available mechanisms for xApp deployment, onboarding, and lifecycle management also vary across RIC implementations~\cite{herrera2025tutorial, polese2023understanding}. These implementation differences complicate the portability of DRL controllers and make results obtained on different software stacks difficult to compare directly. Therefore, reproducible evaluation therefore requires sufficient information about the interaction trajectory and experimental configuration~\cite{henderson2018deep, agarwal2021deep}. Future software stacks should therefore provide standardized hooks. Greater standardization at this level would facilitate cross-platform comparison and provide a more reproducible path from DRL algorithm development to O-RAN deployment.

\begin{comment}
    \input{tables/sec10/sec10_table_benchmark_suite}
\end{comment}

\subsection{Datasets and Data Generation Pipelines}
\label{subsec:datasets-traces}

% \hz{need references. The whole subsection has no reference. this is weird.}

Datasets are central to DRL for two reasons\cite{polese2022colo, herrera2025tutorial}. First, Because unconstrained online exploration can disrupt service, pre-deployment/offline training is an important option for learning-based RAN control. Second, trace-driven evaluation provides a common basis for comparing policies. However, useful DRL datasets differ from ordinary telemetry datasets. A static set of KPMs is not enough to reconstruct a sequential decision problem\cite{kumar2020conservative}. A DRL dataset should describe how the environment evolves after each action.

Existing studies rely on three broad data sources. 

\begin{itemize}[leftmargin=0.15in]
\item 
\textit{Simulation-generated data:}
Simulation is scalable and controllable: researchers can vary topology, traffic load, mobility, channel models, slice requirements, and service types. It is therefore useful for early-stage studies on slicing, resource allocation, traffic steering, scheduler reconfiguration, and energy saving. The limitation is the sim-to-real gap. A policy trained under simplified assumptions may fail when measurements are noisy, actions are delayed, hardware imposes constraints, or traffic demand differs from the training distribution \cite{lacava2023ns,herrera2025tutorial}.

\item 
\textit{Emulation or testbed-generated data:} 
Colosseum-based environments, ColO-RAN, OpenRAN Gym, Arena, POWDER, and similar platforms can generate traces that include the interaction among wireless channels, protocol stacks, RIC services, xApps, and traffic generators \cite{bonati2021colosseum,bonati2023openran,polese2022colo,bertizzolo2020arena,breen2020powder}. Such traces are more expensive to collect, but they are valuable because they contain system effects that a standalone simulator may miss: measurement delay, control-loop timing, interface bottlenecks, and the behavior of real software components.

\item 
\textit{Production or near-production data:} 
These traces are valuable because they reflect real traffic demand, mobility, interference, and operational constraints. Large public datasets containing production-grade O-RAN decision trajectories remain limited\cite{polese2022colo, herrera2025tutorial}. Even when such traces are available, they may contain only counters and KPIs. This makes them useful for system identification, offline model calibration, or traffic generation, but insufficient as a complete DRL benchmark unless they are paired with decision logs.
\end{itemize}

A data-generation pipeline for DRL-based O-RAN should  be designed around decisions. At minimum, each data trajectory should include
\begin{equation}
\tau = \{(s_t, a_t, r_t, s_{t+1}, \Delta_t, z_t)\}_{t=0}^{T-1},
\end{equation}
where $s_t$ is the network state/observation exposed to the agent, $a_t$ is the selected action, $r_t$ is the reward or cost signal, $s_{t+1}$ is the next state/observation, $\Delta_t$ records the control delay or timestamp information, and $z_t$ stores metadata such as topology, traffic profile, mobility state, slice mix, channel scenario, and agent placement. For offline RL, the dataset should also include the behavior policy when available, because out-of-distribution actions are a major source of error in offline policy learning\cite{kumar2020conservative}.

The dataset should also specify the O-RAN control location. A trajectory collected for an xApp at the near-RT RIC is not equivalent to a trajectory collected for an rApp at the non-RT RIC. The observation granularity, action space, reward horizon, and timing constraints differ \cite{polese2023understanding}. Similarly, a dataset for a proposed dApp or in-node control loop should include local MAC/PHY state that may not be exposed through E2. Without placement information, the same data can be misused to formulate an unrealizable control problem.

Diversity is equally important. A useful dataset suite should cover normal operation and stressed conditions. Normal operation should include several load levels, mobility patterns, channel conditions, slice requirements, and user distributions. Stressed operation should include congestion, mobility bursts, sudden slice demand changes, radio failures, delayed or missing telemetry, noisy observations, jamming, and adversarial perturbations, the latter being particularly relevant to robustness evaluation of learning-based xApps \cite{hassan2025advo}.

\subsection{Benchmarks for DRL in O-RAN}
\label{subsec:benchmark-gaps}

To the best of our knowledge, the DRL-for-O-RAN literature does not yet have a widely adopted benchmark suite analogous to Atari\cite{mnih2013playing} or MuJoCo\cite{todorov2012mujoco} in the broader DRL community. This gap matters because DRL performance is sensitive to design choices that are often underreported: state construction, action granularity, reward shaping, traffic generation, mobility assumptions, control interval, random seeds, and baseline tuning. Without common benchmarks, it is difficult to know whether a proposed method improves the state of the art or simply benefits from a favorable setup.
In what follows, we present the requirements of a credible benchmark for DRL in O-RAN.

\begin{itemize}[leftmargin=0.15in]
\item 
\textbf{Task Standardization}. Existing work covers network slicing, traffic steering, scheduler reconfiguration, MIMO and MCS adaptation, resource allocation, energy saving, jamming mitigation, robustness, and explainability. However, each paper often defines its own version of the task. For example, a slicing benchmark may allocate PRBs per slice, tune scheduling weights, admit flows, or jointly configure multiple resource dimensions. These are related but not identical decision problems. A benchmark suite should separate them and define the observation, action, reward, constraint, and timescale for each.

\item 
\textbf{Metric Standardization}. Papers report throughput, latency, packet delivery ratio, SLA violation, spectral efficiency, energy consumption, fairness, resource utilization, convergence speed, inference time, training cost, and robustness. This diversity is appropriate because O-RAN use cases are diverse, but it complicates comparison. A slicing policy optimized for resource utilization cannot be compared directly with another optimized for tail latency unless both are evaluated on a shared set of primary and secondary metrics. Benchmarks should therefore distinguish between optimization objectives and reporting metrics. The reward may focus on a primary objective, but the paper should still report the full metric vector.

\item 
\textbf{Baseline Quality}. Some studies compare against simple heuristics, while others compare multiple DRL algorithms such as DQN, PPO, DDPG, TD3, SAC, CQL, or MARL variants. Some studies compare against simple heuristics, while others evaluate multiple DRL architectures and configurations \cite{tsampazi2024pandora}. A credible benchmark should include at least three baseline classes: a classical heuristic or rule-based controller, a non-DRL optimization or model-based baseline when feasible, and standard DRL baselines with documented hyperparameters. Ablation studies should also be included to isolate the effect of history, encoder design, reward terms, safety filters, or action masking.

\item 
\textbf{Reward Transparency}. Reward design can dominate the result. Many O-RAN rewards combine throughput, latency, fairness, energy, and SLA penalties through weighted sums. Different weights can change the learned policy substantially, yet the sensitivity of the result to those weights is rarely reported. Different reward definitions and weights can substantially change the learned policy and the conclusions drawn from an experiment \cite{tsampazi2024pandora}. Benchmarks should either fix the reward for comparability or provide a standardized sensitivity analysis. When SLAs are central, a constrained formulation should be reported alongside any scalar reward so that the operator-facing thresholds are explicit.
\end{itemize}

\begin{comment}
\input{tables/sec10/sec10_table_benchmark_suite}

Table~\ref{tab:benchmark-suite} proposes a minimal benchmark suite. The purpose is not to freeze the research community into a single set of scenarios, but to provide a common base layer. Researchers can still add richer scenarios, but the shared layer would allow direct comparison across algorithms, platforms, and papers.
\end{comment}

% \subsection{Reproducibility and Reporting Guidelines}
\subsection{Evaluation Reporting for Reproducibility}
\label{subsec:reproducibility}

% \hz{need references. The whole subsection has no reference. this is weird.}

Reproducibility is a persistent challenge in DRL-based networking research, particularly when learning algorithms interact with wireless environments, protocol stacks, and practical control loops \cite{bonati2023openran,tsampazi2024pandora}. In O-RAN, the challenge is amplified by the interaction of learning algorithms, wireless environments, protocol stacks, and control-loop implementation. A policy may perform well under one channel model, traffic trace, RIC implementation, or xApp interval, but fail under another. Reproducibility should therefore be treated as an evaluation requirement rather than an optional artifact.
In what follows, we provide some evaluation reporting requirements for promoting reproducibility of DRL in O-RAN. 

\begin{itemize}[leftmargin=0.15in]
\item 
\textbf{Stochastic Learning}. Random seeds, neural-network initialization, replay-buffer sampling, exploration noise, environment randomness, user arrival, traffic generation, and mobility all affect the learned policy\cite{henderson2018deep, colas2018many}. Reporting only the best run is insufficient. Studies should report multiple random seeds, confidence intervals, and a clear distinction between training performance and evaluation performance.

\item 
\textbf{MDP Description}. Studies that report only network KPIs do not provide enough information to reproduce a DRL agent. The sensitivity of DRL performance to reward definitions, action spaces, architectures, and decision timescales further motivates complete reporting of the MDP and training configuration \cite{tsampazi2024pandora}. 
Therefore, studies should report the complete observation space, action space, reward function, discount factor, episode definition, control interval, policy architecture, optimizer, learning rate, batch size, replay buffer, training steps, exploration schedule, target-network update rule when applicable, and policy update frequency. 
For multi-agent DRL, studies should also report whether training is centralized, whether execution is decentralized, what information agents exchange, and how non-stationarity is handled.

\item 
\textbf{O-RAN Context}. Because O-RAN defines specific architectural placements, interfaces, service models, and timing constraints, a DRL study should explicitly report its control context \cite{polese2023understanding,lacava2023ns}. A DRL study should report where the agent is deployed, whether it is implemented as an xApp, rApp, dApp, or external controller, what KPMs are collected, which service model or interface is assumed, how often observations are reported, how actions are delivered, and whether the RIC-RAN interaction follows an O-RAN-compliant control path. If a Near-RT RIC is used, inference latency and action-delivery latency should be measured, not merely assumed. If the action is abstracted, the abstraction should be stated explicitly.

\item 
\textbf{Robustness Evaluation}. A policy that works only under nominal traffic is not deployment-ready. Evaluations should include distribution shift, traffic bursts, mobility changes, unseen slice demand, noisy measurements, and delayed telemetry. Adversarial perturbations should also be considered when evaluating learning-based xApps \cite{hassan2025advo}. Explainability can further support robustness analysis and policy debugging \cite{fiandrino2023explora}. For example, an explainability framework can reveal whether two policies that achieve similar throughput rely on different state features, which helps operators identify brittle behavior and debug reward design.
\end{itemize}

\begin{comment}
\input{tables/sec10/sec10_table_reporting_checklist}

Table~\ref{tab:reporting-checklist} gives a concise checklist for future papers. The goal is to make results comparable across simulators, emulators, and testbeds, and to make it possible to determine whether a DRL policy is actually deployable in O-RAN.
\end{comment}

\subsection{Remark and Outlook}
\label{subsec:platform-lessons}

The current platform landscape shows clear progress. O-RAN researchers can now train and test DRL controllers with O-RAN stacks, RIC implementations, software-in-the-loop simulators, and hardware-in-the-loop emulators. Colosseum, OpenRAN Gym, ColO-RAN, ns-O-RAN, OAIC, and related platforms have moved the field beyond isolated algorithmic evaluation. They make it possible to test whether DRL policies can be packaged as xApps, consume KPMs, interact with a RIC, and operate under realistic timing and radio conditions.

The remaining gap is not the absence of platforms, but the absence of shared experimental contracts. A credible result should state what is realistic and what is abstracted. It should report both learning performance and system feasibility. It should compare against well-defined baselines. It should release enough configuration detail to reproduce the environment. Most importantly, it should evaluate robustness and trust, not only average reward. The next stage of DRL-for-O-RAN research should therefore move toward shared benchmark tasks, public trajectory datasets, common logging schemas, open-source xApp implementations, and reporting templates aligned with O-RAN control loops. These practices would make the literature more cumulative and would help distinguish policies that are genuinely deployable from policies that perform well only in narrow and under-specified simulations.

% \clearpage

\section{Standards and Industry Adoption}
\label{11_Standards_Industry}
%\lj{A figure in each section to summarize all references.}

% The previous section examined how DRL-based O-RAN controllers should be evaluated through platforms, datasets, benchmarks, and reproducibility practices. Experimental validity, however, is only one part of deployability. A DRL policy that performs well in a simulator, emulator, or research testbed still needs to fit the standard interfaces, software packaging rules, security expectations, certification processes, and procurement constraints that shape real O-RAN deployments. This section therefore shifts the perspective from experimental evaluation to standardization and industrial adoption.

This section shifts the perspective from experimental evaluation to standardization and industrial adoption.
This shift is necessary because O-RAN is not simply a research architecture for running intelligent algorithms. It is an interoperability framework in which multi-vendor RAN functions, RIC platforms, SMO services, O-Cloud resources, and testing facilities must operate across organizational boundaries. 
This shift brings two new requirements for DRL design and deployment. 
First, the DRL agent is constrained by the standard interface through which it observes and acts. An xApp cannot use arbitrary state variables if the corresponding E2 node does not expose them, and it cannot apply arbitrary actions if the intended service model or control procedure does not support them. 
Second, the DRL agent must be managed as an operational software artifact. It needs a deployment package, a model version, a monitoring path, a rollback strategy, a safety envelope, and an audit trail.

\begin{comment}

The central question of this section is therefore: what must happen for a DRL-based xApp, rApp, or future dApp to move from a paper prototype into an interoperable and certifiable O-RAN ecosystem? We answer this question from five angles as shown in Figure~\ref{fig:structure11}.

\begin{figure*}[t]
    \centering
    \includegraphics[width=0.9\linewidth]{figures/11_Structure.pdf}
    \caption{Roadmap: Evolution Toward DRL-enabled O-RAN}.
    % \hz{redraw this figure; make it compact and concise.}
    \hz{poor figure. Editable file is missing.}
    \label{fig:structure11}
\end{figure*}

\end{comment}

%Section~\ref{subsec:oran-ai-ml-specs} reviews O-RAN Alliance AI/ML specifications and explains how they shape the lifecycle of learning-based controllers. Section~\ref{subsec:sdo-alignment} discusses the relationship between O-RAN, 3GPP, ETSI, IEEE, and related standards bodies. Section~\ref{subsec:regulatory-pressure} analyzes regulatory and policy drivers, with emphasis on critical-infrastructure accountability and trustworthy automation. Section~\ref{subsec:research-projects} reviews open-source and research projects that bridge standardization and experimentation. Section~\ref{subsec:otic-certification} discusses certification, PlugFests, and OTICs, and argues that DRL-based controllers will require an assurance profile that goes beyond conventional interface conformance.

\subsection{O-RAN Alliance AI/ML Specifications}
\label{subsec:oran-ai-ml-specs}

The O-RAN Alliance\cite{oran_architecture_v16} provides the most direct standardization context for DRL in O-RAN. Its objective is to define an open, intelligent, virtualized, and interoperable RAN architecture built around disaggregated network functions, RICs, O-Cloud infrastructure, and open interfaces \cite{polese2023understanding}. For DRL, the most important aspect of this architecture is that intelligence is not an external add-on. It is embedded into the RIC hierarchy and the SMO framework, and it is connected to the RAN through interfaces that determine what learning agents can observe, what actions they can apply, and how their models can be managed.

\textbf{O-RAN Work Groups (WGs).}
The O-RAN Alliance has organized this work across technical groups. 
WG2\cite{oran_wg2} is especially important for learning-based control because it defines the Non-RT RIC, the A1 interface, the R1 interface, use cases and requirements related to the Non-RT RIC, and the AI/ML workflow. WG3\cite{oran_wg3} defines the Near-RT RIC and the E2 interface, including the service models that expose telemetry and control functions to xApps. WG6\cite{oran_wg6} defines O-Cloud-related functions and the O2 interface. WG10 addresses OAM and the O1 interface. 
WG11\cite{oran_wg11} addresses security. The nGRG studies how O-RAN should evolve toward 6G, including AI-native RAN functions, new control loops, and extensions that may eventually support real-time or in-node intelligence.

From a DRL perspective, these work groups collectively define an execution envelope. An rApp in the Non-RT RIC can operate on long time horizons, access SMO data services through R1, use O1 telemetry, and influence lower-level control through A1 policies or enrichment information. An xApp in the Near-RT RIC can subscribe to KPMs over E2 and send control actions through E2 service models, typically under a 10\;ms -- 1\;s loop budget. A future dApp-like entity may act closer to the O-DU or O-CU for sub-10 ms control, but this remains less mature in normative O-RAN specifications\cite{oran_ngrg}. This hierarchy is attractive for DRL because it naturally maps long-term policy learning, near-real-time adaptation, and local fast control to different agents. It is also restrictive because the observation and action spaces are not arbitrary; they are bounded by what the relevant interface exposes.

\textbf{AI/ML Workflow.}
The O-RAN AI/ML workflow is the key lifecycle mechanism for standardizing learning-based controllers. It covers data collection and preparation, training, validation, publication, deployment, inference, and continuous operation \cite{oran_wg2, oran-aiml-workflow,polese2023understanding}. For DRL, this workflow has three important implications. 
First, it encourages offline training or pretraining before deployment, which is consistent with the safety concerns discussed earlier in this survey. Direct online exploration in a live RAN can violate SLAs\cite{herrera2025tutorial}, so a policy should be trained and validated before it is allowed to affect operational control. 
Second, the workflow separates model production from model execution. A DRL model may be trained in a central environment, published to a catalog, deployed as part of an xApp or rApp, and monitored during inference. 
Third, continuous operation provides a standard place to discuss drift detection, retraining, rollback, and model replacement.

\textbf{A1 Interface.}
The A1 interface is especially relevant to hierarchical DRL. A1 Policy Management can carry high-level policy guidance from the Non-RT RIC to the Near-RT RIC. A1 Enrichment Information can provide context such as traffic forecasts, mobility predictions, or external information that is not directly available to the RAN. Recent O-RAN specifications include A1 ML model-management services and procedures. The 2026 public A1 specifications define updated A1-ML procedures and an A1-ML/v1 model-management service, while R1 also exposes AI/ML model lifecycle services\cite{oran_wg2_a1}. In a DRL system, A1 can therefore serve as the interface between a slow-timescale rApp and a fast-timescale xApp. For example, an rApp may learn a long-term policy that specifies slice-level priorities or traffic-steering preferences, while an xApp learns a near-real-time policy that translates those priorities into E2 control actions.

\textbf{E2 Interface.}
The E2 interface is the central interface for xApp-based DRL. E2SM KPM\cite{oran_e2sm_kpm_v8} provides measurement reporting, while E2SM RC\cite{oran_e2sm_rc_v10} provides control capabilities for radio-resource allocation, mobility, bearer control, access control, and related RAN functions. The E2 setup procedure also exposes an important practical constraint\cite{oran_e2ap_v8}: an E2 node advertises the RAN functions it supports, and an xApp can only subscribe to or control what the node exposes. A DRL paper that assumes arbitrary control over PRBs, handovers, MCS, or power should therefore state whether that action is expressible through an existing or proposed service model. Otherwise, the work may be algorithmically interesting but not deployable as an interoperable O-RAN xApp.

\begin{table*}[!t]
\centering
\caption{Standards alignment for DRL-based O-RAN controllers.}
\label{tab:sdo-alignment}
\scriptsize
\setlength{\tabcolsep}{2.2pt}
\renewcommand{\arraystretch}{1.18}

\begin{tabular}{
|>{\centering\arraybackslash}m{0.110\textwidth}
|m{0.275\textwidth}
|m{0.255\textwidth}
|m{0.320\textwidth}|}
\hline

\rowcolor[gray]{0.9}
\multicolumn{1}{|c|}{
    \parbox[c][6ex][c]{0.110\textwidth}{
        \centering\textbf{Organization}
    }
}
&
\multicolumn{1}{c|}{
    \parbox[c][6ex][c]{0.275\textwidth}{
        \centering\textbf{Relevant scope}
    }
}
&
\multicolumn{1}{c|}{
    \parbox[c][6ex][c]{0.255\textwidth}{
        \centering\textbf{Impact on DRL controllers}
    }
}
&
\multicolumn{1}{c|}{
    \parbox[c][6ex][c]{0.320\textwidth}{
        \centering\textbf{Typical evidence expected in a deployable study}
    }
}
\\
\hline

% ===================== O-RAN ALLIANCE =====================

\textbf{O-RAN Alliance} \cite{oran_architecture_v16}
&
RICs, SMO, O-Cloud, E2, A1, R1, O1, O2, fronthaul, testing profiles
&
Defines where xApps and rApps run, what they observe, and how they act
&
Interface assumptions, service-model support, xApp/rApp packaging, RIC integration, lifecycle behavior
\\
\hline

% ===================== 3GPP =====================

\textbf{3GPP} \cite{3gpp_ts_38401}
&
RAN protocol stack, QoS, mobility, slicing, NG-RAN architecture, AI/ML for NG-RAN and NR
&
Defines the protocol semantics of many states, actions, and constraints
&
Mapping of actions to 3GPP procedures, QoS and SLA interpretation, compatibility with RAN protocol behavior
\\
\hline

% ===================== ETSI =====================

\textbf{ETSI} \cite{etsi_gs_nfv_006_v451,etsi_ts_103859_v1710}
&
NFV, MANO, OSM, virtualization and orchestration concepts
&
Shapes deployment, workload placement, lifecycle management, and service orchestration
&
Model deployment plan, scaling behavior, rollback, resource requirements, O-Cloud or SMO integration
\\
\hline

% ===================== IEEE =====================

\textbf{IEEE} \cite{ieee_8021cm_2018,ieee_1588_2019}
&
Fronthaul, xHaul, timing, synchronization, Ethernet and packet transport
&
Affects latency, jitter, synchronization, and reliability assumptions for closed-loop control
&
Delay model, timing budget, transport assumptions, robustness to jitter and packet loss
\\
\hline

% ===================== REGULATORY AND SECURITY =====================

\textbf{Regulatory and security}
\cite{eu_ai_act_2024,eu_nis2_2022,eu_gdpr_2016,oran_wg11}
&
Cybersecurity, AI governance, privacy, auditability, accountability, and software supply-chain security
&
Sets assurance and risk-management requirements for autonomous control
&
Security/event logs, traceability, explainability, human oversight, privacy safeguards, SBOM/provenance, and secure update/rollback controls
\\
\hline

\end{tabular}
\end{table*}

\textbf{Missing Pieces.}
To the best of our knowledge, after reviewing the current O-RAN architecture, AI/ML workflow, A1/R1 application-management specifications, there is no DRL-specific descriptor that jointly declares. A deployable DRL controller needs more than a container image and an interface endpoint. It should also declare its observation requirements, action schema, reward or objective interpretation, safety constraints, latency budget, model provenance, training dataset, intended deployment scope, fallback policy, monitoring metrics, and conflict domain\cite{oran_architecture_v16, oran_wg2}. Without such a description, portability across RIC implementations and E2-node vendors will remain limited.
We therefore argue that future O-RAN AI/ML specifications should include a controller-assurance descriptor for learning-based xApps and rApps. Such a descriptor would not standardize the internal DRL algorithm, but it would standardize the operational claims that must be tested.

\begin{comment}
    \input{tables/sec11/sec11_table_oran_drl_specs}
\end{comment}

\subsection{IEEE, ETSI, 3GPP, and Other SDO Alignment}
\label{subsec:sdo-alignment}

O-RAN does not replace the broader mobile-network standardization ecosystem. Instead, it extends and complements it. The O-RAN Alliance is an industry consortium rather than a formal SDO, so standard-related activity must align with organizations such as 3GPP\cite{3gpp_ts_38401}, ETSI\cite{etsi_gs_nfv_006_v451, etsi_ts_103859_v1710}, ITU\cite{itu_r_m2150_3}, and IEEE\cite{ieee_8021cm_2018, ieee_1588_2019}. This alignment is essential for DRL because the learning agent sits at the intersection of radio protocols, O-RAN interfaces, virtualization infrastructure, transport networks, and security processes.

\textbf{3GPP Specifications.}
3GPP TS 38.401\cite{3gpp_ts_38401}, introduced with Release 15 and maintained in later releases, defines the NG-RAN architecture including gNB-CU/gNB-DU splits; O-RAN builds on this foundation and further specifies the O-RU and 7.2x Open Fronthaul split. These 3GPP-defined functions determine where measurements originate and which protocol-layer parameters have operational meaning. A DRL agent that controls mobility, for example, must respect 3GPP mobility procedures and the semantics of handover-related information elements. A DRL agent that controls slicing or bearer behavior must respect QoS-flow and slice definitions. Therefore, even when a DRL policy is implemented through O-RAN interfaces, the meaning of many states and actions remains grounded in 3GPP.

3GPP is also increasingly relevant because 5G-Advanced has introduced AI/ML into RAN and system discussions. Release 18 \cite{3gpp_ts_38401_v18} includes AI/ML for NG-RAN work based on earlier study items, including data collection and signaling support over existing 5G network interfaces \cite{3gpp-aiml-ngran}. AI/ML has also been studied for NR air-interface functions such as CSI feedback, beam management, and positioning \cite{lin2023overview}. Release 19\cite{3gpp_ts_38401} discussions further move toward broader AI-native networking and 6G-oriented functionality \cite{lin2025bridge}. For DRL in O-RAN, this means that the boundary between O-RAN intelligence and 3GPP intelligence will become more important. If a learning function is standardized directly in 3GPP, an O-RAN controller may need to coordinate with it rather than duplicate it. If O-RAN exposes an interface to a 3GPP-defined function, the DRL agent must respect the 3GPP semantics of that function.

\textbf{ETSI Specifications.}
ETSI\cite{etsi_gs_nfv_006_v451, etsi_ts_103859_v1710} contributes through virtualization, management, and orchestration. O-RAN O-Cloud and O2 discussions naturally intersect with ETSI NFV and MANO concepts, because xApps, rApps, RAN functions, and supporting services are deployed as software workloads over cloud infrastructure. ETSI has also adopted O-RAN fronthaul user, control, and synchronization plane specifications as an ETSI standard, which illustrates how O-RAN technical work can move into a formal standardization channel \cite{polese2023understanding}. For DRL, ETSI alignment matters because model placement and lifecycle management are infrastructure problems as much as algorithmic problems. A near-RT xApp with a large neural network may require GPU or accelerator support. A distributed multi-agent DRL system may require edge placement. A safe rollout may require blue-green deployment, rollback, or canary testing. These are orchestration problems that cannot be solved only at the algorithm layer.

\textbf{IEEE Specifications.}
IEEE is relevant primarily through the IEEE 802.1 Time-Sensitive Networking standards and IEEE 1588 Precision Time Protocol\cite{ieee_8021cm_2018, ieee_1588_2019}. These transport and timing constraints should therefore be reflected when evaluating latency-sensitive O-RAN control loops. Open fronthaul and xHaul networks depend on packet transport, timing accuracy, and synchronization behavior. These properties directly affect DRL when the agent controls functions with tight timing constraints. For example, a policy that adjusts beam or scheduling behavior may perform well in an ideal simulator but fail when fronthaul jitter, synchronization error, or transport congestion affects the timing of measurements and actions. IEEE-related transport and timing work therefore shapes the delay and reliability assumptions under which DRL controllers must be tested.

\textbf{Key Takeaways.}
Table~\ref{tab:sdo-alignment} summarizes the standards for DRL-based O-RAN controllers. 
The interaction among these standard organizations leads to a practical rule: a DRL-for-O-RAN contribution should state its standardization layer. Some works are primarily algorithmic and assume an abstract RAN environment. Some are O-RAN-aware and model E2 or A1 constraints. Some are 3GPP-aware and respect protocol semantics. Some are orchestration-aware and include O1, O2, and O-Cloud placement. Some are certification-oriented and provide interface traces, conformance assumptions, safety logs, and fallback behavior. Without this statement, it is difficult to assess how close the contribution is to industrial adoption.

\subsection{Research Projects and Open-Source Ecosystem}
\label{subsec:research-projects}

Open-source software and research platforms are important enablers for bridging algorithms, interoperability testing, and deployment. This is because the gap between an algorithm and a deployable controller is large. Researchers need RAN stacks, RIC platforms, interface libraries, xApp and rApp SDKs, data pipelines, emulators, and orchestration tools. Operators and vendors need shared reference implementations to test interoperability before committing to proprietary integration. Open-source projects provide the common substrate on which these communities can meet.\cite{polese2023understanding, herrera2025tutorial}

\textbf{Open Source for RAN.}
srsRAN/OCUDU \cite{srsran} and OpenAirInterface \cite{kaltenberger2020openairinterface} are the most widely used open-source cellular stacks for research prototypes. They allow researchers to run softwarized LTE and 5G functions, expose measurements, and modify parts of the protocol stack that would be inaccessible in commercial base stations. For DRL, these stacks are useful because they can be instrumented to collect trajectories and to test whether a control action affects the protocol stack as expected. They are also frequently connected to SDR hardware or emulation platforms.

\textbf{Open Source for RIC.}
The O-RAN Software Community provides reference implementations of O-RAN components, including RIC and SMO functions. ONF SD-RAN provides an ONOS-based near-RT RIC and a set of xApps and SDK components for programmable RAN control \cite{onos-sdran}. FlexRIC provides a lightweight RIC implementation and a RAN agent that can connect to OpenAirInterface and support service-model experimentation \cite{schmidt2021flexric}. ColO-RAN and OpenRAN Gym connect RIC-oriented control with experimental platforms, allowing xApps to be trained and tested in controlled environments \cite{polese2022colo,bonati2023openran}. ns-O-RAN extends ns-3-based simulation with O-RAN concepts and has become useful for scalable software-in-the-loop experimentation \cite{lacava2023ns}. OAIC focuses on AI-enabled cellular control and O-RAN research, including security and xApp experimentation \cite{upadhyaya2022prototyping}.

\textbf{Open Source for SMO and Orchestration.}
ONAP \cite{onap} and OSM \cite{osm} provide open-source management and orchestration environments that are relevant to O-RAN integration \cite{polese2023understanding}. They are not DRL frameworks by themselves, but they matter because DRL controllers need lifecycle management. A model must be deployed, monitored, updated, scaled, and rolled back. A multi-agent system may require coordinated placement across edge and regional sites. A policy that fails validation must be removed or replaced. These tasks are orchestration tasks, not only learning tasks.

\begin{table*}[!t]
\centering
\caption{A sample certification-oriented assurance profile for DRL-based O-RAN controllers.}
\label{tab:drl-certification}
\scriptsize
\setlength{\tabcolsep}{2.2pt}
\renewcommand{\arraystretch}{1.18}

\begin{tabular}{
|>{\centering\arraybackslash}m{0.150\textwidth}
|m{0.365\textwidth}
|m{0.430\textwidth}|}
\hline

\rowcolor[gray]{0.9}
\multicolumn{1}{|c|}{
    \parbox[c][6ex][c]{0.150\textwidth}{
        \centering\textbf{Assurance Dimension}
    }
}
&
\multicolumn{1}{c|}{
    \parbox[c][6ex][c]{0.365\textwidth}{
        \centering\textbf{Test Objective}
    }
}
&
\multicolumn{1}{c|}{
    \parbox[c][6ex][c]{0.430\textwidth}{
        \centering\textbf{Representative Evidence}
    }
}
\\
\hline

% ===================== INTERFACE CONFORMANCE =====================

\textbf{Interface conformance}
&
Verify that the xApp or rApp uses E2, A1, R1, O1, or O2 procedures correctly
&
Valid message traces, service-model compatibility, policy-schema validation
\\
\hline

% ===================== LIFECYCLE MANAGEMENT =====================

\textbf{Lifecycle management}
&
Verify onboarding, configuration, scaling, update, rollback, and removal
&
Descriptor validation, container image metadata, version logs, rollback tests
\\
\hline

% ===================== LATENCY COMPLIANCE =====================

\textbf{Latency compliance}
&
Verify that inference and action delivery fit the declared control loop
&
Measured inference latency, action latency, tail latency, and resource usage under load
\\
\hline

% ===================== SAFETY ENVELOPE =====================

\textbf{Safety envelope}
&
Verify that the policy cannot issue unsafe or infeasible actions
&
Action bounds, runtime shielding, constraint-violation rates, and fallback activation logs
\\
\hline

% ===================== ROBUSTNESS =====================

\textbf{Robustness}
&
Verify behavior under traffic bursts, mobility shifts, noisy KPMs, missing data, and attacks
&
Stress-test results, distribution-shift evaluation, and adversarial or fault-injection traces
\\
\hline

% ===================== EXPLAINABILITY AND AUDIT =====================

\textbf{Explainability and audit}
&
Verify that decisions can be inspected after normal and abnormal events
&
Decision logs, feature attribution, reward decomposition, and counterfactual examples
\\
\hline

% ===================== CONFLICT HANDLING =====================

\textbf{Conflict handling}
&
Verify coexistence with other xApps and policy sources
&
Multi-xApp tests, conflict-resolution traces, priority declarations, and scope declarations
\\
\hline

% ===================== PRIVACY AND PROVENANCE =====================

\textbf{Privacy and provenance}
&
Verify that telemetry and models comply with privacy and software supply-chain requirements
&
Data-use declarations, anonymization or federated-learning mechanisms, model provenance, and software BOM
\\
\hline

\end{tabular}
\end{table*}

\subsection{Certification, PlugFests, and OTICs}
\label{subsec:otic-certification}

Certification is the point where the research view of DRL meets the industrial view of trust. O-RAN already has a testing and integration ecosystem. The O-RAN Alliance organizes Global PlugFests to bring vendors, operators, OTICs, and independent laboratories together for testing and integration. It also cooperates with OTICs on certification and badging, and its TIFG defines testing, integration, and certification-related processes \cite{oran-testing,oran-tifg,oran-plugfest}. This process is essential for multi-vendor interoperability: components must not only implement an interface, but also work correctly with components from other vendors under realistic integration conditions.

For conventional O-RAN components, certification can focus on interface conformance, interoperability, performance, and security. For DRL-based xApps and rApps, these criteria remain necessary but are not sufficient. A learning-based controller can be syntactically compliant with E2 or A1 and still be unsafe. It can pass a basic interface test and still make unstable decisions under distribution shift. It can satisfy average throughput and still violate fairness or slice-level SLAs. Therefore, DRL requires an additional layer of intelligence assurance.

To fill this gap, the certification endeavors for DRL-based O-RAN controllers may be separated to three levels. 

\begin{itemize}[leftmargin=0.15in]
\item 
The first level is conformance. The controller must use standard interfaces correctly. An xApp must subscribe to KPMs through an appropriate E2 procedure, interpret the service-model payload correctly, and send control messages that conform to the expected schema. An rApp must use R1 and A1 procedures correctly and express its guidance in valid policy or enrichment-information structures. This level ensures that the controller can talk to the O-RAN system.

\item 
The second level is interoperability and integration. The controller must work with different RIC implementations, E2 nodes, and SMO environments, at least within a declared scope. This requires testing with multiple vendors or reference stacks, not only with the environment in which the policy was trained. It also requires testing the software lifecycle: onboarding, configuration, startup, scaling, termination, update, rollback, and logging. A DRL controller that cannot be safely updated or removed is not operationally deployable.

\item 
The third level is intelligence assurance. This is the level that is specific to DRL-based control. It should test whether the policy respects safety constraints, meets latency budgets, handles missing or delayed measurements, avoids invalid actions, remains robust under traffic and channel shifts, and triggers fallback behavior under anomaly or low-confidence conditions. It should also test whether the controller can coexist with other xApps. Conflict mitigation is particularly important because independent xApps may optimize different objectives and act on overlapping RAN functions. A certification profile for DRL should therefore include multi-controller scenarios rather than only single-agent tests.
\end{itemize}

Table~\ref{tab:drl-certification} outlines a possible DRL-specific OTIC profile. 
% The profile should be viewed as a research proposal rather than a description of an existing O-RAN certification program. 
Its purpose is to connect the trustworthiness requirements discussed in this survey with the industrial testing mechanisms that already exist in O-RAN. 
% Such a profile would also improve academic evaluation: if researchers know what evidence a controller would need for certification, they can design experiments that are closer to deployment requirements.

%\input{tables/sec11/sec11_table_drl_certification}

\subsection{Remark}
\label{subsec:standards-lessons}

The standardization and adoption path for DRL in O-RAN is promising yet incomplete. It is promising because O-RAN already provides many of the ingredients that learning-based control needs: programmable RICs, telemetry interfaces, policy interfaces, model-lifecycle concepts, cloud-native deployment, open-source reference implementations, and testing institutions. These components make O-RAN a more suitable substrate for DRL than a closed monolithic RAN.
It is incomplete because current specifications and certification mechanisms were not designed specifically around DRL policies that learn from interaction and generalize imperfectly under uncertainty. To the best of our knowledge, the current O-RAN specification and certification ecosystem does not yet define a unified DRL-specific profile covering for learning-based xApps and rApps\cite{oran_architecture_v16, oran_wg2, oran_wg2_a1}. 
% Without these elements, interoperability may remain limited to software interfaces while the learned control behavior remains difficult to compare, audit, and trust.

For industry adoption, a DRL-based controller should be treated not merely as a trained model, but as an operational O-RAN software artifact. Deployment-ready DRL requires DevOps/RLOps support, including automated testing, compatibility validation, runtime monitoring, policy rollback, and cross-platform verification across simulators, emulators, and real testbeds\cite{herrera2025tutorial}.

\begin{comment}
\hz{????????????}
The next stage should therefore connect four communities that have often worked separately. DRL researchers should formulate policies in terms of realistic O-RAN observation and action paths. O-RAN standardization groups should expose enough AI/ML lifecycle metadata to support safe model deployment and rollback. Open-source projects should provide portable xApp and rApp templates for learning-based control. OTICs and PlugFests should extend testing from interface conformance to intelligence assurance. If these efforts converge, DRL can move from a promising optimization technique to a certifiable and interoperable component of future O-RAN systems.
\end{comment}

% \clearpage

\section{Research Gaps and Directions}
\label{12_Challenges}
\begin{figure}[!t]
    \centering
    \includegraphics[width=0.98\linewidth]{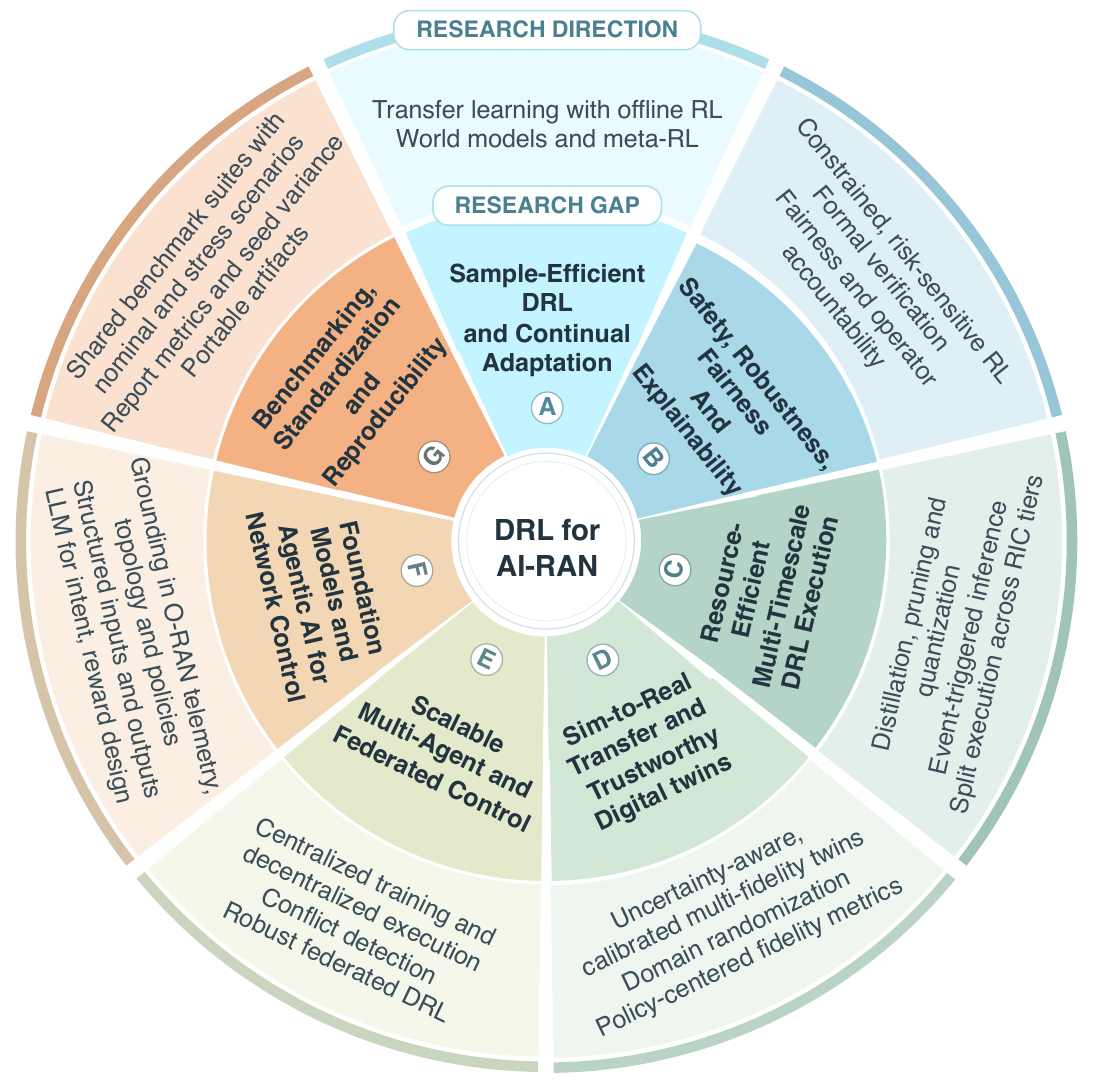}
    \caption{Open research gaps and future directions for DRL-enabled AI-RAN.
  Sectors A-G correspond to Sections~\ref{challenge_a} to \ref{challenge_g}, respectively. Each sector summarizes the research gap and representative research directions.}
    \label{fig:gaps}
\end{figure}

Despite the rapid progress in advancing DRL and O-RAN, deploying DRL as a dependable control mechanism in large-scale AI-RAN systems remains challenging. Fig.~\ref{fig:gaps} organizes the remainder of this section into seven gaps
between algorithmic performance and operational deployment, together with the directions we consider most promising for closing them. Each sector states a challenge area, the research gap underlying it, and representative directions.

\subsection{Sample-Efficient DRL and Continual Adaptation}
\label{challenge_a}

\textbf{Research Gap.}
DRL typically requires extensive interaction with its environment, but direct exploration in an operational RAN is expensive and can temporarily degrade performance or violate SLAs. Historical telemetry provides a safer source of experience, yet it may not cover the states and actions encountered by a newly learned policy. Moreover, traffic demand, mobility patterns, interference, topology, and user populations continuously change, causing the deployment environment to depart from the distribution used for training. A policy that performs well initially may become ineffective or unsafe over time.

\textbf{Research Directions.}
One direction is to combine transfer learning with conservative offline RL. Policies can first be pretrained using logged trajectories from related cells, deployments, or network configurations and then adapted using a limited amount of target-domain data. Conservative value estimation and behavior-constrained policy improvement can restrict the learned controller to actions sufficiently supported by historical experience, reducing the risk of unsafe extrapolation. The resulting policy may subsequently be evaluated in shadow mode or introduced through constrained online fine-tuning.

Another direction is model-based learning and trajectory models, which provide additional opportunities to improve sample efficiency. Learned dynamics or world models can generate imagined rollouts, evaluate candidate actions, and expose the policy to rare operating conditions (corner cases) without repeatedly interacting with the physical network. Meta-RL can further enable rapid adaptation from a small number of observations when traffic, channel, or service conditions change. These methods can be integrated with change-detection mechanisms that trigger adaptation only when statistically significant drift is observed. Each update should pass offline validation and runtime safety checks, retain a verified rollback policy, and incorporate replay or regularization mechanisms that prevent catastrophic forgetting.

\subsection{Safety, Robustness, Fairness, and Explainability}

\textbf{Research Gap.}
Safety is essential for DRL deployment in operational O-RAN systems because inappropriate control actions can violate SLAs, destabilize services, generate excessive interference, or exceed hardware and regulatory limits. However, deep policies are commonly treated as black boxes and may produce unsafe or difficult-to-interpret actions under unseen conditions, corrupted telemetry, adversarial attacks, or poorly specified rewards. A policy may also maximize aggregate performance while systematically disadvantaging cell-edge users, congested cells, or low-priority slices. These limitations make it difficult for operators to trust, diagnose, and certify DRL-based xApps and rApps.

\textbf{Research Directions.}
Future systems should combine design-time and runtime assurance. Risk-sensitive and constrained RL can optimize performance while explicitly bounding SLA violations, power consumption, interference, or tail latency. Runtime safety shields can inspect proposed actions and project, modify, or reject those that violate predefined constraints. Formal verification, barrier or Lyapunov certificates, and reachability analysis can provide stronger evidence that a policy satisfies critical properties over specified operating regions. Adversarial training, telemetry anomaly detection, uncertainty estimation, and certified robustness should also be developed to protect controllers against distribution shifts and malicious manipulation.

Trustworthy DRL also needs to account for fairness and operator accountability. Fairness-aware rewards and constraints can regulate performance across users, slices, cells, tenants, or operators rather than optimizing only aggregate utility. Interpretable policy structures, symbolic approximations, feature-attribution methods, and counterfactual explanations can reveal which observations led to a control decision and how alternative actions would affect network outcomes. Explanations should be presented through operator-facing tools that support inspection, human override, and policy approval. Runtime monitors should continuously compare predicted and observed outcomes and automatically invoke rollback or fallback controllers when safety, robustness, or fairness requirements are no longer satisfied.

\subsection{Resource-Efficient Multi-Timescale DRL Execution}

\textbf{Research Gap.}
O-RAN control spans multiple timescales, ranging from long-term optimization at the Non-RT RIC to sub-second control at the Near-RT RIC and potentially sub-10-ms execution at dApps. Meeting these timing requirements is difficult when policies must process high-dimensional telemetry from many users, cells, beams, and slices. Edge platforms also have limited compute, memory, communication bandwidth, and energy budgets. Large neural policies may therefore produce high inference latency, delayed actions, or excessive resource consumption, even when their control performance is strong in simulation.

\textbf{Research Directions.}
Policy compression should become an integral part of DRL design rather than a post-training optimization. Policy distillation can transfer the behavior of a large teacher model into a compact policy suitable for deployment. Pruning, quantization, low-rank adaptation, and hardware-aware neural architecture design can further reduce model size and inference time. Adaptive observation selection can process only the KPMs relevant to the current decision, while action-space reduction can eliminate infeasible or redundant actions before inference. Event-triggered policies may execute only when the network state changes sufficiently, thereby avoiding unnecessary computation during stable periods.

Execution should also be distributed according to control timescale and available resources. Computationally intensive representation learning, planning, and policy adaptation can be placed in the SMO, Non-RT RIC, or O-Cloud, while compact policies and precomputed safety checks execute closer to the radio loop. Split inference can partition a model across these components when communication latency permits. Future work should jointly optimize control reward, inference latency, memory footprint, communication overhead, and energy consumption. Reporting these deployment metrics together with conventional networking KPIs is necessary to determine whether a policy can satisfy the operational budget of its intended rApp, xApp, or dApp host.

\subsection{Sim-to-Real Transfer and Trustworthy Digital Twins}

\textbf{Research Gap.}
Simulation and digital twins provide scalable environments for DRL training and exploration, but policies trained in these environments can degrade after deployment. Simulators cannot fully reproduce site-specific propagation, mobility, traffic distributions, protocol implementations, scheduling behavior, hardware imperfections, processing delays, and software interactions. Even small modeling errors can accumulate through sequential decisions and lead a policy to exploit simulator-specific behavior that is absent in the physical RAN. A digital twin may therefore appear accurate at the level of individual channel or traffic statistics while still producing misleading policy evaluations.

\textbf{Research Direction.}
Future digital twins should be uncertainty-aware and continuously calibrated using measurements collected from the corresponding physical network. Instead of representing uncertain parameters through fixed nominal values, the twin should maintain distributions or confidence intervals for channel, mobility, traffic, hardware, and protocol parameters. Online system identification can update these parameters as deployment conditions change. Multi-fidelity twins can further combine analytical models, packet-level simulation, emulation, hardware-in-the-loop experimentation, and real-network traces to balance scalability and realism.

Domain randomization can expose policies to a range of plausible environments, while system identification can narrow this range using deployment-specific measurements. Limited real-network fine-tuning, performed under safety shields or in shadow mode, can then correct the remaining mismatch. Twin fidelity should not be evaluated solely through prediction errors for individual physical variables. It should also be assessed through task-level KPIs, action rankings, constraint violations, state-occupancy distributions, and the relative performance of candidate policies. This policy-centered evaluation would determine whether the twin is sufficiently accurate for the control task it is intended to support.

\subsection{Scalable Multi-Agent and Federated Control}

\textbf{Research Gap.}
Large O-RAN deployments may contain many independently developed xApps and rApps that share radio, transport, compute, and spectrum resources. These applications may operate at different timescales and optimize different objectives, producing non-stationarity and conflicting actions. Centralized coordination becomes increasingly difficult as the number of cells and agents grows, while full exchange of observations and trajectories introduces communication, privacy, and data-governance concerns. Federated learning avoids centralizing raw data, but heterogeneous vendors, network domains, hardware capabilities, and local data distributions can slow convergence and bias the resulting policy.

\textbf{Research Directions.}
Hierarchical MARL can decompose large control problems according to network topology, administrative domain, or timescale. High-level agents can determine policies, budgets, or coordination targets, while lower-level agents perform local radio control. Graph-based MARL is particularly promising because O-RAN deployments naturally form dynamic graphs of users, cells, interference relations, slices, and transport links. Graph representations can support variable-size deployments and allow agents to exchange information primarily with relevant neighbors. Centralized training with decentralized execution can use broader network information during training while preserving low-latency local execution.

Coordination mechanisms should also explicitly detect and resolve conflicts among xApps and rApps rather than relying only on shared rewards. Arbitration policies can evaluate action compatibility, priority, predicted impact, and SLA risk before actions reach the E2 nodes. Communication-efficient federated DRL should support partial participation, compressed updates, asynchronous aggregation, personalized policies, and robustness to unreliable or malicious participants. Future formulations must remain effective under non-identically distributed experience and heterogeneous state, action, and reward definitions. Standard semantics for observations, actions, constraints, and policy metadata will be essential for coordinating agents developed by different vendors.

\subsection{Foundation Models and Agentic AI for Network Control}

\textbf{Research Gap.}
Foundation models and agentic AI can improve network-state representation, intent interpretation, reward and constraint design, tool use, and long-horizon planning. However, their high computational cost, hallucinations, weak grounding in current network conditions, and nondeterministic behavior prevent unconstrained deployment in critical control loops. Their outputs may be syntactically plausible but inconsistent with radio constraints, current configurations, or available control capabilities. Their large inference latency and limited auditability also conflict with the requirements of Near-RT RIC and dApp control.

\textbf{Research Directions.}
Foundation models should be grounded in standardized O-RAN telemetry, configuration models, interface specifications, topology information, and operator policies. Retrieval mechanisms and tool interfaces can provide current network context rather than relying only on information encoded during pretraining. Structured input and output representations should constrain the model to valid network entities, parameters, and actions. Uncertainty estimates and consistency checks should accompany generated plans so that poorly grounded recommendations can be rejected before execution.

A promising architecture is to restrict foundation models to high-level roles such as intent decomposition, reward generation, scenario generation, policy selection, or world-model construction. Deterministic DRL policies or conventional controllers can then implement time-critical actions. Candidate plans generated by an LLM or agentic system should pass symbolic constraints, formal policy checks, and DRL safety layers before deployment. Behaviors that are repeatedly validated can be distilled into smaller, deterministic policies for execution at the edge. This separation allows foundation models to contribute reasoning and generalization without placing an unconstrained generative model directly in a critical radio-control loop.

\subsection{Benchmarking, Standardization, and Reproducibility}
\label{challenge_g}

\textbf{Research Gap.}
Progress in DRL for O-RAN is difficult to assess because existing studies use different MDP formulations, observations, action spaces, rewards, datasets, topologies, traffic models, baselines, and evaluation protocols. Many evaluations report average reward or throughput but omit safety violations, inference cost, generalization, or sim-to-real performance. Proprietary platforms and incomplete artifact releases further limit reproducibility and make it difficult to transfer policies across software stacks and vendors. The absence of common representations for agents and their lifecycle metadata also complicates integration, testing, and certification.

\textbf{Research Directions.}
The research community should establish shared O-RAN benchmark suites covering representative control tasks, network scales, traffic and mobility patterns, uncertainty levels, and failure conditions. Benchmark scenarios should include both nominal operation and stress cases involving distribution shifts, telemetry loss, adversarial manipulation, and conflicts among control applications. Standard reporting should include reward, networking KPIs, SLA violations, fairness, robustness, sample efficiency, inference latency, model size, energy consumption, target hardware, and statistical variation across random seeds. Evaluations should also document behavior-policy provenance, training data, software versions, hyperparameters, and the complete path from simulation to deployment.

Reproducible xApp and rApp artifacts should package policy weights, model architecture, observation and action schemas, normalization parameters, safety constraints, training provenance, and rollback information. Portable agent-description and model-metadata formats could enable the same policy to be validated across RIC implementations and vendor platforms. Standardization efforts should clarify how DRL observations, actions, constraints, model updates, monitoring data, and lifecycle events map onto the A1, E2, O1, and O2 interfaces. 
OTICs should extend testing from interface conformance toward closed-loop policy evaluation, including safety, interoperability, resource consumption, conflict handling, and controlled rollback. These workflows would provide a practical path from research prototypes to certifiable O-RAN applications.

\section{Conclusion}
\label{13_Conclusion}

This survey presented a comprehensive examination of DRL for 6G AI-RAN, connecting DRL methodology with the architecture, control loops, interfaces, and operational requirements of O-RAN. We reviewed the foundations of value-based, policy-gradient, model-based, offline, safe, multi-agent, federated, and transfer learning methods, and explained how RAN control problems can be formulated through observations, states, actions, rewards, constraints, and temporal structure. We then surveyed DRL applications across radio resource management, mobility management, interference control, traffic steering, energy efficiency, network slicing, ISAC, and security. Beyond individual use cases, we examined multi-agent and federated coordination, foundation models and agentic AI, trustworthy DRL, training and deployment strategies, experimental platforms, benchmarks, and relevant standardization activities.

The surveyed literature demonstrates that DRL offers a promising framework for adaptive, automated, and long-term optimization in O-RANs. 
However, operational adoption requires DRL-based xApps, rApps, and dApps to be sample-efficient, adaptive, trustworthy, resource-efficient, generalizable, interoperable, and supported by digital twins, real-world testbeds, reproducible benchmarks, standardized lifecycle management, and O-RAN-aligned certification.
Addressing these challenges will enable DRL to evolve from an experimental optimization technique to a dependable control foundation for open, intelligent, and autonomous 6G RANs.

% \clearpage

\section*{Appendix: List of Acronyms}
\label{tab_acronyms}
%\section*{List of Acronyms}

\begin{IEEEdescription}[\IEEEsetlabelwidth{Near-RT RIC}]
\item[3GPP] The 3rd Generation Partnership Project
\item[6G] Sixth-Generation
\item[A1-EI] A1 Enrichment Information
\item[A1-ML] A1 Machine Learning
\item[A1-P] A1 Policy Management
\item[AI] Artificial Intelligence
\item[AI-RAN] Artificial Intelligence Radio Access Network
\item[AP] Acess Point 
\item[BCQ] Behavior-Constrained Q-Learning
\item[BS] Base Station 
\item[CMDP] Constrained Markov Decision Process
\item[CQL] Conservative Q-Learning
\item[CTDE] Centralized Training, Decentralized Execution 
\item[DDPG] Deep Deterministic Policy Gradient
\item[DecT] Decision Transformer
\item[DNN] Deep Neural Network
\item[DQN] Deep Q-Network
\item[DRL] Deep Reinforcement Learning
\item[E2SM] E2 Service Model
\item[ETSI] European Telecommunications Standards Institute
\item[FL] Federated Learning
\item[FLOP] floating-point operation
\item[FM] Foundation Model
\item[FRL] Federated Reinforcement Learning
\item[GAE] Generalized Advantage Estimation
\item[GenAI] Generative Artificial Intelligence
\item[gNB] the Next Generation NodeB
\item[GNN] Graph Neural Network
\item[GRPO] Group Relative Policy Optimization
\item[HFRL] Horizontal Federated Reinforcement Learning
\item[IBN] Intent-Based Networking
\item[IQL] Implicit Q-Learning
\item[I/Q] In-phase/Quadrature
\item[ISAC] Integrated Sensing and Communication
\item[ITU] International Telecommunication Union
\item[KPI] Key Performance Indicator
\item[KPM] Key Performance Measurement
\item[LLM] Large Language Model
\item[LTE] Long-Term Evolution
\item[MAC] Medium Access Control
\item[MANO] Management and Orchestration
\item[MARL] Multi-Agent Reinforcement Learning
\item[MDP] Markov Decision Process
\item[MIMO] Multiple-Input and Multiple-Output
\item[ML] Machine Learning
\item[Near-RT RIC] Near-Real-Time RAN Intelligent Controller
\item[NFV] Network Functions Virtualization
\item[NG-RAN] Next Generation Radio Access Network
\item[NIC] Network Interface Card
\item[Non-RT RIC] Non-Real-Time RAN Intelligent Controller
\item[NR] New Radio
\item[OAI] Open Air Interface
\item[OAIC] Open AI Cellular
\item[OAM] Operations, Administration and Maintenance
\item[O-CU] O-RAN Central Unit
\item[O-CU-CP] O-RAN Central Unit-Control Plane
\item[O-CU-UP] O-RAN Central Unit-User Plane
\item[O-DU] O-RAN Distributed Unit
\item[OFH] Open Fronthaul
\item[ONAP] Open Network Automation Platform
\item[ONF] Open Networking Foundation
\item[ONOS] Open Network Operating System
\item[O-RAN] Open Radio Access Network
\item[O-RU] O-RAN Radio Unit
\item[OSC] O-RAN Software Community
\item[OSM] Open Source MANO
\item[OTIC] Open Testing and Integration Centre 
\item[PER] Prioritized Experience Replay
\item[PF] Proportional Fair
\item[PRB] Physical Resource Block
\item[POMDP] Partially Observable Markov Decision Process
\item[PPO] Proximal Policy Optimization
\item[RAG] Retrieval-Augmented Generation
\item[RAN] Radio Access Network
\item[rApp] Non-Real-Time RAN Intelligent Controller Application
\item[RAT] Radio Access Technology 
\item[RC] RAN Control
\item[RF] Radio Frequency
\item[RIC] RAN Intelligent Controller
\item[RL] Reinforcement Learning
\item[RLOps] Reinforcement Learning Operations 
\item[RRM] Radio Resource Management
\item[RTG] Return-to-Go
\item[RU] Radio Unit
\item[SAC] Soft Actor-Critic
\item[SDAP] Service Data Adaptation Protocol 
\item[SDK] Software Development Kit
\item[SDO] Standards Development Organization
\item[SDR] Software-Defined Radio
\item[SLA] Service Level Agreement
\item[SLICES] Scientific Large-Scale Infrastructure for Computing/Communication Experimental Studies
\item[SMO] Service Management and Orchestration
\item[SSM] State-Space Model
\item[TD3] Twin Delayed Deep Deterministic Policy Gradient
\item[TIFG] Test and Integration Focus Group
\item[TRPO] Trust Region Policy Optimization
\item[UE] User Equipment
\item[URLLC] Ultra-Reliable Low-Latency Communication
\item[VAE] Variational Autoencoder
\item[VFRL] Vertical Federated Reinforcement Learning
\item[VNF] Virtual Network Functions
\item[WG] O-RAN ALLIANCE Work Group
\item[XAI] Explainable Artificial Intelligence
\item[xApp] Near-Real-Time RAN Intelligent Controller Application
\end{IEEEdescription}

\bibliographystyle{IEEEtran}
\bibliography{ref.bib}

@article{jiang2021road,
  title={The road towards 6G: A comprehensive survey},
  author={Jiang, Wei and Han, Bin and Habibi, Mohammad Asif and Schotten, Hans Dieter},
  journal={IEEE Open Journal of the Communications Society},
  volume={2},
  pages={334--366},
  year={2021},
  publisher={IEEE}
}

@article{wang2023road,
  title={On the road to 6G: Visions, requirements, key technologies, and testbeds},
  author={Wang, Cheng-Xiang and You, Xiaohu and Gao, Xiqi and Zhu, Xiuming and Li, Zixin and Zhang, Chuan and Wang, Haiming and Huang, Yongming and Chen, Yunfei and Haas, Harald and others},
  journal={IEEE Communications Surveys \& Tutorials},
  volume={25},
  number={2},
  pages={905--974},
  year={2023},
  publisher={IEEE}
}

@techreport{ORAN_OAD_17_2026,
  author      = {{O-RAN Alliance}},
  title       = {{O-RAN Architecture Description 17.0}},
  institution = {{O-RAN Alliance Working Group 1}},
  type        = {Technical Specification},
  number      = {O-RAN.WG1.TS.OAD-R005-v17.00},
  year        = {2026},
  month       = jun,
  note        = {Release R005}
}

@article{krizhevsky2012imagenet,
  title={Imagenet classification with deep convolutional neural networks},
  author={Krizhevsky, Alex and Sutskever, Ilya and Hinton, Geoffrey E},
  journal={Advances in neural information processing systems},
  volume={25},
  year={2012}
}

@inproceedings{chen2020simple,
  title={A simple framework for contrastive learning of visual representations},
  author={Chen, Ting and Kornblith, Simon and Norouzi, Mohammad and Hinton, Geoffrey},
  booktitle={International conference on machine learning},
  pages={1597--1607},
  year={2020},
  organization={PmLR}
}

@article{grill2020bootstrap,
  title={Bootstrap your own latent-a new approach to self-supervised learning},
  author={Grill, Jean-Bastien and Strub, Florian and Altch{\'e}, Florent and Tallec, Corentin and Richemond, Pierre and Buchatskaya, Elena and Doersch, Carl and Avila Pires, Bernardo and Guo, Zhaohan and Gheshlaghi Azar, Mohammad and others},
  journal={Advances in neural information processing systems},
  volume={33},
  pages={21271--21284},
  year={2020}
}

@article{arnaz2022toward,
  title={{Toward integrating intelligence and programmability in open radio access networks: A comprehensive survey}},
  author={Arnaz, Azadeh and Lipman, Justin and Abolhasan, Mehran and Hiltunen, Matti},
  journal={Ieee Access},
  volume={10},
  pages={67747--67770},
  year={2022},
  publisher={IEEE}
}

@article{wu2025mogul,
  title={{MOGUL: A Model-Guided Learning Approach for Scheduling in {5G} {O-RAN}}},
  author={Wu, Yubo and Zeng, Huacheng and Lou, Wenjing and Hou, Y Thomas},
  journal={IEEE Internet of Things Journal},
  year={2025},
  publisher={IEEE}
}

@article{barker2025real,
  title   = {{REAL: Reinforcement Learning-Enabled {xApps} for Experimental Closed-Loop Optimization in {O-RAN} with {OSC} {RIC} and {srsRAN}}},
  author  = {Barker, Ryan and Dorcheh, Alireza Ebrahimi and Seyfi, Tolunay and Afghah, Fatemeh},
  journal = {arXiv preprint arXiv:2502.00715},
  year    = {2025},
  doi     = {10.48550/arXiv.2502.00715}
}

@article{brik2022deep,
  title={{Deep learning for {B5G} open radio access network: Evolution, survey, case studies, and challenges}},
  author={Brik, Bouziane and Boutiba, Karim and Ksentini, Adlen},
  journal={IEEE Open Journal of the Communications Society},
  volume={3},
  pages={228--250},
  year={2022},
  publisher={IEEE}
}

@article{abdalla2022toward,
  title={{Toward next generation open radio access networks: What {O-RAN} can and cannot do!}},
  author={Abdalla, Aly S and Upadhyaya, Pratheek S and Shah, Vijay K and Marojevic, Vuk},
  journal={IEEE Network},
  volume={36},
  number={6},
  pages={206--213},
  year={2022},
  publisher={IEEE}
}

@article{polese2023understanding,
  title={{Understanding {O-RAN}: Architecture, interfaces, algorithms, security, and research challenges}},
  author={Polese, Michele and Bonati, Leonardo and D’oro, Salvatore and Basagni, Stefano and Melodia, Tommaso},
  journal={IEEE Communications Surveys \& Tutorials},
  volume={25},
  number={2},
  pages={1376--1411},
  year={2023},
  publisher={IEEE}
}

@article{polese2023empowering,
  title={{Empowering the {6G} cellular architecture with open {RAN}}},
  author={Polese, Michele and Dohler, Mischa and Dressler, Falko and Erol-Kantarci, Melike and Jana, Rittwik and Knopp, Raymond and Melodia, Tommaso},
  journal={IEEE Journal on Selected Areas in Communications},
  volume={42},
  number={2},
  pages={245--262},
  year={2023},
  publisher={IEEE}
}

@article{masaracchia2023digital,
  title={{Digital twin for open {RAN}: Toward intelligent and resilient {6G} radio access networks}},
  author={Masaracchia, Antonino and Sharma, Vishal and Fahim, Muhammad and Dobre, Octavia A and Duong, Trung Q},
  journal={IEEE Communications Magazine},
  volume={61},
  number={11},
  pages={112--118},
  year={2023},
  publisher={IEEE}
}

@inproceedings{abubakar2023energy,
  title={{Energy efficiency of open radio access network: A survey}},
  author={Abubakar, Attai Ibrahim and Onireti, Oluwakayode and Sambo, Yusuf and Zhang, Lei and Ragesh, GK and Imran, Muhammad Ali},
  booktitle={2023 IEEE 97th Vehicular Technology Conference (VTC2023-Spring)},
  pages={1--7},
  year={2023},
  organization={IEEE}
}

@article{liang2024energy,
  title={{Energy consumption of machine learning enhanced open {RAN}: A comprehensive review}},
  author={Liang, Xuanyu and Wang, Qiao and Al-Tahmeesschi, Ahmed and Chetty, Swarna B and Grace, David and Ahmadi, Hamed},
  journal={IEEE Access},
  volume={12},
  pages={81889--81910},
  year={2024},
  publisher={IEEE}
}

@article{wani2024open,
  title={{Open {RAN}: A concise overview}},
  author={Wani, Mohamad and Kretschmer, Mathias and Schr{\"o}der, Bernd and Grebe, Andreas and Rademacher, Michael},
  journal={IEEE Open Journal of the Communications Society},
  volume={6},
  pages={13--28},
  year={2024},
  publisher={IEEE}
}

@article{brik2024explainable,
  title={{Explainable {AI} in {6G} {O-RAN}: A tutorial and survey on architecture, use cases, challenges, and future research}},
  author={Brik, Bouziane and Chergui, Hatim and Zanzi, Lanfranco and Devoti, Francesco and Ksentini, Adlen and Siddiqui, Muhammad Shuaib and Costa-P{\'e}rez, Xavier and Verikoukis, Christos},
  journal={IEEE Communications Surveys \& Tutorials},
  volume={27},
  number={5},
  pages={2826--2859},
  year={2024},
  publisher={IEEE}
}

@article{amachaghi2024survey,
  title={{A survey for intrusion detection systems in open {RAN}}},
  author={Amachaghi, Emmanuel N and Shojafar, Mohammad and Foh, Chuan Heng and Moessner, Klaus},
  journal={IEEE access},
  volume={12},
  pages={88146--88173},
  year={2024},
  publisher={IEEE}
}

@article{couto2024survey,
  title={{A survey of public datasets for {O-RAN}: fostering the development of machine learning models}},
  author={Couto, Rodrigo S and Cruz, Pedro and Pacheco, Roberto G and Souza, Vivian Maria S and Campista, Miguel Elias M and Costa, Lu{\'\i}s Henrique MK},
  journal={Annals of Telecommunications},
  volume={79},
  number={9},
  pages={649--662},
  year={2024},
  publisher={Springer}
}

@article{herrera2025tutorial,
  title={{A tutorial on {O-RAN} deployment solutions for {5G}: From simulation to emulated and real testbeds}},
  author={Herrera, Juan Luis and Montebugnoli, Sofia and Scotece, Domenico and Foschini, Luca and Bellavista, Paolo},
  journal={IEEE Communications Surveys \& Tutorials},
  year={2025},
  publisher={IEEE}
}

@article{agarwal2025open,
  title={{Open {RAN} for {6G} networks: Architecture, use cases and open issues}},
  author={Agarwal, Bharat and Irmer, Ralf and Lister, David and Muntean, Gabriel-Miro},
  journal={IEEE Communications Surveys \& Tutorials},
  year={2025},
  publisher={IEEE}
}

@article{alam2025comprehensive,
  title={{A comprehensive tutorial and survey of {O-RAN}: Exploring slicing-aware architecture, deployment options, use cases, and challenges}},
  author={Alam, Khurshid and Habibi, Mohammad Asif and Tammen, Matthias and Krummacker, Dennis and Saad, Walid and Di Renzo, Marco and Melodia, Tommaso and Costa-P{\'e}rez, Xavier and Debbah, M{\'e}rouane and Dutta, Ashutosh and others},
  journal={IEEE Communications Surveys \& Tutorials},
  year={2025},
  publisher={IEEE}
}

@article{mehrban2025integrating,
  title={{Integrating Zero Trust Architecture in {O-RAN}: A Comprehensive Survey and Analysis}},
  author={Mehrban, Ali and Abou El Houda, Zakaria and Moudoud, Hajar and Le, Long Bao},
  journal={IEEE Open Journal of the Communications Society},
  volume={6},
  pages={10465--10495},
  year={2025},
  publisher={IEEE}
}

@inproceedings{lee2025survey,
  title={{A Survey on Intelligent Traffic Steering Techniques in {O-RAN}}},
  author={Lee, Seungchan and Choi, Seongin and Won, Dongwook and Do, Quang Tuan and Kim, Jaemin and Cho, Sungrae},
  booktitle={2025 Sixteenth International Conference on Ubiquitous and Future Networks (ICUFN)},
  pages={29--31},
  year={2025},
  organization={IEEE}
}

@article{deng2026ai,
  title={{{AI}-Native Open {RAN} for Non-Terrestrial Networks: An Overview}},
  author={Deng, Jikang and Hasan, S Fizza and Zhou, Hui and Al-Ahmadi, Saad and Alouini, Mohamed-Slim and Da Costa, Daniel B},
  journal={IEEE Open Journal of the Communications Society},
  year={2026},
  publisher={IEEE}
}

@article{kirana2026ml,
  title={{{ML}-Enabled Open {RAN}: A Comprehensive Survey of Architectures, Challenges, and Opportunities}},
  author={Kirana, Mira Chandra and Keyela, Patatchona and Rostamian, Fatemeh and Tashman, Deemah H and Cherkaoui, Soumaya},
  journal={IEEE Communications Surveys \& Tutorials},
  year={2026},
  publisher={IEEE}
}

@article{mahmoud2026review,
  title={{A Review of Open-{RAN} Intelligence: Opportunities, Challenges, Real-Life Applications and Impacts}},
  author={Mahmoud, Haitham Hassan and Mi, De and Aneiba, Adel and Idrissi, Moad and Sharma, Gaurav and Foh, Chuan Heng and Xiao, Pei},
  journal={ACM Computing Surveys},
  volume={58},
  number={8},
  pages={1--36},
  year={2026},
  publisher={ACM New York, NY}
}

@article{liang2026security,
  title={{Security and privacy in {O-RAN} for {6G}: A comprehensive review of threats and mitigation approaches}},
  author={Liang, Lujia and Zhang, Lei},
  journal={IEEE Communications Surveys \& Tutorials},
  year={2026},
  publisher={IEEE}
}

@article{hamdan2023recent,
  title={{Recent advances in machine learning for network automation in the {O-RAN}}},
  author={Hamdan, Mutasem Q and Lee, Haeyoung and Triantafyllopoulou, Dionysia and Borralho, R{\'u}ben and Kose, Abdulkadir and Amiri, Esmaeil and Mulvey, David and Yu, Wenjuan and Zitouni, Rafik and Pozza, Riccardo and others},
  journal={Sensors},
  volume={23},
  number={21},
  pages={8792},
  year={2023},
  publisher={MDPI}
}

@article{santos2025managing,
  title={{Managing {O-RAN} networks: {xApp} development from zero to hero}},
  author={Santos, Joao F and Huff, Alexandre and Campos, Daniel and Cardoso, Kleber V and Both, Cristiano B and DaSilva, Luiz A},
  journal={IEEE Communications Surveys \& Tutorials},
  year={2025},
  publisher={IEEE}
}

@article{hoffmann2023open,
  title={{Open {RAN} {xApps} design and evaluation: Lessons learnt and identified challenges}},
  author={Hoffmann, Marcin and Janji, Salim and Samorzewski, Adam and Ku{\l}acz, {\L}ukasz and Adamczyk, Cezary and Dryja{\'n}ski, Marcin and Kryszkiewicz, Pawel and Kliks, Adrian and Bogucka, Hanna},
  journal={IEEE Journal on Selected Areas in Communications},
  volume={42},
  number={2},
  pages={473--486},
  year={2023},
  publisher={IEEE}
}

@article{puterman1990markov,
  title={{Markov decision processes}},
  author={Puterman, Martin L},
  journal={Handbooks in operations research and management science},
  volume={2},
  pages={331--434},
  year={1990},
  publisher={Elsevier}
}

@article{wiering2012reinforcement,
  title={{Reinforcement learning}},
  author={Wiering, Marco A and Van Otterlo, Martijn},
  journal={Adaptation, learning, and optimization},
  volume={12},
  number={3},
  pages={729},
  year={2012},
  publisher={Springer}
}

@article{kaelbling1998planning,
  title={{Planning and acting in partially observable stochastic domains}},
  author={Kaelbling, Leslie Pack and Littman, Michael L and Cassandra, Anthony R},
  journal={Artificial intelligence},
  volume={101},
  number={1-2},
  pages={99--134},
  year={1998},
  publisher={Elsevier}
}

@article{esslinger2022deep,
  title={{Deep transformer q-networks for partially observable reinforcement learning}},
  author={Esslinger, Kevin and Platt, Robert and Amato, Christopher},
  journal={arXiv preprint arXiv:2206.01078},
  year={2022}
}

@article{watkins1992q,
  title={{Q-learning}},
  author={Watkins, Christopher JCH and Dayan, Peter},
  journal={Machine learning},
  volume={8},
  number={3},
  pages={279--292},
  year={1992},
  publisher={Springer}
}

@article{mnih2015human,
  title={{Human-level control through deep reinforcement learning}},
  author={Mnih, Volodymyr and Kavukcuoglu, Koray and Silver, David and Rusu, Andrei A and Veness, Joel and Bellemare, Marc G and Graves, Alex and Riedmiller, Martin and Fidjeland, Andreas K and Ostrovski, Georg and others},
  journal={nature},
  volume={518},
  number={7540},
  pages={529--533},
  year={2015},
  publisher={Nature Publishing Group}
}

@inproceedings{van2016deep,
  title={{Deep reinforcement learning with double {Q-learning}}},
  author={Van Hasselt, Hado and Guez, Arthur and Silver, David},
  booktitle={Proceedings of the AAAI conference on artificial intelligence},
  volume={30},
  number={1},
  year={2016}
}

@inproceedings{wang2016dueling,
  title={{Dueling network architectures for deep reinforcement learning}},
  author={Wang, Ziyu and Schaul, Tom and Hessel, Matteo and Hasselt, Hado and Lanctot, Marc and Freitas, Nando},
  booktitle={International conference on machine learning},
  pages={1995--2003},
  year={2016},
  organization={PMLR}
}

@article{schaul2015prioritized,
  title={{Prioritized experience replay}},
  author={Schaul, Tom and Quan, John and Antonoglou, Ioannis and Silver, David},
  journal={arXiv preprint arXiv:1511.05952},
  year={2015}
}

@inproceedings{bellemare2017distributional,
  title={{A distributional perspective on reinforcement learning}},
  author={Bellemare, Marc G and Dabney, Will and Munos, R{\'e}mi},
  booktitle={International conference on machine learning},
  pages={449--458},
  year={2017},
  organization={Pmlr}
}

@article{sutton1999policy,
  title={{Policy gradient methods for reinforcement learning with function approximation}},
  author={Sutton, Richard S and McAllester, David and Singh, Satinder and Mansour, Yishay},
  journal={Advances in neural information processing systems},
  volume={12},
  year={1999}
}

@article{williams1992simple,
  title={{Simple statistical gradient-following algorithms for connectionist reinforcement learning}},
  author={Williams, Ronald J},
  journal={Machine learning},
  volume={8},
  number={3},
  pages={229--256},
  year={1992},
  publisher={Springer}
}

@article{konda1999actor,
  title={{Actor-critic algorithms}},
  author={Konda, Vijay and Tsitsiklis, John},
  journal={Advances in neural information processing systems},
  volume={12},
  year={1999}
}

@inproceedings{mnih2016asynchronous,
  title={{Asynchronous methods for deep reinforcement learning}},
  author={Mnih, Volodymyr and Badia, Adria Puigdomenech and Mirza, Mehdi and Graves, Alex and Lillicrap, Timothy and Harley, Tim and Silver, David and Kavukcuoglu, Koray},
  booktitle={International conference on machine learning},
  pages={1928--1937},
  year={2016},
  organization={PmLR}
}

@misc{wu2017baselines,
  author       = {Wu, Yuhuai and Mansimov, Elman and Liao, Shun and Radford, Alec and Schulman, John},
  title        = {{{OpenAI} Baselines: {ACKTR} \& {A2C}}},
  howpublished = {\url{https://openai.com/index/openai-baselines-acktr-a2c/}},
  year         = {2017},
  month        = aug
}

@inproceedings{schulman2015trust,
  title={{Trust region policy optimization}},
  author={Schulman, John and Levine, Sergey and Abbeel, Pieter and Jordan, Michael and Moritz, Philipp},
  booktitle={International conference on machine learning},
  pages={1889--1897},
  year={2015},
  organization={PMLR}
}

@article{schulman2017proximal,
  title={{Proximal policy optimization algorithms}},
  author={Schulman, John and Wolski, Filip and Dhariwal, Prafulla and Radford, Alec and Klimov, Oleg},
  journal={arXiv preprint arXiv:1707.06347},
  year={2017}
}

@article{shao2024deepseekmath,
  title={{{DeepSeekMath}: Pushing the limits of mathematical reasoning in open language models}},
  author={Shao, Zhihong and Wang, Peiyi and Zhu, Qihao and Xu, Runxin and Song, Junxiao and Bi, Xiao and Zhang, Haowei and Zhang, Mingchuan and Li, YK and Wu, Yang and others},
  journal={arXiv preprint arXiv:2402.03300},
  year={2024}
}

@article{lillicrap2015continuous,
  title={{CONTINUOUS CONTROL WITH DEEP REINFORCEMENT LEARNING}},
  author={Lillicrap, Timothy P and Hunt, Jonathan J and Pritzel, Alexander and Heess, Nicolas and Erez, Tom and Tassa, Yuval and Silver, David and Wierstra, Daan},
  journal={arXiv preprint arXiv:1509.02971},
  year={2015}
}

@inproceedings{silver2014deterministic,
  title={{Deterministic policy gradient algorithms}},
  author={Silver, David and Lever, Guy and Heess, Nicolas and Degris, Thomas and Wierstra, Daan and Riedmiller, Martin},
  booktitle={International conference on machine learning},
  pages={387--395},
  year={2014},
  organization={Pmlr}
}

@inproceedings{fujimoto2018addressing,
  title={{Addressing function approximation error in actor-critic methods}},
  author={Fujimoto, Scott and Hoof, Herke and Meger, David},
  booktitle={International conference on machine learning},
  pages={1587--1596},
  year={2018},
  organization={PMLR}
}

@inproceedings{haarnoja2018soft,
  title={{Soft actor-critic: Off-policy maximum entropy deep reinforcement learning with a stochastic actor}},
  author={Haarnoja, Tuomas and Zhou, Aurick and Abbeel, Pieter and Levine, Sergey},
  booktitle={International conference on machine learning},
  pages={1861--1870},
  year={2018},
  organization={Pmlr}
}

@inproceedings{henderson2018deep,
  title={{Deep reinforcement learning that matters}},
  author={Henderson, Peter and Islam, Riashat and Bachman, Philip and Pineau, Joelle and Precup, Doina and Meger, David},
  booktitle={Proceedings of the AAAI conference on artificial intelligence},
  volume={32},
  number={1},
  year={2018}
}

@article{m2023model,
  title={{Model-based reinforcement learning: A survey}},
  author={M. Moerland, Thomas and Broekens, Joost and Plaat, Aske and M. Jonker, Catholijn},
  journal={Foundations and Trends in Machine Learning},
  volume={16},
  number={1},
  pages={1--118},
  year={2023},
  publisher={Emerald Publishing Limited}
}

@article{shapley1953stochastic,
  title={{Stochastic games}},
  author={Shapley, Lloyd S},
  journal={Proceedings of the national academy of sciences},
  volume={39},
  number={10},
  pages={1095--1100},
  year={1953},
  publisher={National Academy of Sciences}
}

@article{bernstein2002complexity,
  title={{The complexity of decentralized control of Markov decision processes}},
  author={Bernstein, Daniel S and Givan, Robert and Immerman, Neil and Zilberstein, Shlomo},
  journal={Mathematics of operations research},
  volume={27},
  number={4},
  pages={819--840},
  year={2002},
  publisher={INFORMS}
}

@book{oliehoek2016concise,
  title={{A concise introduction to decentralized {POMDPs}}},
  author={Oliehoek, Frans A and Amato, Christopher and others},
  volume={1},
  year={2016},
  publisher={Springer}
}

@article{lowe2017multi,
  title={{Multi-agent actor-critic for mixed cooperative-competitive environments}},
  author={Lowe, Ryan and Wu, Yi I and Tamar, Aviv and Harb, Jean and Pieter Abbeel, OpenAI and Mordatch, Igor},
  journal={Advances in neural information processing systems},
  volume={30},
  year={2017}
}

@article{rashid2020monotonic,
  title={{Monotonic value function factorisation for deep multi-agent reinforcement learning}},
  author={Rashid, Tabish and Samvelyan, Mikayel and De Witt, Christian Schroeder and Farquhar, Gregory and Foerster, Jakob and Whiteson, Shimon},
  journal={Journal of Machine Learning Research},
  volume={21},
  number={178},
  pages={1--51},
  year={2020}
}

@article{yu2022surprising,
  title={{The surprising effectiveness of {PPO} in cooperative multi-agent games}},
  author={Yu, Chao and Velu, Akash and Vinitsky, Eugene and Gao, Jiaxuan and Wang, Yu and Bayen, Alexandre and Wu, Yi},
  journal={Advances in neural information processing systems},
  volume={35},
  pages={24611--24624},
  year={2022}
}

@inproceedings{mcmahan2017communication,
  title={{Communication-efficient learning of deep networks from decentralized data}},
  author={McMahan, Brendan and Moore, Eider and Ramage, Daniel and Hampson, Seth and y Arcas, Blaise Aguera},
  booktitle={Artificial intelligence and statistics},
  pages={1273--1282},
  year={2017},
  organization={Pmlr}
}

@article{qi2021federated,
  title={{Federated reinforcement learning: Techniques, applications, and open challenges}},
  author={Qi, Jiaju and Zhou, Qihao and Lei, Lei and Zheng, Kan},
  journal={arXiv preprint arXiv:2108.11887},
  year={2021}
}

@article{zhuo2019federated,
  title={{Federated deep reinforcement learning}},
  author={Zhuo, Hankz Hankui and Feng, Wenfeng and Lin, Yufeng and Xu, Qian and Yang, Qiang},
  journal={arXiv preprint arXiv:1901.08277},
  year={2019}
}

@article{li2020federated,
  title={{Federated optimization in heterogeneous networks}},
  author={Li, Tian and Sahu, Anit Kumar and Zaheer, Manzil and Sanjabi, Maziar and Talwalkar, Ameet and Smith, Virginia},
  journal={Proceedings of Machine learning and systems},
  volume={2},
  pages={429--450},
  year={2020}
}

@article{geiping2020inverting,
  title={{Inverting gradients-how easy is it to break privacy in federated learning?}},
  author={Geiping, Jonas and Bauermeister, Hartmut and Dr{\"o}ge, Hannah and Moeller, Michael},
  journal={Advances in neural information processing systems},
  volume={33},
  pages={16937--16947},
  year={2020}
}

@inproceedings{khodadadian2022federated,
  title={{Federated reinforcement learning: Linear speedup under markovian sampling}},
  author={Khodadadian, Sajad and Sharma, Pranay and Joshi, Gauri and Maguluri, Siva Theja},
  booktitle={International conference on machine learning},
  pages={10997--11057},
  year={2022},
  organization={PMLR}
}

@inproceedings{espeholt2018impala,
  title={{{IMPALA}: Scalable distributed deep-{RL} with importance weighted actor-learner architectures}},
  author={Espeholt, Lasse and Soyer, Hubert and Munos, Remi and Simonyan, Karen and Mnih, Vlad and Ward, Tom and Doron, Yotam and Firoiu, Vlad and Harley, Tim and Dunning, Iain and others},
  booktitle={International conference on machine learning},
  pages={1407--1416},
  year={2018},
  organization={PMLR}
}

@article{espeholt2019seed,
  title={{{SEED RL}: Scalable and efficient deep-{RL} with accelerated central inference}},
  author={Espeholt, Lasse and Marinier, Rapha{\"e}l and Stanczyk, Piotr and Wang, Ke and Michalski, Marcin},
  journal={arXiv preprint arXiv:1910.06591},
  year={2019}
}

@article{DBLP:journals/corr/abs-1803-00933,
  author       = {Dan Horgan and
                  John Quan and
                  David Budden and
                  Gabriel Barth{-}Maron and
                  Matteo Hessel and
                  Hado van Hasselt and
                  David Silver},
  title        = {{Distributed Prioritized Experience Replay}},
  journal      = {CoRR},
  volume       = {abs/1803.00933},
  year         = {2018},
  url          = {http://arxiv.org/abs/1803.00933},
  eprinttype   = {arXiv},
  eprint       = {1803.00933},
  bibsource    = {dblp computer science bibliography, https://dblp.org}
}

@article{levine2020offline,
  title={{Offline reinforcement learning: Tutorial, review, and perspectives on open problems}},
  author={Levine, Sergey and Kumar, Aviral and Tucker, George and Fu, Justin},
  journal={arXiv preprint arXiv:2005.01643},
  year={2020}
}

@inproceedings{fujimoto2019off,
  title={{Off-policy deep reinforcement learning without exploration}},
  author={Fujimoto, Scott and Meger, David and Precup, Doina},
  booktitle={International conference on machine learning},
  pages={2052--2062},
  year={2019},
  organization={PMLR}
}

@article{kumar2020conservative,
  title={{Conservative {Q-learning} for offline reinforcement learning}},
  author={Kumar, Aviral and Zhou, Aurick and Tucker, George and Levine, Sergey},
  journal={Advances in neural information processing systems},
  volume={33},
  pages={1179--1191},
  year={2020}
}

@article{kostrikov2021offline,
  title={{Offline reinforcement learning with implicit {Q-learning}}},
  author={Kostrikov, Ilya and Nair, Ashvin and Levine, Sergey},
  journal={arXiv preprint arXiv:2110.06169},
  year={2021}
}

@article{garcia2015comprehensive,
  title={{A comprehensive survey on safe reinforcement learning}},
  author={Garc{\i}a, Javier and Fern{\'a}ndez, Fernando},
  journal={Journal of Machine Learning Research},
  volume={16},
  number={1},
  pages={1437--1480},
  year={2015}
}

@book{altman2021constrained,
  title={{Constrained Markov decision processes}},
  author={Altman, Eitan},
  year={2021},
  publisher={Routledge}
}

@article{chow2018risk,
  title={{Risk-constrained reinforcement learning with percentile risk criteria}},
  author={Chow, Yinlam and Ghavamzadeh, Mohammad and Janson, Lucas and Pavone, Marco},
  journal={Journal of Machine Learning Research},
  volume={18},
  number={167},
  pages={1--51},
  year={2018}
}

@techreport{ray2019benchmarking,
  title        = {{Benchmarking Safe Exploration in Deep Reinforcement Learning}},
  author       = {Ray, Alex and Achiam, Joshua and Amodei, Dario},
  year         = {2019},
  institution  = {OpenAI},
  url          = {https://cdn.openai.com/safexp-short.pdf}
}

@inproceedings{tamar2015optimizing,
  title={{Optimizing the {CVaR} via sampling}},
  author={Tamar, Aviv and Glassner, Yonatan and Mannor, Shie},
  booktitle={Proceedings of the AAAI Conference on Artificial Intelligence},
  volume={29},
  number={1},
  year={2015}
}

@article{iyengar2005robust,
  title={{Robust dynamic programming}},
  author={Iyengar, Garud N},
  journal={Mathematics of Operations Research},
  volume={30},
  number={2},
  pages={257--280},
  year={2005},
  publisher={INFORMS}
}

@inproceedings{pinto2017robust,
  title={{Robust adversarial reinforcement learning}},
  author={Pinto, Lerrel and Davidson, James and Sukthankar, Rahul and Gupta, Abhinav},
  booktitle={International conference on machine learning},
  pages={2817--2826},
  year={2017},
  organization={PMLR}
}

@inproceedings{finn2017model,
  title={{Model-agnostic meta-learning for fast adaptation of deep networks}},
  author={Finn, Chelsea and Abbeel, Pieter and Levine, Sergey},
  booktitle={International conference on machine learning},
  pages={1126--1135},
  year={2017},
  organization={PMLR}
}

@article{duan2016rl,
  title={{{{RL}}$^2$: Fast reinforcement learning via slow reinforcement learning}},
  author={Duan, Yan and Schulman, John and Chen, Xi and Bartlett, Peter L and Sutskever, Ilya and Abbeel, Pieter},
  journal={arXiv preprint arXiv:1611.02779},
  year={2016}
}

@inproceedings{rakelly2019efficient,
  title={{Efficient off-policy meta-reinforcement learning via probabilistic context variables}},
  author={Rakelly, Kate and Zhou, Aurick and Finn, Chelsea and Levine, Sergey and Quillen, Deirdre},
  booktitle={International conference on machine learning},
  pages={5331--5340},
  year={2019},
  organization={PMLR}
}

@article{zhu2023transfer,
  title={{Transfer learning in deep reinforcement learning: A survey}},
  author={Zhu, Zhuangdi and Lin, Kaixiang and Jain, Anil K and Zhou, Jiayu},
  journal={IEEE Transactions on Pattern Analysis and Machine Intelligence},
  volume={45},
  number={11},
  pages={13344--13362},
  year={2023},
  publisher={IEEE}
}

@article{rusu2015policy,
  title={{Policy distillation}},
  author={Rusu, Andrei A and Colmenarejo, Sergio Gomez and Gulcehre, Caglar and Desjardins, Guillaume and Kirkpatrick, James and Pascanu, Razvan and Mnih, Volodymyr and Kavukcuoglu, Koray and Hadsell, Raia},
  journal={arXiv preprint arXiv:1511.06295},
  year={2015}
}

@article{barreto2017successor,
  title={{Successor features for transfer in reinforcement learning}},
  author={Barreto, Andr{\'e} and Dabney, Will and Munos, R{\'e}mi and Hunt, Jonathan J and Schaul, Tom and Van Hasselt, Hado P and Silver, David},
  journal={Advances in neural information processing systems},
  volume={30},
  year={2017}
}

@inproceedings{tobin2017domain,
  title={{Domain randomization for transferring deep neural networks from simulation to the real world}},
  author={Tobin, Josh and Fong, Rachel and Ray, Alex and Schneider, Jonas and Zaremba, Wojciech and Abbeel, Pieter},
  booktitle={2017 IEEE/RSJ international conference on intelligent robots and systems (IROS)},
  pages={23--30},
  year={2017},
  organization={IEEE}
}

@article{yan2025near,
  title={{Near-Real-Time Resource Slicing for {QoS} Optimization in {5G} {O-RAN} Using Deep Reinforcement Learning}},
  author={Yan, Peihao and Lu, Jie and Zeng, Huacheng and Hou, Y Thomas},
  journal={IEEE Transactions on Networking},
  volume={34},
  pages={1596--1611},
  year={2025},
  publisher={IEEE}
}

@article{lu2026eexapp,
  title={{Eexapp: {GNN}-based reinforcement learning for radio unit energy optimization in {5G} {O-RAN}}},
  author={Lu, Jie and Yan, Peihao and Zeng, Huacheng},
  journal={arXiv preprint arXiv:2602.09206},
  year={2026}
}

@inproceedings{chebotar2019closing,
  title={{Closing the sim-to-real loop: Adapting simulation randomization with real world experience}},
  author={Chebotar, Yevgen and Handa, Ankur and Makoviychuk, Viktor and Macklin, Miles and Issac, Jan and Ratliff, Nathan and Fox, Dieter},
  booktitle={2019 international conference on robotics and automation (ICRA)},
  pages={8973--8979},
  year={2019},
  organization={IEEE}
}

@article{tan2018sim,
  title={{Sim-to-real: Learning agile locomotion for quadruped robots}},
  author={Tan, Jie and Zhang, Tingnan and Coumans, Erwin and Iscen, Atil and Bai, Yunfei and Hafner, Danijar and Bohez, Steven and Vanhoucke, Vincent},
  journal={arXiv preprint arXiv:1804.10332},
  year={2018}
}

@article{sadeghi2016cad2rl,
  title={{{CAD2RL}: Real single-image flight without a single real image}},
  author={Sadeghi, Fereshteh and Levine, Sergey},
  journal={arXiv preprint arXiv:1611.04201},
  year={2016}
}

@article{zhu2018reinforcement,
  title={{Reinforcement and imitation learning for diverse visuomotor skills}},
  author={Zhu, Yuke and Wang, Ziyu and Merel, Josh and Rusu, Andrei and Erez, Tom and Cabi, Serkan and Tunyasuvunakool, Saran and Kram{\'a}r, J{\'a}nos and Hadsell, Raia and de Freitas, Nando and others},
  journal={arXiv preprint arXiv:1802.09564},
  year={2018}
}

@inproceedings{ganin2015unsupervised,
  title={{Unsupervised domain adaptation by backpropagation}},
  author={Ganin, Yaroslav and Lempitsky, Victor},
  booktitle={International conference on machine learning},
  pages={1180--1189},
  year={2015},
  organization={PMLR}
}

@article{ganin2016domain,
  title={{Domain-adversarial training of neural networks}},
  author={Ganin, Yaroslav and Ustinova, Evgeniya and Ajakan, Hana and Germain, Pascal and Larochelle, Hugo and Laviolette, Fran{\c{c}}ois and March, Mario and Lempitsky, Victor},
  journal={Journal of machine learning research},
  volume={17},
  number={59},
  pages={1--35},
  year={2016}
}

@inproceedings{bousmalis2017unsupervised,
  title={{Unsupervised pixel-level domain adaptation with generative adversarial networks}},
  author={Bousmalis, Konstantinos and Silberman, Nathan and Dohan, David and Erhan, Dumitru and Krishnan, Dilip},
  booktitle={Proceedings of the IEEE conference on computer vision and pattern recognition},
  pages={3722--3731},
  year={2017}
}

@techreport{3gpp_38_801,
  author      = {{3GPP}},
  title       = {{Study on New Radio Access Technology;
                 Radio Access Architecture and Interfaces}},
  type        = {Technical Specification},
  number      = {TR 38.801, V14.0.0},
  institution = {3rd Generation Partnership Project (3GPP)},
  year        = {2017},
}

@techreport{3gpp_38_463,
  author      = {{3GPP}},
  title       = {{{NG-{RAN}}; {E1 Application Protocol (E1AP)}}},
  institution  = {3rd Generation Partnership Project (3GPP)},
  type        = {Technical Specification},
  number      = {TS 38.463, V16.3.0},
  year        = {2020},
}

@techreport{3gpp_38_473,
  author       = {{3GPP}},
  title        = {{{NG-{RAN}}; {F1} Application Protocol ({F1AP})}},
  institution  = {3rd Generation Partnership Project (3GPP)},
  type         = {Technical Specification},
  number       = {TS 38.473, V19.1.0},
  year         = {2025},
}

@techreport{oran_wg4_cus,
  author      = {{O-RAN Alliance}},
  title       = {{{{O-RAN}} Working Group 4 (Open Fronthaul Interfaces {WG});
                 Control, User and Synchronization Plane Specification}},
  type        = {Technical Specification},
  number      = {O-RAN.WG4.CUS.0-R005-v20.00},
  institution = {O-RAN Alliance e.V.},
  year        = {2026},
}

@techreport{oran_wg2_nonrtric_arch,
  author       = {{O-RAN Alliance}},
  title        = {{{O-RAN} {Non-RT} {RIC}: Architecture}},
  institution  = {O-RAN Alliance Working Group 2},
  type         = {Technical Specification},
  number       = {O-RAN.WG2.TS.Non-RT-RIC-ARCH-R004-v07.00},
  year         = {2025},
}

@techreport{oran_wg3_ricarch,
  author       = {{O-RAN Alliance}},
  title        = {{{O-RAN} {Near-RT} {RIC} Architecture}},
  institution  = {O-RAN Alliance Working Group 3},
  type         = {Technical Specification},
  number       = {O-RAN.WG3.TS.RICARCH-R005-v08.00},
  year         = {2026},
}

@techreport{oran_wg6_o2,
  author       = {{O-RAN Alliance}},
  title        = {{{O-RAN} O2 Interface General Aspects and Principles}},
  institution  = {O-RAN Alliance Working Group 6},
  type         = {Technical Specification},
  number       = {O-RAN.WG6.TS.O2-GA\&P-R005-v10.00},
  year         = {2026},
}

@techreport{oran_wg2_a1,
  author       = {{O-RAN Alliance}},
  title        = {{{{O-RAN}} A1 interface: General Aspects and Principles}},
  institution  = {O-RAN Alliance},
  type         = {Technical Specification},
  number       = {O-RAN.WG2.TS.A1GAP-R005-v05.03},
  year         = {2026},
}

@techreport{oran_wg2_r1,
  author       = {{O-RAN Alliance}},
  title        = {{{{O-RAN}} {R1} Interface: General Aspects and Principles}},
  institution  = {O-RAN Alliance},
  type         = {Technical Specification},
  number       = {O-RAN.WG2.TS.R1GAP-R005-v13.00},
  year         = {2026},
}

@techreport{oran_wg3_y1,
  author       = {{O-RAN Alliance}},
  title        = {{{{O-RAN}} {Y1} Interface: General Aspects and Principles}},
  institution  = {O-RAN Alliance},
  type         = {Technical Specification},
  number       = {O-RAN.WG3.TS.Y1GAP-R005-v01.02},
  year         = {2026},
}

@techreport{oran_wg1_o1,
  author       = {{O-RAN Alliance}},
  title        = {{{{O-RAN}} Operations and Maintenance Interface}},
  institution  = {O-RAN Alliance},
  type         = {Technical Specification},
  number       = {O-RAN.WG1.O1-Interface.0-v04.00},
  year         = {2021}
}

@article{d2022dapps,
  title={{{dApps}: Distributed applications for real-time inference and control in {O-RAN}}},
  author={D'Oro, Salvatore and Polese, Michele and Bonati, Leonardo and Cheng, Hai and Melodia, Tommaso},
  journal={IEEE Communications Magazine},
  volume={60},
  number={11},
  pages={52--58},
  year={2022},
  publisher={IEEE}
}

@techreport{3gpp_38_300,
  author       = {{3GPP}},
  title        = {{NR; NR and NG-{RAN} Overall Description; Stage 2}},
  institution  = {3rd Generation Partnership Project (3GPP)},
  type         = {Technical Specification},
  number       = {TS 38.300, V18.3.0},
  year         = {2024},
}

@techreport{3gpp_38_331,
  author       = {{3GPP}},
  title        = {{NR; Radio Resource Control (RRC) Protocol Specification}},
  institution  = {3rd Generation Partnership Project (3GPP)},
  type         = {Technical Specification},
  number       = {TS 38.331, V18.4.0},
  year         = {2024},
}

@techreport{hexax_ai_native,
  author       = {{Hexa-X-II Consortium}},
  title        = {{Deliverable D3.5: Final Architectural Framework and Analysis}},
  institution  = {Hexa-X-II Project, European Union Horizon Europe SNS JU},
  year         = {2025},
  url          = {https://hexa-x-ii.eu/wp-content/uploads/2025/03/Hexa-X-II_D3.5_v1.0.pdf}
}

@misc{ngmn_6g_position,
  author       = {{NGMN Alliance}},
  title        = {{{6G} Position Statement: An Operator View}},
  howpublished = {White Paper},
  year         = {2023},
  url          = {https://www.ngmn.org/highlight/ngmn-publishes-6g-position-statement.html}
}

@article{liu2022integrated,
  title={{Integrated sensing and communications: Toward dual-functional wireless networks for {6G} and beyond}},
  author={Liu, Fan and Cui, Yuanhao and Masouros, Christos and Xu, Jie and Han, Tony Xiao and Eldar, Yonina C and Buzzi, Stefano},
  journal={IEEE journal on selected areas in communications},
  volume={40},
  number={6},
  pages={1728--1767},
  year={2022},
  publisher={IEEE}
}

@article{kodheli2020satellite,
  title={{Satellite communications in the new space era: A survey and future challenges}},
  author={Kodheli, Oltjon and Lagunas, Eva and Maturo, Nicola and Sharma, Shree Krishna and Shankar, Bhavani and Montoya, Jesus Fabian Mendoza and Duncan, Juan Carlos Merlano and Spano, Danilo and Chatzinotas, Symeon and Kisseleff, Steven and others},
  journal={IEEE Communications Surveys \& Tutorials},
  volume={23},
  number={1},
  pages={70--109},
  year={2020},
  publisher={IEEE}
}

@techreport{oran_ntn_2025,
  author       = {{O-RAN Alliance}},
  title        = {{Deployments of {O-RAN}-Based Non-Terrestrial Networks}},
  institution  = {O-RAN Alliance},
  type         = {White Paper},
  year         = {2025},
  month        = feb,
  url          = {https://mediastorage.o-ran.org/ecosystem-resources/O-RAN-2025.04.02.WP.O-RAN_NTN_Deployments-v08.4.pdf}
}

@article{baena2025space,
  title={{Space-{O-RAN}: Enabling intelligent, open, and interoperable non terrestrial networks in {6G}}},
  author={Baena, Eduardo and Testolina, Paolo and Polese, Michele and Koutsonikolas, Dimitrios and Jornet, Josep and Melodia, Tommaso},
  journal={IEEE Communications Magazine},
  year={2025},
  publisher={IEEE}
}

@techreport{oran_ngrg,
  author       = {{O-RAN Alliance next Generation Research Group (nGRG)}},
  title        = {{{dApps} for Real-Time {RAN} Control: Use Cases and Requirements}},
  institution  = {O-RAN Alliance},
  type         = {Contributed Research Report},
  number       = {nGRG-RR-2024-10},
  year         = {2024}
}

@techreport{oran_wg3_e2sm_kpm,
  author       = {{O-RAN Alliance}},
  title        = {{{O-RAN} Working Group 3, Near-Real-time {RAN} Intelligent 
                   Controller, E2 Service Model (E2SM), {KPM}}},
  institution  = {O-RAN Alliance},
  number       = {O-RAN.WG3.TS.E2SM-KPM-R004-v07.00},
  type          = {Technical Specification},
  year         = {2025}
}

@techreport{3gpp_28_552,
  author       = {{3GPP}},
  title        = {{Management and orchestration; {5G} performance measurements}},
  institution  = {3rd Generation Partnership Project (3GPP)},
  number       = {TS 28.552, V19.1.0},
  year         = {2024}
}

@techreport{3gpp_28_554,
  author       = {{3GPP}},
  title        = {{Management and orchestration; {5G} end-to-end 
                   Key Performance Indicators ({KPI})}},
  number       = {TS 28.554, V19.1.0},
  institution  = {3rd Generation Partnership Project (3GPP)},
  year         = {2024}
}

@article{ammar2026towards,
  title={Towards Intelligent and Adaptive Multi-Agent Slicing in Maritime Networks},
  author={Ammar, Sahar and Abderrahim, Wiem and Shihada, Basem},
  journal={IEEE Transactions on Mobile Computing},
  year={2026},
  publisher={IEEE}
}

@article{ammar2025maritime,
  title={Maritime-oriented network slicing in O-RAN integrated aerial-terrestrial networks},
  author={Ammar, Sahar and Abderrahim, Wiem and Shihada, Basem},
  journal={IEEE Transactions on Mobile Computing},
  year={2025},
  publisher={IEEE}
}

@article{qazzaz2026oreo,
  title={OREO: Open RAN Energy Optimisation via Deep Reinforcement Learning for 6G Networks},
  author={Qazzaz, Mohammed MH and Salama, Abdelaziz and Hafeez, Maryam and Zaidi, Syed Ali Raza},
  journal={IEEE Open Journal of the Communications Society},
  year={2026},
  publisher={IEEE}
}

@ARTICLE{xdiff,
  author={Yan, Peihao and Zeng, Huacheng and Hou, Y. Thomas},
  journal={IEEE Transactions on Networking}, 
  title={{xDiff: Online Diffusion Model for Collaborative Inter-Cell Interference Management in {5G} {O-RAN}}}, 
  year={2026},
  volume={34},
  number={},
  pages={1363-1376},
  doi={10.1109/TON.2025.3622520}}

@article{reinders2026aiim,
  title={{AIIM: Adaptive Inter-cell Interference Mitigation for Heterogeneous Multi-vendor {5G} {O-RAN} Networks}},
  author={Reinders, Samuel and Dorcheh, Alireza Ebrahimi and Barker, Ryan and Seyfi, Tolunay and Afghah, Fatemeh},
  journal={arXiv preprint arXiv:2605.01112},
  year={2026}
}

@inproceedings{gopal2025adapshare,
  title={{AdapShare: An {RL}-based dynamic spectrum sharing solution for {O-RAN}}},
  author={Gopal, Sneihil and Griffith, David and Rouil, Richard A and Liu, Chunmei},
  booktitle={2025 IEEE 22nd Consumer Communications \& Networking Conference (CCNC)},
  pages={1--7},
  year={2025},
  organization={IEEE}
}

@article{abedin2022elastic,
  title={{Elastic {O-RAN} slicing for industrial monitoring and control: A distributed matching game and deep reinforcement learning approach}},
  author={Abedin, Sarder Fakhrul and Mahmood, Aamir and Tran, Nguyen H and Han, Zhu and Gidlund, Mikael},
  journal={IEEE Transactions on Vehicular Technology},
  volume={71},
  number={10},
  pages={10808--10822},
  year={2022},
  publisher={IEEE}
}

@article{ghafouri2024multi,
  title={{A multi-level deep {RL}-based network slicing and resource management for {O-RAN}-based {6G} cell-free networks}},
  author={Ghafouri, Navideh and Vardakas, John S and Ramantas, Kostas and Verikoukis, Christos},
  journal={IEEE Transactions on Vehicular Technology},
  volume={73},
  number={11},
  pages={17472--17484},
  year={2024},
  publisher={IEEE}
}

@article{sohaib2025optimizing,
  title={{Optimizing {URLLC} in open {RAN}: A deep reinforcement learning-based trade-off analysis}},
  author={Sohaib, Rana Muhammad and Shah, Syed Tariq and Jamshed, Muhammad Ali and Onireti, Oluwakayode and Yadav, Poonam},
  journal={IEEE Communications Standards Magazine},
  year={2025},
  publisher={IEEE}
}

@article{li2025toward,
  title={{Toward Practical Operation of Deep Reinforcement Learning Agents in Real-World Network Management at Open {RAN} Edges}},
  author={Li, Haiyuan and Madhukumar, Hari and Li, Peizheng and Liu, Yuelin and Teng, Yiran and Wu, Yulei and Wang, Ning and Yan, Shuangyi and Simeonidou, Dimitra},
  journal={IEEE Communications Magazine},
  year={2025},
  publisher={IEEE}
}

@article{yan2026tarmm,
  title={{TARMM: Scaling Delay-Critical Edge {AI} Offloading in {5G} {O-RAN} via Temporal Graph Mobility Management}},
  author={Yan, Peihao and Chen, Yun and Lu, Jie and Wang, Qijun and Zeng, Huacheng},
  journal={arXiv preprint arXiv:2604.24501},
  year={2026}
}

@article{qazzaz2026xapp,
  title={{{xApp} Empowered Resource Management for Non-Terrestrial Users in {5G} {O-RAN} Networks}},
  author={Qazzaz, Mohammed MH and Zaidi, Syed Ali and Al-Hameed, Aubida A and Salama, Abdelaziz and Mclernon, Des},
  journal={IEEE Transactions on Machine Learning in Communications and Networking},
  year={2026},
  publisher={IEEE}
}

@article{kalntis2026meta,
  title={{Meta-Learning-Based Handover Management in {NextG} {O-RAN}}},
  author={Kalntis, Michail and Iosifidis, George and Su{\'a}rez-Varela, Jos{\'e} and Lutu, Andra and Kuipers, Fernando A},
  journal={IEEE Journal on Selected Areas in Communications},
  year={2026},
  publisher={IEEE}
}

@inproceedings{dai2024intelligent,
  title={{Intelligent handover management enabled by {O-RAN} and deep reinforcement learning}},
  author={Dai, Jiongyu and Mahboob, Shadab and Wang, Haining and Liu, Lingjia},
  booktitle={2024 IEEE 100th Vehicular Technology Conference (VTC2024-Fall)},
  pages={1--6},
  year={2024},
  organization={IEEE}
}

@article{wadud2026ai,
  title={{{AI}-Driven Multi-Modal Adaptive Handover Control Optimization for {O-RAN}}},
  author={Wadud, Abdul and Golpayegani, Fatemeh and Afraz, Nima},
  journal={arXiv preprint arXiv:2603.17158},
  year={2026}
}

@article{lacava2023programmable,
  title={{Programmable and customized intelligence for traffic steering in {5G} networks using open {RAN} architectures}},
  author={Lacava, Andrea and Polese, Michele and Sivaraj, Rajarajan and Soundrarajan, Rahul and Bhati, Bhawani Shanker and Singh, Tarunjeet and Zugno, Tommaso and Cuomo, Francesca and Melodia, Tommaso},
  journal={IEEE Transactions on Mobile Computing},
  volume={23},
  number={4},
  pages={2882--2897},
  year={2023},
  publisher={IEEE}
}

@article{habib2024machine,
  title={{Machine learning-enabled traffic steering in {O-RAN}: A case study on hierarchical learning approach}},
  author={Habib, Md Arafat and Zhou, Hao and Iturria-Rivera, Pedro Enrique and Ozcan, Yigit and Elsayed, Medhat and Bavand, Majid and Gaigalas, Raimundas and Erol-Kantarci, Melike},
  journal={IEEE Communications Magazine},
  volume={63},
  number={1},
  pages={100--107},
  year={2024},
  publisher={IEEE}
}

@inproceedings{erdol2022federated,
  title={{Federated meta-learning for traffic steering in {O-RAN}}},
  author={Erdol, Hakan and Wang, Xiaoyang and Li, Peizheng and Thomas, Jonathan D and Piechocki, Robert and Oikonomou, George and Inacio, Rui and Ahmad, Abdelrahim and Briggs, Keith and Kapoor, Shipra},
  booktitle={2022 IEEE 96th Vehicular Technology Conference (VTC2022-Fall)},
  pages={1--7},
  year={2022},
  organization={IEEE}
}

@article{kavehmadavani2024empowering,
  title={{Empowering traffic steering in {6G} open {RAN} with deep reinforcement learning}},
  author={Kavehmadavani, Fatemeh and Nguyen, Van-Dinh and Vu, Thang X and Chatzinotas, Symeon},
  journal={IEEE Transactions on Wireless Communications},
  volume={23},
  number={10},
  pages={12782--12798},
  year={2024},
  publisher={IEEE}
}

@article{nguyen2023network,
  title={{Network-aided intelligent traffic steering in {6G} {O-RAN}: A multi-layer optimization framework}},
  author={Nguyen, Van-Dinh and Vu, Thang X and Nguyen, Nhan Thanh and Nguyen, Dinh C and Juntti, Markku and Luong, Nguyen Cong and Hoang, Dinh Thai and Nguyen, Diep N and Chatzinotas, Symeon},
  journal={IEEE Journal on Selected Areas in Communications},
  volume={42},
  number={2},
  pages={389--405},
  year={2023},
  publisher={IEEE}
}

@inproceedings{tamim2023intelligent,
  title={{Intelligent {O-RAN} traffic steering for {URLLC} through deep reinforcement learning}},
  author={Tamim, Ibrahim and Aleyadeh, Sam and Shami, Abdallah},
  booktitle={ICC 2023-IEEE International Conference on Communications},
  pages={112--118},
  year={2023},
  organization={IEEE}
}

@article{sroka2024policy,
  title={{Policy-based traffic steering and load balancing in {O-RAN}-based vehicle-to-network communications}},
  author={Sroka, Pawe{\l} and Ku{\l}acz, {\L}ukasz and Janji, Salim and Dryja{\'n}ski, Marcin and Kliks, Adrian},
  journal={IEEE Transactions on Vehicular Technology},
  volume={73},
  number={7},
  pages={9356--9369},
  year={2024},
  publisher={IEEE}
}

@inproceedings{sharma2025adaptive,
  title={{Adaptive Traffic Steering in Open {RAN}: Integrating Rule-Based Policies with Reinforcement Learning}},
  author={Sharma, Utkarsh and Wei, Hua and Chen, Mingzhe and Xu, Jie and Liu, Yuchen},
  booktitle={IEEE INFOCOM 2025-IEEE Conference on Computer Communications Workshops (INFOCOM WKSHPS)},
  pages={1--6},
  year={2025},
  organization={IEEE}
}

@inproceedings{truong2025reinforcement,
  title={{Reinforcement Learning for {RAN} Intelligent Controller: A Case Study on Traffic Steering}},
  author={Truong, Thanh Phung and Van, Hieu Hoang and Canh, Trung Nguyen and Nguyen, Hieu V and Cho, Sungrae},
  booktitle={International Conference on Intelligence of Things},
  pages={31--42},
  year={2025},
  organization={Springer}
}

@inproceedings{kefalas2025traffic,
  title={{Traffic Steering for {O-RAN} Multi-RAT {5G} Networks Using {ML}-Based Demand Forecasting}},
  author={Kefalas, Dimitris and Makris, Nikos and Korakis, Thanasis and Fdida, Serge},
  booktitle={2025 IEEE Conference on Network Function Virtualization and Software-Defined Networking (NFV-SDN)},
  pages={1--6},
  year={2025},
  organization={IEEE}
}

@article{polese2026enabling,
  title={{Enabling Programmable Inference and {ISAC} at the 6GR Edge with {dApps}}},
  author={Polese, Michele and Gangula, Rajeev and Melodia, Tommaso},
  journal={arXiv preprint arXiv:2603.29146},
  year={2026}
}

@article{baena2026toward,
  title={{Toward Native {ISAC} Support in {O-RAN} Architectures for {6G}}},
  author={Baena, Eduardo and Krishnan, Rajesh and Vu, Mai and Zussman, Gil and Koutsonikolas, Dimitrios},
  journal={arXiv preprint arXiv:2603.03607},
  year={2026}
}

@article{villegas2026isac,
  title={{{ISAC}-Assisted {DRL} for Dynamic {MAC} Scheduler Reconfiguration in {O-RAN}}},
  author={Villegas, Neco and Herrera, Juan Luis and Diez, Luis and Scotece, Domenico and Foschini, Luca and Ag{\"u}ero, Ram{\'o}n},
  journal={IEEE Open Journal of the Communications Society},
  year={2026},
  publisher={IEEE}
}

@inproceedings{nikbakht2024memory,
  title={{A memory-based reinforcement learning approach to integrated sensing and communication}},
  author={Nikbakht, Homa and Wigger, Mich{\`e}le and Shitz, Shlomo Shamai and Poor, H Vincent},
  booktitle={2024 58th Asilomar Conference on Signals, Systems, and Computers},
  pages={433--437},
  year={2024},
  organization={IEEE}
}

@article{agarwal2021deep,
  title={Deep reinforcement learning at the edge of the statistical precipice},
  author={Agarwal, Rishabh and Schwarzer, Max and Castro, Pablo Samuel and Courville, Aaron C and Bellemare, Marc},
  journal={Advances in neural information processing systems},
  volume={34},
  pages={29304--29320},
  year={2021}
}

@incollection{hamidi2025ran,
  title={{{RAN} architecture and implementation requirements for {ISAC}}},
  author={Hamidi-Sepehr, Fatemeh and Hewavithana, Thushara and Vannithamby, Rath and Merwaday, Arvind},
  booktitle={Integrated Sensing and Communications for Future Wireless Networks},
  pages={325--346},
  year={2025},
  publisher={Elsevier}
}

@inproceedings{todorov2012mujoco,
  title={{{MuJoCo}: A physics engine for model-based control}},
  author={Todorov, Emanuel and Erez, Tom and Tassa, Yuval},
  booktitle={2012 IEEE/RSJ international conference on intelligent robots and systems},
  pages={5026--5033},
  year={2012},
  organization={IEEE}
}

@article{mnih2013playing,
  title={{Playing atari with deep reinforcement learning}},
  author={Mnih, Volodymyr and Kavukcuoglu, Koray and Silver, David and Graves, Alex and Antonoglou, Ioannis and Wierstra, Daan and Riedmiller, Martin},
  journal={arXiv preprint arXiv:1312.5602},
  year={2013}
}

@article{oh2023decentralized,
  title={{A decentralized pilot assignment algorithm for scalable {O-RAN} cell-free massive {MIMO}}},
  author={Oh, Myeung Suk and Das, Anindya Bijoy and Hosseinalipour, Seyyedali and Kim, Taejoon and Love, David J and Brinton, Christopher G},
  journal={IEEE Journal on Selected Areas in Communications},
  volume={42},
  number={2},
  pages={373--388},
  year={2023},
  publisher={IEEE}
}

@inproceedings{eskandari2025network,
  title={{Network slicing in {O-RAN}-enabled cell-free massive {MIMO}: A {DRL}-based power control}},
  author={Eskandari, Mahdi and Rahmani, Mostafa and Burr, Alister G},
  booktitle={2025 IEEE Wireless Communications and Networking Conference (WCNC)},
  pages={1--7},
  year={2025},
  organization={IEEE}
}

@article{abdelmoaty2025enabling,
  title={{Enabling seamless connectivity in consumer electronics: A {DRL} approach for scalable cell-free {mMIMO} handover under {O-RAN}}},
  author={Abdelmoaty, Ahmed and Naboulsi, Diala and Kaddoum, Georges},
  journal={IEEE Transactions on Consumer Electronics},
  year={2025},
  publisher={IEEE}
}

@inproceedings{sohaib2024drl,
  title={{{DRL}-based joint resource scheduling of {eMBB} and {URLLC} in {O-RAN}}},
  author={Sohaib, Rana M and Shah, Syed Tariq and Onireti, Oluwakayode and Sambo, Yusuf and Abbasi, Qammer H and Imran, Muhammad Ali},
  booktitle={2024 IEEE International Conference on Communications Workshops (ICC Workshops)},
  pages={1523--1528},
  year={2024},
  organization={IEEE}
}

@inproceedings{tan2025deep,
  title={{Deep Reinforcement Learning based Resource Allocation in {O-RAN}}},
  author={Tan, Xin Yi and Tham, Mau Luen and Lee, Ying Loong and Chow, Chee Onn and Wong, Yi Jie and Pu, Chuan-Hsian},
  booktitle={2025 IEEE Industrial Electronics and Applications Conference (IEACon)},
  pages={92--97},
  year={2025},
  organization={IEEE}
}

@article{filali2024open,
  title={{Open {RAN} slicing for MVNOs with deep reinforcement learning}},
  author={Filali, Abderrahime and Mlika, Zoubeir and Cherkaoui, Soumaya},
  journal={IEEE Internet of Things Journal},
  volume={11},
  number={10},
  pages={18711--18725},
  year={2024},
  publisher={IEEE}
}

@INPROCEEDINGS{mhatre2024aiaas1,
  author={Mhatre, Suvidha and Adelantado, Ferran and Ramantas, Kostas and Verikoukis, Christos},
  booktitle={ICC 2024 - IEEE International Conference on Communications}, 
  title={{{AIaaS} for ORAN-based {6G} Networks: Multi-time Scale Slice Resource Management with {DRL}}}, 
  year={2024},
  volume={},
  number={},
  pages={5407-5412},
  doi={10.1109/ICC51166.2024.10622601}}

@INPROCEEDINGS{villegas2025drl2,
  author={Villegas, N. and Herrera, J. L. and Diez, L. and Scotece, D. and Foschini, L. and Agüero, R.},
  booktitle={ICC 2025 - IEEE International Conference on Communications}, 
  title={{{DRL}-Based Dynamic {MAC} Scheduler Reconfiguration in {O-RAN}}}, 
  year={2025},
  volume={},
  number={},
  pages={5023-5028},
  doi={10.1109/ICC52391.2025.11160805}}

@inproceedings{an2024dragon4,
author = {An, Qing and Doost-Mohammady, Rahman and Yang, Roy and Sridhar, Kamakshi},
title = {{DRAGON: A {DRL}-based {MIMO} Layer and {MCS} Adapter in Open {RAN} {5G} Networks}},
year = {2024},
isbn = {9798400704895},
publisher = {Association for Computing Machinery},
address = {New York, NY, USA},
url = {https://doi.org/10.1145/3636534.3701549},
doi = {10.1145/3636534.3701549},
booktitle = {Proceedings of the 30th Annual International Conference on Mobile Computing and Networking},
pages = {2323–2328},
numpages = {6},
location = {Washington D.C., DC, USA},
series = {ACM MobiCom '24}
}

@ARTICLE{dai2024ran6,
  author={Dai, Jiongyu and Li, Lianjun and Safavinejad, Ramin and Mahboob, Shadab and Chen, Hao and Ratnam, Vishnu V and Wang, Haining and Zhang, Jianzhong and Liu, Lingjia},
  journal={IEEE Transactions on Mobile Computing}, 
  title={{{O-RAN}-Enabled Intelligent Network Slicing to Meet Service-Level Agreement ({SLA})}}, 
  year={2025},
  volume={24},
  number={2},
  pages={890-906},
  doi={10.1109/TMC.2024.3476338}}

@ARTICLE{filali2023communication7,
  author={Filali, Abderrahime and Nour, Boubakr and Cherkaoui, Soumaya and Kobbane, Abdellatif},
  journal={IEEE Communications Standards Magazine}, 
  title={{Communication and Computation {O-RAN} Resource Slicing for {URLLC} Services Using Deep Reinforcement Learning}}, 
  year={2023},
  volume={7},
  number={1},
  pages={66-73},
  doi={10.1109/MCOMSTD.0002.2100078}}

@ARTICLE{rezazadeh2024intelligible10,
  author={Rezazadeh, Farhad and Chergui, Hatim and Siddiqui, Shuaib and Mangues, Josep and Song, Houbing and Saad, Walid and Bennis, Mehdi},
  journal={IEEE Wireless Communications}, 
  title={{Intelligible Protocol Learning for Resource Allocation in {6G} {O-RAN} Slicing}}, 
  year={2024},
  volume={31},
  number={5},
  pages={192-199},
  doi={10.1109/MWC.015.2300552}}

@ARTICLE{ergu2024efficient14,
  author={Ergu, Yared Abera and Nguyen, Van-Linh and Hwang, Ren-Hung and Lin, Ying-Dar and Cho, Chuan-Yu and Yang, Hui-Kuo and Shin, Hyundong and Duong, Trung Q.},
  journal={IEEE Transactions on Vehicular Technology}, 
  title={{Efficient Adversarial Attacks Against {DRL}-Based Resource Allocation in Intelligent {O-RAN} for {V2X}}}, 
  year={2025},
  volume={74},
  number={1},
  pages={1674-1686},
  doi={10.1109/TVT.2024.3466511}}

@ARTICLE{rezazadeh2022specialization16,
  author={Rezazadeh, Farhad and Zanzi, Lanfranco and Devoti, Francesco and Chergui, Hatim and Costa-Pérez, Xavier and Verikoukis, Christos},
  journal={IEEE Transactions on Vehicular Technology}, 
  title={{On the Specialization of {FDRL} Agents for Scalable and Distributed {6G} {RAN} Slicing Orchestration}}, 
  year={2023},
  volume={72},
  number={3},
  pages={3473-3487},
  doi={10.1109/TVT.2022.3218158}}

@INPROCEEDINGS{hammami2022policy18,
  author={Hammami, Nessrine and Nguyen, Kim Khoa},
  booktitle={2022 IEEE Wireless Communications and Networking Conference (WCNC)}, 
  title={{On-Policy vs. Off-Policy Deep Reinforcement Learning for Resource Allocation in Open Radio Access Network}}, 
  year={2022},
  volume={},
  number={},
  pages={1461-1466},
  doi={10.1109/WCNC51071.2022.9771605}}

@ARTICLE{mhatre2024intelligent22,
  author={Mhatre, Suvidha and Adelantado, Ferran and Ramantas, Kostas and Verikoukis, Christos},
  journal={IEEE Transactions on Vehicular Technology}, 
  title={{Intelligent {QoS}-Aware Slice Resource Allocation With User Association Parameterization for Beyond {5G} {O-RAN}-Based Architecture Using {DRL}}}, 
  year={2025},
  volume={74},
  number={2},
  pages={3096-3109},
  doi={10.1109/TVT.2024.3483288}}

@INPROCEEDINGS{lotfi2025meta30,
  author={Lotfi, Fatemeh and Afghah, Fatemeh},
  booktitle={2025 IEEE Wireless Communications and Networking Conference (WCNC)}, 
  title={{Meta Reinforcement Learning Approach for Adaptive Resource Optimization in {O-RAN}}}, 
  year={2025},
  volume={},
  number={},
  pages={1-6},
  doi={10.1109/WCNC61545.2025.10978365}}

@ARTICLE{boateng2022consortium33,
  author={Boateng, Gordon Owusu and Sun, Guolin and Mensah, Daniel Ayepah and Doe, Daniel Mawunyo and Ou, Ruijie and Liu, Guisong},
  journal={IEEE Transactions on Mobile Computing}, 
  title={{Consortium Blockchain-Based Spectrum Trading for Network Slicing in {5G} {RAN}: A Multi-Agent Deep Reinforcement Learning Approach}}, 
  year={2023},
  volume={22},
  number={10},
  pages={5801-5815},
  doi={10.1109/TMC.2022.3190449}}

@inproceedings{martinez2024drl,
  title={{{DRL}-based {xApps} for Dynamic {RAN} and {MEC} Resource Allocation and Slicing in {O-RAN}}},
  author={Mart{\'\i}nez-Morfa, Mario and De Mendoza, Carlos Ruiz and Cervell{\'o}-Pastor, Cristina and Sallent, Sebasti{\`a}},
  booktitle={2024 15th International Conference on Network of the Future (NoF)},
  pages={106--114},
  year={2024},
  organization={IEEE}
}

@article{filali2026drl,
  title={{{DRL}-based {RAN} slicing with efficient inter-slice isolation in tactical wireless networks}},
  author={Filali, Abderrahime and Naboulsi, Diala and Kaddoum, Georges},
  journal={IEEE Open Journal of Vehicular Technology},
  year={2026},
  publisher={IEEE}
}

@ARTICLE{qiao2025resource,
  author={Qiao, Kai and Wang, Hongchao and Zhang, Weiting and Yang, Dong and Zhang, Yuming and Zhang, Ning},
  journal={IEEE Transactions on Cognitive Communications and Networking},
  title={{Resource Allocation for Network Slicing in Open {RAN}: A Hierarchical Learning Approach}},
  year={2025},
  volume={11},
  number={4},
  pages={2584-2600},
  doi={10.1109/TCCN.2024.3524641},
  url={https://ieeexplore.ieee.org/document/10820044}
}

@INPROCEEDINGS{lotfi2023attention,
  author={Lotfi, Fatemeh and Afghah, Fatemeh and Ashdown, Jonathan},
  booktitle={GLOBECOM 2023 - 2023 IEEE Global Communications Conference},
  title={{Attention-Based Open {RAN} Slice Management Using Deep Reinforcement Learning}},
  year={2023},
  pages={6328-6333},
  doi={10.1109/GLOBECOM54140.2023.10436850},
  url={https://ieeexplore.ieee.org/document/10436850}
}

@INPROCEEDINGS{lotfi2025llm,
  author={Lotfi, Fatemeh and Rajoli, Hossein and Afghah, Fatemeh},
  booktitle={ICC 2025 - IEEE International Conference on Communications},
  title={{{LLM}-Augmented Deep Reinforcement Learning for Dynamic {O-RAN} Network Slicing}},
  year={2025},
  pages={3827-3832},
  doi={10.1109/ICC52391.2025.11161572},
  url={https://ieeexplore.ieee.org/document/11161572}
}

@INPROCEEDINGS{motalleb2023moving,
  author={Motalleb, Mojdeh Karbalaee and Benzaid, Chafika and Taleb, Tarik and Shah-Mansouri, Vahid},
  booktitle={GLOBECOM 2023 - 2023 IEEE Global Communications Conference},
  title={{Moving Target Defense based Secured Network Slicing System in the {O-RAN} Architecture}},
  year={2023},
  pages={6358-6363},
  doi={10.1109/GLOBECOM54140.2023.10437795},
  url={https://ieeexplore.ieee.org/document/10437795}
}

@INPROCEEDINGS{habib2023hierarchical,
  author={Habib, Md Arafat and Zhou, Hao and Iturria-Rivera, Pedro Enrique and Elsayed, Medhat H. M. and Bavand, Majid and Gaigalas, Raimundas and Ozcan, Yigit and Erol-Kantarci, Melike},
  booktitle={ICC 2023 - IEEE International Conference on Communications},
  title={{Hierarchical Reinforcement Learning Based Traffic Steering in Multi-RAT {5G} Deployments}},
  year={2023},
  pages={100-105},
  doi={10.1109/ICC45041.2023.10278983},
  url={https://ieeexplore.ieee.org/document/10278983}
}

@ARTICLE{joda2022deep,
  author={Joda, Rana and Pamuklu, Taha and Iturria-Rivera, Pedro Enrique and Erol-Kantarci, Melike},
  journal={IEEE Transactions on Network and Service Management},
  title={{Deep Reinforcement Learning-Based Joint User Association and CU-DU Placement in {O-RAN}}},
  year={2022},
  volume={19},
  number={4},
  pages={4097-4110},
  doi={10.1109/TNSM.2022.3219411},
  url={https://ieeexplore.ieee.org/document/9946423}
}

@ARTICLE{ergu2025radar,
  author={Ergu, Yared Abera and Nguyen, Van-Linh},
  journal={IEEE Transactions on Green Communications and Networking},
  title={{RADAR: Robust {DRL}-Based Resource Allocation Against Adversarial Attacks in Intelligent {O-RAN}}},
  year={2025},
  volume={9},
  number={4},
  pages={2305-2318},
  doi={10.1109/TGCN.2025.3562895},
  url={https://ieeexplore.ieee.org/document/10971998}
}

@ARTICLE{seid2025multiagent,
  author={Seid, Abegaz Mohammed and Abishu, Hayla Nahom and Hevesli, Muhammet and Elbiaze, Halima and Erbad, Aiman and Guizani, Mohsen},
  journal={IEEE Transactions on Communications},
  title={{A Multi-Agent {DRL}-Based Dynamic Resource Allocation in {O-RAN}-Enabled TN-{NTN} Metaverse Services}},
  year={2025},
  volume={73},
  number={12},
  pages={14243-14259},
  doi={10.1109/TCOMM.2025.3597810},
  url={https://ieeexplore.ieee.org/document/11122487}
}

@article{metahrl2025,
  author  = {Lotfi, S. and others},
  title   = {{Meta-Hierarchical Reinforcement Learning for Scalable Resource Management in {O-RAN}}},
  journal = {arXiv preprint arXiv:2512.13715},
  year    = {2025}
}

@inproceedings{riggio2026deployable,
  author    = {Riggio, R.},
  title     = {{Deployable Hierarchical {ML} Traffic Steering for {O-RAN} RICs}},
  booktitle = {Proceedings of the IEEE/IFIP Network Operations and Management Symposium (NOMS)},
  year      = {2026}
}

@article{lotfi2024open,
  author  = {Lotfi, S. and others},
  title   = {{Open {RAN} resource allocation via distributed and hierarchical deep reinforcement learning}},
  journal = {Computer Networks},
  volume  = {240},
  pages   = {110150},
  year    = {2024}
}

@inproceedings{he2025heterogeneous,
  title={{Heterogeneous-Agent {PPO} {RL} for {xApps} Coordination in Digital Twin Enabled {O-RAN}}},
  author={He, Zhizhou and Luo, Yang and Shojafar, Mohammad and Mi, De},
  booktitle={2025 IEEE/CIC International Conference on Communications in China (ICCC)},
  pages={1--6},
  year={2025},
  organization={IEEE}
}

@article{he2025digital,
  title={{Digital Twin-Enhanced Reinforcement Learning for Intelligent {xApps} Management in {O-RAN} Systems}},
  author={He, Zhizhou and Al-Tahmeesschi, Ahmed and Foh, Chuan Heng and Ahmadi, Hamed and Shojafar, Mohammad},
  journal={IEEE Internet of Things Magazine},
  year={2025},
  publisher={IEEE}
}

@article{kouchaki2025federated,
  title={{Federated neuroevolution {O-RAN}: Enhancing the robustness of deep reinforcement learning {xApps}}},
  author={Kouchaki, Mohammadreza and Abdalla, Aly Sabri and Marojevic, Vuk},
  journal={IEEE Communications Magazine},
  year={2025},
  publisher={IEEE}
}

@article{giannopoulos2026interoperable,
  title={{Interoperable {rApp}/{xApp} Control over {O-RAN} for Mobility-aware Dynamic Spectrum Allocation}},
  author={Giannopoulos, Anastasios and Spantideas, Sotirios and Bartsioka, Maria Lamprini and Trakadas, Panagiotis},
  journal={arXiv preprint arXiv:2601.13769},
  year={2026}
}

@article{erdol2025machine,
  title={Machine Learning-based Applications for Open Radio Access Networks in B5G},
  author={Erdol, Hakan and Erdol, Hakan},
  journal={simulation},
  volume={3},
  pages={1--1},
  year={2025}
}

@inproceedings{orhan2021connection,
  title={{Connection management {xApp} for {O-RAN} {RIC}: A graph neural network and reinforcement learning approach}},
  author={Orhan, Oner and Swamy, Vasuki Narasimha and Tetzlaff, Thomas and Nassar, Marcel and Nikopour, Hosein and Talwar, Shilpa},
  booktitle={2021 20th IEEE international conference on machine learning and applications (ICMLA)},
  pages={936--941},
  year={2021},
  organization={IEEE}
}

@article{montebugnoli2026manatee,
  title={{MANATEE: A DevOps Platform for {xApp} Lifecycle Management and Testing in Open {RAN}}},
  author={Montebugnoli, Sofia and Bonati, Leonardo and Sabbioni, Andrea and Foschini, Luca and Bellavista, Paolo and D'Oro, Salvatore and Polese, Michele and Melodia, Tommaso},
  journal={arXiv preprint arXiv:2601.14009},
  year={2026}
}

@article{zafar2024ric,
  title={{{RIC}-Apps-Conflict Management}},
  author={Zafar, Hammad and Tohidi, Ehsan and Kasparick, Martin and Lorbeer, Boris and Lehmann, Heiko and Weh, Matthias and Rastogi, Gunja and Charaf, Jonas and Tarwala, Monika and Kliks, Adrian and others},
  journal={Tech Rep.},
  year={2024}
}

@article{zhang2026optimized,
  title={{Optimized Traffic Scheduling in Distributed Multi-Tier Edge-Cloud Open-{RAN} for {5G} based on Multi-Agent Deep Reinforcement Learning}},
  author={Zhang, Zhifu and Liu, Yucheng and Sliva, Bruno and Liu, Hao and Zhou, Mo and Hancke, Gerhard Petrus},
  journal={IEEE Transactions on Consumer Electronics},
  year={2026},
  publisher={IEEE}
}

@inproceedings{rezazadeh2023multi,
  title={{A multi-agent deep reinforcement learning approach for {RAN} resource allocation in {O-RAN}}},
  author={Rezazadeh, Farhad and Zanzi, Lanfranco and Devoti, Francesco and Barrachina-Mu{\~n}oz, Sergio and Zeydan, Engin and Costa-P{\'e}rez, Xavier and Mangues-Bafalluy, Josep},
  booktitle={IEEE INFOCOM 2023-IEEE Conference on Computer Communications Workshops (INFOCOM WKSHPS)},
  pages={1--2},
  year={2023},
  organization={IEEE}
}

@article{shokouhi2025distributedkasi2025risk,
  title={{Distributed Precoding for Cell-free Massive {MIMO} in {O-RAN}: A Multi-agent Deep Reinforcement Learning Framework}},
  author={Shokouhi, Mohammad Hossein and Wong, Vincent WS},
  journal={arXiv preprint arXiv:2510.27069},
  year={2025}
}

@article{iturria2022multi,
  title={{Multi-agent team learning in virtualized open radio access networks ({O-RAN})}},
  author={Iturria-Rivera, Pedro Enrique and Zhang, Han and Zhou, Hao and Mollahasani, Shahram and Erol-Kantarci, Melike},
  journal={Sensors},
  volume={22},
  number={14},
  pages={5375},
  year={2022},
  publisher={MDPI}
}

@inproceedings{shokouhi2025distributed,
  title={{Distributed Precoding for {eMBB} and {URLLC} Traffic in Cell-Free {O-RAN}: A Multi-Agent Reinforcement Learning Framework}},
  author={Shokouhi, Mohammad Hossein and Wong, Vincent WS},
  booktitle={ICC 2025-IEEE International Conference on Communications},
  pages={2114--2119},
  year={2025},
  organization={IEEE}
}

@article{lotfi2025task,
  title={{Task-Specific Sharpness-Aware {O-RAN} Resource Management Using Multi-Agent Reinforcement Learning}},
  author={Lotfi, Fatemeh and Rajoli, Hossein and Afghah, Fatemeh},
  journal={IEEE Transactions on Machine Learning in Communications and Networking},
  volume={4},
  pages={98--114},
  year={2025},
  publisher={IEEE}
}

@inproceedings{giarre2025hierarchical,
  title={{Hierarchical multi agent {DRL} for soft handovers between edge clouds in open {RAN}}},
  author={Giarr{\`e}, Federico and Meer, Irshad A and Masoudi, Meysam and Ozger, Mustafa and Cavdar, Cicek},
  booktitle={2025 IEEE International Conference on Machine Learning for Communication and Networking (ICMLCN)},
  pages={1--6},
  year={2025},
  organization={IEEE}
}

@article{zangooei2023flexible,
  title={{Flexible {RAN} slicing in open {RAN} with constrained multi-agent reinforcement learning}},
  author={Zangooei, Mohammad and Golkarifard, Morteza and Rouili, Mohamed and Saha, Niloy and Boutaba, Raouf},
  journal={IEEE Journal on Selected Areas in Communications},
  volume={42},
  number={2},
  pages={280--294},
  year={2023},
  publisher={IEEE}
}

@article{raftopoulos2024drl,
  title={{{DRL}-based latency-aware network slicing in {O-RAN} with time-varying {SLAs}}},
  author={Raftopoulos, Raoul and D'Oro, Salvatore and Melodia, Tommaso and Schembra, Giovanni},
  journal={arXiv preprint arXiv:2401.05042},
  year={2024}
}

@article{hazarika2024enhancing,
  title={{Enhancing vehicular networks with hierarchical {O-RAN} slicing and federated {DRL}}},
  author={Hazarika, Bishmita and Saikia, Prajwalita and Singh, Keshav and Li, Chih-Peng},
  journal={IEEE Transactions on Green Communications and Networking},
  volume={8},
  number={3},
  pages={1099--1117},
  year={2024},
  publisher={IEEE}
}

@inproceedings{ahmed2025federated,
  title={{Federated Deep Reinforcement Learning-Driven {O-RAN} for Automatic Multirobot Reconfiguration}},
  author={Ahmed, Faisal and Lee, Myungjin and Lien, Shao-Yu and Subramaniam, Suresh and Matsuura, Motoharu and Hasegawa, Hiroshi and Lin, Shih-Chun},
  booktitle={NOMS 2025-2025 IEEE Network Operations and Management Symposium},
  pages={1--7},
  year={2025},
  organization={IEEE}
}

@article{ndikumana2023federated,
  title={{Federated learning assisted deep {Q-learning} for joint task offloading and fronthaul segment routing in open {RAN}}},
  author={Ndikumana, Anselme and Nguyen, Kim Khoa and Cheriet, Mohamed},
  journal={IEEE Transactions on Network and Service Management},
  volume={20},
  number={3},
  pages={3261--3273},
  year={2023},
  publisher={IEEE}
}

@article{alsenwi2025ran,
  title={{{O-RAN} Architecture-based Distributed Learning Framework for Multi-{RIS}-aided Vehicular Networks}},
  author={Alsenwi, Madyan and Abolhasan, Mehran and Lipman, Justin},
  journal={Computer Networks},
  pages={111940},
  year={2025},
  publisher={Elsevier}
}

@inproceedings{kouchaki2023openai,
  title={{{OpenAI} {dApp}: An open {AI} platform for distributed federated reinforcement learning apps in {O-RAN}}},
  author={Kouchaki, Mohammadreza and Abdalla, Aly Sabri and Marojevic, Vuk},
  booktitle={2023 IEEE Future Networks World Forum (FNWF)},
  pages={1--6},
  year={2023},
  organization={IEEE}
}

@article{amiri2023edge,
  title={{Edge-{AI} empowered dynamic {VNF} splitting in {O-RAN} slicing: A federated {DRL} approach}},
  author={Amiri, Esmaeil and Wang, Ning and Shojafar, Mohammad and Tafazolli, Rahim},
  journal={IEEE Communications Letters},
  volume={28},
  number={2},
  pages={318--322},
  year={2023},
  publisher={IEEE}
}

@article{abouaomar2022federated,
  title={{Federated deep reinforcement learning for open {RAN} slicing in {6G} networks}},
  author={Abouaomar, Amine and Taik, Afaf and Filali, Abderrahime and Cherkaoui, Soumaya},
  journal={IEEE Communications Magazine},
  volume={61},
  number={2},
  pages={126--132},
  year={2022},
  publisher={IEEE}
}

@article{abou2024federated,
  title={{Federated deep reinforcement learning for efficient jamming attack mitigation in {O-RAN}}},
  author={Abou El Houda, Zakaria and Moudoud, Hajar and Brik, Bouziane},
  journal={IEEE Transactions on Vehicular Technology},
  volume={73},
  number={7},
  pages={9334--9343},
  year={2024},
  publisher={IEEE}
}

@article{el2025secure,
  title={{Secure and trustworthy open radio access network ({O-RAN}) optimization: A zero-trust and federated learning framework for {6G} networks}},
  author={El-Hajj, Mohammed},
  journal={Future Internet},
  volume={17},
  number={6},
  pages={233},
  year={2025},
  publisher={MDPI}
}

@inproceedings{zhang2022federated,
  title={{Federated deep reinforcement learning for resource allocation in {O-RAN} slicing}},
  author={Zhang, Han and Zhou, Hao and Erol-Kantarci, Melike},
  booktitle={GLOBECOM 2022-2022 IEEE Global Communications Conference},
  pages={958--963},
  year={2022},
  organization={IEEE}
}

@article{guo2025towards,
  title={{Towards Transparent {6G} {AI}-{RAN}: A Survey on Explainable Deep Reinforcement Learning for Intelligent Network Slicing}},
  author={Guo, Shuaishuai and Zhong, Yutong and Feng, Zhenyu and Kang, Shengqi and Chen, Jichao},
  journal={Journal of Information and Intelligence},
  year={2025},
  publisher={Elsevier}
}

@article{nouri2024generative,
  title={{Generative {AI} for {O-RAN} Slicing: A Semi-Supervised Approach with {VAE} and Contrastive Learning}},
  author={Nouri, Salar and Motalleb, Mojdeh Karbalaee and Shah-Mansouri, Vahid and Shariatpanahi, Seyed Pooya},
  journal={arXiv preprint arXiv:2401.08861},
  year={2024}
}

@article{lotfi2025prompt,
  title={{Prompt-tuned {LLM}-augmented {DRL} for dynamic {O-RAN} network slicing}},
  author={Lotfi, Fatemeh and Rajoli, Hossein and Afghah, Fatemeh},
  journal={arXiv preprint arXiv:2506.00574},
  year={2025}
}

@article{lotfi2025oran,
  title={{{ORAN-GUIDE}: {RAG}-driven prompt learning for {LLM}-augmented reinforcement learning in {O-RAN} network slicing}},
  author={Lotfi, Fatemeh and Rajoli, Hossein and Afghah, Fatemeh},
  journal={arXiv preprint arXiv:2506.00576},
  year={2025}
}

@article{mashaal2026iqfm,
  title={{Iqfm--a wireless foundation model for i/q streams in {AI}-native {6G}}},
  author={Mashaal, Omar and Abou-Zeid, Hatem},
  journal={IEEE Open Journal of the Communications Society},
  year={2026},
  publisher={IEEE}
}

@article{wei2025large,
  title={{Large language models for next-generation wireless network management: A survey and tutorial}},
  author={Wei, Bisheng and Jiang, Ruihong and Zhang, Ruichen and Liu, Yinqiu and Niyato, Dusit and Sun, Yaohua and Lu, Yang and Li, Yonghui and Mao, Shiwen and Yuen, Chau and others},
  journal={arXiv preprint arXiv:2509.05946},
  year={2025}
}

@inproceedings{oluwaseyi2025llm,
  title={{The {LLM} as a Network Operator: A Vision for Generative {AI} in the {6G} Radio Access Network}},
  author={Oluwaseyi, Giwa and Adewole, Michael and Awodumila, Tobi and Aderinto, Pelumi},
  booktitle={NeurIPS 2025 Workshop: AI and ML for Next-Generation Wireless Communications and Networking},
  year={2025}
}

@article{rezazadeh2025agentic,
  title={{Agentic World Modeling for {6G}: Near-Real-Time Generative State-Space Reasoning}},
  author={Rezazadeh, Farhad and Chergui, Hatim and Debbah, Merouane and Song, Houbing and Niyato, Dusit and Liu, Lingjia},
  journal={arXiv preprint arXiv:2511.02748},
  year={2025}
}

@article{cai2025tutorial,
  title={{Tutorial on Large Language Model-Enhanced Reinforcement Learning for Wireless Networks}},
  author={Cai, Lingyi and Fu, Wenjie and Huang, Yuxi and Zhang, Ruichen and Liu, Yinqiu and Kang, Jiawen and Xiong, Zehui and Jiang, Tao and Niyato, Dusit and Wang, Xianbin and others},
  journal={arXiv preprint arXiv:2512.03722},
  year={2025}
}

@article{hu2025reflection,
  title={{Reflection-Driven Self-Optimization {6G} Agentic {AI} {RAN} via Simulation-in-the-Loop Workflows}},
  author={Hu, Yunhao and Lyu, Xinchen and Ren, Chenshan and Chen, Keda and Cui, Qimei and Tao, Xiaofeng},
  journal={arXiv preprint arXiv:2512.20640},
  year={2025}
}

@article{sun2026igaa,
  title={{IGAA: Intent-Driven General Agentic {AI} for Edge Services Scheduling using Generative Meta Learning}},
  author={Sun, Yan and Liu, Yinqiu and Guo, Shaoyong and Zhang, Ruichen and Qi, Feng and Qiu, Xuesong and Gong, Weifeng and Niyato, Dusit and Wu, Qihui},
  journal={arXiv preprint arXiv:2601.13702},
  year={2026}
}

@article{navidan2026toward,
  title={{Toward autonomous {O-RAN}: A multi-scale agentic {AI} framework for real-time network control and management}},
  author={Navidan, Hojjat and Cheraghinia, Mohammad and Fontaine, Jaron and Seif, Mohamed and De Poorter, Eli and Poor, H Vincent and Moerman, Ingrid and Shahid, Adnan},
  journal={arXiv preprint arXiv:2602.14117},
  year={2026}
}

@article{shokouhi2026agentic,
  title={{Agentic {AI} for Intent-driven Optimization in Cell-free {O-RAN}}},
  author={Shokouhi, Mohammad Hossein and Wong, Vincent WS},
  journal={arXiv preprint arXiv:2602.22539},
  year={2026}
}

@article{he2026agentic,
  title={{Agentic {AI}-{RAN}: Enabling Intent-Driven, Explainable and Self-Evolving Open {RAN} Intelligence}},
  author={He, Zhizhou and Luo, Yang and Liu, Xinkai and Mashhadi, Mahdi Boloursaz and Shojafar, Mohammad and Debbah, Merouane and Tafazolli, Rahim},
  journal={arXiv preprint arXiv:2602.24115},
  year={2026}
}

@article{habib2026reimagining,
  title={{Reimagining {RAN} Automation in {6G}: An Agentic {AI} Framework with Hierarchical Online Decision Transformer}},
  author={Habib, Md Arafat and Elsayed, Medhat and Bavand, Majid and Rivera, Pedro Enrique Iturria and Ozcan, Yigit and Erol-Kantarci, Melike},
  journal={arXiv preprint arXiv:2604.03908},
  year={2026}
}

@article{zou2026telecom,
  title={{Telecom World Models: Unifying Digital Twins, Foundation Models, and Predictive Planning for {6G}}},
  author={Zou, Hang and Yang, Yuzhi and Bariah, Lina and Tian, Yu and Lu, Yuhuan and Wang, Bohao and Bara, Anis and Mefgouda, Brahim and Liu, Hao and Tao, Yiwei and others},
  journal={arXiv preprint arXiv:2604.06882},
  year={2026}
}

@article{ferrag20266g,
  title={{{6G} Needs Agents: Toward Agentic {AI}-Native Networks for Autonomous Intelligence}},
  author={Ferrag, Mohamed Amine and Lakas, Abderrahmane and Debbah, Merouane},
  journal={arXiv preprint arXiv:2605.01546},
  year={2026}
}

@article{wang2026bridging,
  title={{Bridging the Cognitive Gap: A Unified Memory Paradigm for {6G} Agentic {AI}-{RAN}}},
  author={Wang, Xijun and Liu, Zhaoyang and Feng, Chenyuan and Chen, Xiang and Yang, Howard H and Quek, Tony QS},
  journal={arXiv preprint arXiv:2605.10036},
  year={2026}
}

@article{gajjar2026agents,
  title={{Agents Should Replace Narrow Predictive {AI} as the Orchestrator in {6G} {AI}-{RAN}}},
  author={Gajjar, Pranshav and Shah, Vijay K},
  journal={arXiv preprint arXiv:2605.11516},
  year={2026}
}

@article{natanzi2026advanced,
  title={{Advanced {AI} Service Provisioning in {O-RAN} through {LLM} Engine Integration}},
  author={Natanzi, Seyed Bagher Hashemi and Gajja, Pranshav and Tang, Bo and Shah, Vijay K},
  journal={arXiv preprint arXiv:2605.23809},
  year={2026}
}

@article{aghayev2026genesis,
  title={{GENESIS: Harnessing {AI} Agents for Autonomous {6G} {RAN} Synthesis, Research, and Testing}},
  author={Aghayev, Tamerlan and Elkael, Maxime and Polese, Michele and Nguyen, Minh Dat and Gemmi, Gabriele and Lacava, Andrea and Saeizadeh, Ali and Prasad, Reshma and Testolina, Paolo and Feraudo, Angelo and others},
  journal={arXiv preprint arXiv:2605.27360},
  year={2026}
}

@inproceedings{baena2026knows,
  title={{Who Knows What? Semantic Negotiation for Human-Supervised {RAN} Agentic Coordination}},
  author={Baena, Eduardo and Mandal, Ankita and Elango, Abhiram and Dinh, Phuc and Koutsonikolas, Dimitrios},
  booktitle={Proceedings of the 27th International Workshop on Mobile Computing Systems and Applications},
  pages={151--156},
  year={2026}
}

@misc{liurobust,
  title={{Robust Foundation Models Empowered {RAN} Intelligence for Reliable Embodied Robot Scenarios}},
  author={Liu, Yijing and Quek, Tony and Jiang, Wenbo and others},
  note={Under review as a conference paper at ICLR 2026},
  year={2026}
}

@article{salmi2026ai,
  title={{{AI}-native {O-RAN} architectures for {6G}: Towards real-time adaptation, conflict resolution, and efficient resource management}},
  author={Salmi, Sifeddine and Ouameur, Messaoud Ahmed and Bagaa, Miloud and Alexandropoulos, George C and Tahenni, Abdellah and Massicotte, Daniel and Ksentini, Adlen},
  journal={IEEE Transactions on Network and Service Management},
  year={2026},
  publisher={IEEE}
}

@article{zheng2026advanced,
  title={{Advanced Deep Reinforcement Learning for Agentic {AI} and Their Applications in Wireless Network}},
  author={Zheng, Jie and Niyato, Dusit and Zhang, Ruichen and Wang, Jiacheng and Nie, Jiangtian and Du, Hongyang and Kang, Jiawen and Zhang, Haijun and Jamalipour, Abbas and Kim, Dong In},
  journal={IEEE Transactions on Cognitive Communications and Networking},
  year={2026},
  publisher={IEEE}
}

@inproceedings{ccimen2025overview,
  title={{An Overview of Large Language Models in {6G} Radio Access Networks}},
  author={{\c{C}}imen, Sedat and Alt{\i}nta{\c{s}}, M{\"u}cahit and Duru, Ismail and Karahan, S{\"u}meye Nur and Yaz{\i}c{\i}, Ibrahim},
  booktitle={2025 International Conference on Electrical, Communication and Computer Engineering (ICECCE)},
  pages={1--6},
  year={2025},
  organization={IEEE}
}

@inproceedings{sidhu2025artificial,
  title={{Artificial Intelligence in Intent-Based Networking: A Comprehensive Survey on Enabling Autonomous Network Management}},
  author={Sidhu, Sneh Kanwar Singh and Sharma, Ankita},
  booktitle={2025 3rd International Conference on Advances in Computation, Communication and Information Technology (ICAICCIT)},
  volume={1},
  pages={579--584},
  year={2025},
  organization={IEEE}
}

@article{altintas2026beyond,
  title={{Beyond Connectivity via Artificial Intelligence Agents Enabling Perceptive, Adaptive, and Anticipatory {6G} Networks}},
  author={Altintas, Mucahit and Duru, Ismail and Yazici, Ibrahim and Karahan, Sumeye Nur and Cimen, Sedat},
  journal={IEEE Access},
  volume={14},
  pages={45977--46023},
  year={2026}
}

@inproceedings{navidan2025closed,
  title={{Closed-loop Intelligence Using Large Language Models in Wireless Networks}},
  author={Navidan, Hojjat and Seif, Mohamed and Poor, H Vincent and Moerman, Ingrid and Shahid, Adnan},
  booktitle={2025 16th IFIP Wireless and Mobile Networking Conference (WMNC)},
  pages={184--185},
  year={2025},
  organization={IEEE}
}

@article{zhou2026convergence,
  title={{Convergence of Reinforcement Learning and Time-Sensitive Networking for Future Industrial {AI} Agent Communication: Fundamentals, Challenges, and Opportunities}},
  author={Zhou, Nan and Yao, Yingfei and Liang, Xiaojun and Yao, Shunchun and Wan, Jiafu and Yang, Chunhua and Gui, Weihua and Gao, Wen},
  journal={IEEE Communications Surveys \& Tutorials},
  year={2026},
  publisher={IEEE}
}

@article{yang2025decision,
  title={{Decision-Making Large Language Model for Wireless Communication: A Comprehensive Survey on Key Techniques}},
  author={Yang, Ning and Fan, Mingrui and Wang, Wentao and Zhang, Haijun},
  journal={IEEE Communications Surveys \& Tutorials},
  year={2025},
  publisher={IEEE}
}

@article{habib2026generative,
  title={{Generative {AI} for Intent-Driven Network Management in {6G} {RAN}: A Case Study on the Mamba Model}},
  author={Habib, Md Arafat and Elsayed, Medhat and Ozcan, Yigit and Iturria-Rivera, Pedro Enrique and Bavand, Majid and Erol-Kantarci, Melike},
  journal={IEEE Wireless Communications},
  year={2026},
  publisher={IEEE}
}

@article{habib2025harnessing,
  title={{Harnessing the Power of {LLMs}, Informers and Decision Transformers for Intent-Driven {RAN} Management in {6G}}},
  author={Habib, Md Arafat and Iturria-Rivera, Pedro Enrique and Ozcan, Yigit and Elsayed, Medhat and Bavand, Majid and Gaigalas, Raimundas and Erol-Kantarci, Melike},
  journal={IEEE Transactions on Network Science and Engineering},
  year={2025},
  publisher={IEEE}
}

@article{habib2026hierarchical,
  title={{Hierarchical Decision Mamba Meets Agentic {AI}: A Novel Approach for {RAN} Slicing in {6G}}},
  author={Habib, Md Arafat and Elsayed, Medhat and Bavand, Majid and Rivera, Pedro Enrique Iturria and Ozcan, Yigit and Erol-Kantarci, Melike},
  journal={IEEE Networking Letters},
  year={2026},
  publisher={IEEE}
}

@article{hong2025comprehensive,
  title={{A Comprehensive Survey on {LLM}-Based Network Management and Operations}},
  author={Hong, Jibum and Tu, Nguyen Van and Hong, James Won-Ki},
  journal={International Journal of Network Management},
  volume={35},
  number={6},
  pages={e70029},
  year={2025},
  publisher={Wiley Online Library}
}

@article{liu2025lameta,
  title={{Lameta: Intent-aware agentic network optimization via a large {AI} model-empowered two-stage approach}},
  author={Liu, Yinqiu and Liu, Guangyuan and Wang, Jiacheng and Zhang, Ruichen and Niyato, Dusit and Sun, Geng and Xiong, Zehui and Han, Zhu},
  journal={IEEE Journal on Selected Areas in Communications},
  year={2025},
  publisher={IEEE}
}

@inproceedings{habib2025llm,
  title={{{LLM}-based intent processing and network optimization using attention-based hierarchical reinforcement learning}},
  author={Habib, Md Arafat and Rivera, Pedro Enrique Iturria and Ozcan, Yigit and Elsayed, Medhat and Bavand, Majid and Gaigalas, Raimundus and Erol-Kantarci, Melike},
  booktitle={2025 IEEE Wireless Communications and Networking Conference (WCNC)},
  pages={1--6},
  year={2025},
  organization={IEEE}
}

@misc{chiaranillm,
  title={{{LLM}-Guided Reinforcement Learning for Adaptive Inter-Slice Resource Prioritization in {6G} {O-RAN}}},
  author={Chiarani, Martino and Roy, Swastika and Ramantas, Kostas and Verikoukis, Christos},
  note={Preprint},
  year={2025}
}

@article{sun2026large,
  title={{Large Language Model-Empowered Resource Allocation in Intent-Driven Wireless Networks}},
  author={Sun, Haofeng and Tian, Hui and Zheng, Jingheng and Ni, Wanli and Cui, Qimei and Niyato, Dusit and Zhang, Ping},
  journal={IEEE Transactions on Cognitive Communications and Networking},
  year={2026},
  publisher={IEEE}
}

@article{zheng2026large,
  title={{Large language model-enabled reinforcement learning for wireless network optimization}},
  author={Zheng, Jie and Zhang, Ruichen and Niyato, Dusit and Zhang, Haijun and Wang, Jiacheng and Du, Hongyang and Kang, Jiawen and Xiong, Zehui},
  journal={IEEE Communications Magazine},
  year={2026},
  publisher={IEEE}
}

@article{zhou2024large,
  title={{Large language model ({LLM}) for telecommunications: A comprehensive survey on principles, key techniques, and opportunities}},
  author={Zhou, Hao and Hu, Chengming and Yuan, Ye and Cui, Yufei and Jin, Yili and Chen, Can and Wu, Haolun and Yuan, Dun and Jiang, Li and Wu, Di and others},
  journal={IEEE Communications Surveys \& Tutorials},
  volume={27},
  number={3},
  pages={1955--2005},
  year={2024},
  publisher={IEEE}
}

@inproceedings{tageldien2025large,
  title={{Large language models in intent-based networking: a comprehensive survey across the intent lifecycle}},
  author={Tageldien, Marwa and Selim, Bassant and Sboui, Lokman},
  booktitle={2025 International Telecommunications Conference (ITC-Egypt)},
  pages={810--817},
  year={2025},
  organization={IEEE}
}

@article{zhang2025multi,
  title={{Multi-Modal Data-Enhanced Foundation Models for Prediction and Control in Wireless Networks: A Survey}},
  author={Zhang, Han and Farzanullah, Mohammad and Ghassemi, Mohammad and Sediq, Akram Bin and Afana, Ali and Erol-Kantarci, Melike},
  journal={IEEE Communications Surveys \& Tutorials},
  volume={28},
  pages={4359--4393},
  year={2025},
  publisher={IEEE}
}

@inproceedings{gemayel2025network,
  title={{Network Function Orchestration with {LLM} based Multi-Agent System}},
  author={Gemayel, Johny and Mokh, Ali},
  booktitle={2025 IEEE International Conference on Communications Workshops (ICC Workshops)},
  pages={262--267},
  year={2025},
  organization={IEEE}
}

@article{gajjar2025oransight,
  title={{Oransight-2.0: Foundational {LLMs} for {O-RAN}}},
  author={Gajjar, Pranshav and Shah, Vijay K},
  journal={IEEE Transactions on Machine Learning in Communications and Networking},
  year={2025},
  publisher={IEEE}
}

@article{zaidi2026reasoning,
  title={{Reasoning and Acting (ReAct) with Multimodal {LLMs}: A Framework for Intent Driven {6G} Networks}},
  author={Zaidi, Syed Ali Raza and Hafeez, Maryam and Qazzaz, Mohammed MH and Tatipamula, Mallik and Chowdary, Kuntal and Win, Moe Z},
  journal={IEEE Open Journal of the Communications Society},
  year={2026},
  publisher={IEEE}
}

@inproceedings{lotfi2026scalable,
  title={{Scalable {LLM}-Augmented {DRL} with Context-Aware Prompt Learning for {O-RAN} Slicing}},
  author={Lotfi, Fatemeh and Rajoli, Hossein and Afghah, Fatemeh},
  booktitle={ICASSP 2026-2026 IEEE International Conference on Acoustics, Speech and Signal Processing (ICASSP)},
  pages={21436--21440},
  year={2026},
  organization={IEEE}
}

@article{alkuwaiti6577052distributed,
  title={{Distributed Deep Reinforcement Learning and Large Language Models for Intelligent Network Management: A Comparative Study}},
  author={Alkuwaiti, Saeed and AbuAli, Najah},
  journal={Available at SSRN 6577052},
  year={2026}
}

@article{soliman2025foundation,
  title={{Foundation Models and Large Language Models in Next Generation Networks: Comparison, Opportunities and Challenges}},
  author={Soliman, Abdelrahman and Hussein, Ahmed Refaey},
  journal={Authorea Preprints},
  year={2025},
  publisher={Authorea}
}

@article{elkael2026agentran,
  title={{AgentRAN: An agentic {AI} architecture for autonomous control of open {6G} networks}},
  author={Elkael, Maxime and D'Oro, Salvatore and Bonati, Leonardo and Polese, Michele and Lee, Yunseong and Furueda, Koichiro and Melodia, Tommaso},
  journal={IEEE Communications Magazine},
  year={2026},
  publisher={IEEE}
}

@article{bao2026llm,
  title={{{LLM}-hRIC: {LLM}-empowered hierarchical {RAN} intelligent control for {O-RAN}}},
  author={Bao, Lingyan and Yun, Sinwoong and Lee, Jemin and Quek, Tony QS},
  journal={IEEE Communications Magazine},
  year={2026},
  publisher={IEEE}
}

@article{li2026agentic,
  title={{Agentic Open {RAN}: A Deterministic and Auditable Framework for Intent-Driven Radio Control}},
  author={Li, Hengxu and Xu, Dongkuan and Chen, Mingzhe and Liu, Yuchen},
  journal={arXiv preprint arXiv:2604.13384},
  year={2026}
}

@article{ding2025ridas,
  title={{RIDAS: A Multi-Agent Framework for {AI}-{RAN} with Representation-and Intention-Driven Agents}},
  author={Ding, Kuiyuan and Guo, Caili and Yang, Yang and Guo, Jianzhang},
  journal={arXiv preprint arXiv:2507.13140},
  year={2025}
}

@article{ngo2026llm,
  title={{{LLM}-Based Net Analyzer {rApp} for Explainable and Safe Automation in {O-RAN} {Non-RT} {RIC}}},
  author={Ngo, Tuan V and Ngo, Mao V and Chen, Binbin and Quek, Tony QS and Kumari, Tejaswita and Nekovee, Maziar},
  journal={arXiv preprint arXiv:2603.13775},
  year={2026}
}

@inproceedings{achiam2017constrained,
  title={{Constrained policy optimization}},
  author={Achiam, Joshua and Held, David and Tamar, Aviv and Abbeel, Pieter},
  booktitle={International conference on machine learning},
  pages={22--31},
  year={2017},
  organization={Pmlr}
}

@inproceedings{alshiekh2018safe,
  title={{Safe reinforcement learning via shielding}},
  author={Alshiekh, Mohammed and Bloem, Roderick and Ehlers, R{\"u}diger and K{\"o}nighofer, Bettina and Niekum, Scott and Topcu, Ufuk},
  booktitle={Proceedings of the AAAI conference on artificial intelligence},
  volume={32},
  number={1},
  year={2018}
}

@inproceedings{cheng2019end,
  title={{End-to-end safe reinforcement learning through barrier functions for safety-critical continuous control tasks}},
  author={Cheng, Richard and Orosz, G{\'a}bor and Murray, Richard M and Burdick, Joel W},
  booktitle={Proceedings of the AAAI conference on artificial intelligence},
  volume={33},
  number={01},
  pages={3387--3395},
  year={2019}
}

@article{berkenkamp2017safe,
  title={{Safe model-based reinforcement learning with stability guarantees}},
  author={Berkenkamp, Felix and Turchetta, Matteo and Schoellig, Angela and Krause, Andreas},
  journal={Advances in neural information processing systems},
  volume={30},
  year={2017}
}

@inproceedings{nagib2025safeslice,
  title={{SafeSlice: Enabling {SLA}-compliant {O-RAN} slicing via safe deep reinforcement learning}},
  author={Nagib, Ahmad M and Abou-Zeid, Hatem and Hassanein, Hossam S},
  booktitle={2025 IEEE International Conference on Machine Learning for Communication and Networking (ICMLCN)},
  pages={1--7},
  year={2025},
  organization={IEEE}
}

@article{nagib2023safe,
  title={{Safe and accelerated deep reinforcement learning-based {O-RAN} slicing: A hybrid transfer learning approach}},
  author={Nagib, Ahmad M and Abou-Zeid, Hatem and Hassanein, Hossam S},
  journal={IEEE Journal on Selected Areas in Communications},
  volume={42},
  number={2},
  pages={310--325},
  year={2023},
  publisher={IEEE}
}

@article{kasi2025risk,
  title={{Risk-aware reinforcement learning framework for user-centric {O-RAN}}},
  author={Kasi, Shahrukh Khan and Khan, Fahd Ahmed and Ekin, Sabit and Imran, Ali},
  journal={IEEE Transactions on Machine Learning in Communications and Networking},
  volume={3},
  pages={195--214},
  year={2025},
  publisher={IEEE}
}

@inproceedings{navarro20252offran,
  title={{2OffRAN: Offline Off-Policy Reinforcement Learning for Safe Handover in {O-RAN}}},
  author={Navarro, Annalisa and Botta, Alessio and Canonico, Roberto and Wang, Yizhou and Fitzek, Frank HP and Nguyen, Giang T},
  booktitle={2025 IEEE International Conference on Machine Learning for Communication and Networking (ICMLCN)},
  pages={1--6},
  year={2025},
  organization={IEEE}
}

@inproceedings{tuerxun2026safe,
  title={{Safe {RAN} Slicing in {O-RAN}: Minimizing {SLA} Violations via Model-Based Reinforcement Learning}},
  author={Tuerxun, Aerman and Nakao, Akihiro},
  booktitle={2026 IEEE 23rd Consumer Communications \& Networking Conference (CCNC)},
  pages={1--6},
  year={2026},
  organization={IEEE}
}

@article{yungaicela2026rslaq,
  title={{RSLAQ-A Robust {SLA}-driven {6G} {O-RAN} {QoS} {xApp} using deep reinforcement learning}},
  author={Yungaicela-Naula, Noe M and Sharma, Vishal and Scott-Hayward, Sandra},
  journal={IEEE Transactions on Mobile Computing},
  year={2026},
  publisher={IEEE}
}

@inproceedings{fairrlsurvey2024,
  title={{Fairness in reinforcement learning: A survey}},
  author={Reuel, Anka and Ma, Devin},
  booktitle={Proceedings of the AAAI/ACM Conference on AI, Ethics, and Society},
  volume={7},
  number={1},
  pages={1218--1230},
  year={2024}
}

@article{mo2002fair,
  title={{Fair end-to-end window-based congestion control}},
  author={Mo, Jeonghoon and Walrand, Jean},
  journal={IEEE/ACM Transactions on networking},
  volume={8},
  number={5},
  pages={556--567},
  year={2002},
  publisher={IEEE}
}

@article{jain1984quantitative,
  title={{A quantitative measure of fairness and discrimination}},
  author={Jain, Rajendra K and Chiu, Dah-Ming W and Hawe, William R and others},
  journal={Eastern Research Laboratory, Digital Equipment Corporation, Hudson, MA},
  volume={21},
  number={1},
  pages={2022--2023},
  year={1984}
}

@inproceedings{siddique2020learning,
  title={{Learning fair policies in multi-objective (deep) reinforcement learning with average and discounted rewards}},
  author={Siddique, Umer and Weng, Paul and Zimmer, Matthieu},
  booktitle={International Conference on Machine Learning},
  pages={8905--8915},
  year={2020},
  organization={PMLR}
}

@article{lopezsanchez2022latency,
  title={{Latency fairness optimization on wireless networks through deep reinforcement learning}},
  author={L{\'o}pez-S{\'a}nchez, Maria and Villena-Rodr{\'\i}guez, Alejandro and G{\'o}mez, Gerardo and Martin-Vega, Francisco J and Aguayo-Torres, Mari Carmen},
  journal={IEEE Transactions on Vehicular Technology},
  volume={72},
  number={4},
  pages={5407--5412},
  year={2022},
  publisher={IEEE}
}

@inproceedings{hasheminezhad2025derric,
  title={{Reinforced Fairness-Aware Multi-Agent Self-Organization for {6G} Radio Access Network Orchestration}},
  author={Nezhad, Elham Hashemi and Di Maio, Antonio and Braun, Torsten},
  booktitle={2025 IEEE 50th Conference on Local Computer Networks (LCN)},
  pages={1--9},
  year={2025},
  organization={IEEE}
}

@inproceedings{comsa2025fairq,
  title={{FAIR-Q: Fairness and Adaptive Intelligent Resource Management with {QoS} Optimization in Dynamic {6G} Radio Access Networks}},
  author={Com{\c{s}}a, Ioan-Sorin and Bergamin, Per and Muntean, Gabriel-Miro and Shah, Purav and Trestian, Ramona},
  booktitle={2025 IEEE International Symposium on Broadband Multimedia Systems and Broadcasting (BMSB)},
  pages={1--7},
  year={2025},
  organization={IEEE}
}

@article{giwa2025hetnet,
  title={{Optimisation of Resource Allocation in Heterogeneous Wireless Networks Using Deep Reinforcement Learning}},
  author={Giwa, Oluwaseyi and Shock, Jonathan and Toit, Jaco Du and Awodumila, Tobi},
  journal={arXiv preprint arXiv:2509.25284},
  year={2025}
}

@article{mondal2023fairauction,
  title={{Fairness guaranteed and auction-based x-haul and cloud resource allocation in multi-tenant O-RANs}},
  author={Mondal, Sourav and Ruffini, Marco},
  journal={IEEE Transactions on Communications},
  volume={71},
  number={6},
  pages={3452--3468},
  year={2023},
  publisher={IEEE}
}

@article{aslan2024fairvran,
  title={{Fair resource allocation in virtualized {O-RAN} platforms}},
  author={Aslan, Fatih and Iosifidis, George and Ayala-Romero, Jose A and Garcia-Saavedra, Andres and Costa-Perez, Xavier},
  journal={Proceedings of the ACM on Measurement and Analysis of Computing Systems},
  volume={8},
  number={1},
  pages={1--34},
  year={2024},
  publisher={ACM New York, NY, USA}
}

@article{tsampazi2024pandora,
  title={{PandORA: Automated design and comprehensive evaluation of deep reinforcement learning agents for open {RAN}}},
  author={Tsampazi, Maria and D'Oro, Salvatore and Polese, Michele and Bonati, Leonardo and Poitau, Gwenael and Healy, Michael and Alavirad, Mohammad and Melodia, Tommaso},
  journal={IEEE Transactions on Mobile Computing},
  volume={24},
  number={4},
  pages={3223--3240},
  year={2024},
  publisher={IEEE}
}

@article{milani2024explainable,
  title={{Explainable reinforcement learning: A survey and comparative review}},
  author={Milani, Stephanie and Topin, Nicholay and Veloso, Manuela and Fang, Fei},
  journal={ACM Computing Surveys},
  volume={56},
  number={7},
  pages={1--36},
  year={2024},
  publisher={ACM New York, NY}
}

@inproceedings{ribeiro2016should,
  title={{Why should i trust you? Explaining the predictions of any classifier}},
  author={Ribeiro, Marco Tulio and Singh, Sameer and Guestrin, Carlos},
  booktitle={Proceedings of the 22nd ACM SIGKDD international conference on knowledge discovery and data mining},
  pages={1135--1144},
  year={2016}
}

@article{lundberg2017unified,
  title={{A unified approach to interpreting model predictions}},
  author={Lundberg, Scott M and Lee, Su-In},
  journal={Advances in neural information processing systems},
  volume={30},
  year={2017}
}

@article{heuillet2021explainability,
  title={{Explainability in deep reinforcement learning}},
  author={Heuillet, Alexandre and Couthouis, Fabien and D{\'\i}az-Rodr{\'\i}guez, Natalia},
  journal={Knowledge-Based Systems},
  volume={214},
  pages={106685},
  year={2021},
  publisher={Elsevier}
}

@article{lu2026demystifying,
  title={{Demystifying Deep Reinforcement Learning: A Neuro-Symbolic Framework for Interpretable Open {RAN} Automation}},
  author={Lu, Jie and Yan, Peihao and Tan, Pang-Ning and Hou, Y Thomas and Zeng, Huacheng},
  journal={arXiv preprint arXiv:2605.10648},
  year={2026}
}

@article{fiandrino2023explora,
  title={{EXPLORA: {AI/ML} explainability for the open {RAN}}},
  author={Fiandrino, Claudio and Bonati, Leonardo and D'Oro, Salvatore and Polese, Michele and Melodia, Tommaso and Widmer, Joerg},
  journal={Proceedings of the ACM on Networking},
  volume={1},
  number={CoNEXT3},
  pages={1--26},
  year={2023},
  publisher={ACM New York, NY, USA}
}

@inproceedings{duttagupta2025symbxrl,
  title={{SymbXRL: symbolic explainable deep reinforcement learning for mobile networks}},
  author={Duttagupta, Abhishek and Jabbari, MohammadErfan and Fiandrino, Claudio and Fiore, Marco and Widmer, Joerg},
  booktitle={IEEE INFOCOM 2025-IEEE Conference on Computer Communications},
  pages={1--10},
  year={2025},
  organization={IEEE}
}

@article{jabbari2026sia,
  title={{SIA: Symbolic Interpretability for Anticipatory Deep Reinforcement Learning in Network Control}},
  author={Jabbari, MohammadErfan and Duttagupta, Abhishek and Fiandrino, Claudio and Bonati, Leonardo and D'Oro, Salvatore and Polese, Michele and Fiore, Marco and Melodia, Tommaso},
  journal={arXiv preprint arXiv:2601.22044},
  year={2026}
}

@article{rezazadeh2024sliceops,
  title={{SliceOps: Explainable MLOps for streamlined automation-native {6G} networks}},
  author={Rezazadeh, Farhad and Chergui, Hatim and Alonso, Luis and Verikoukis, Christos},
  journal={IEEE Wireless Communications},
  volume={31},
  number={5},
  pages={224--230},
  year={2024},
  publisher={IEEE}
}

@inproceedings{rezazadeh2023explanation,
  title={{Explanation-guided deep reinforcement learning for trustworthy {6G} {RAN} slicing}},
  author={Rezazadeh, Farhad and Chergui, Hatim and Mangues-Bafalluy, Josep},
  booktitle={2023 ieee international conference on communications workshops (icc workshops)},
  pages={1026--1031},
  year={2023},
  organization={IEEE}
}

@article{fatehi2026interpretable,
  title={{Interpretable Attention-Based Multi-Agent {PPO} for Latency Spike Resolution in {6G} {RAN} Slicing}},
  author={Fatehi, Kavan and Ghourtani, Mostafa Rahmani and Sonee, Amir and Yadav, Poonam and Russo, Alessandra M and Ahmadi, Hamed and Calinescu, Radu},
  journal={arXiv preprint arXiv:2602.11076},
  year={2026}
}

@inproceedings{sun2025explainable,
  title={{An Explainable {AI} Framework for Dynamic Resource Management in Vehicular Network Slicing}},
  author={Sun, Haochen and Liu, Yifan and Al-Tahmeesschi, Ahmed and Chetty, Swarna and Zaidi, Syed Ali Raza and Nag, Avishek and Ahmadi, Hamed},
  booktitle={2025 IEEE 36th International Symposium on Personal, Indoor and Mobile Radio Communications (PIMRC)},
  pages={1--6},
  year={2025},
  organization={IEEE}
}

\end{document}